\documentclass[11pt]{emulateapj}
\usepackage{natbib}
\usepackage{graphicx}
\usepackage{amsmath}
\usepackage{amssymb}
\def\jgr{J. Geophys. Res. }

\def\aj{Astron. J. }
\def\apj{Astrophys. J. }
\def\aap{Astron. \& Astrophys. }
\def\chf{CH$_4$ }
\def\chfx{CH$_4$}
\def\chisq{$\chi^2$ }
\def\chisqx{$\chi^2$}

\def\icmx{cm$^{-1}$}

\def\Wm2{W/m$^2$}
\def\Wpm2sr{Wm$^{-2}sr^{-1}$}
\def\deg{$^\circ$ }
\def\degx{$^\circ$}

\def\mum{$\mu$m }
\def\mumx{$\mu$m}

\def\nhfshx{NH$_4$SH}
\def\hts{H$_2$S }
\def\htsx{H$_2$S}

\def\planss{Plan. \& Sp. Sci. }
\renewcommand{\baselinestretch}{1.3}  % turned off for single-spaced text

\begin{document}
%\setpagewiselinenumbers
%\linenumbers 
\title{The methane distribution and
    polar brightening on Uranus based on HST/STIS\footnotemark[\dag], Keck/NIRC2, and IRTF/SpeX observations through 2015}
\author{L.~A. Sromovsky\altaffilmark{1}, E. Karkoschka\altaffilmark{2}, P.~M. Fry\altaffilmark{1},
 I. de Pater\altaffilmark{3}, H.~B. Hammel\altaffilmark{4} \altaffilmark{5}}
\altaffiltext{1}{University of Wisconsin - Madison, Madison WI 53706, USA}
\altaffiltext{2}{University of Arizona, Tucson AZ 85721, USA}
\altaffiltext{3}{University of California, Berkeley, CA 94720, USA}
\altaffiltext{4}{AURA, 1212 New York Ave. NW, Suite 450, Washington, DC 20005, USA}
\altaffiltext{5}{Space Science Institute, Boulder, CO 80303, USA}
\altaffiltext{\dag}{Based in part on observations with the NASA/ESA Hubble Space
Telescope obtained at the Space Telescope Science Institute, which is
operated by the Association of Universities for Research in Astronomy,
Incorporated under NASA Contract NAS5-26555.}

\slugcomment{Journal reference: Icarus (2018) Under review.}

\newpage
%% \pagenumbering{arabic}
%% \setcounter{page}{3}

%%\section*{Abstract} 2089 words
\begin{abstract} Space Telescope Imaging Spectrograph (STIS) observations of Uranus in
2015 show that the depletion of upper tropospheric methane has 
been relatively stable and that the polar region has been brightening over time
as a result of increased aerosol scattering.  This interpretation is confirmed
by near-IR imaging from HST and from the Keck telescope using NIRC2
adaptive optics imaging.  Our analysis of the
2015 spectra, as well as prior spectra from 2012,
 shows that there is a factor of
three decrease in the effective upper tropospheric methane mixing
ratio between 30\deg N and 70\deg N.
The absolute value of the deep methane mixing
ratio, which probably does not vary with latitude,
 is lower than our previous estimate, and depends significantly on
the style of aerosol model that we assume, ranging from a high of
3.5$\pm$0.5\% for conservative non-spherical particles with a simple
Henyey-Greenstein phase function to a low of 2.7\%$\pm$0.3\% for
conservative spherical particles.  Our previous higher estimate of
4$\pm$0.5\% was a result of a forced consistency
 with occultation results of Lindal et al. (1987, JGR 92, 14987-15001).
That requirement was abandoned in our new analysis because new work by  
Orton et al. (2014, Icarus 243, 494-513) and by
 Lellouch et al. (2015, Astron. \& AstroPhys. 579, A121), called into question
the occultation results.  For the main cloud layer in our models we found
 that both large and small particle
solutions are possible for spherical particle models. At low latitudes the
 small-particle solution has a mean
particle radius near 0.3 \mumx, a real refractive index near 1.65, and
a total column mass of 0.03 mg/cm$^2$, while the large-particle
solution has a particle radius near 1.5 \mumx, a real index near 1.24,
and a total column mass 30 times larger.  The pressure boundaries of
the main cloud layer are between about 1.1 and 3 bars, within which
\hts is the most plausible condensable.
\end{abstract}


\maketitle
\shortauthors{Sromovsky et al.} 
\shorttitle{Methane distribution and polar brightening on Uranus}

\section{Introduction}

% \citep{Sro2009eqdyn}.
%From 1994 through its 2007 equinox, the southern hemisphere, recently
%coming out of a long summer, was generally brighter than the northern
%hemisphere. Most obvious was the presence of a southern bright cloud
%band centered near 45\deg S, while no corresponding feature was
%present in the northern hemisphere. The presence of a significant
%asymmetry near equinox is consistent with a delayed response to the
%seasonal forcing.
%We also saw
%surprisingly rapid changes in the
%the beginning of changes that were occurring faster than
%expected. The 
%bright band at 45\deg S, which was fading at the same time a new
%bright band was forming at 45\deg N.  

The visible and infrared spectra of Uranus are both dominated by the
absorption features of methane, its third most abundant gas. In some
spectral regions, however, the effects of collision-induced absorption
(CIA) by hydrogen can be seen competing with methane. Those
wavelengths provide constraints on the number density of methane with
respect to hydrogen, and thus on the volume mixing ratio of methane. From
analysis of HST/STIS spatially resolved spectra of Uranus obtained in
2002, 5 years before equinox and limited to latitudes south of 30\deg
N, \cite{Kark2009IcarusSTIS}, henceforth referred to as {\bf KT2009},
discovered that good fits to the latitudinal variation of these
spectra required a latitudinal variation in the effective volume
mixing ratio of methane.  They inferred that the southern polar region
was depleted in methane with respect to low latitudes by about a
factor of two.  This suggested a possible meridional circulation in
which upwelling methane-rich gas at low latitudes was dried out by
condensation then moved to high latitudes, where descending motions
brought the methane-depleted gas downward, with a return flow at
deeper levels.

Because post equinox groundbased observations revealed numerous small
``convective'' features in the north polar region that had not been
seen in the south polar region just prior to equinox,
\cite{Sro2012polar} suggested that the downwelling movement of
methane-depleted gas would suppress convection in the south polar
region, providing a plausible explanation for the lack of discrete
cloud features there, and further suggested that the presence of
discrete cloud features at high northern latitudes might mean that
methane is not depleted there.  However, using 2012 HST/STIS
observations designed to test that hypothesis, \cite{Sro2014stis}
showed that the depletion was indeed symmetric, with both polar
regions depleted by similar amounts, and from imaging observations
taken near equinox using discrete narrow band filters that sampled
methane-dominated and hydrogen-dominated wavelengths, they showed that
the symmetry was also present at equinox and thus probably a stable
feature of the Uranian atmosphere.

In spite of the apparent general stability of the latitudinal
distribution of methane, there were significant post-equinox increases
in the brightness of the north polar region, as well as some evidence
for brightening at low latitudes.  New HST/STIS observations were
obtained in 2015 to further investigate possible changes in the
methane distribution and the nature of the polar brightening that was
taking place.

Other relevant developments occurred since our last analysis of the
HST/STIS spectra of Uranus.  The first is the inference of new mean
temperature and methane profiles for Uranus by
\cite{Orton2014uratemp}, which are in disagreement with the
occultation-based profiles of \cite{Lindal1987} and also with those
using a reduced He/H$_2$ volume mixing ratio derived by
\cite{Sro2011occult}.  The second development is an independent
determination of the methane volume mixing ratio in the lower
stratosphere and upper troposphere by \cite{Lellouch2015} using
Herschel far infrared and sub-mm observations.  Since both of these
results question the validity of Uranus occultation results in general, and
more specifically the validity of using them at all times and all
latitudes, we decided to modify our analysis so that models would be
constrained by STIS spectral observations alone, and abandoned the
requirement that our low-latitude methane profiles also be consistent
with occultation constraints.  We also took a fresh look at how to
best model the aerosol structure, and found that a much simpler
2-3 layer structure could produce fits as accurate as the more complex
five-layer structure we had used in our previous analysis of the 2012
STIS observations.

%% In addition, we deviated from our approach to the 2012 analysis, which used much of
%% the cloud model KT2009 developed to fit the 2002 STIS observations
%% covering mainly the southern hemisphere.  Considering that cloud
%% scattering properties might be different in the northern hemisphere or
%% changed with time, we also took a fresh look at how to best model the
%% aerosol structure, and found that a much simpler 2-layer structure
%% could produce fits as accurate as the more complex five-layer structure we
%% had used in our previous analysis.

In the following we first describe the approach to constraining
the methane mixing ratio on Uranus, then discuss the new thermal profile
for Uranus and its implications. We follow that with a discussion of our
STIS, WFC3, and Keck supporting observations, the calibration of the
STIS spectra, direct comparisons of STIS spectra, comparisons of imaging
observations at key wavelengths, description of our approach to radiative
transfer modeling, results of modeling cloud structure and the distribution
of methane, interpretation of those results, how models can be
extended to longer wavelengths to match spectra obtained
at NASA's Infrared Telescope Facility (IRTF), comparisons
with other models, and a final summary and conclusions.

\section{How observations constrain the methane mixing ratio on Uranus}

\begin{figure*}\centering
\includegraphics[width=6.2in]{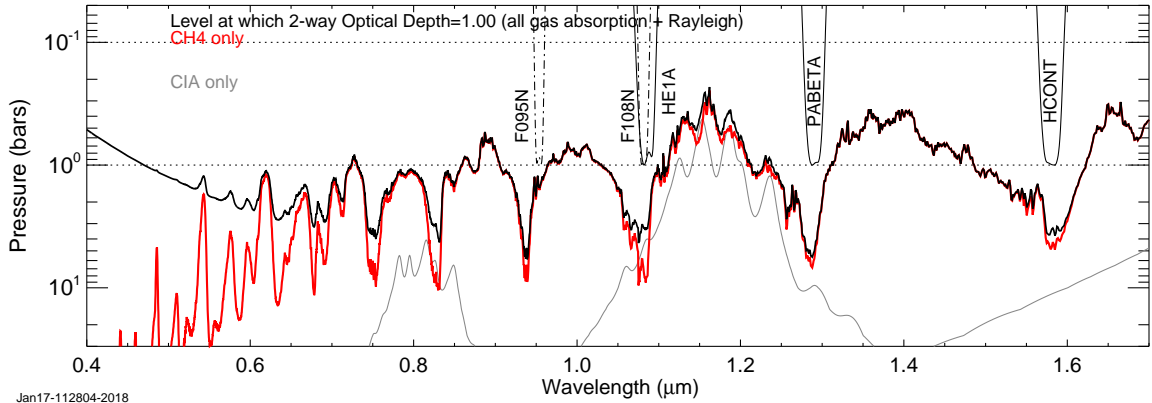}
\caption{Penetration depth vs. wavelength as limited by different
  opacity sources assuming the \cite{Orton2014uratemp} temperature profile and
 a 3.5\% deep methane
mixing ratio. Where absorption dominates, penetration is about
one optical depth, but when Rayleigh scattering dominates, light can penetrate many 
optical depths. Transmission profiles of key HST/NICMOS (dot-dash)
  and Keck NIRC2 (solid) narrow-band filters are also plotted.}
 \label{Fig:pendepth}
%SOURCE: plot_pendepth_from_ura_pendepthfile_Jan_2018 in /home/sro/ir_ms_ura on puck:
% and also p 109, Log L
\end{figure*}

Constraining the mixing ratio of \chf on Uranus is based on
differences in the spectral absorption of \chf and H$_2$, illustrated
by the penetration depth plot of Fig.\ \ref{Fig:pendepth}. There
methane absorption can be seen to dominate at most wavelengths, while
hydrogen's Collision Induced Absorption (CIA) is relatively more
important in narrow spectral ranges near 825 nm, which is covered by
our STIS observations, and also near 1080 nm, which we were able to
sample with imaging observations using the NICMOS F108N filter and
Keck He1\_A filter.  Model calculations that don't have the correct
ratio of methane to hydrogen lead to a relative reflectivity mismatch
near these wavelengths.
\cite{Kark2009IcarusSTIS} used the 825-nm spectral constraint to infer a
methane mixing ratio of 3.2\% at low latitudes, but dropping to 1.4\%
at high southern latitudes.
\cite{Sro2011occult} analyzed the same data set, but used only
temperature and mixing ratio profiles that were consistent with the
\cite{Lindal1987} refractivity profiles.  They confirmed the depletion
but inferred a somewhat higher mixing ratio of 4\% at low latitudes
and found that better fits were obtained if the high latitude
depletion was restricted to the upper troposphere
(down to $\sim$2-4 bars).  Subsequently, 2009 groundbased spectral observations at the
NASA Infrared Telescope Facility (IRTF) using the SpeX instrument,
which provided coverage of the key 825-nm spectral region, were used
by \cite{Tice2013} to infer that both northern and southern mid latitudes
 were weakly depleted in methane, relative to the near equatorial region, which
was enriched by at least 9\%. Their I-band analysis yielded a broad peak centered at 6\degx S,
which was 32$\pm$24\% above the minimum found at 44\degx N.
 These low IRTF-based values for the latitudinal variation might be partly a result of
lower spatial resolution combined with worse view angles into higher
latitudes than obtained by HST observations.

%%  A probably more
%% important contributor
%% to the lower absolute value is that \cite{Tice2013} did not constrain
%% their results to be occultation consistent.  A discussion of
%% alternative profiles is provided in Sec. \ref{Sec:RT}.

\section{HST/STIS 2015 Observations}

Our 2015 spectral observations of Uranus (Cycle 23 HST program 14113, L. Sromovsky P.I.) used four
HST orbits, three of them devoted to STIS spatial mosaics and one
devoted to Wide Field Camera 3 (WFC3) support imaging. The STIS
observations were taken on 10 October 2015 and the WFC3 observations
on 11 October 2015. Observing conditions and exposures are summarized
in Table\ \ref{Tbl:sci_obs}.

\begin{table*} \centering
%\fontfamily{phv}\selectfont
% turned off for single-spaced text
\caption{Science exposures from 2015 HST program 14113.\label{Tbl:sci_obs}}
\vspace{0.1in}
\begin{tabular}{|c | c | c | c | c | r | r | r|}
\hline
Relative &Start      & Start      & Instrument & Filter or & Exposure        & No. of & Phase\\[0.in]
Orbit &Date (UT) & Time (UT) &            & Grating   & (sec)  & Exp.      & Angle (\degx)\\
\hline
  1 & 2015-10-10 & 13:54:06 & STIS & MIRVIS &   5.0 &  1 & 0.09 \\
  1 & 2015-10-10 & 14:09:48 & STIS &  G430L &  70.0 & 13 & 0.09 \\
  2 & 2015-10-10 & 15:28:02 & STIS &  G750L &  84.0 & 19 & 0.09 \\
  3 & 2015-10-10 & 17:03:25 & STIS &  G750L &  84.0 & 19 & 0.09 \\
 19 & 2015-10-11 & 18:30:59 & WFC3 &  F336W &  30.0 &  1 & 0.04 \\
 19 & 2015-10-11 & 18:32:44 & WFC3 &  F467M &  16.0 &  1 & 0.04 \\
 19 & 2015-10-11 & 18:34:24 & WFC3 &  F547M &   6.0 &  1 & 0.04 \\
 19 & 2015-10-11 & 18:35:48 & WFC3 &  F631N &  65.0 &  1 & 0.04 \\
 19 & 2015-10-11 & 18:38:17 & WFC3 &  F665N &  52.0 &  1 & 0.04 \\
 19 & 2015-10-11 & 18:40:24 & WFC3 &  F763M &  26.0 &  1 & 0.04 \\
 19 & 2015-10-11 & 18:42:11 & WFC3 &  F845M &  35.0 &  1 & 0.04 \\
 19 & 2015-10-11 & 18:44:05 & WFC3 &  F953N & 250.0 &  1 & 0.04 \\
 19 & 2015-10-11 & 18:50:43 & WFC3 & FQ889N & 450.0 &  1 & 0.04 \\
 19 & 2015-10-11 & 19:02:25 & WFC3 & FQ937N & 150.0 &  1 & 0.04 \\
 19 & 2015-10-11 & 19:09:25 & WFC3 & FQ727N & 210.0 &  1 & 0.04 \\
\hline
\end{tabular}
%% This table was created by triton:~patf/uranus/paper_stis15/obstable_stis15ura.pro.
\renewcommand{\baselinestretch}{1.}\normalsize  % turned off for single-spaced text
\parbox{6.in}{\vspace{0.1in}On October 10 the sub-observer planetographic latitude
  was 31.7\deg S, the observer range was 18.984 AU
  (2.8400$\times$10$^9$ km), and the equatorial angular diameter of
  Uranus was 3.7126 arcseconds. The first two STIS orbits used the
  52$''\times$0.1$''$ slit and the third inadvertently used the
  52$''\times$0.05$''$ slit}
\end{table*}
\renewcommand{\baselinestretch}{1.3}  % turned off for single-spaced text

\vspace{-0.1in}
\subsection{STIS spatial mosaics.} 
\vspace{-0.05in}
Our STIS observations used the G430L and G750L gratings and the CCD
detector, which has $\sim$0.05 arcsecond square pixels covering a
nominal 52$''\times$52$''$ square field of view (FOV) and a
spectral range from $\sim$200 to 1030 nm \citep{Hernandez2012}.
Using the 52$''\times$0.1$''$ slit, the resolving power varies from 500
to 1000 over each wavelength range due to fixed wavelength dispersion
of the gratings. Observations had to be carried out within a few days of Uranus
opposition (12 October 2015) when the telescope roll angle could be
set 
%to 300\deg or 120\deg 
to orient the STIS slit parallel
to the spin axis of Uranus. 

One STIS orbit produced a mosaic of half of Uranus using the CCD
detector, the G430L grating, and the 52$''\times$0.1$''$ slit. The G430L
grating covers 290 to 570 nm with a 0.273 nm/pixel dispersion. The
slit was aligned with Uranus' rotational axis, and stepped from the
evening limb to the central meridian in 0.152 arcsecond increments
(because the planet has no high spatial resolution center-to-limb
features at these wavelengths we used interpolation to fill in missing
columns of the mosaic). Two additional STIS orbits were used to mosaic
the planet with the G750L grating. We intended to use the 52$''\times$0.1$''$ slit
(524-1027 nm coverage with 0.492 nm/pixel dispersion) for both
orbits, but an error in the program resulted in half of the half-disk
covered with the nominal 0.05$''$ slit.
%, In one email Erich said this actually has a width of
% 0.035$''$ according to the STIS Instrument Handbook, but I could
% not find it in the handbook, and Erich in a later email said it was 0.05$''$. 
This produced a higher spectral resolution
at the cost of a significant reduction in signal to noise ratio. The limb to
central meridian stepping was at 0.0562 arcsecond intervals for the G750L
grating. Aside from the slit width error, this was the
same procedure that was used successfully for HST program 9035 in 2002
(E. Karkoschka, P.I.) and for HST program 12894 in 2012 (L. Sromovsky, P.I.). 
 As Uranus' equatorial radius was 1.85
arcseconds when observations were performed, stepping from one step
off the limb to the central meridian required 13 positions for the
G430L grating and 36 for the G750L grating. 
Two orbits were needed to complete the
G750L grating observations, spanning a total time of 2 h 17 m, during which Uranus rotated
47\degx. This rotation was not a problem because of the high degree of
zonal symmetry of Uranus and because our analysis rejected any small
scale deviations from it.

Exposure times were similar to those used in the 2002 and 2012
programs, with 70-second exposures for G430L and 84-second exposures
for G750L gratings, using the 1 electron/DN gain setting. These
exposures yielded single-pixel signal-to-noise ratios of around 10:1
at 300 nm, $>$ 40:1 from around 400 to 700 nm, and decreasing to
around 20:1 (methane windows) to $<$ 10:1 (methane absorption bands)
at 1000 nm.  For the G750L grating, the part of the planet covered by
the narrow slit setting (from 0.9$''$ from the planet center to the
central meridian) the signal to noise at continuum wavelengths was reduced by a factor of
$\sim$1.7 at short wavelengths to $\sim$2.1 at the longest wavelengths.  In methane
bands, where readout noise dominates, the degradation factor was $\sim$4.5.
%% SNR reductions are based on sqrt(throughput reduction) Erich's email of 4/19/2016

\subsection{Supporting WFC3 imaging.}\label{Sec:synth}

\begin{figure*}[!h]\centering
%SOURCE Pat, in /usr5/stis12ura/calibrated/
%\includegraphics[width=6in]{wfc3_stis15_ratio_bilin_comp_smaller.ps}
\includegraphics[width=6in]{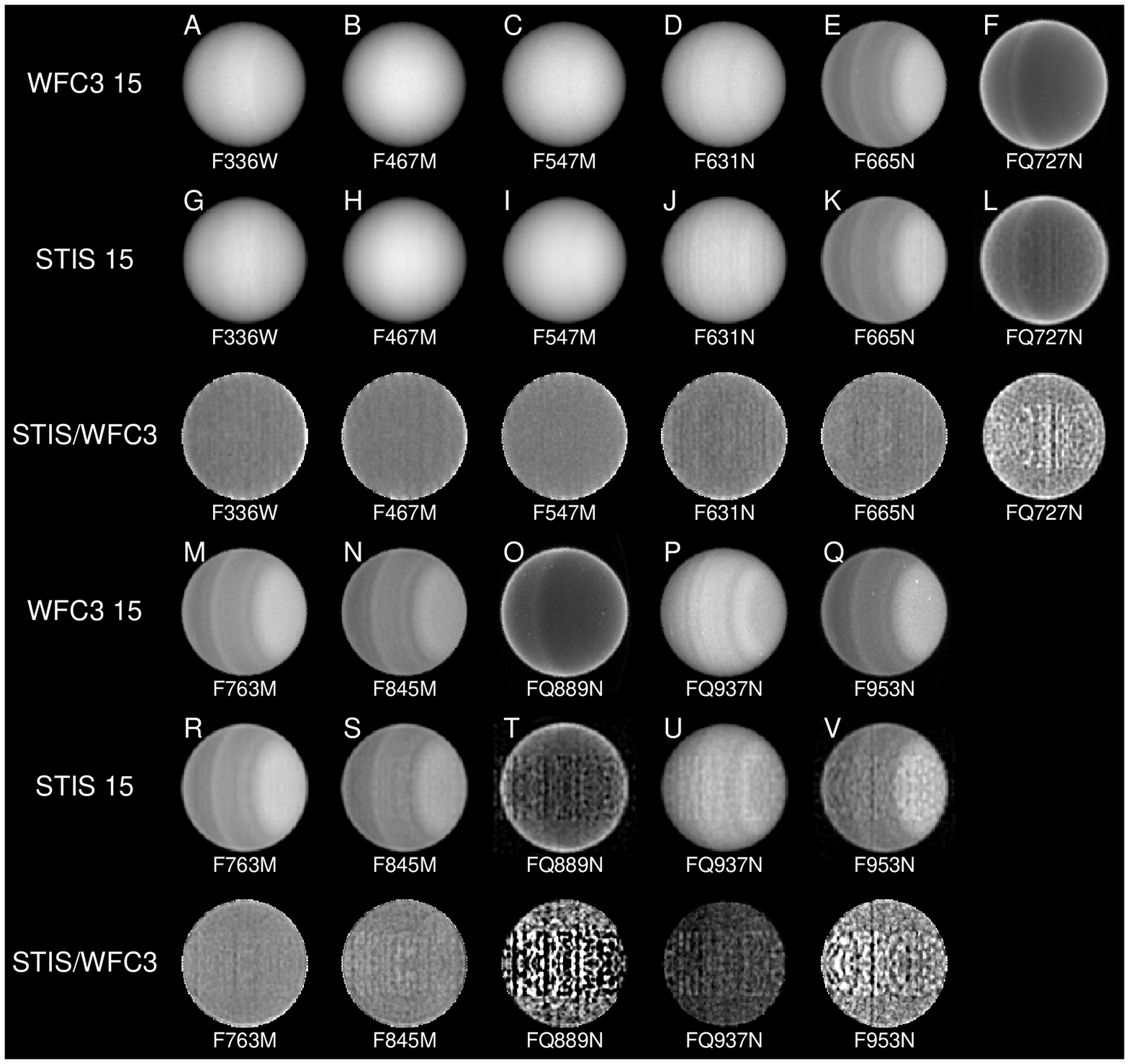}
\caption{WFC3 images of Uranus taken on 11 October 2015 (A-F and
  M-Q) compared to synthetic band-pass filter images (G-L and R-V)
  created from weighted averages of STIS spectral data cubes using
  WFC3 throughput and solar spectral weighting. The north pole is at
  at the right. Portions of the synthetic images east of the central
  meridian are obtained by reflection of the images west of the
  central meridian. 
The ratio images are stretched to make 0.8 black and 1.2 white.}
\label{Fig:stisimages}
\end{figure*}

The complex radiometric calibration of the STIS spectra relies on
calibrated WFC3 images to provide the final wavelength dependent
correction functions.   To
ensure that this function was determined as well as possible for the
Cycle 23 observations in 2015, and to cross check the extensive
spatial and spectral corrections that are required for STIS
observations, we used one additional orbit of WFC3 imaging at a pixel
scale of 0.04 arcseconds with eleven different filters spread over the
300-1000 nm range of the STIS spectra. These WFC3 images are displayed
in Fig. \ref{Fig:stisimages}, along with synthetic images with the
same spectral weighting constructed from STIS spatially resolved
spectra, as described in the following section.  The filters and
exposures are provided in Table\ \ref{Tbl:sci_obs}.

\subsection{Supporting near-IR imaging}

\begin{figure*}\centering
\includegraphics[width=6.2in]{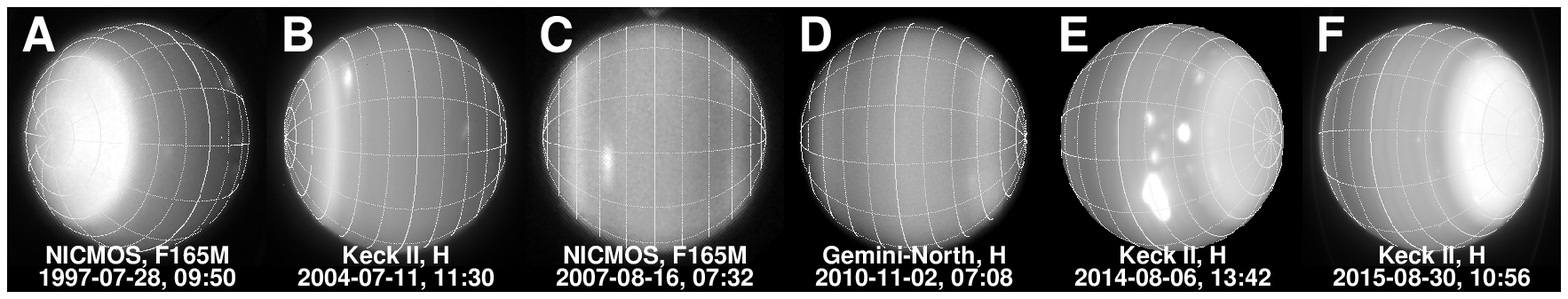}
%SOURCE: made by plot_targsr_stis15.pro in ../paper_stis15/ on triton
% using control file near_ir_1997-2015_6column_control
% documented on page 3, Uranus Log I
\caption{H-band (1.6-\mumx) images of Uranus from 1997 through 2015,
  from observatories/instruments given in the legends. The bright
  south polar region seen in 1997 (A), 10 years before equinox, is
  similar to the bright north polar region seen in 2015 (F), eight
  years after equinox. Images taken during the 2007 equinox year (C)
  found that neither polar region was bright. The longitude and planetographic
latitude grid lines are at 30\deg and 15\deg intervals respectively.}
\label{Fig:Hband}
\end{figure*}

HST/NICMOS and groundbased Keck and Gemini imaging at near-IR
wavelengths help to extend and fill gaps in the temporal record of changes occurring in the
atmosphere of Uranus. Fig.\ \ref{Fig:Hband} shows that the difference between polar and low latitude
cloud structures has evolved over time. The relatively rapid decline of the bright
``polar cap'' in the south and its reformation in the north is faster
than seems consistent with the long radiative time constants of the
Uranian atmosphere \citep{Conrath1990}.  In following sections we will
show that the polar brightness in 2015 (and presumably also in 1997)
is not due very much to latitudinal variations in aerosol scattering,
but is mainly due to a much lower degree of methane absorption at high
latitudes. This latitudinal variation of methane absorption appears
to be stable over time according to infrared observations
\citep{Sro2014stis}.  Thus, at times when the polar region was as dark
as low latitudes (compared at the same view angles), it appears not
that methane absorption was greater then, but instead that aerosol
scattering was reduced, a causal relationship we will here further
confirm regarding polar brightness increases between 2012 and 2015.

\begin{figure*}\centering
\includegraphics[width=6.2in]{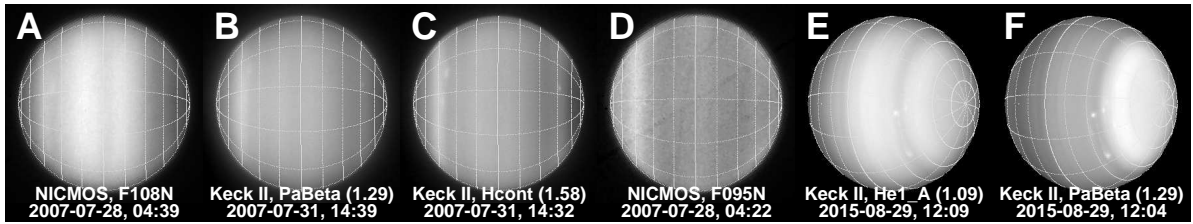}
%SOURCE: made by plot_targsr_stis15.pro in ../paper_stis15/ on triton
% using control file near_ir_narrowband_6column_control
% documented on page 6, Uranus Log I, improved readability on P110, Log L
\caption{Near-IR narrow-band images of Uranus obtained near the 2007 equinox
(A-D) and in August 2015 (E, F)  from observatories/instruments given in the legends. 
The NICMOS F108N image (A) and the Keck II He1\_A image (E) sample wavelengths
at which hydrogen absorption dominates.  The remaining images sample wavelengths
of comparable absorption but due entirely to methane. Note that low latitudes
are relatively darker than high latitudes at methane dominated wavelengths (e.g. in B and F), which
is not the case for wavelengths dominated by hydrogen absorption (as in A and E). Grid lines
are the same as in Fig.\ \ref{Fig:Hband}.}
\label{Fig:nirmulti}
\end{figure*}

The aforementioned interpretation of the bright polar region on Uranus
 can be partly inferred from the characteristics
of near-IR images at key wavelengths that have different sensitivities
to methane and hydrogen absorption, as illustrated in Fig.\ \ref{Fig:nirmulti}.
Images at hydrogen dominated wavelengths (panels A and E) reveal relatively
bright low latitudes, and high latitudes that were either darker, as
at equinox (A), or comparably bright, as in 2015 (E). At methane dominated
wavelengths, low latitudes are relatively darker, especially in 2015, where
the excess methane absorption at low latitudes is obvious from comparing
panels E and F.

\begin{table*}[!hbt]\centering
\caption{Near-IR observations from HST, Keck, and Gemini observatories.}\label{Tbl:ground}
\renewcommand{\baselinestretch}{1.}\normalsize  % turned off for single-spaced text
\vspace{0.15in}
\begin{tabular}{l l c r r r r l}
Telescope &       &           &   Obs.    &         & Phase  & S.O.     & \\
/Instrument & PID & Obs. Date &   Time &  Filter & Angle  & CLat & PI, Notes\\
\hline\\[-0.18in]
HST/NICMOS & 7429 & 1997-07-28 & 09:50:24 & F165M &    &-40.3 & Tomasko, 1 \\ 
Keck II/NIRC2 &   & 2003-10-06 &07:14:51  & H &  &-18.1 & Hammel, 2 \\
Keck II/NIRC2 &   & 2004-07-11 &11:30:32 & H &  &-11.1 & de Pater, 2 \\
HST/NICMOS & 11118 & 2007-07-28 & 04:39:xx & F095N & 2.0  & 0.61 & Sromosvky, 3 \\ 
HST/NICMOS & 11118 & 2007-07-28 & 04:22:30 & F095N &   & 0.58 & Sromovsky, 3 \\ 
HST/NICMOS & 11118 & 2007-07-28 & 04:39:13 & F108N &    & 0.58 & Sromovsky, 3 \\ 
Keck II/NIRC2 & & 2007-07-31 & 14:39:28 & PaBeta &  1.87  & 0.51  & Sromovsky, 2\\
Keck II/NIRC2 & & 2007-07-31 & 14:32:33 & Hcont &   & 0.49  & Sromovsky, 2\\
HST/NICMOS & 11190 & 2007-08-16 & 07:32:32 & F165M &   &-0.0 & Trafton, 3\\
Gemini-N/NIRI & \parbox{0.5in}{2010B-Q-110}  & 2010-11-02 & 07:08:57 &H\_G0203 &  &9.3 &Sromovsky, 4\\
Keck II/NIRC2 & &2014-08-06 & 13:42:06 &H&   & 28.4 & de Pater, 2 \\
Keck II/NIRC2 & & 2015-08-29 & 12:09:05 & He1A  &    & 31.96  & de Pater, 2\\
Keck II/NIRC2 & & 2015-08-29 & 12:04:41 & PaBeta &   & 31.96  & de Pater, 2\\
Keck II/NIRC2 & & 2015-08-30 & 10:56:55 & H &   & 31.9 & de Pater, 2\\
\hline\\[-0.16in]
\end{tabular}
\vspace{0.14in}
\parbox{5.76in}{NOTES: 1: pscale = 0.0431 as/pixel; 2: pscale = 0.009942
  as/pixel ; 3: pscale = 0.0432 as/pixel ; 4: pscale= 0.02138 as/pixel}
\end{table*}
%%  pscale = 0.0431 as/pixel, M. Tomasko, PI; 6: pscale = 0.009942 as/pixel, L. Sromovsky, PI
%% ;7: pscale =  0.0432 as/pixel,  L. Trafton, PI; 8: pscale= 0.02138 as/pixel, 
%% Obserever=K. Chiboucas ; 5: pscale = 0.009942 as/pixel, de Pater, PI; 6:  pscale = 0.009942 as/pixel,   de Pater, PI}
\renewcommand{\baselinestretch}{1.3}\normalsize  % turned off for single-spaced text

%\vspace{-0.14in}
\section{STIS data reduction and calibration.}\label{Sec:cal}

% does not yield suitably
%calibrated two-dimensional spectral images for an object like Uranus.
%Considerable additional effort was required to reach a final calibration of
%these data, using techniques developed by KT2009 and closely followed in the
%calibration of the 2012 STIS observations.
%The calibration procedure is 
%, as described by [Erich's paper in preparation, as stated in 
%4/13/16 email].  

The STIS pipeline processing used at STScI is just the first step of a rather
complex calibration procedure, which is 
described by KT2009 for the 2002 observations,
and by \cite{Sro2014stis} for the 2012 observations. Essentially the same procedure
was followed for the 2015 observations.  Additionally, 2002 and 2012 STIS
cubes were recalibrated using WFPC2 and WFC3 images newly reduced using
the best available detector responsivity functions and filter throughput
functions.
All three calibrated STIS cubes
and related information can be found online in the HST MAST archive as described
in the supplemental material section at the end of the paper.   In the
discussion that follows, we first describe the processing of
supporting WFC3 imaging. In the case of 2012 recalibration, WFC3
imaging was also utilized, but for the 2002 recalibration, WFPC2
images were used. We then describe the creation of our calibration
correction function, describe our spectral cube construction,
and finally our comparison of STIS synthetic images with bandpass
filter images.

Each WFC3 image was deconvolved with an appropriate Point-Spread
Function (PSF) obtained from the tiny tim code of \cite{Krist1995},
optimized to result in data values close to zero in the space view just off the limb of Uranus. To
match the spatial resolution of the STIS images, the WFC3 images were
then reconvolved with an approximation of the PSF given in the
analysis supplemental file of \cite{Sro2014stis}.
The images were then converted to I/F using the best available header
PHOTFLAM values [given in WFC3 ISR 2016-001] and the \cite{Colina1996} solar flux spectrum,
averaged over the WFC3 filter band passes. (PHOTFLAM is a multiplier used
to convert instrument counts of electrons/second to flux units of ergs/s/cm$^2$/\AA.) To obtain a
disk-averaged I/F, the planet's light was integrated out to 1.15 times the equatorial radius,
 then averaged over the planet's cross section in pixels, which was computed using
NAIF ephemerides \citep{Acton1996} and SPICELIB limb ellipse model
(SPICELIB is NAIF toolkit software used in generating navigation and
ancillary instrument information files.) The disk-averaged I/F
(using the initial calibration) was also computed for each of the STIS
monochromatic images, and the filter- and solar flux-weighted I/F was
computed for each of the WFC3 filter pass bands that we used.

%% not sure what Erich used so removing this: out to 1.1 equatorial radii 
%THE FOLLOWING PARAGRAPH NEEDS TO BE UPDATED: 
%[MORE INFO FROM Erich NEEDED], 
% [Erich, DO THESE RMS DEVIATIONS INCLUDE FQ937N?].  

\begin{figure*}[!ht]\centering
%% figure produced by /usr5/stis15ura/wfc3/compare_stis_cals_revA.pro
%% PS file from /usr5/stis15ura/wfc3/compare_stis_cals_revA.ps
%% See PMF log Uranus & Neptune 4/2016->, p. 105
%\includegraphics[width=5in]{compare_stis_cals_revA.ps}
\includegraphics[width=5in]{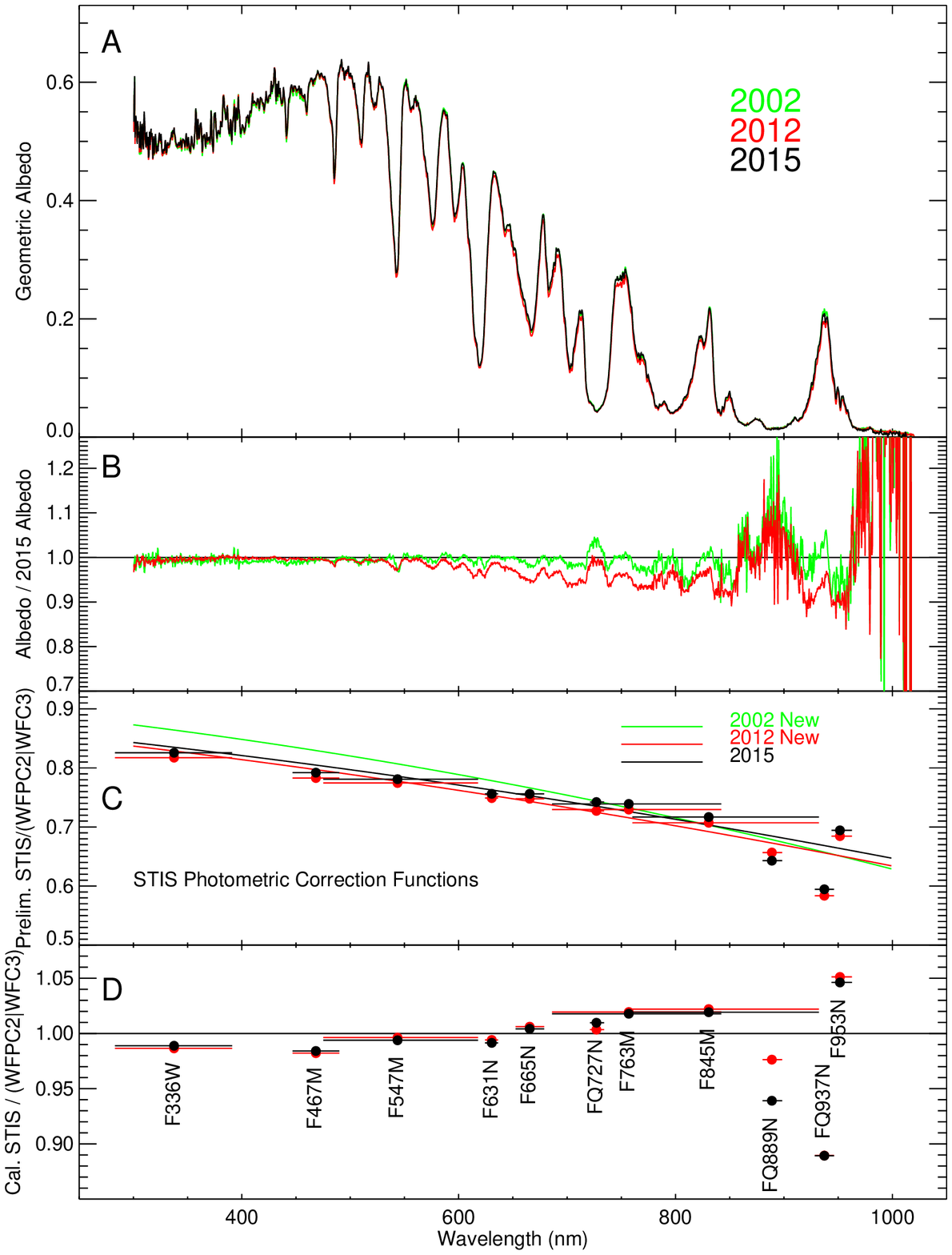}
\caption{{\bf A:} Radiometrically calibrated disk-averaged I/F spectra
  for three year of STIS observations. STIS observations from 2002 and
  2012 have been recalibrated from previous incarnations
  \citep{Kark2009IcarusSTIS,Sro2014stis}.  {\bf B:} Ratio of each
    year's disk-averaged I/F to 2015. {\bf C:} STIS photometric
      calibration functions (raw stis albedo divided by WFC3
      albedo). The functions are a fit to ratios constructed using
      synthetic band-pass filter disk-integrated I/F values
      (preliminary-calibration) divided by corresponding I/F values
      obtained from WFC3 measurements (circles and horizontal bars
      indicate filter effective wavelength and full-width half maximum
      transmission for 2012 and 2015). {\bf D:} Ratio of final
        calibrated STIS synthetic disk-averaged I/F values to WFC3
        reflectivities, showing scatter of ratios relative to the
        fits, for 2012 and 2015.}
\label{Fig:pat}
\end{figure*}

By comparing the synthetic disk-averaged STIS I/F, using the initial
calibration, to the corresponding
WFC3 values, we constructed a correction function to improve the radiometric
calibration of the STIS cube. Figure \ref{Fig:pat}C shows the ratios of
STIS to WFC3 disk-integrated brightness, and the quadratic function
that we fit to these ratios as a function of wavelength, for the 2015
data set and recalibrations of the previous two data sets.  We heavily
weighted the broadband filters, 
and computed an effective wavelength
weighted according to the product of the solar spectrum and the I/F
spectrum of Uranus.  The RMS deviation of individual filters from the
calibration curve given in Fig. \ref{Fig:pat} is about 1\% RMS for
2012 and 2015 correction curves, but about 2 \% RMS for the 2002
calibration curve (fit points not shown).
For narrow
filters such as FQ937N, typical deviations are some three times
larger.  The difference between the 2002 and the later calibration
curves is mostly due to use of different slit locations, which result
in different light paths through the monochrometer.  The 2012 and 2015
curves use the same slit location and thus should be the same, and
indeed they are consistent to about 1\%.

If one assumes that the spectrum of Uranus varies
only slowly with time, one can add many other filters to plots of
Fig.\ \ref{Fig:pat} where images are available somewhat close to the
time of STIS spectroscopy.  The medium and wide filters plot quite
consistently near the same curve while many narrow filters show
significant offsets, suggesting that an improved calibration weights
the narrow filters much less in the fitted curve.  This consideration
changes the calibration by about 1\% and thus does not make a big
difference with respect to our previous adopted calibration, but our
new calibration is more reliable because it is less dependent on
unreliable data from narrow filters. 

The final calibrated cubes contain 150 pixels parallel to the spin
axis of Uranus and 75 pixels perpendicular to its spin axis, with a
spatial sampling interval of of 0.015$\times R_U$ km/pixel (384
km/pixel), which is equivalent to 0.028 arc seconds per pixel for 2015
observations. (Here $R_U$ is the equatorial radius of Uranus).
 The center of Uranus is located at coordinates (74, 74),
where (0, 0) is the lower left corner pixel.  The spatial resolution of
the final cube is defined by a point-spread function with a FWHM of 3
pixels.  The cube contains an image for each of 1800 wavelengths
sampled at a spacing of 0.4 nm, with a uniform spectral resolution of
1 nm. Navigation backplanes are provided, in which the center of each
pixel is given a planetographic latitude and longitude, solar and
observer zenith angle cosines, and an azimuth angle.

As a sanity check on the STIS processing we compared WFC3 images to synthetic
WFC3 images created from our calibrated STIS data cubes, as shown in Fig.\ \ref{Fig:stisimages}.
Ratio plots of STIS/WFC3 show the desired flat behavior, except very close
to the limbs, where STIS I/F values exceed WFC3 values.
The most significant discrepancy is in the overall I/F level computed
for the FQ937N filter (note the dark ratio plot in the bottom row),
 a consequence of our calibration curve being
10\% high for that filter.

\section{Center-to-limb fitting}

The low frequency of prominent discrete cloud features on Uranus and
its zonal uniformity make it possible to characterize the smooth
center-to-limb profiles of the background cloud structure without much
concern about longitudinal variability, even though we observed only
half the disk of Uranus.  These profiles provide important constraints
on the vertical distribution of cloud particles and the vertical
variation of methane compared to hydrogen.  Because our observations
were taken very close to zero phase, these profiles are a function of
just one angular parameter, which we take to be $\mu$, the cosine of the
zenith angle (the observer and solar zenith angles are
essentially equal).  They also have a relatively simple structure that
we characterized using the same 3-parameter function KT2009 used to
analyze the 2002 STIS observations, and which we also used to fit the
2012 observations.  For each 1\deg of latitude from 30\deg S to 87\deg
N, all image samples within 1\deg of the selected latitude and with
$\mu > $ 0.175 are collected and fit to the empirical function
\begin{eqnarray}
 I(\mu)= a + b \mu + c/\mu \label{Eq:ctl},
\end{eqnarray}
 assuming all samples were collected at the desired latitude and using
 the $\mu$ value for the center of each pixel of the image samples.
 Fitting this function to center-to-limb (CTL) variations at high latitudinal
 resolution makes it possible to separate latitudinal variations
 from those associated with view angle variations.

%% \paragraph{(Erich's version).}  For each latitude and wavelength,  a
%% three-parameter center-to-limb fit was done and evaluated at cos(theta) values of
%% 0.4, 0.6, 0.8, and 1.0.  The cos(theta) = 0.2
%% I/F was estimated with data right at the limb.  
%% with 1800 wavelengths, 140 latitudes (from 54 South to 85 North spaced
%% every degree) and five scattering geometries, 0.2 to 1.0.  All of the
%% 1.0 data and most of the 0.8 data are extrapolations, of course, and
%% should better not be used.  
%% Center limb fits gives a data cube 
%% 

\begin{figure*}\centering
%% Figure produced by triton:~patf/uranus/ctlfit/extract_stis_ctlfit2016.pro.
\hspace{-0.1in}
\includegraphics[width=3.15in]{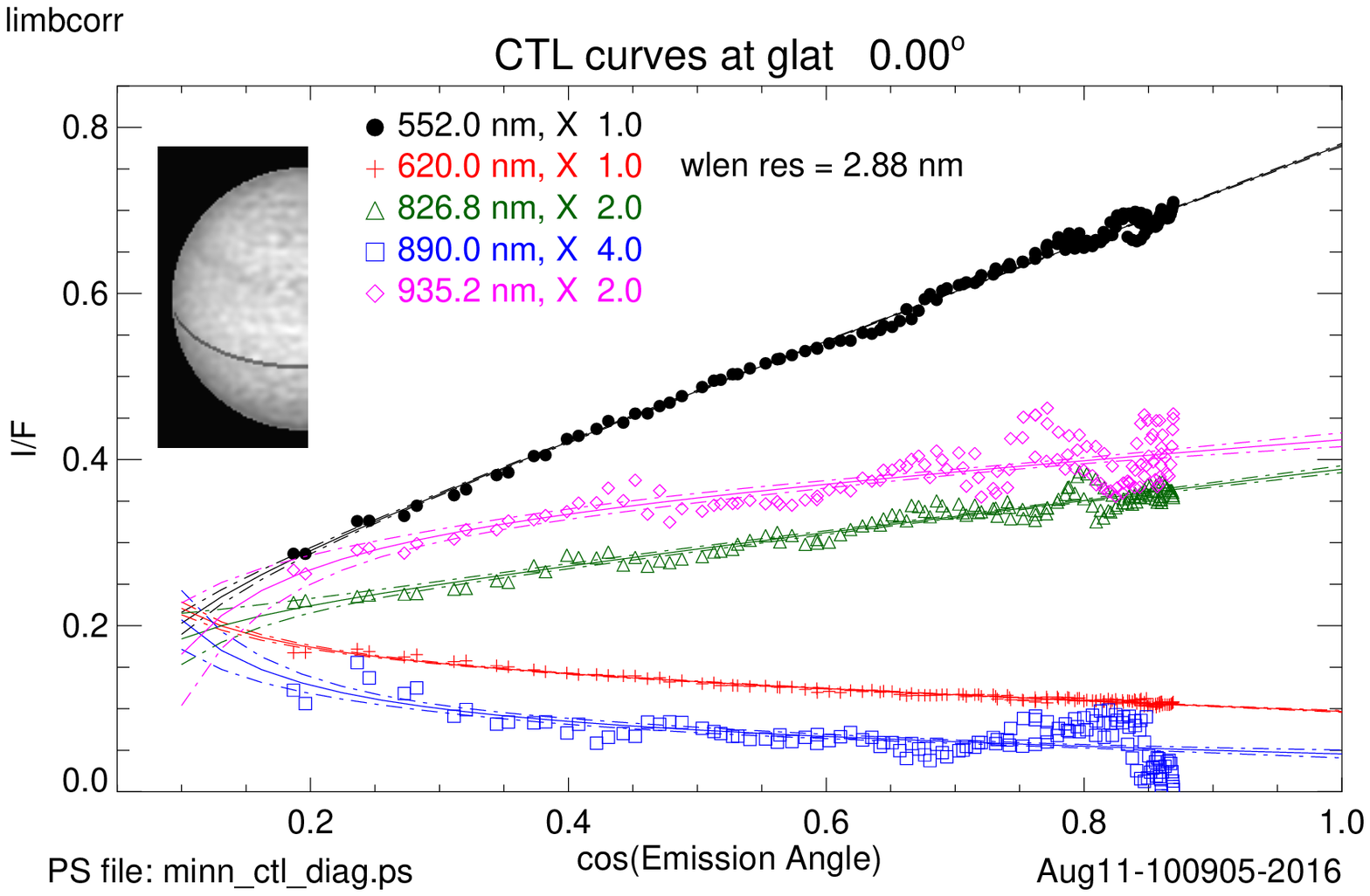}
\hspace{-0.2in}
\includegraphics[width=3.15in]{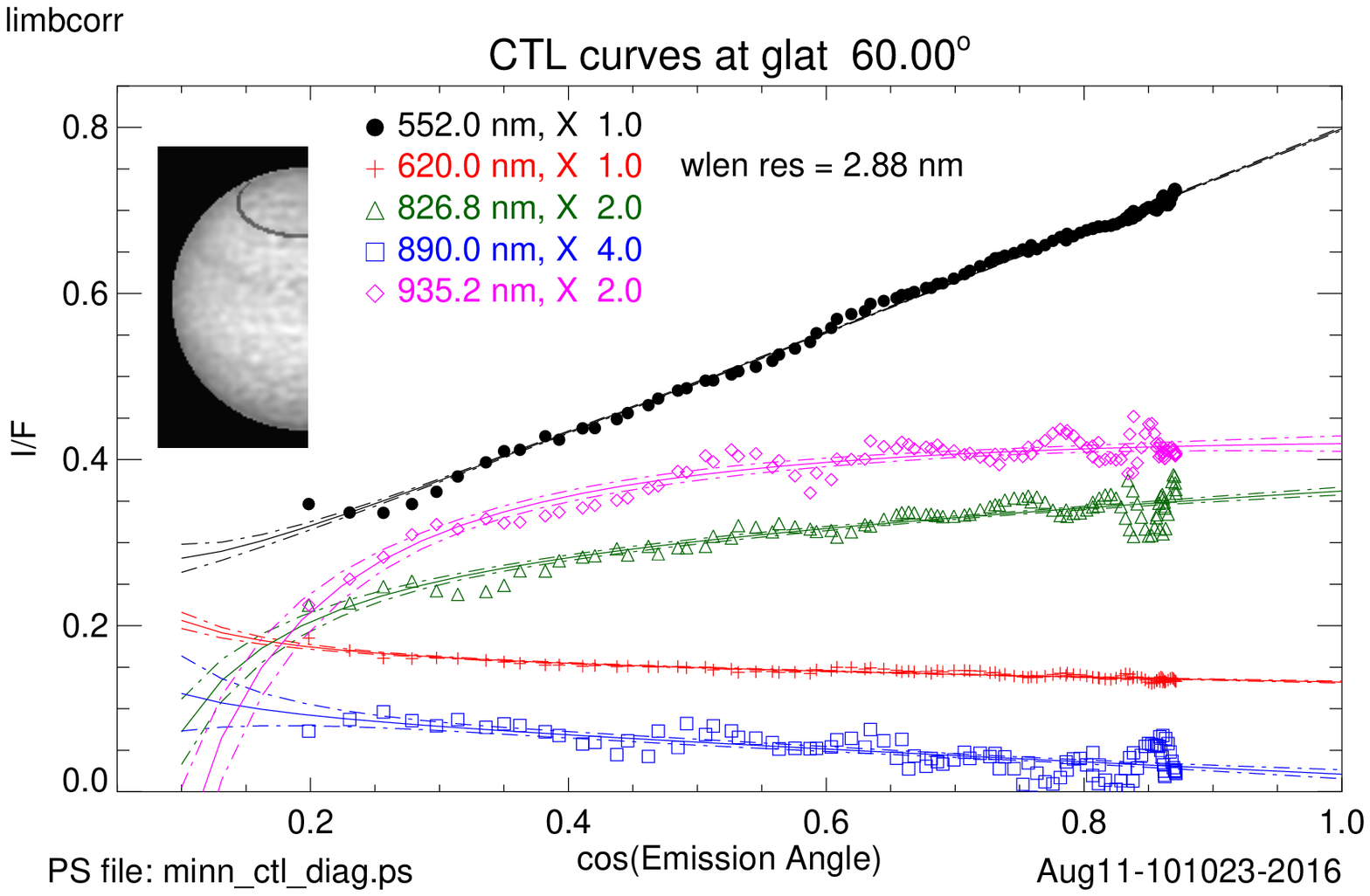}
\caption{Sample center-to-limb fits for 0\deg N (left) and 60\degx N (right), as described in the
  main text.  STIS I/F samples and fit lines with uncertainty bands
  are shown for five different wavelengths indicated in the
  legends. The latitude bands sampled for these fits are darkened in the
  inset images of the half-disk of Uranus.}
\label{Fig:ctlfits}
\end{figure*}

Before fitting the CTL profile for each wavelength,
the spectral data were smoothed to a resolution of 2.88-nm to improve
signal to noise ratios without significantly blurring key spectral
features.  (Our prior analysis was conducted in the wavenumber domain
and used smoothing to a resolution of 36 \icmx.)  Sample fits are
provided in Fig.\ \ref{Fig:ctlfits}. Most of the scatter about the
fitted profiles is due to noise, which is often amplified by the
deconvolution process.  Because the range of observed $\mu$ values
decreases away from the equator at high southern and northern
latitudes, we chose a moderate value of $\mu$ = 0.7 as the maximum
view-angle cosine to provide a reasonably large unextrapolated range
of 16\deg S to 77\deg N.  Ranges for other years and for a $\mu$ range
of 0.3 to 0.6 are given in Table\ \ref{Tbl:latranges}.
%of  [UPDATE THESE NUMBERS THAT WERE COMPUTED FOR mu=0.6: 33.6\deg S to
%  72.6\deg N for 2012 observations and 74.5\deg S to 31.7\deg N for
%  2002].
 Unless otherwise noted all our results are derived without
extrapolation.

\begin{table}\centering
\caption{Latitude ranges for two different view-angle cosine ranges.}\label{Tbl:latranges}
\vspace{0.1in}
\begin{tabular}{c c c}
Year  & 0.3 $\le \mu \le$ 0.6 &  0.3 $\le \mu \le$ 0.7 \\[0.03in]
\hline
 \\[-0.1in]
2002 & 74\degx S -- 33\degx N & 67\degx S -- 26\degx N\\[0.1in] 
2012 & 35\degx S -- 72\degx N & 28\degx S -- 65\degx N\\[0.1in] 
2015 & 23\degx S -- 77\degx N & 16\degx S -- 77\degx N\\[0.1in] 
\hline
\end{tabular}
\end{table}

The CTL fits can also be used to create zonally smoothed images by
replacing the observed I/F for each pixel by the fitted value. Results
of that procedure are displayed in a later section.  
%% Note that the image for the H$_2$ CIA
%% dominated wavelength (826.8 nm) shows much more equator-to-pole
%% darkening than the image for the methane-dominated wavelength of 833.7
%% nm, implying that there is relatively less methane absorption
%% (compared to H$_2$ absorption) at high latitudes. This implication is
%% based on the fact that the two wavelengths are roughly sensitive
%% to the same pressure range and thus are not sensing grossly different
%% aerosol contributions.

%% \begin{figure}\centering
%% %2002\includegraphics[width=5in]{STIS2002_FittedImages619.2-930.0.eps}
%% %2012\includegraphics[width=5in]{STIS_FittedImages619.2-930.0.eps}
%% \includegraphics[width=6in]{sample_ctl_images_0212.ps}
%% \caption{Comparison of 2002 and 2012 synthetic images created from CTL fits for
%%   six sample wavelengths (these are autoscaled)). Note that the
%%   images for the H$_2$ CIA-dominated wavelength (826.8 nm) have relatively
%% bright low latitudes and darker polar regions, while the images for the methane dominate
%%   wavelength of 930 nm do not. This implies that there is relatively less
%%   methane absorption (compared to H$_2$ absorption) at high latitudes. Note that the longitudinal
%% structure seen near the poles, a region where CTL fits did not
%% replace the original image data, is mostly due to noise.}
%% \label{Fig:ctlimages}
%% \end{figure}

\section{Direct comparisons of STIS spectra}

%For 2015 there is little difference between 10\deg N and
%60\deg N spectra at continuum and hydrogen dominated wavelengths,
%while in regions of intermediate methane absorption there are
%significant differences (panels E and F, described in
%Sec. \ref{Sec:scanstis}).

\begin{figure*}\centering
\includegraphics[width=3.1in]{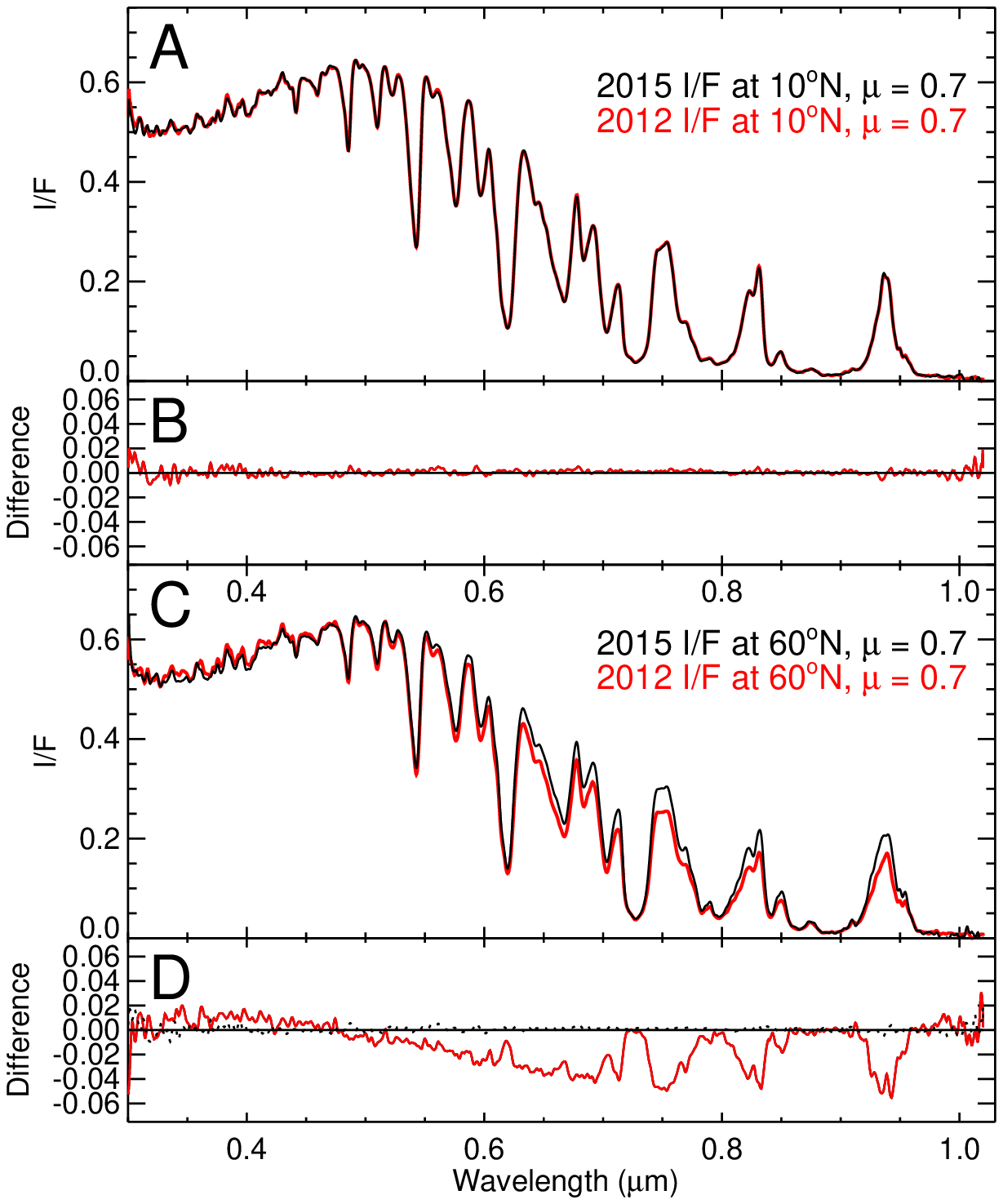}
\includegraphics[width=3.1in]{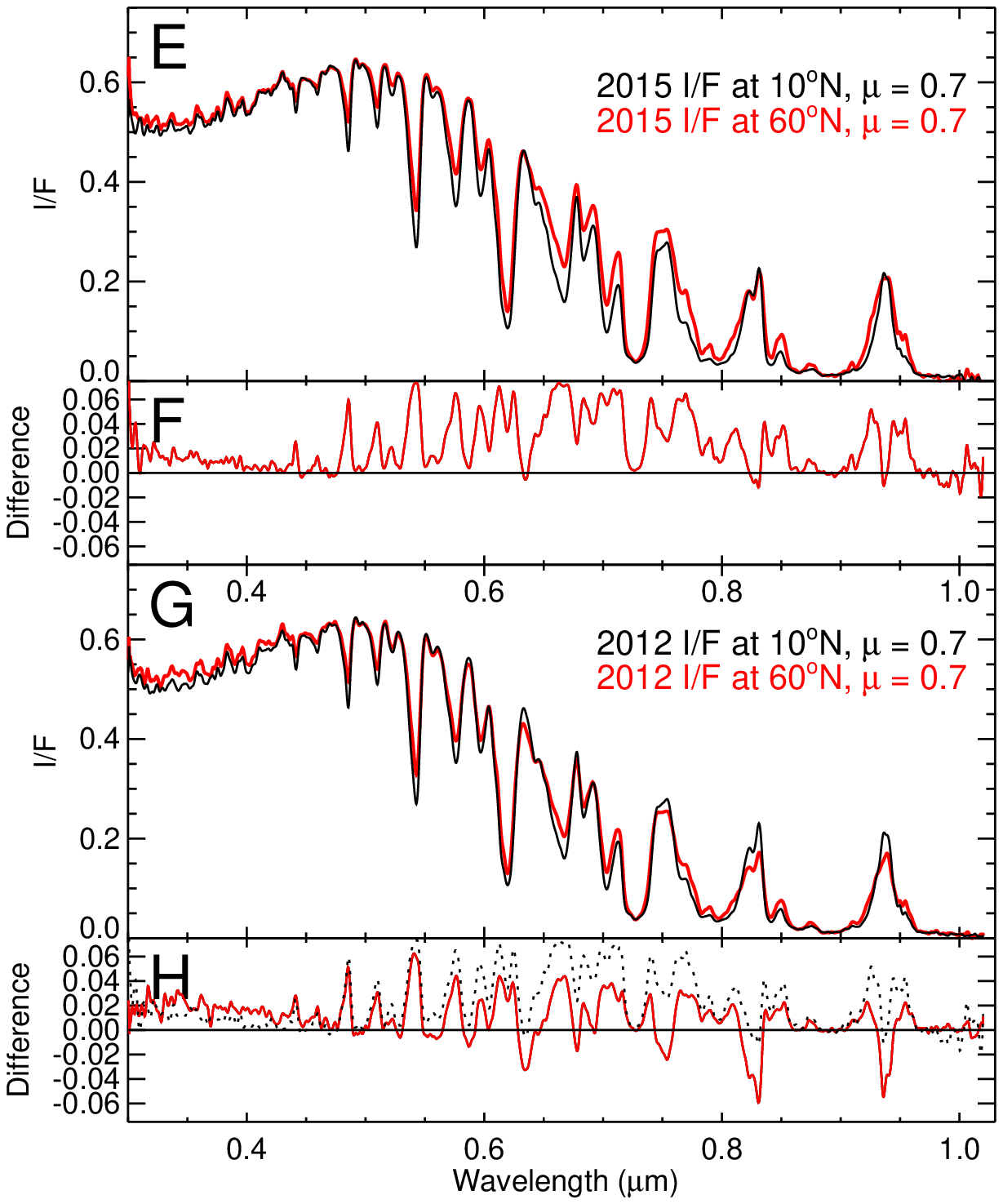}
%SOURCE: plots made by program plot_polarcapspecdif.pro on puck in ir_ms_ura
% original version documented on P 122-3, Uranus log H and on P 4, Uranus log I
% latest version on page 43 of Uranus Log I
\caption{Comparison of 2012 and 2015 STIS spectra at 10\deg N (A) and
  60\deg N (C), and comparison of STIS 10\deg N and 60\deg N spectra
  from 2015 (E) and 2012 (G), with difference plots shown in panels B,
  D, F, and H respectively.  The dotted curve in panel H is a copy of
  the 2015 difference curve from panel F. Latitudes are
  planetographic.  Note the nearly exact equality (in A) of 10\deg N
  spectra from 2012 and 2015.}
\label{Fig:speccomp}
\end{figure*}

A rough assessment of the changes between 2012 and 2015 and the
differences between high and low latitudes in these two years can be
made with the help of direct comparisons of STIS spectra, as
in Fig.\ \ref{Fig:speccomp}.  Note that at 10\deg N there is almost no
difference between 2012 and 2015 spectra (panels A and B).  This is
also the case for $\mu$ values of 0.3 and 0.5, which are not shown in
the figure.  For 2015, (see panel E) the lack of any I/F difference between 10\deg N and 60\deg
  N at 0.83 \mumx, which is a wavelength at which hydrogen absorption
  dominates, suggests that there is not much difference in aerosol
  scattering between these two latitudes.  A similar lack of difference
  at 0.93 \mumx, a wavelength of weak (but dominant) methane
  absorption, suggests that at very deep levels, there may not be much
  of a latitudinal difference in methane mixing ratios, or that there
  is an aerosol layer blocking visibility down to levels that might
  sense such a difference.  Yet the fact that wavelengths of
  intermediate methane absorption do show a significant increase in
  I/F with latitude suggests that at upper tropospheric levels the
  methane mixing ratio does decline with latitude, which is a known
 result from previous work, and is refined by the analysis presented
in following sections.
 Somewhat different results are seen, for 2012 (in panel G). The 10\deg N
and 60\deg N I/F values at 0.83 \mum and 0.93 \mum do differ (panels G and
H), which we will later show is a result of differences in aerosol
scattering.  The small size of continuum differences between 2012 and
2015 (panels D and H) is partly a result of the relatively smaller
impact of particulates at short wavelengths where Rayleigh scattering
is more significant.  At absorbing wavelengths for which gas absorption
is important, the optical depth and
vertical distribution of particulates have a greater fractional effect
on I/F and thus small secular changes in these parameters can be more
easily noticed.

\section{Direct comparison of methane and hydrogen absorptions vs. latitude.}

If methane and hydrogen absorptions had the same dependence on
pressure, then it would be simple to estimate the latitudinal
variation in their relative abundances by looking at the relative
variation in I/F values with latitude for wavelengths that produce
similar absorption at some reference latitude.  Although this idea is
compromised by different vertical variations in absorption, which
means that latitudinal variation in the vertical distribution of
aerosols can also play a role, it is nevertheless useful in a
semi-quantitative sense.  Thus we explore several cases below.

\subsection{Image comparisons at key near-IR wavelengths in 2007 and 2015}
%came from HST
%  program 11118 (L. Sromovsky PI) and was taken at 4:39 UT on 28 July
%  2007. 
%, at a phase angle of 2.00\deg and a sub-observer latitude of
%  0.61\degx. 
%The Keck II image was taken at 14:39 UT on 31 July 2007
%  (I. de Pater, PI), at a phase angle of 1.87\deg and a sub-observer
%  latitude of 0.51\degx.]  

\begin{figure*}[!ht]\centering
%\includegraphics[width=6.2in]{lat_profile_pair_F108N_PaBeta_mu0.4-0.6.eps}
%\includegraphics[width=6.2in]{fig07_rev.eps}
%\includegraphics[width=6.2in]{latscans_2015_keck.eps}
%SOURCE: plot15scans_revC18.pro, in /home/patf/uranus/paper_stis15 on puck
%\includegraphics[width=6in]{scans_0715_mu86_match10N.ps}
\includegraphics[width=6in]{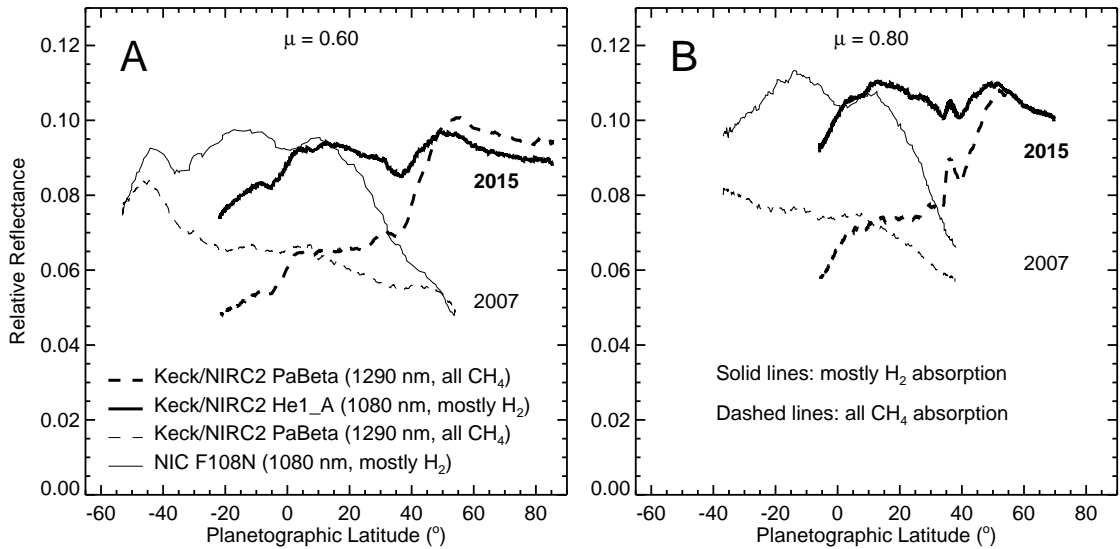}
% see page 111 in Uranus log L
\caption{Latitudinal profiles at fixed zenith angle cosines of 0.6 (A)
  and 0.8 (B) for F108N (HST/NICMOS) and PaBeta (Keck/NIRC2) filters
  (light solid and dashed lines respectively) taken near the Uranus
  equinox in 2007, and Keck/NIRC2 filter He1\_A and PaBeta filters
  (thick solid and dashed lines respectively) in 2015.  In 2007 the
  southern hemisphere was still generally brighter than the northern
  hemisphere and the 38\deg S - 58\deg S southern bright band was
  still better defined and considerably brighter than the
  corresponding northern bright band. The relatively low equatorial
  I/F values for the methane-dominated PaBeta filter (1290 nm)
  implies greater CH$_4$/H$_2$ absorption at low latitudes. We
  scaled the Keck 2015 observations to approximately match the 2007
  observations at 10\degx N, where we found almost no change between
  2012 and 2015 at CCD wavelengths (Fig.\ \ref{Fig:speccomp}A).  Note
  that between 2007 and 2015, the north polar region has brightened by
  comparable amounts at both hydrogen-dominated and methane-dominated
  wavelengths, indicating that it the brightening is due to increased
  aerosol scattering, not a temporal change in the methane mixing
  ratio at high latitudes. More information about the observations is
  given in Table\ \ref{Tbl:ground}.}
\label{Fig:nearircomp}
\end{figure*}

Our first example compares an HST/NICMOS image made with an F108N
filter (centered at 1080 nm), which is dominated by H$_2$ CIA, to a
KeckII/NIRC2 image made with a PaBeta filter (centered at 1290 nm),
which is dominated by methane absorption.  The images are shown in
panels A and B in Fig.\ \ref{Fig:nirmulti} and latitude scans at fixed view angles are
shown in Fig.\ \ref{Fig:nearircomp}.  The NICMOS observation was taken
on 28 July 2007 at 4:39 UT and the Keck observation on 31 July 2007 at
14:39 UT (see Table\ \ref{Tbl:ground} for more information).  That
these two observations sense roughly the same level in the atmosphere
is demonstrated by the penetration depth plot in
Fig.\ \ref{Fig:pendepth}, which also displays the filter transmission
functions.  The absolute (unscaled) I/F profiles for these two images
near the 2007 Uranus equinox are displayed for $\mu$ = 0.6 and $\mu$ =
0.8 by thinner lines in Fig.\ \ref{Fig:nearircomp}.  At high latitudes
in both hemispheres, profiles at the two wavelengths agree closely,
and both increase towards the equator.  But as low latitudes are
approached the two profiles diverge dramatically, with the I/F for the
hydrogen-dominated wavelength ending up 50\% greater than for the
methane-dominated wavelength, indicating much greater methane
absorption at low latitudes than at high latitudes.  As noted by
\cite{Sro2014stis} this suggests that upper tropospheric methane
depletion (relative to low latitudes) was present at both northern and
southern high latitudes in 2007, at least roughly similar to the
pattern that was inferred by \cite{Tice2013} from analysis of 2009
IRTF SpeX observations.  Latitudinal variations in aerosol scattering
could distort these results somewhat, but because they affect both
wavelengths to similar degrees, most of the effect is likely due to
methane variations.

A second example is shown by the thicker lines in
Fig.\ \ref{Fig:nearircomp}, which display latitudinal scans of 2015
images shown in panels E and F of Fig.\ \ref{Fig:nirmulti}. These 
were made by the KeckII/NIRC2 camera with He\_1A and PaBeta filters
on 29 August 2015 (see Table\ \ref{Tbl:ground}).  The He\_1A filter is
similar to the NICMOS F108N filter, as shown in
Fig.\ \ref{Fig:pendepth}.  The 2015 observations present a picture
that is somewhat different from the 2007 observations, with high
northern latitudes much brighter (at the same view angles) than in
2007.  This change appears to be entirely due to increased aerosol
scattering.  This conclusion is supported by the characteristics of
images obtained with H$_2$-dominated filters (NICMOS F108N filter and
Keck He1\_A filter).  In 2015 the I/F in the He1\_A filter is
relatively independent of latitude as shown by the image in panel E of
Fig.\ \ref{Fig:nirmulti}, indicating that aerosol scattering must have
a relatively weak latitudinal dependence. Note that latitude scans at
fixed view angles for these filters (shown in
Fig.\ \ref{Fig:nearircomp}) exhibit a low-latitude divergence of the
hydrogen-dominated and methane-dominated wavelengths which has about
the same magnitude in 2015 as seen for the 2007 observations,
indicating a similar increase of methane absorption at low latitudes.
  
\subsection{Direct comparison of key STIS wavelength scans}\label{Sec:scanstis}

\begin{figure*}[!h]\centering
\includegraphics[width=5.in]{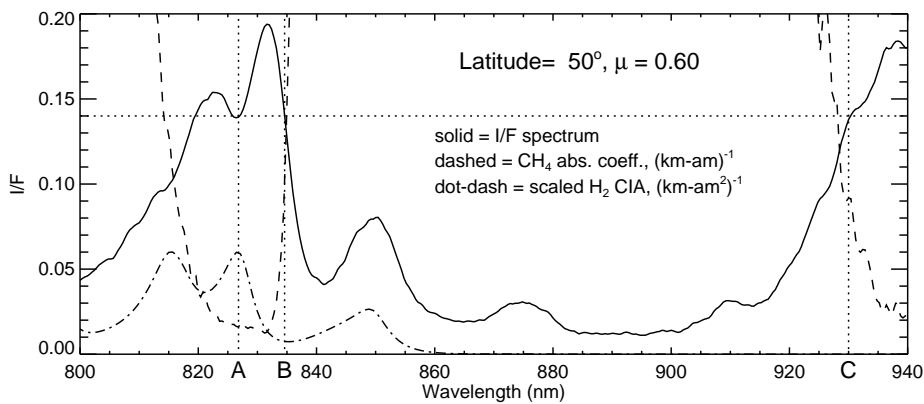}
%SOURCE: plotted by plot_latdep_stis_sm2015.pro in ..uranus/rt_raman on triton
\caption{I/F and absorption spectra comparing the equilibrium H$_2$
  CIA coefficient spectrum (divided by 1.2$\times 10^{-7}$, shown as dot-dash
  curve) and methane absorption coefficient spectrum (dashed). Note
  that the I/F spectrum has nearly equal I/F values at 826.8 nm (A),
  834.6 nm (B), and 930 nm (C), but H$_2$ absorption is much greater
  at A than at B, while the opposite is true of methane absorption,
  and at C only methane absorption is present. In a reflecting layer
  model, changes in cloud reflectivity should affect wavelengths A-C
  by the same factor, but changes in methane mixing ratio would affect
  C most and A least. From \cite{Sro2014stis}.}
\label{Fig:absprofiles}
\end{figure*}

A comparison of the STIS latitude scans at methane dominated
wavelengths with scans at  H$_2$ CIA dominated wavelengths is also
 informative. By selecting wavelengths that at one latitude provide
similar I/F values but very different contributions by H$_2$ CIA and
methane, one can then make comparisons at other latitudes to see how
I/F values at the two wavelengths vary with latitude.  If aerosols did
not vary at all with latitude, then any observed I/F variation would be a clear indicator of
variation in the ratio of \chf to H$_2$. Fig.\ \ref{Fig:absprofiles} displays a
detailed view of I/F in the spectral region where hydrogen CIA exceeds
methane absorption (see Fig.\ \ref{Fig:pendepth} for penetration
depths). Near 827 nm (A) and 930 nm (C) the I/F values are similar but the
former is dominated by hydrogen absorption (dot-dash curve) and the latter by methane
absorption (dashed curve). Near 835 nm (B) there is a relative minimum in hydrogen
absorption, while methane absorption is still strong.  For the latitude
and view angle of this figure (50\deg N and $\mu$ = 0.6), I/F values are nearly the same at all three
wavelengths, suggesting that they all produce roughly the same
attenuation of the vertically distributed aerosol scattering.  

\begin{figure*}[!h]\centering
\includegraphics[width=6.2in]{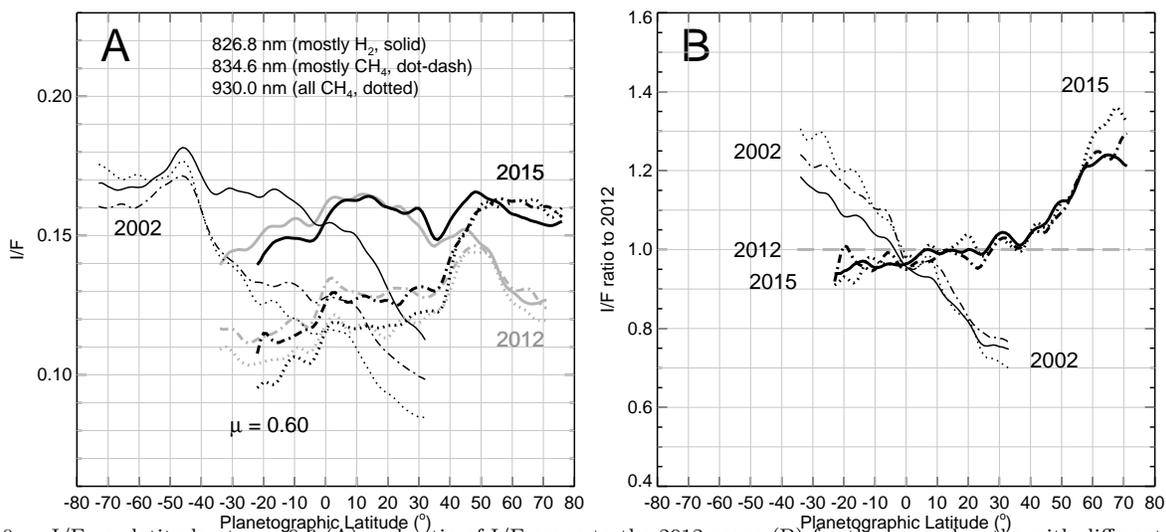}
%SOURCE: plotted with plot_stis_latscan2018.pro in /home/home2/sro/uranus/rt_raman on triton
\vspace{-0.15in}
\caption{I/F vs. latitude at $\mu=0.6$ (A) and ratio of I/F scans to
  the 2012 scans (B) for three wavelengths with different amounts of
  methane and hydrogen absorption.  Thin curves are from 2002, thick
  gray curves from 2012, and thick black curves from 2015. These are
  plots of center-to-limb fitted values instead of raw image data.  In
  all cases the methane-dominated wavelengths have much reduced I/F at
  low latitudes, compared to hydrogen-dominated wavelengths,
  indicating higher \chf/H$_2$ ratios at low latitudes.  Also note the
  increased north polar brightness at all displayed wavelengths
  between 2012 and 2015, indicating that the temporal change is due to
  increased aerosol scattering.}
\label{Fig:latscan3}
\end{figure*}

Figure\ \ref{Fig:latscan3} displays the latitudinal scans for the
three wavelengths highlighted in Fig.\ \ref{Fig:absprofiles} for the
STIS observations in 2002 (shown by thin lines), 2012 (thick gray
lines), and 2015 (thick black lines).  This is for a view angle cosine
of $\mu = 0.6$, chosen as a compromise between amplitude of variation
and coverage in latitude.  The 2012 I/F for the hydrogen-dominated
wavelength increases towards low latitudes, while the I/F for the
methane-dominated wavelength decreases substantially, indicating an
increase in the amount of methane relative to hydrogen at low
latitudes.  Similar effects are seen in 2002 (providing the best view
of southern latitudes) and 2015 (providing the best view of the
northern latitudes).  The hydrogen-dominated wavelengths have
relatively flat latitudinal profiles of I/F in the southern hemisphere
in 2002 and in the northern hemisphere in 2015, while the
methane-dominated wavelengths show strong decreases towards the equator,
beginning at about 45\deg S and 50\deg N.  For $\mu$=0.8 (not shown), which probes more
deeply, the latitudinal variation for the methane dominated
wavelengths is somewhat greater (a 30\% decrease in I/F
at low latitudes vs. a 20\% decrease for $\mu$ = 0.6).

The spectral comparisons in Fig.\ \ref{Fig:latscan3} also reveal
substantial secular changes between 2002 and 2012 and between 2012 and
2015. At wavelengths for which methane and/or hydrogen absorption are
important, the northern low-latitudes have brightened substantially,
while the southern low latitudes have darkened. The bright band
between 38\deg and 58\deg N continued to brighten.  Its brightening
and the darkening of the corresponding southern band was already
apparent from a comparison of 2004 and 2007 imaging
\citep{Sro2009eqdyn}.  The most dramatic change between 2012 and 2015
is the increased brightness of the polar region.  The nearly identical
brightening at all wavelengths, shown by the ratio plot in panel B of
Fig.\ \ref{Fig:latscan3}, argues that the brightening is due to
aerosol scattering rather than a decrease in the amount of methane. We
will confirm this with radiation transfer modeling in Section 9.

%MORE WORDS
%ABOUT 2015 RESULTS NEED TO BE ADDED.  
%Aside from
%what appears to be a small calibration disagreement between 2002 and
%2012, in which corrected 2002 I/F values are about 2\% higher (at
%continuum wavelengths), the latitudinal variations are very similar
%for both epochs. [The corrected 2002 calibration we use here is based
%  on WFPC2 comparisons and leads to I/F values that are 3\% smaller
%  than published by KT2009.]  

\begin{figure*}[!h]\centering
\includegraphics[width=6.0in]{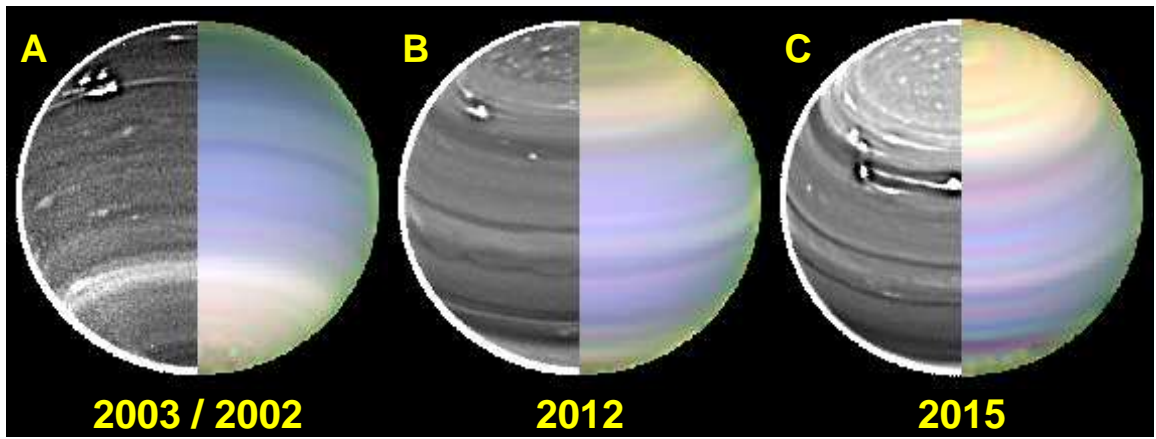}
% This version is 4x smaller than original (300K instead of 1.2 MB)
%ORIGINAL SOURCE: stis15_compare_epochs_keck_revA.pro in /home/patf/uranus/ctlfit on triton.
% See page 44 Uranus Log I
\caption{Color composites of fitted center-to-limb smoothed images for
2002 (A, right), 2012 (B, right), and 2015 (C, right), using color assignments R=930 nm (all methane)
G=834.6 nm (methane and hydrogen), and B=826.8 nm (mostly hydrogen). The blue
tint at low latitudes in all years is due to locally increased methane absorption.
We also show near-IR NIRC2 H-band images for 2003 (A, left), 2012 (B, left), and 2015 (C, left).
The NIRC2 images are rotation removed averages following \cite{Fry2012} and
 processed to enhance the contrast of small spatial scales, by adding k times
the difference between the original image and a 0.13-arcsecond smoothed version, where k was
taken to be 30 for 2003 and 2012 images, but only 22.5 for the 2015 image because of 
better seeing conditions during its acquisition.}
\label{Fig:colorcomps}
\end{figure*}

A color composite of the highlighted wavelengths (using R = 930 nm, G
= 834.6 nm, and B = 826.8 nm) is shown in Fig.\ \ref{Fig:colorcomps},
where the three components are balanced to produce comparable dynamic
ranges for each wavelength. This results in nearly blue low latitudes
where absorption at the two methane dominated wavelengths is
relatively high and green/orange polar regions as a result of the
decreased absorption by methane there.  The spatial structure in
high-resolution near-IR H-band Keck II images is also shown in each
panel, revealing that small discrete cloud features remain visible in
the north polar regions even in the 2015 images, taken after the polar
region brightened significantly between 2012 and 2015. Also
noteworthy, is the lack of such features in the south polar region
(panel A).  This asymmetry is somewhat surprising.  As noted by
\cite{Sro2014stis}, because both polar regions are depleted in
methane, the suggested downwelling motions that could produce such
depletion would be expected to inhibit convection in both polar
regions.

\section{Radiative transfer modeling of methane and aerosol distributions}\label{Sec:RT}

\subsection{Radiation transfer calculations}\label{Sec:radtran}

\begin{figure*}\centering
\includegraphics[width=3.1in]{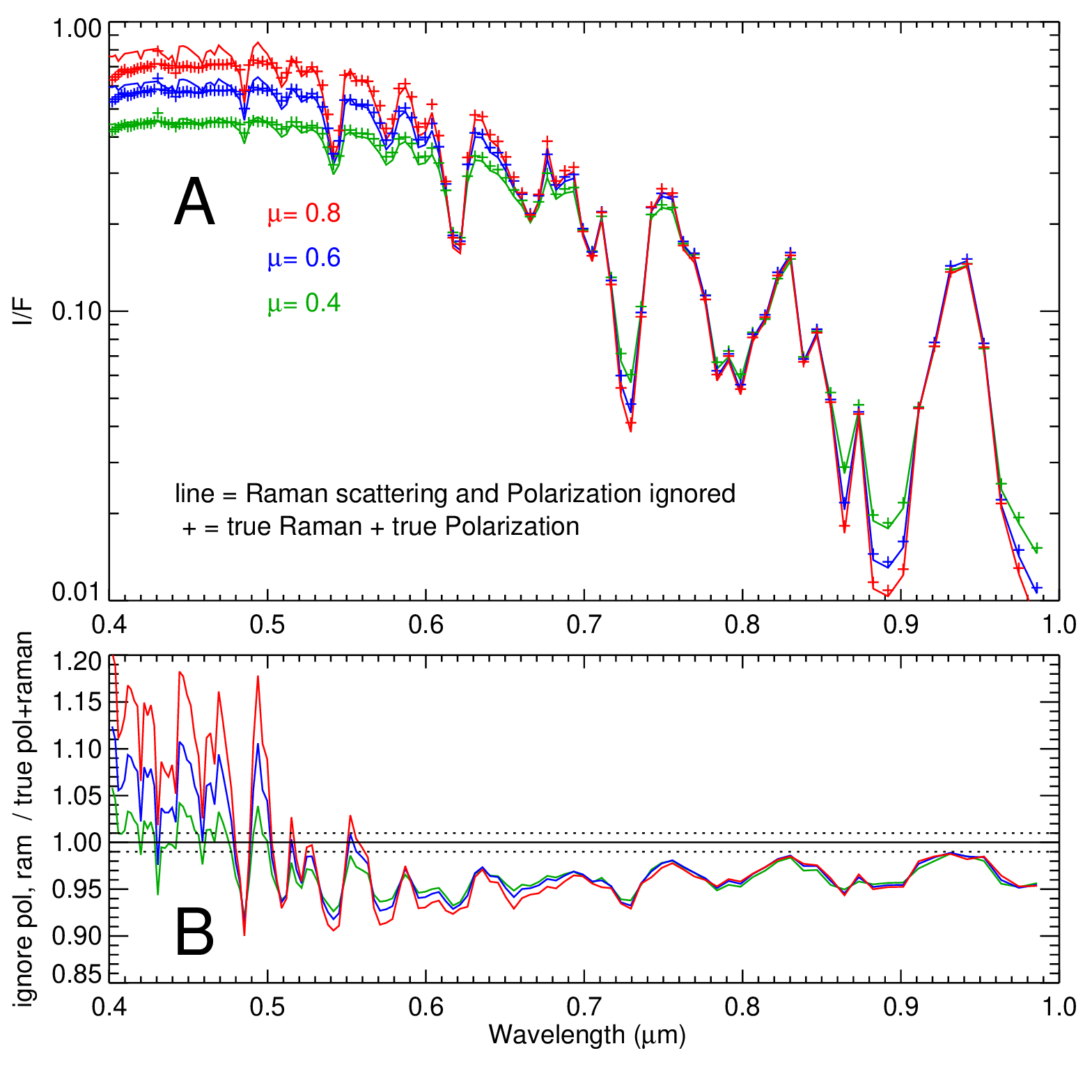}
\includegraphics[width=3.1in]{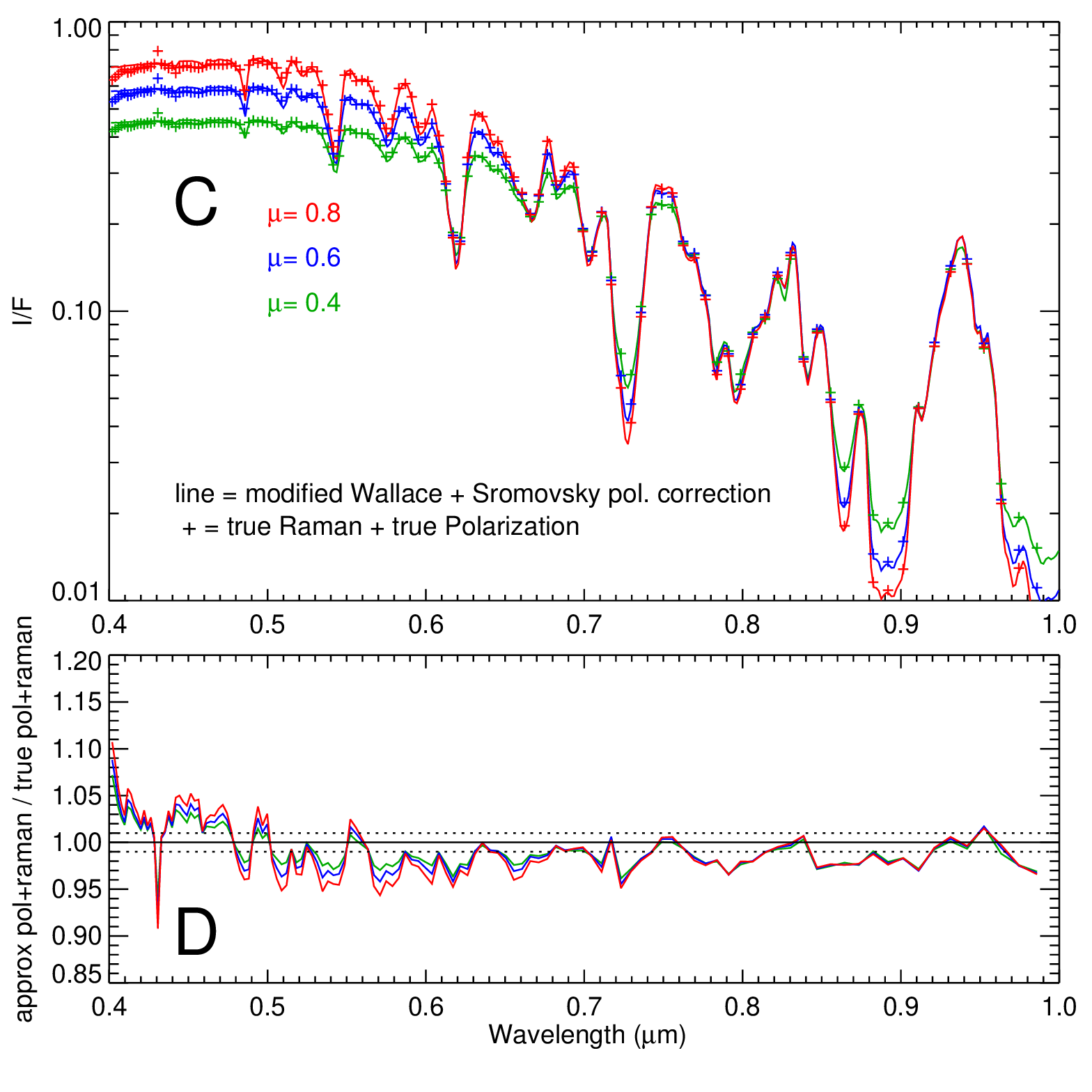}
% SOURCE plotted by compare_ccd_ir_calcs.pro on puck in /home/sro/ir_ms_ura
% documented on P 42-47 of Uranus Log H
\caption{Trial calculations showing errors produced by ignoring Raman scattering
and polarization (A and B) and greatly reduced errors achieved by
employing the modified Wallace approximation of Raman scattering \citep{Sro2005raman} and
the approximation of polarization effects following  \cite{Sro2005pol}.}\label{Fig:trialcalc}
\end{figure*}

In contrast to our prior analysis \citep{Sro2014stis}, which was
carried out in the wavenumber domain to accommodate our Raman
scattering code, here we worked in the wavelength domain, which is
better suited to the uniform wavelength resolution of our calibrated
STIS data cubes. We also used an approximation for the effects of
Raman scattering rather than carrying out the full Raman scattering
calculations.  We again used the accurate polarization correction
described by \cite{Sro2005pol} instead of carrying out the time
consuming rigorous polarization calculations.  To assess the adequacy
of our approximations, we did sample calculations that included Raman
scattering and polarization effects on outgoing intensity using the
radiation transfer code described by \cite{Sro2005raman}.  Examples in
Fig.\ \ref{Fig:trialcalc} show that at most wavelengths the errors
from these approximations do not exceed a few percent.

We improved our characterization of methane absorption at CCD
wavelengths by using correlated-k model fits by \cite{Irwin2010Icar}, which are available at
http://users.ox.ac.uk/$\sim$atmp0035/ktables/ in files
ch4\_karkoschka\_IR.par.gz and ch4\_karkoschka\_vis.par.gz. 
These fits are based on band model results of \cite{Kark2010ch4}.
% [{\it Using later
%    versions seemed to produce slightly worse fits.}]  
To model collision-induced opacity of H$_2$-H$_2$ and He-H$_2$
interactions, we interpolated tables of absorption coefficients as a
function of pressure and temperature that were computed with a program
provided by Alexandra Borysow \citep{Borysow2000}, and available at
the Atmospheres Node of NASA'S Planetary Data System. We assumed
equilibrium hydrogen, following
KT2009 and \cite{Sro2011occult}.
% but did look into the
%effects of non-equilibrium distributions, which is discussed in
%Sec. \ref{Sec:noneqh2}. [IT MIGHT also be useful to provide spectral
%  comparisons with and without Raman and polarization effects, or
%  provide a more detailed reference.]

After trial
calculations to determine the effect of different quadrature schemes
on the computed spectra, we selected 12 zenith angle quadrature points
per hemisphere and 12 azimuth angles. Test calculations with 10 and 14 quadrature
points in each variable changed fit parameters by only about 1\%, which is much less than
their uncertainties.  

\subsection{Thermal profiles for Uranus}\label{Sec:thermal}

Assuming the helium volume mixing ratio (VMR) of 0.152 inferred by
\cite{Conrath1987JGR}, \cite{Lindal1987} used radio occultation
measurements of refractivity versus altitude to infer a family of
thermal and methane profiles, with each profile distinguished by the
assumed methane relative humidity above the cloud level, and the
resultant deep volume mixing ratio (VMR) of methane below the cloud
level.  The cloud was positioned at the point where the refractivity
profile had a sharp change in slope. None of these profiles achieved
methane saturation inside the cloud layer, even the profile with the
highest physically realistic humidity level (limited by the
requirement that lapse rates could not be superadiabatic). This
high-humidity profile also had the highest deep temperatures and the
largest deep methane VMR of 4\%.  By allowing the He/H$_2$ ratio to
take on values near the low end of the uncertainty range given by
\cite{Conrath1987JGR}, \cite{Sro2011occult} were able to find
solutions that achieved methane saturation inside the cloud layer, as
well as deep methane mixing ratios somewhat greater than 4\%.  

The above results are based on the Voyager ingress profile, which sampled
latitudes from 2\degx S to 6\degx S.  As to whether this local sample
can be taken to roughly represent a global mean profile, some guidance
is provided by the results that \cite{Hanel1986Sci} derived from the
Voyager 2 Infrared Interferometer Spectrometer (IRIS) observations.
Inversion of spectral samples near both poles and near the equator
yielded temperature profiles that differed by less than 1 K from about
150 mbar to 600 mbar, and the equator and south pole profiles remained
within 2 K from 60 mbar to 150 mbar, with the north polar profile
deviating up to 4 K above the tropopause.  More significant variations
can be seen at middle latitudes, however, especially in the 60 -- 200
mbar range where average temperatures are 3.5 K higher than the
latitudinal average near the equator and 4.5 K lower near 30\degx S
\citep{Conrath1991urabook}.  In the 200 -- 1000 mbar range latitudinal
excursions are within 1--1.5 K. Thus it appears that in the most
important region of the atmosphere for our applications, the thermal
structure was not strongly variable with latitude, at least in 1986.
Models of seasonal temperature variations on Uranus by
\cite{Friedson1987} suggest that the effective temperature variation
at low latitudes will be extremely small, only 0.2 K peak-to-peak at
the equator, increasing to a still relatively small 2.5 K at the
poles.  Thus, it is plausible to analyze observations during the 2012
-- 2015 period with thermal profiles obtained as far back as 1986,
even though they are local, but probably more appropriate to use
thermal structures derived from observations in 2007, averaging over a
wide range of latitudes, such as those inferred by
\cite{Orton2014uratemp} from nearly disk-integrated spectral
observations.

\begin{figure*}[!htb]\centering
\includegraphics[width=6.2in]{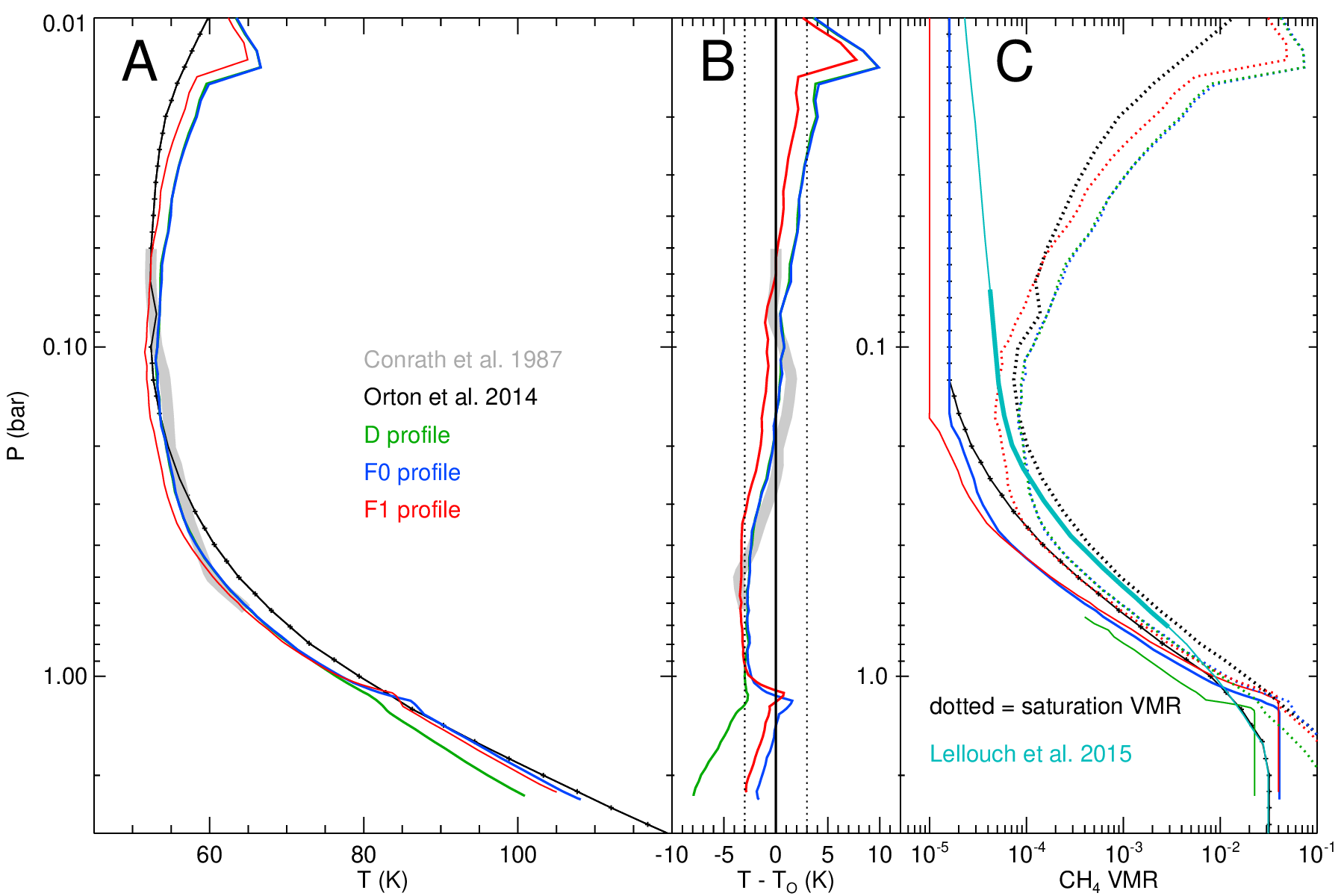}
%SOURCE:  plot_orton2014.pro revised, documentation on p42 Uranus Log I
% Revision includes T-difference plot and fewer profiles
% (further documentation on pages 140-148, Uranus Log G)
\caption{{\bf A:} Alternate Uranus T(P) profiles. F0 and D profiles
  were derived from radio occultation measurements \citep{Lindal1987}
  assuming a helium VMR of 0.152.  The F1 profile was also derived
  from radio occultation measurements, but using a lower helium VMR
  0.1155, following \cite{Sro2011occult}. The Voyager IRIS profile of
  \cite{Conrath1987JGR} (thick gray curve) is in best agreement with
  the F0, D, and F1 profiles.  The \cite{Orton2014uratemp} profile
  (solid black curve), is based on Spitzer Space Telescope spectral
  observations. {\bf B:} Temperatures relative to the
  \cite{Orton2014uratemp} profile, which strongly disagrees with the
  radio occultation profiles in 500 mbar -- 1 bar region, where it is
  3 K warmer. {\bf C:} Methane VMR profiles corresponding to
  temperature profiles shown in A, using the same line styles, with an
  additional estimated profile by \cite{Lellouch2015}, based on
  Herschel far-IR and sub-mm observations.}\label{Fig:tch4}
\end{figure*}

Sample thermal and methane profiles are displayed in
Fig.\ \ref{Fig:tch4}. The profile of \cite{Orton2014uratemp},
hereafter referred to as {\bf O14}, is based on nearly
disk-integrated spectral observations obtained with the Spitzer Space
Telescope near the Uranus equinox in 2007.  Among the occultation
profile sets, it is only those with high methane VMR values that
provide decent agreement with the O14 deep temperature structure,
but none of the occultation profiles are compatible with the O14
profile in the 0.30 -- 1.0 bar range.  One might argue that if radio
occultation results agree with O14 at 100 mb and at pressures beyond
1 bar, then the disagreement in temperatures at intermediate pressure
levels is more likely due to an error in the Orton et al. profile
because that profile is inferred from different spectrometers in
different spectral regions that sample different altitudes, which
might suffer from differences in calibrations, while the radio
occultation uses the same measurement (the frequency of a radio
signal) throughout the pressure range. It seems more likely that the
errors in the radio profile would be in the altitude scale or in
offsets due to uncertain He/H$_2$ ratios, rather than varying in the
way the differences between the radio and Orton et al. profile do. A
similar argument might be made in favor of the \cite{Conrath1987JGR}
profile over the O14 profile because the former is based on
interferometric measurements using the same instrument over the entire
spectral range.  And the former profile is in good agreement with the
occultation-based profiles in the 300--600 mbar range, where the
latter is not.  On the other hand, the Orton et al.  profile allows
higher \chf mixing ratios without saturation in the 0.3 -- 1 bar
region (Fig.\ \ref{Fig:tch4}) and are thus more compatible with the
recent \cite{Lellouch2015} \chf VMR profile derived from Herschel
far-IR and sub mm observations.

The methane VMR in the stratosphere was estimated to be no greater
than $10^{-5}$ by \cite{Orton1987spectra}.  A best fit estimate for
the tropopause value of the methane VMR, based on more recent Spitzer
observations, is $(1.6^{+0.2}_{-0.1})\times 10^{-5}$ according to
Fig. 4 of \cite{Orton2014uracomp}, which is the value we assumed
here in
deriving the new F0 profile. However, the even more recent
\cite{Lellouch2015} result is three times larger.  In terms of
relative humidity (ratio of vapor pressure to saturation vapor
pressure) these stratospheric mixing ratios correspond to humidities
of 25\% and 75\% at the Orton et al. tropopause temperature of 52.4 K.
The F profile of \cite{Lindal1987} was derived assuming a constant
methane relative humidity of 53\% above the cloud tops and a constant
stratospheric mixing ratio equal to the tropopause value. In deriving
the F0 profile we used linear-in-altitude interpolation of the methane
humidity values between the cloud top and tropopause.  The F1 profile
of \cite{Sro2011occult} followed the same procedure except that the
value of the tropopause mixing ratio was taken to be the earlier upper
bound of $10^{-5}$ and the He VMR was taken to be 0.1155 instead of
0.152.  The lower He VMR was chosen to produce a saturated methane
mixing ratio inside the cloud layer.  

Our analysis for this paper is primarily based on the
O14 thermal profile, although we did consider the
effects of using these alternative profiles.  From trial retrievals we
found no significant difference in the absolute mixing ratios inferred
for different thermal profiles.  The main differences occurred when
these mixing ratios were converted to relative humidities.  We often
found supersaturation above the condensation level for the cooler
occultation profiles, whereas the same mixing ratios did not lead to
supersaturation for the warmer O14 profile, or for the F0
profile.  The F0 profile would also have been a decent baseline
choice, as long as we did not also use the F0 methane profile, and
instead let the STIS spectra constrain the methane mixing ratios without
 regard to occultation consistency.

\subsection{Vertical Profiles of Methane}

\begin{figure*}\centering
\hspace{-0.15in}
\includegraphics[width=3.2in]{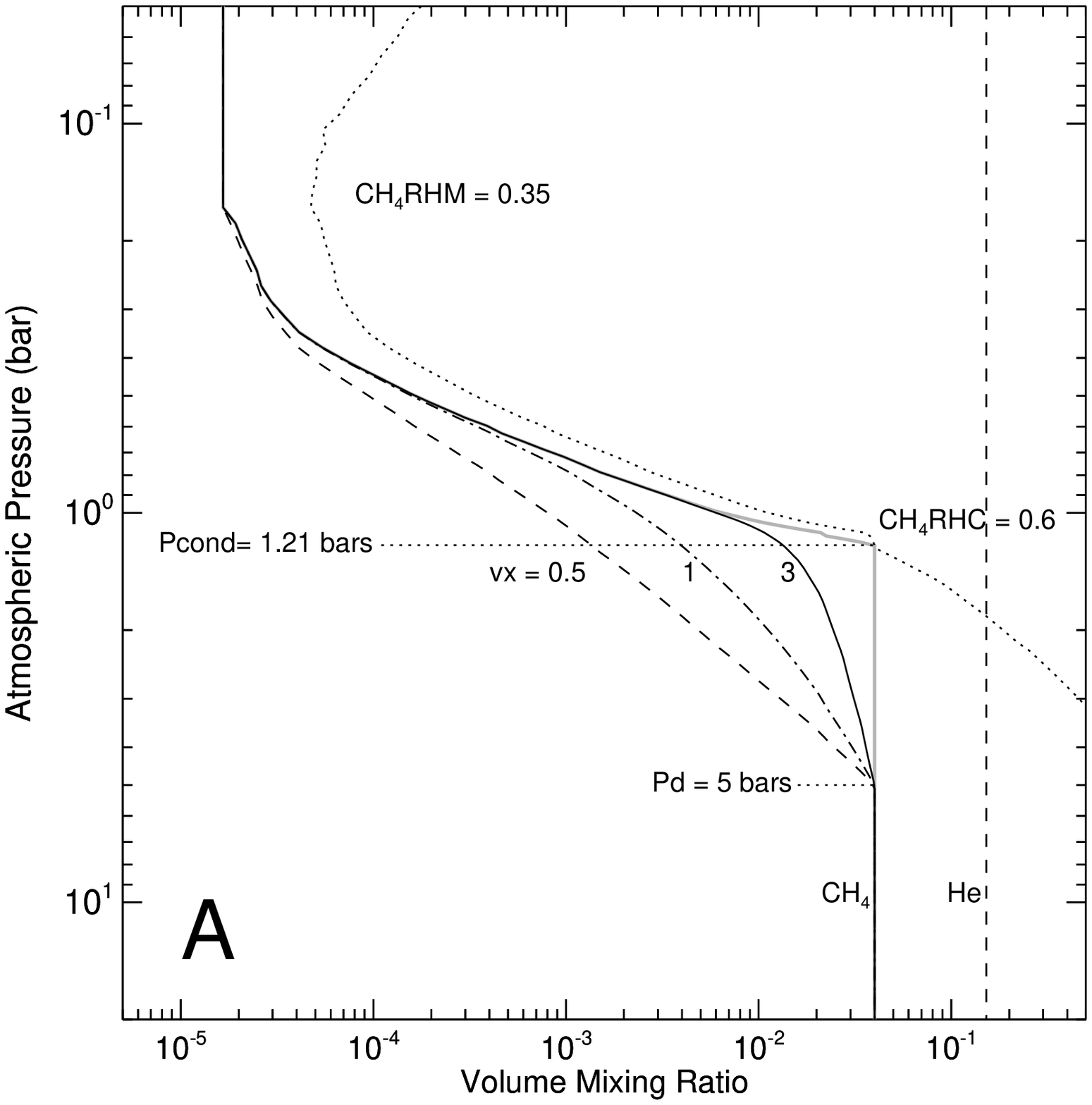}
\hspace{-0.15in}
\includegraphics[width=3.2in]{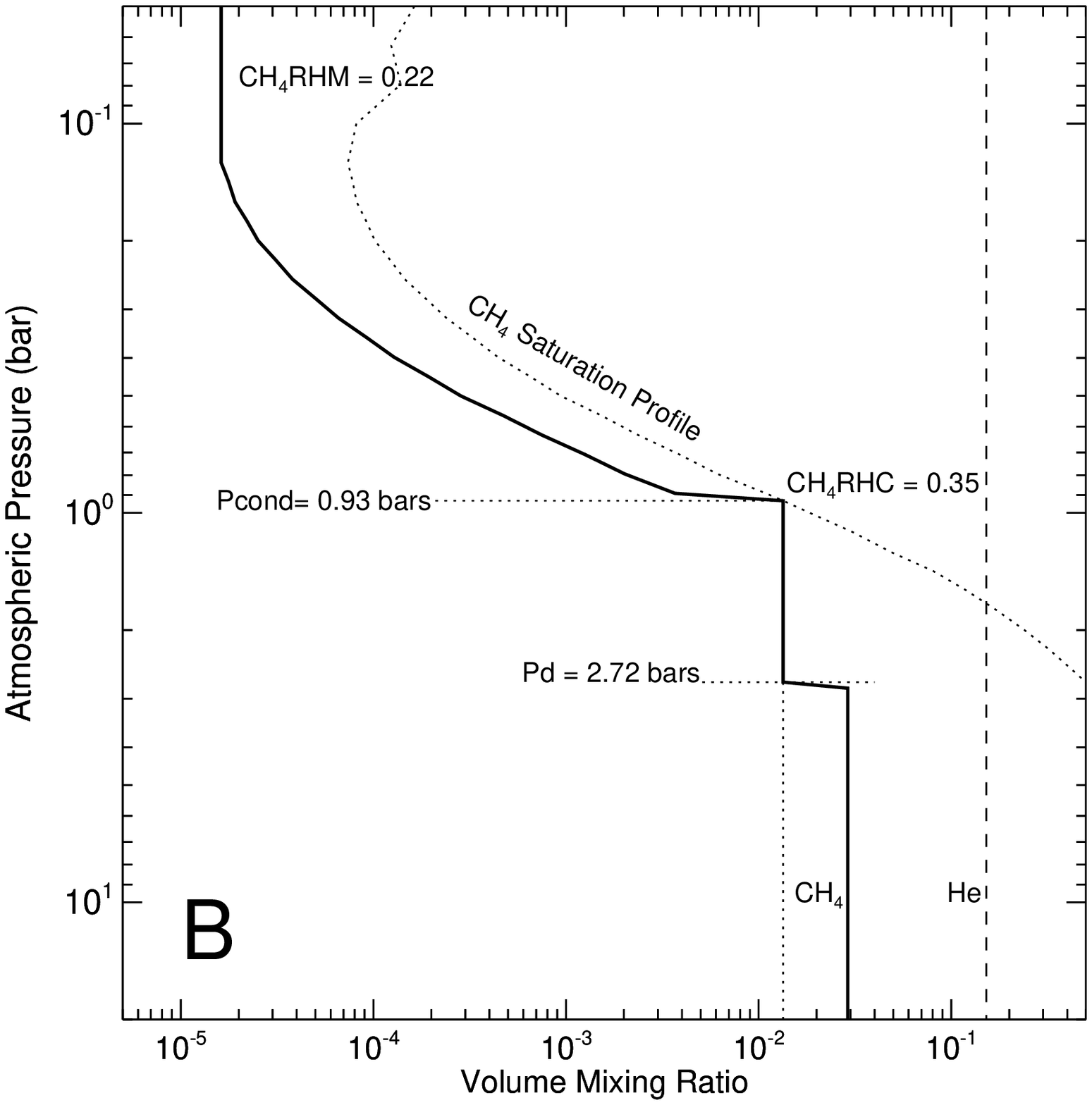}
%SOURCE plotcldgasura_nonsub.pro on puck in /home/sro/uranus/ir_ms_ura
% documented on Uranus Log I, page 45.
\caption{{\bf A:} Sample ``descended gas'' methane profiles with $pd$ =
  5 bars and $vx=$ 0.5 (dashed), 1 (dot-dash), and 3 (solid). The
  starting profile before descent is shown in solid gray and is based
  on the F1 T(P) profile with methane constrained by its deep mixing
  ratio and the humidities above the condensation level (CH$_4$RHC) and
  at the tropopause (CH$_4$RHM), with linear in log P
  interpolation between these levels.  {\bf B:} Sample step-function vertical
  methane profile using the T(P) profile of \cite{Orton2014uratemp} to
  define the saturation vapor pressure profile (dotted curve). This
  particular example fits the 2015 spectra at 40\degx N. See text for
  further explanation.}\label{Fig:descend}
%SOURCE: plotcldgasura_driver.pro was used, documented on page 26-27, Uranus Log H.
\end{figure*}

In our prior analysis the vertical profile of methane was generally coupled to the
vertical temperature profile so that the vertical variation of atmospheric refractivity
was consistent with occultation measurement of refractivity.  In our current analysis
we uncoupled temperature and methane profiles because of questions
raised about the reliability of the occultation results, especially by the
new temperature structure results of \cite{Orton2014uratemp}, and by the
new methane measurements of \cite{Lellouch2015}, which imply supersaturation
for the cooler temperature profiles obtained from occultation analysis.
Another difference in our current analysis is that we included the
parameters describing the methane vertical distribution as adjustable
parameters in the fitting process.
We first
carry out fits of spectra at different latitudes assuming a vertically
invariant (but adjustable) methane VMR ($\alpha_0$) below the condensation level. Slightly
above the condensation level we fit a relative humidity $rh_c$, and assume a
minimum relative humidity of $rh_m$ at the tropopause between 20\% and
60\%, which yields mixing ratios within a factor of two of
\cite{Orton2014uratemp}). The high end of this range is in better
agreement with \cite{Lellouch2015}.  The STIS spectra themselves are not very
sensitive to the exact value at the tropopause, as evident from
Fig.\ \ref{Fig:pendepth}.  Between the tropopause and the condensation
level we interpolate relative humidity between $rh_c$ and $rh_m$ using
the function
\begin{eqnarray}
rh(P) = rh_m + (rh_c - rh_m)\times \\ \nonumber
 \big[ 1 - \log(P_c/P)/\log(P_c/P_m)\big],
\label{Eq:rh}
\end{eqnarray}
where $P_c$ is the pressure at which \chf condensation would occur for the
given thermal profile and a given uniform deep methane VMR, and $P_m$
is the pressure at which the relative humidity attains its minimum
value near the tropopause. Given a deep methane VMR ($\alpha_0$) and a
temperature profile from which a condensation pressure can be defined,
Eq.\ \ref{Eq:rh} then defines a methane VMR as a function of pressure for $P < P_c$,
denoted by $\alpha(P)$.  That profile is generated prior to
application of the \cite{Sro2011occult} ``descended profile'' function
in which the initial mixing ratio profile $\alpha (P)$ is dropped down
to increased pressure levels $P'(\alpha)$ using the equation
\begin{eqnarray}
 P' = P\times [1 + (\alpha(P)/\alpha_{0})^{vx}(P_{d}/P_{c}-1)] \\ \nonumber
      \mathrm{for}\quad P_{tr} <P<P_{d},\label{Eq:deplete}
\end{eqnarray}
where $P_{d}$ is the pressure depth at which the revised mixing ratio
$\alpha'(P)=\alpha(P')$ equals the uniform deep mixing ratio
$\alpha_{0}$, $P_{c}$ is the methane condensation pressure before
methane depletion, $P_{tr}$ is the tropopause pressure (100 mb), and
the exponent $vx$ controls the shape of the profile between 100 mb and
$P_{d}$.  Sample plots of descended profiles are displayed in Fig.\ \ref{Fig:descend}.
The profiles with $vx=1$ are similar in
form to those adopted by \cite{Kark2011nep}.  Our prior analysis obtained
the best fits with $vx$=3, while our current analysis obtains a latitude
dependent value ranging from $\ge$ 9 at low latitudes to 2.4$\pm$0.7 at 70\deg N.

Fig.\ \ref{Fig:descend}B displays an alternative step-function depletion model in which
the methane mixing ratio decreases from the deep value to a lower vertically uniform
value beginning at pressure $P_d$ and continuing upward until the condensation
level is reached for that mixing ratio. This is parameterized by four variables:
the deep mixing ratio $\alpha_{0}$, the pressure break-point  $P_d$ ,
 the upper mixing ratio  $\alpha_1$, and
the relative humidity immediately above the condensation level $rh_c$.  The parameters
of all three of these vertical profile models are summarized in Table\ \ref{Tbl:gasparams}.

\begin{table*}\centering
\caption{Methane vertical profile model parameters.}
\vspace{0.15in}
\begin{tabular}{c | l l }
\hline
Model Type &  Parameter (description)                  &  Value        \\ 
\hline
              &   $\alpha_0$ (deep mixing ratio)    & adjustable \\
              &   $P_c$ (condensation pressure)  & derived from  $\alpha_0$, $P(T)$ profile\\
uniform deep  &   $P_t$ (tropopause pressure)   & derived from  $P(T)$ profile\\
              &   $rh_c$ (relative humidity at $0.95\times P_c$) & adjustable\\
              &   $rh_m$ (relative humidity at $P_t$) & adjustable, or from \cite{Orton2014uratemp}\\
 \hline
              &    $\alpha_0$ (mixing ratio for $P>P_d$)    & adjustable \\
              &    $\alpha_1$ (mixing ratio for $P_c< P<P_d$)    & adjustable \\
              &    $P_d$ (transition pressure) & adjustable \\
\raisebox{1.5ex}[0pt]{2-step uniform}  &   $P_c$ (condensation pressure)  & derived from  $\alpha_1$, $P(T)$ profile\\
              &   $rh_c$ (relative humidity at $0.95\times P_c$) & adjustable\\
              &   $rh_m$ (relative humidity at $P_t$) & fixed at various values\\
 \hline
              &    $\alpha_0$ (mixing ratio for $P>P_d$)    & adjustable \\
               &    $P_d$ (transition pressure) & adjustable \\
\raisebox{1.5ex}[0pt]{descended} &    $vx$ (exponent of shape function) & adjustable\\
              &   $\alpha'(P)$ (descended VMR profile) & derived by inverting Eq. \ref{Eq:deplete}\\
              &   $rh_c$ (relative humidity at $0.95\times P_c$) & adjustable\\
              &   $rh_m$ (relative humidity at $P_t$) & fixed at various values\\
 \hline
\end{tabular}\label{Tbl:gasparams}
\parbox[]{5.8in}{\vspace{0.05in}NOTE: we assumed the same mixing ratio for $P<P_t$ as for $P=P_t$.  For the 1 and
2-step uniform models $rh(P)$ for $P_t<P<0.95\times P_c$ is obtained from Eq. \ref{Eq:rh}.}
\end{table*}

\subsection{Cloud models}

\subsubsection{Prior cloud models}

\begin{figure*}[!htb]\centering
\includegraphics[width=6.2in]{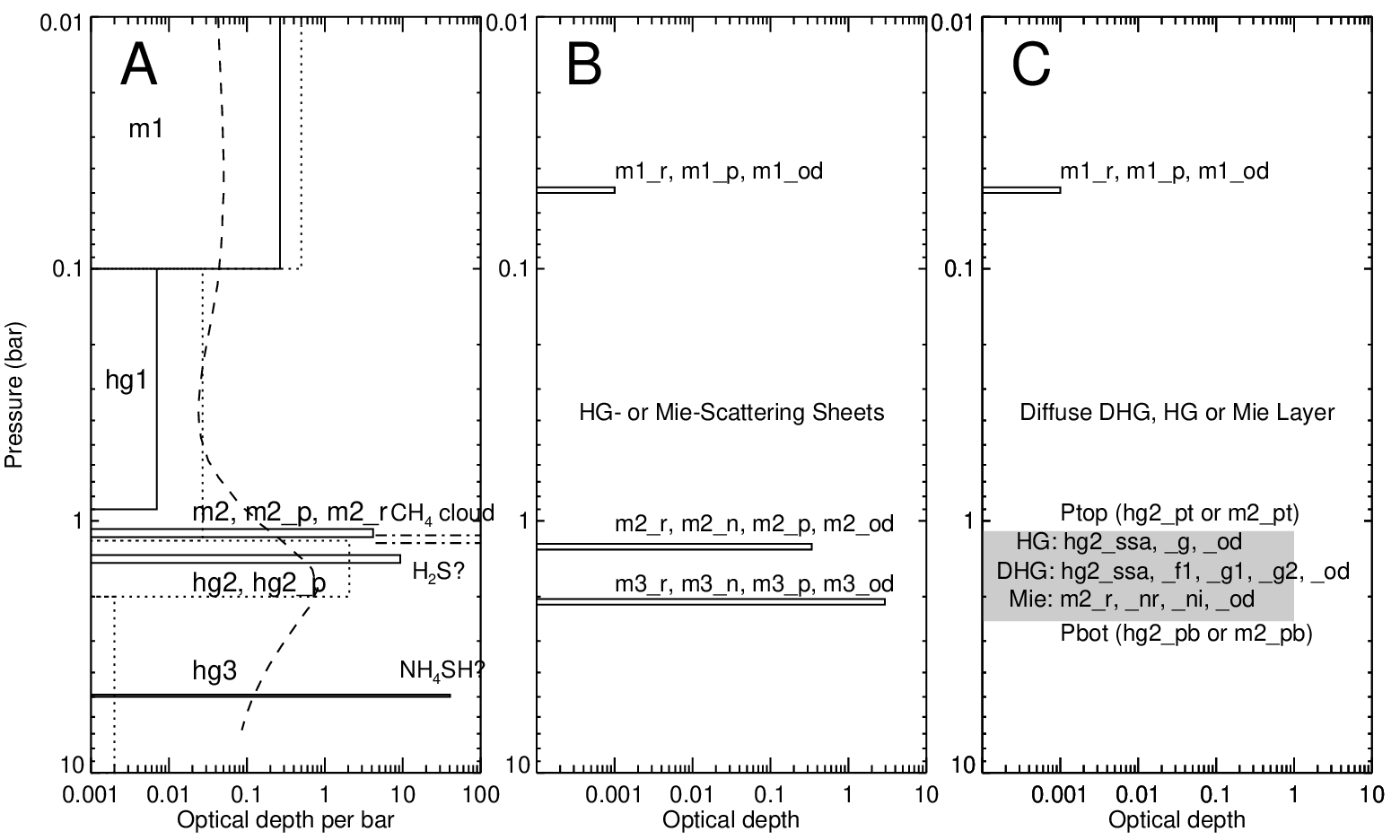}
%SOURCE: plot by plot_diffuse_compact_model_structures.pro
% Updated on 29 Dec 2017, 
% previously documented on page 38, Log L.  Previous version documented
% on page 46 of Uranus Log I
\caption{{\bf A:} Comparison of the KT2009 model (dotted) with the
  similar but more complex 5-layer model used by \cite{Sro2014stis}, which 
replaced two diffuse layers with 3 compact layers. {\bf B:} A
  preliminary simplified model with three compact layers, mostly
  defined by the two lower layers. This model was constructed
with the possibility in mind that $m2$ might be formed from
  methane and $m3$ from \htsx. {\bf C:} Our
baseline simplest model in which the tropospheric cloud is uniformly
mixed between top and bottom pressures and has the same particle
properties throughout layer 2. }
\label{Fig:cloudmodels}
\end{figure*}

Our prior analysis used an overly complex five-layer model that was
based on the KT2009 four-layer model, with the main difference being
replacement of their main Henyey-Greenstein (HG) layer with two
layers, the higher of which was a Mie-scattering layer that was a
putative methane condensation cloud, as illustrated in
Fig.\ \ref{Fig:cloudmodels}A.  In this model the scattering properties
of the three remaining Henyey-Greenstein layers were taken from
KT2009, with no adjustment to improve fit quality.  This model was
partly based on parameters tuned to fit the 2002 STIS observations,
taken 13 years before our most recent ones, and thus it was
appropriate to reconsider the aerosol structure.  In addition, the
five-layer model actually has too many parameters to meaningfully
constrain independently with STIS observations. Our starting point
consisted of three Mie-scattering sheet clouds, as illustrated in
Fig.\ \ref{Fig:cloudmodels}B.  But we obtained fits of comparable
quality using for the tropospheric aerosols a simpler single cloud of
uniform scattering properties and uniformly mixed with the gas between
top and bottom boundaries, as in in Fig.\ \ref{Fig:cloudmodels}C.  As
a result, that simpler model became our baseline model.
\cite{Tice2013}, \cite{Irwin2015reanalysis}, and \cite{DeKleer2015}
were successful in using a similar model structure to fit near-IR spectra.

\subsubsection{Simplified Mie-scattering aerosol models}

We have two options for our 2-cloud baseline model displayed in
Fig.\ \ref{Fig:cloudmodels}C.  Both options use a sheet cloud of
spherical Mie-scattering particles to approximate the stratospheric
haze contribution.  The parameters defining a sheet cloud of spherical
particles are the size distribution of particles, their refractive
index, effective pressure, and optical depth.  We chose the
\cite{Hansen1971JAScircpol} gamma distribution, characterized by an
effective radius and effective dimensionless variance.  As spectra are
not very sensitive to the variance, we chose an arbitrary value of
0.1.  Based on preliminary fits we chose a particle size of 0.06
\mumx. Other researchers have selected a slightly larger size of 0.1
\mumx. We also found generally low sensitivity to the effective
pressure as long as it was sufficiently low.  We thus chose a somewhat
arbitrary value of 50 mbar, putting the haze above the tropopause.  We
made an arbitrary choice of 1.4 for the layer's refractive index. At
wavelengths shorter than our lower limit, the haze undoubtedly
provides some absorption, as noted by KT2009, but we did not need to
include stratospheric haze absorption to model its effects in our
spectral range.  Usually the only adjustable parameter for this layer
we took to be the optical depth. Test calculations showed that an
extended haze spanning pressures from 1 mbar to 200 mbar worked almost
as well as our sheet cloud model.  It did produce a slightly larger
\chisq but using a diffuse stratospheric haze model had little effect
on derived parameter values.  The optical depth of the haze only
increased by 2\%, and the fractional changes in all the other fitted
parameters were less than 0.4\%, putting these changes well below their
estimated uncertainties.  In any case, our aim with the haze model was
to account for its spectral effects, not to accurately describe the
physical characteristics of the haze itself.

For a tropospheric sheet cloud of conservative particles the fitted
parameters would be particle size, real refractive index, effective
pressure, and optical depth (4 parameters).  For a pair of
tropospheric sheet clouds, as in Fig.\ \ref{Fig:cloudmodels}B, there
would be 8 parameters to constrain.  Assuming both layers had the same
scattering properties, that would drop the number of fitted parameters
to 6.  Replacing the pair of sheet clouds with a single diffuse layer
with uniform scattering properties, as in
Fig.\ \ref{Fig:cloudmodels}C, reduces the number of optical depths to
one, but keeps the number of pressure parameters to two, this time
used for top and bottom boundaries, yielding a new total of 5
parameters for the tropospheric aerosols.  Instead of fitting top and
bottom pressures to control the vertical distribution, \cite{Tice2013}
chose to fit the base pressure and the particle to gas scale height
ratio.  Which approach is more realistic remains to be determined.  At this
point we have a nominal total of 6 adjustable parameters to describe
our aerosol particles, one for the stratospheric sheet, and five for
the vertically extended tropospheric layer.  These are named $m1\_r$,
$m2\_r$, $m2\_pb$, $m2\_pt$, $m2\_od$, and $m2\_nr$, where the
characters preceding the number indicate the type of particle ($m$
denotes Mie scattering spherical particle), the number is the layer
number, and the type of parameter is indicated after the underscore
($r$ for radius, $pb$ for bottom pressure, $pt$ for top pressure, $od$
for optical depth, and $nr$ for real refractive index).

For these spherical (Mie-scattering) particles, wavelength dependent
properties are controlled by particle size and refractive index. Even
if both of these are wavelength-independent, scattering cross section
(or optical depth) and phase function do have a wavelength dependence
because of the physical interaction of light with spherical
particles. Where our chosen parameters fail to provide sufficient
wavelength dependence, we will also add another parameter, namely the
imaginary refractive index $m2\_ni$, which will in general be
wavelength dependent, and have its main influence over the
single-scattering albedo $\varpi$. We also have between two and four
parameters chosen to constrain the vertical methane profile, yielding
generally between eight and ten total parameters to constrain by the
non-linear regression routine.

\subsubsection{Non-spherical aerosol models.}

Because the particles in the atmosphere of Uranus are thought to be
mostly solid particles, they are unlikely to be perfect spheres, and
thus we also considered a more generalized description of their
scattering properties. To investigate non-spherical scattering, we
employed the commonly used double Henyey-Greenstein phase function, in
which three generally wavelength-dependent parameters need to be
defined: the scattering asymmetry parameter ($g_1 > 0$) of a mainly
forward scattering term, the asymmetry parameter ($g_2 < 0$) of the
mainly backscattering term, and their respective fractional weights
($f_1$ and $1-f_1$ respectively).  An additional fourth parameter is
the single-scattering albedo ($\varpi$), which might also be
wavelength dependent.  The double Henyey-Greenstein (DHG) phase function is
given by \begin{eqnarray} P(\theta)=\! f_1\!\times \! (1 - g_1^2)/(1 + g_1^2-2
  g_1\cos(\theta))^{3/2}\quad \\ \nonumber
 + (1-f_1)\times
  (1-g_2^2)/(1+g_2^2-2g_2\cos(\theta))^{3/2},\label{Eq:dhg}
\end{eqnarray}
where $\theta$ is the scattering angle. KT2009 modeled their results
assuming $g_1=0.7$ and $g_2=-0.3$ and used a wavelength-dependent $f_1$
to adjust the phase function of their tropospheric cloud layers so
that they would appear relatively bright enough at short wavelengths.
For haze layers composed of fractal aggregate particles, as inferred
to exist in Titan's atmosphere, one would expect both phase function
and optical depth to be wavelength dependent, and modeling the fractal
aggregate phase function variation with double Henyey-Greenstein
functions would require wavelength dependence in $g_1$ and $g_2$ as
well as $f_1$, judging from the aggregate models of \cite{Rannou1999}.
An alternate approach to matching observed spectra with spherical
particles is to make the particles absorbing at longer wavelengths and
conservative at shorter wavelengths.  

The simplest DHG particle is just an HG particle characterized by an
asymmetry parameter g, and a single scattering albedo $\varpi$, and
for a limited spectral range, a wavelength dependence parameter, which
can be taken as a linear slope in optical depth, which amounts to
three parameters ($g$, $\varpi$, d$\tau$/d$\lambda$).  This is the same
number needed to characterize scattering by a Mie particle ($r$, $nr$,
$ni$).  However, if we use a full DHG formulation, then there are five
particle parameters to constrain ($f_1$, $g_1$, $g_2$, $\varpi$, and
d$\tau$/d$\lambda$).

An alternative way to produce the wavelength dependence of a spherical particle
without its potentially complex phase function, containing features like
glories and rainbows, which would not be seen in randomly oriented solid particles, is to 
follow the procedure of \cite{Irwin2015reanalysis}.  They computed scattering
properties of spherical particles to determine the wavelength dependence 
of the scattering cross section, but fit the phase function to a double HG
function to smooth out the spherical particle features.  The refractive
index they assumed was the typical value of 1.40 at short wavelengths, but was
modified by the Kramers-Kronig relation to be consistent with the fitted
variation of the imaginary index.  Whether there are any cases of
 randomly oriented solid particles actually displaying these modified
Mie scattering characteristics remains to be determined.

\subsubsection{Fractal aggregate particles}

For those layers that are produced by photochemistry, it is also plausible
that the hazes might consist of fractal aggregates, which have phase
functions that are strongly peaked in the forward direction, but are
shaped at other angles by the scattering properties of the monomers
from which the aggregates are assembled.  It is a convenience to
assume identical monomers, and to parameterize the aggregate
scattering in terms of the number of monomers, the fractal dimension
of the aggregate, and the potentially wavelength dependent real and
imaginary refractive index of the monomers \citep{Rannou1999}.  If the
refractive index were wavelength independent, this would require
fitting of potentially five parameters (rm, Nm, dim, nr, ni), the
same number as for the most general DHG particle.  Assuming
ni = 0, rm = fixed size, this would require fitting just three parameters (Nm, dim, nr),
a tractable task, but one which we have not so far implemented in our fitting code.

%An example of fractal aggregate scattering properties as a function
%of wavelength is provided in Fig.\ \ref{Fig:aggregate} for an aggregate
%of 100 monomers 0.05 \mum in radius with a real refractive index
%of 1.4, and for a fractal dimension of 2.01. 
%Note in the right panel of Fig.\ \ref{Fig:aggregate} that 

To better understand the wavelength dependent properties of aggregates
we made some sample calculations. We first considered an aggregate of
100 monomers 0.05 \mum in radius with a real refractive index of 1.4,
and a fractal dimension of 2.01. These particles have the mass of 
a particle of 0.23 \mum in radius.  This provides a physical connection
between monomer parameters and the wavelength dependent aggregate
phase function and scattering and absorption cross sections. We found
that it is possible to at least roughly characterize the fractal
aggregate phase functions with double Henyey-Greenstein functions,
although this provides no physical connection to a wavelength
dependent cross-section and single-scattering albedo unless DHG fits
to the fractal aggregates are done for each wavelength.  We found for
this example that the backscatter phase function amplitude declines as
wavelength decreases, opposite to the model of KT2009, while the
scattering efficiency (and thus optical depth) has a strong wavelength
dependence, also contradicting the KT2009 model, which assumed
wavelength independence for optical depth.  By increasing the number
of monomers from 100 to 500 (mass equivalent to a particle 0.4 \mum in
radius), the asymmetry parameter can be made
relatively flat over the 0.5 \mum to 1 \mum range, but the strong
wavelength dependence of the extinction efficiency remains, suggesting
that it is optical depth dependence on wavelength that offers the best
lever for adjusting model I/F spectra, rather than the phase function.
It is also clear that no spherical particle can simultaneously
reproduce both the fractal phase function and scattering efficiency and their
wavelength dependencies.

\subsubsection{Photochemical vs. condensation cloud models}

According to \cite{Tomasko2005}, the dominant aerosol in Titan's 
atmosphere is a deep photochemical haze
extending from at least 150 km all the way to the surface, with a smoothly
increasing optical depth reaching a total vertical optical depth of 4-5 at 531 nm,
with no evident layers of significant concentration that might suggest
condensation clouds  (only a thin
layer of 0.001 optical depths was seen at 21 km).  KT2009
argued for a similar origin for the dominant aerosols on Uranus. The
fact that the main aerosol opacity on Uranus is found somewhat deeper than would be expected
for a methane condensation cloud certainly suggests that the aerosols in the 1.2-2 bar
region are either H$_2$S, which might condense as deep as the 5-bar level or higher,
or some photochemical product, or both. And residual haze particles might serve
as condensation nuclei for \htsx. This putative deeper photochemical haze is apparently not
the haze modeled by \cite{Rages1991}, which is produced at very high
levels of the atmosphere and has UV absorbing properties that do
not seem to be characteristic of the deeper haze.  In fact, it is
not clear that there is enough penetration of UV light to the 1.2-bar
level to produce significant photochemical production of any haze material.
Ignoring the issue of production rate, the main arguments for a photochemical haze  
are based on the following expected characteristics of such a haze: (1)
a strong north-south asymmetry before the 2007 equinox, with more haze in the
south compared to the north; (2) a declining
haze near the south pole as solar insolation decreased towards the 2007 equinox (this
assumes that the lag between production and insolation is only a few years); (3)
an increasing haze near the north pole as it starts to receive sunlight after
the 2007 equinox; (4)  slow changes because the sub-solar latitude changes
by only 4\degx/year; (5)  a time lag with respect to solar insolation
because haze particles accumulate after production but do not exist at the
beginning of production (equilibrium would be reached when the fall rate
of particles equals the production rate). All five characteristics are
indeed observed for Uranus, at least qualitatively, while these changes
are not obvious expectations for condensation clouds.  

Given our preferred explanation for the polar methane depletion,
namely that there is a downwelling flow from above the methane
condensation level, the mixing ratio of methane would be too low to
allow any methane condensation in the polar region at pressures
greater than about 1 bar.  Thus it is challenging to explain the
increase in haze in the polar region after equinox as an increase in
the mass of condensed particles in that region.  One possibility is
that the clouds are formed below the region of downwelling methane,
and instead in a region of upwelling H$_2$S.  But microwave
observations suggest that the polar subsidence extends deeper than the
deepest aerosol layers that we detect, which would seem to inhibit all
cloud formation by condensation.  Another possibility is that
meridional transport of condensed H$_2$S particles at the observed
pressures, if it occurred at a sufficiently high rate, could resupply
the falling particles.

One odd feature of the putative tropospheric
photochemical haze in the KT2009 model, is the concentration of optical depth within the
1.2-2 bar region, which has about 2 optical depths per bar, which
far exceeds the density of any of the other four layers in the KT2009 model.
A possible explanation of this effect is that the photochemical aerosols
absorb significant quantities of methane, as appears to have occurred
in Titan's atmosphere \citep{Tomasko2008}, growing larger and also diluting the UV absorption
of the particles originating from the stratosphere. The bottom boundary of this
region of enhanced opacity may be where the methane that was adsorbed into the
photochemical aerosols is released and evaporated. A problem with this concept is that
it is also hard to explain the growth of the haze following equinox in a region of greatly
reduced methane abundance.

Another mystery is why the methane mixing ratio is so stable over
time, if methane is involved in fattening the photochemical particles
that have a time varying production.  This might just be due to the
fact that it takes very small amounts of condensed material to produce
a significant optical depth of particulates.  The rate limiting factor
might be the arrival rate of UV photons, rather than the amount of
methane either as the parent molecule of the photochemical chain of
events in the stratosphere, or as the adsorbed material needed to enhance the optical
depth of the haze particles in the troposphere.  We can hope that some clues can be
gleaned from the characteristics of the time dependence and latitude
dependence observed in the model parameters.

\begin{table*}\centering
\caption{Summary of 2 layer cloud model parameters}
\vspace{0.15in}
\begin{tabular}{c | l l l }
\hline
Layer & Description       &  Parameter (function)                  &  Value        \\ 
\hline
  &Stratospheric haze  &  $m1\_p$ (bottom pressure)   & fixed at 60 mb \\
  &of Mie particles    &  $m1\_r$ (particle radius)    & fixed at 0.06 \mum \\
1 &with gamma size     &  $m1\_b$ (variance)           & fixed at 0.1 \\
  &distribution (m1)   &  $n1$ (refractive index)      & nr=1.4, ni=0\\
  &                    &  $m1\_od$ (optical depth)     & adjustable \\
\hline
\hline
  &                   & $m2\_pt$ (top pressure)            & adjustable \\
  &Upper tropospheric & $m2\_pb$ (bottom pressure)         & adjustable \\
  &haze layer of Mie  &  $m2\_r$ (particle radius)         & adjustable \\
  &particles (m2)     &   $m2\_b$ (variance)               & fixed at 0.1 \\
  &                   &  $m2\_nr$ (real refractive index)  & adjustable\\
  &                   &  $m2\_ni$ (imag. refractive index) & adjustable\\
2 &                   &  $m2\_od$ (optical depth)          & adjustable \\
\cline{2-4}
  &                             & $hg2\_pt$ (top pressure)  & adjustable \\
  &Alternate upper trop. & $hg2\_pb$ (bottom pressure)  & adjustable \\
  &haze of HG particles  & $\varpi_2 (\lambda)$ (single-scatt. albedo) & adjustable or fixed\\
  &(hg2)                 &  g (defines HG phase func.) & adjustable\\
   &                     & $hg2\_od$ (optical depth) & adjustable\\
   &                     & $hg2\_kod$ (optical depth slope) & adjustable\\
\cline{2-4}
  & Second alternate            & $hg2\_pt$ (top pressure)  & adjustable \\
  &upper tropospheric           & $hg2\_pb$ (bottom pressure)  & adjustable \\
  &haze of double-HG particles  & $\varpi_2 (\lambda)$ (single-scatt. albedo) & adjustable or fixed\\
  &(hg2)                        & $P_2(\theta,\lambda)$ (phase function) & DHG function of KT2009\\
   &                            & $hg2\_od$ (optical depth) & adjustable\\
\hline
\end{tabular}\label{Tbl:extendedmodel}
%% \parbox[]{6.2in}{NOTE: KT2009 equations defining wavelength dependent
%%   parameters are reproduced in the analysis supplement file. Usually
%%   $hg1\_odpb$ was found to be too small to bother including in our
%%   fits.}
\end{table*}

\subsection{Fitting procedures}

To avoid errors in our approximations of Raman scattering and the
effects of polarization on reflected intensity, we did not
fit wavelengths less than 0.54 \mumx. An upper limit of 0.95 \mum
was selected because of significant uncertainty in
characterization of noise at longer wavelengths. To increase S/N
without obscuring key spectral features, we smoothed the STIS spectra
to a FWHM value of 2.88 nm.  We chose three spectral samples of the
CTL variation, at view and solar zenith angle cosines of 0.3, 0.5, and
0.7, which are fit simultaneously.  In its simplest form our
multi-layer Mie model has three adjustable parameters per layer
(pressure, particle radius, and optical depth). Each layer is assumed
to be a sheet cloud of insignificant vertical thickness.  

We also fit adjustable gas parameters, illustrated in
Fig.\ \ref{Fig:descend} and described in Table\ \ref{Tbl:gasparams}.
For the vertically uniform mixing ratio model (up to the \chf
condensation level) we have two adjustable parameters: the deep
methane volume mixing ratio and the relative methane humidity above
the condensation level (methane relative humidity is the ratio of its partial
pressure to its saturation pressure). For the 2-layer
Mie-scattering aerosol model, this yields a total of 8-9 adjustable
parameters (the top Mie layer has a fixed pressure and often a fixed
particle size as well, with optical depth remaining as the only
adjustable parameter because the others are so poorly constrained).
For the step-function 2-mixing ratio gas model, we use three
adjustable gas parameters: the break point pressure, and the upper
\chf mixing ratio, and the relative methane humidity above the
condensation level, for a total of nine adjustable parameters.  The
third parameterization of the methane distribution, the descended gas
model, also uses three adjustable parameters: the pressure limit of
the descent, the methane relative humidity above the condensation
level (prior to descent), and the shape exponent $vx$.

We used a modified Levenberg-Marquardt non-linear fitting algorithm
\citep{Sro2010iso} to adjust the fitted parameters to minimize
$\chi^2$ and to estimate uncertainties in the fitted
parameters. Evaluation of \chisq requires an estimate of the expected
difference between a model and the observations due to the
uncertainties in both. We used a relatively complex noise
model following \cite{Sro2011occult}, which combined measurement noise
(estimated from comparison of individual measurements with smoothed
values), modeling errors of 1\%, relative calibration errors of 1\%
(larger absolute calibration errors were treated as scale factors),
and effects of methane absorption coefficient errors, taken to be
random with RMS value of 2\% plus an offset uncertainty of
5$\times10^{-4}$ (km-amagat)$^{-1}$.  This is referred to in the
following as the COMPLX2 error model.

\section{Fit results for 2012 and 2015 STIS observations}

Here we first consider conservative fits over a wide 540-980 nm spectral range,
which identifies a problem in matching the needed particle properties
to fit such a wide range.  That problem is then deferred by fitting
the critical 730-900 nm wavelength range that provides the strongest constraints
on the methane/hydrogen ratio, first using Mie scattering particles for
all cloud layers, then using an alternative model in which the
main two tropospheric layers are characterized by adjustable DHG
phase functions. If we assume that the methane mixing ratio is uniform
up to the condensation level, we find that it must decrease with
latitude by factors of 2-3 from equator to pole with different
absolute levels, depending on whether particles are modeled as spheres
or with DHG phase functions. We then consider two models that restrict
methane depletions to an upper tropospheric layer, and find that
improved fits are obtained with models that restrict depletions
to the region above the 5-bar level.

\subsection{Initial conservative fits to the 540-980 nm spectrum.}

\begin{figure*}\centering
\includegraphics[width=6in]{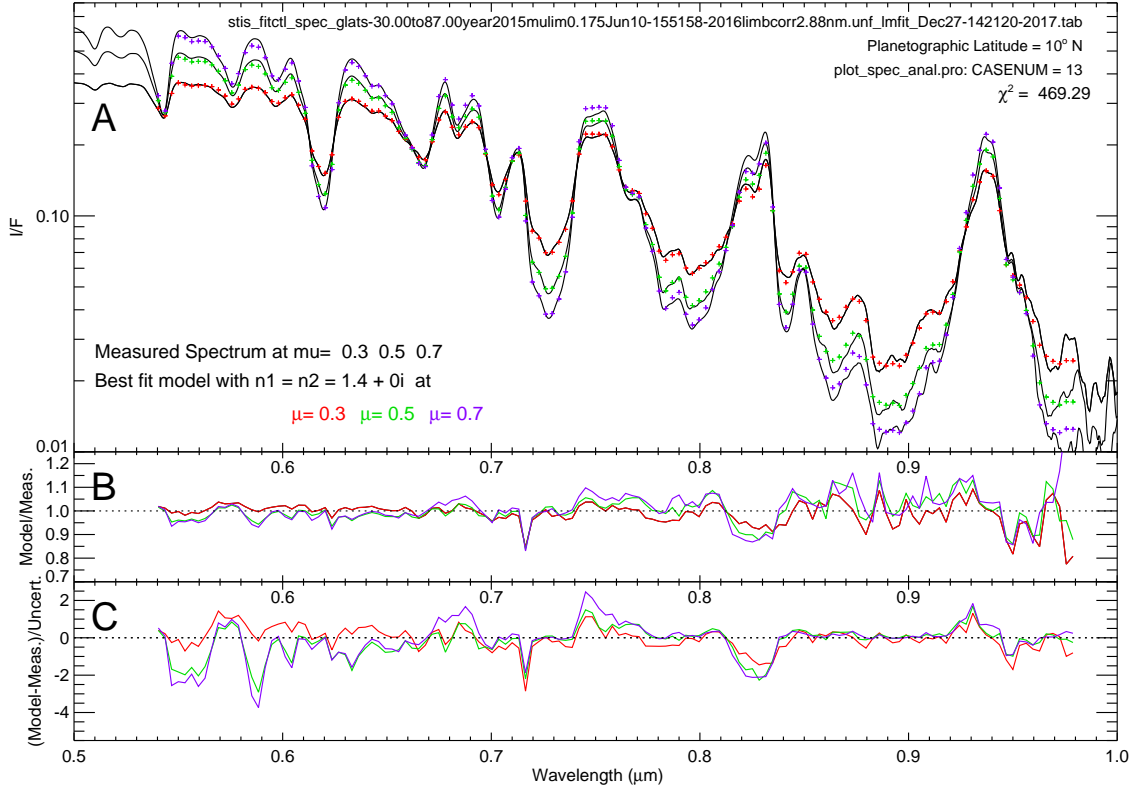}
%SOURCE latest 2-cloud model fit on page 101, Uranus Log L, numcase=13 for plot_spec_anal.pro
%SOURCE original 3-sheet fit produced by plot_spec_anal.pro on puck, see Page 17, Uranus Log J
% parameters are from fit on page 10.
\caption{{\bf Top:} Model spectra at three view angle cosines (colored
  as noted in the legend) compared to the 10\deg N 2015 STIS spectra
  (black curves).  {\bf Middle:} Ratio of model to measured spectra.
  {\bf Bottom:} Difference between model and measured spectra divided
  by expected uncertainty. The aerosol model used the baseline 2-cloud
  model, except that $m2\_nr$ was fixed at a value of 1.4.  Best fit
  parameter values are given in Table\ \ref{Tbl:initfit}.  Note the
  significant discrepancies at short wavelength continuum peaks, near
  740 nm, and within the critical region near 830 nm which is most
  sensitive to the methane to hydrogen ratio. Better fits were
  obtained with $m2\_nr$ allowed to adjust as part of the fitting
  process.}
\label{Fig:initfit}
\end{figure*}

Assuming a real refractive index of $m2\_nr$ = 1.4, and an imaginary
index of zero, we fit our simplified 2-layer model to spectra covering
the 540-980 nm range by adjusting the seven remaining parameters.  We
obtained a best fit model spectrum with significant flaws that are
illustrated in Fig.\ \ref{Fig:initfit}.  The parameter values and
uncertainties are listed in Table\ \ref{Tbl:initfit}.  The best-fit
value for the methane mixing ratio was a remarkably low
1.27$\pm$0.05\%, but is not credible because the region near 830 nm,
which is most sensitive to the \chfx/H$_2$ ratio is very poorly fit.
Additional flaws are seen near 750 nm, as well as at other continuum
features at shorter wavelengths.  Almost exactly the same fit quality
and the same specific flaws were obtained when we replaced the single
tropospheric cloud with two sheet clouds with two more adjustable
parameters.

Better results were obtained by letting the real refractive index be a
fitted parameter as well.  This is in contrast to the common procedure
of fixing the refractive index, most often at a value of 1.4, as we
also did in our initial fit. \cite{Irwin2015reanalysis}, for example,
justified their choice of 1.4 by noting that most plausible
condensables have real indexes between 1.3 (methane) and 1.4
(ammonia).  Other simple hydrocarbons are also in this range.
However, at the levels where we see significant aerosol optical depth,
ammonia is not very plausible, and methane is in doubt because most
particles are found at pressures exceeding the condensation level. On
the other hand, the plausible condensable H$_2$S has a significantly larger
real index of 1.55 \citep{Havriliak1954} at the 80 K -- 100 K temperatures  
characteristic of the main cloud layer on Uranus.  Another possible
cloud particle is a complex photochemical product, one example of
which is the tholin material described by \cite{Khare1993}, which has
a real index near 1.5.  Thus, it seems premature to settle on a fixed
value at this point.
  
When the initial fit is redone with starting values of $m2\_r$ = 1 \mum
and $m2\_nr$ = 1.4, as documented in Table\ \ref{Tbl:initfit}, we
obtain a final large particle solution of $m2\_r$ = 1.918$\pm$0.33
\mum and $m2\_nr$ = 1.184$\pm$0.02.  Although this is an improved fit,
there are still the same significant, though slightly smaller, local
flaws and the inferred methane mixing ratio is again at a quite low
value, this time 1.20$\pm$0.15\%. A considerably better fit is
obtained with the small particle solution, which produced a decrease
in \chisq /N to 0.91.  This solution was obtained by using an initial
guess of $m2\_r$ = 0.5 \mum and $m2\_nr$ = 1.4.  As also shown in
Table\ \ref{Tbl:initfit}, these parameters adjusted to best-fit values
of $m2\_r$ = 0.235$\pm$0.03 \mum and $m2\_nr$ = 1.83$\pm$0.09. The real
index in this case is even larger than the expected value for H$_2$S, and the
methane VMR has increased to a more credible 1.90$\pm$0.13\%. However,
even this fit has a few significant local flaws, near 550 nm, 590 nm,
and 750 nm.  Our interpretation of this situation is that there are
wavelength dependent properties to the particle scattering that are
not captured by conservative spherical particle models.  This suggests
 that problems in fitting the wavelength dependent I/F over a wide
range interfere with attempts to constrain the methane mixing
ratio. Thus we decided to separate these problems.  Leaving
wavelength-dependence for the moment, we next focus on a narrower
spectral region that provides the best constraint on the methane
mixing ratio.

\begin{table*}[!hbt]\centering
\caption{Preliminary fits to the 540-900 nm part of 2015 STIS 10\deg N spectra.}\label{Tbl:initfit}
\vspace{0.1in}
\begin{tabular}{ r c c c }
Parameter  &  Value    &   Value for LP soln.  &  Value for SP soln.        \\[-0.03in]
Name       & $m2\_nr$ fixed at 1.4 & with $m2\_nr$ fitted  & with $m2\_nr$ fitted \\
\hline
  $m1\_od$ at $\lambda$ = 0.5 \mum&  0.046$\pm$  0.01&  0.050$\pm$  0.01&  0.048$\pm$  0.01\\
  $m2\_od$ at $\lambda$ = 0.5 \mum&  5.155$\pm$  0.46&  7.536$\pm$  1.42&  2.437$\pm$  0.24\\
      $m2\_pt$ (bar)&  1.149$\pm$  0.04&  1.054$\pm$  0.04&  0.962$\pm$  0.04\\
      $m2\_pb$ (bar)&  4.137$\pm$  0.25&  4.102$\pm$  0.25&  3.696$\pm$  0.22\\
     $m2\_r$ (\mumx)&  0.382$\pm$  0.03&  1.918$\pm$  0.33&  0.235$\pm$  0.03\\
             $m2\_nr$&  1.400      &  1.184$\pm$  0.02&  1.828$\pm$  0.09\\
 $\alpha_0$ (\%)&  1.270$\pm$  0.05&  1.380$\pm$  0.07&  1.900$\pm$  0.13\\
            $ch4rhc$&  0.986$\pm$  0.13&  1.200$\pm$  0.15&  1.200$\pm$  0.15\\
            \chisq& 469.29 &434.53& 371.03\\
         \chisq /NF&   1.16 &  1.07&   0.91\\
\hline \\[-0.15in]
\end{tabular}
\parbox[]{5.2in}{NOTES: In the last two columns LP soln. denotes large particle solution
and SP soln. denotes small particle solution. The \chisq values
given here are based on fitting points spaced 3.2 nm apart.}
\end{table*}

\subsection{Fitting the 730--900 nm region}

Our next step was to concentrate on the spectral region where the ratio of methane to hydrogen
is best constrained, i.e. the 730--900 nm region.  As shown if Fig.\ \ref{Fig:pendepth}, the
short-wavelength side is free of CIA and sensitive to the deep methane mixing ratio, while the middle
 region from about 810 to 835 nm is strongly affected by hydrogen CIA, and the long-wavelength side of
the region is sensitive to the methane mixing ratio at pressure less than 1 bar.   By using this
entire region we expect to obtain good constraints on both the ratio of methane to hydrogen as well
as on the vertical cloud structure.  Results from fitting this region should not be strongly affected by
wavelength-dependent particle properties, given the relative modest spectral range we
are considering here. If the assumption of Mie scattering over this limited range is seriously flawed, 
that should show up in an inability to get high quality fits.  This relatively narrow spectral
range also weakens constraints on particle size, as might be expected.

\subsubsection{Effects of different aerosol models}

We were somewhat surprised to find that the kind of aerosol model chosen
to fit the observations has a significant effect on the
derived vertical and latitudinal distribution of methane. To investigate these effects 
 we did model
fits at two key latitudes: 10\degx N and 60\degx N.  From more detailed
latitudinal profiles discussed later, we know that the apparent methane
mixing ratio peaks near 10\deg N and is approaching its polar minimum
near 60\deg N.  These are also two latitudes for which 2012 and 2015 observations
provide good samples at the three view angle cosines we selected.

\subsubsection{Fitting the spherical particle 2-cloud model assuming a uniform \chf distribution.}\label{Sec:2mieu}

We first consider a methane vertical distribution that has a constant
mixing ratio from the deep atmosphere to the condensation level.  Above
that level (at lower pressures) we assume a drop in relative humidity
to an adjustable fraction of the saturation vapor pressure, and from
there to the tropopause we interpolate from the above cloud value
to the tropopause minimum as described in Section\ \ref{Sec:thermal}.  The key
parameters describing the methane distribution are then the
above cloud relative humidity and the deep mixing ratio.

We first consider a simple aerosol model in which the tropospheric
contribution is characterized by an adjustable optical depth and a
single layer of spherical particles bounded by top and bottom
pressures and uniformly mixed with the gas.  We assume initially that
these particles scatter light conservatively, but allow the real
refractive index to be constrained by the spectral observations.

\begin{figure*}[!htb]\centering
%\hspace{-0.15in}\includegraphics[width=3.2in]{case22_spectra.eps}
\hspace{-0.15in}\includegraphics[width=3.2in]{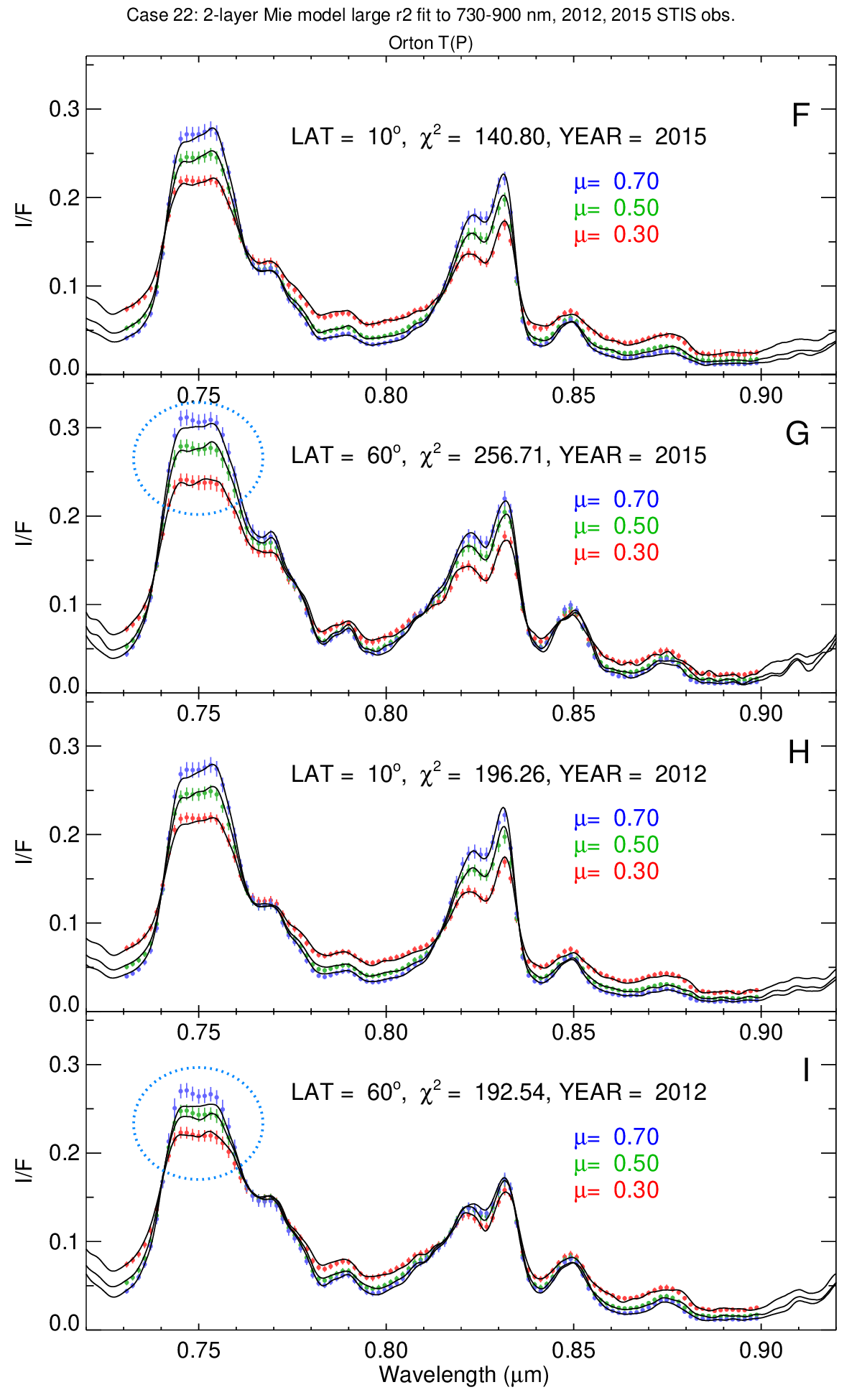}
%\hspace{-0.15in}\includegraphics[width=3.2in]{case23_spectra.eps}
\hspace{-0.15in}\includegraphics[width=3.2in]{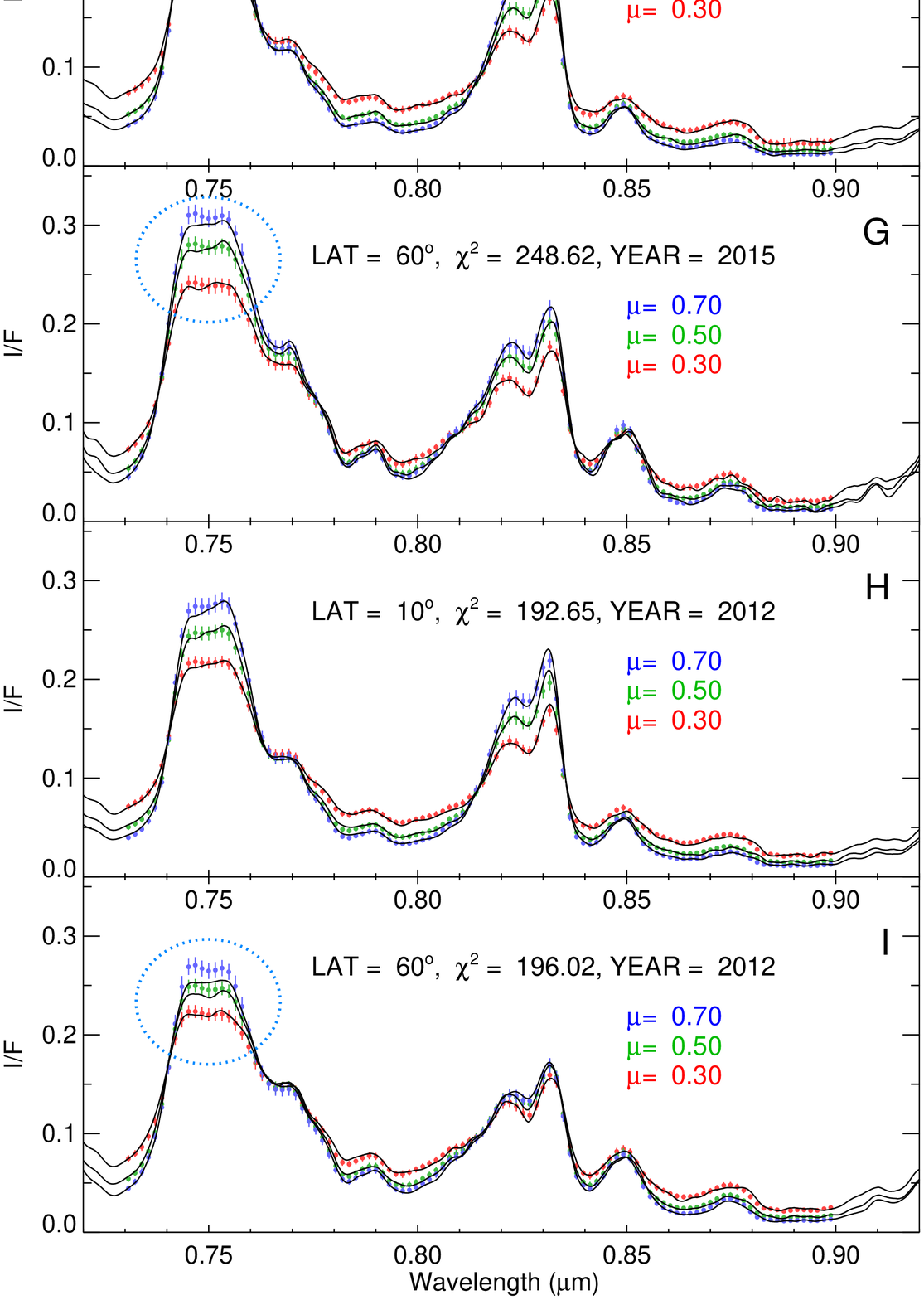}
%SOURCE: produced by running case 23 and case 22 of plot_fitctl2015mie_5panel17.pro in ir_ms_ura on puck:
%\vspace{-0.15in}
\caption{Comparison of observed spectra (curves) with model fits
  (points) for the large particle solutions (left) and the small particle
  solutions (right), both using the model parameterization defined in
  Table\ \ref{Tbl:mie_ls}.  Fits to 2015 STIS observations are shown
  in the top pair of panels and fits to 2012 observations in the
  bottom pair of panels.  Blue dotted ovals identify regions of
  high-latitude fitting errors, which can be greatly reduced by using
  a non-uniform vertical distribution of
  methane.\label{Fig:specmie_ls}}
\end{figure*}

The results of this series of fits for both 2015 and 2012 observations are given
in Table\ \ref{Tbl:mie_ls} where small particle solutions are given in
the first four rows and large-particle solutions in the remaining four rows.  The model
spectra are compared to the observations in Fig.\ \ref{Fig:specmie_ls}.
These fits do achieve their intended result of providing more precise constraints
on the above-cloud methane humidity, which is high at 10\degx N and about
50\% of those levels at 60\deg N.  The temporal change between 2012 and 2015 in the effective methane 
mixing ratios is very small and well within uncertainty limits. The low latitude
values of 3.14$\pm$0.45\% and 3.16$\pm$0.50\% are consistent with no change, as are the 60\degx N
values, which are 0.99$\pm$0.08\% and 0.93$\pm$0.08\%, for 2015 and 2012 respectively.
The factors by which the effective methane mixing ratio declines with latitude are 3.17 and 3.40
for 2015 and 2012 respectively.  

%The methane mixing ratio values for 2015 are larger than
%given in Table\ \ref{Tbl:conserv}, by 31\% at 10\degx N and by 37\% at 60\degx N.

%\hspace{-1in}
\begin{table*}[!hbt]\centering
\caption{Single tropospheric Mie layer fits to 10\degx N and 60\degx N STIS 730 - 900 nm spectra.\label{Tbl:mie_ls}}
% SOURCE: This table is produced by running case 23 and case 22 of plot_fitctl2015mie_5panel17.pro in ir_ms_ura on puck:
\small
\vspace{0.1in}
\setlength\tabcolsep{2pt}
\begin{tabular}{ c c c c c c c c c c c}
   Lat.    &  $m1\_od$        &           &      $m2\_pt$ &    $m2\_pb$ &        $m2\_r$ &          &      $\alpha_0$    &              &        & \\[-0.03in]
   (\degx) &  $\times$100 &  $m2\_od$ &       (bar) &     (bar) &       (\mumx) &   $m2\_nr$  &      (\%)     &       $ch4rhc$ & \chisq & YR\\[0.05in]
\hline
\\[-.15in]
%   LAT &    m1\_od &    m2\_od &    m2\_pt &    m2\_pb &     m2\_r &     m2\_n &    ch4v0 &   ch4rhc &    chisq &  YEAR\\[0.05in]
   10 &  2.8$\pm$0.8 &   3.07$\pm$0.9 &   1.13$\pm$0.04 &   2.46$\pm$0.22 &   0.34$\pm$0.10 &   1.55$\pm$0.16 &   3.14$\pm$0.45 &   0.68$\pm$0.13 &   148.39 &  2015\\[0.05in]
   60 &  0.1$\pm$70.7 &   1.45$\pm$0.3 &   1.02$\pm$0.02 &   2.53$\pm$0.13 &   0.25$\pm$0.09 &   1.86$\pm$0.30 &   0.99$\pm$0.08 &   0.31$\pm$0.18 &   248.62 &  2015\\[0.05in]
\hline
\\[-.15in]
   10 &  3.0$\pm$0.7 &   2.52$\pm$0.6 &   1.07$\pm$0.04 &   2.37$\pm$0.20 &   0.25$\pm$0.09 &   1.74$\pm$0.26 &   3.16$\pm$0.50 &   0.95$\pm$0.16 &   192.65 &  2012\\[0.05in]
   60 &  2.2$\pm$1.6 &   1.10$\pm$0.2 &   1.02$\pm$0.04 &   2.22$\pm$0.13 &   0.24$\pm$0.07 &   1.81$\pm$0.25 &   0.93$\pm$0.08 &   0.42$\pm$0.20 &   196.02 &  2012\\[0.05in]
\hline
\hline
\\[-.15in]
10 &  2.8$\pm$0.8 &   4.95$\pm$1.4 &   1.11$\pm$0.04 &   2.69$\pm$0.19 &   1.09$\pm$0.48 &   1.28$\pm$0.07 &   2.69$\pm$0.28 &   0.67$\pm$0.14 &   140.80 &  2015\\[0.05in]
60 &  0.8$\pm$4.5 &   4.28$\pm$1.1 &   1.07$\pm$0.03 &   2.96$\pm$0.16 &   1.75$\pm$0.52 &   1.23$\pm$0.05 &   0.81$\pm$0.05 &   0.39$\pm$0.24 &   256.71 &  2015\\[0.05in]
%   PGLAT &    m1_od &    m2_od &    m2_pt &    m2_pb &     m2_r &     m2_n &    ch4v0 &   ch4rhc &    chisq &  YEAR\\[0.05in]
\hline
\\[-.15in]
10 &  2.8$\pm$0.7 &   6.09$\pm$1.9 &   1.09$\pm$0.04 &   2.67$\pm$0.19 &   1.54$\pm$0.58 &   1.23$\pm$0.06 &   2.56$\pm$0.26 &   0.88$\pm$0.16 &   196.26 &  2012\\[0.05in]
60 &  3.3$\pm$1.5 &   3.21$\pm$0.8 &   1.08$\pm$0.04 &   2.51$\pm$0.14 &   1.44$\pm$0.53 &   1.22$\pm$0.05 &   0.74$\pm$0.05 &   0.56$\pm$0.26 &   192.54 &  2012\\[0.05in]
\hline
\end{tabular}
\normalsize  % turned off for single-spaced text
\parbox{6.in}{\vspace{0.1in} NOTE: The optical depths are given for a wavelength of 0.5 \mumx. These fits used 318 points of comparison and fit 8 parameters, for a nominal value of NF=310, for which the normalized \chisq /NF
ranged from 0.48 to 0.802.}
\end{table*}

For these fits, the refractive index results for the small-particle solution have a
weighted average for both latitudes and both years of 1.68$\pm$0.11,
which is much closer to the 1.55 value expected for H$_2$S, although
the 60\degx N values exceed that value by slightly more than their
uncertainties.  Perhaps this is an indication of a cloud composition
difference between the two latitudes. A quite different
result is obtained for the large-particle fit.  In this case the
average index is 1.23$\pm$0.03, which is lower than that of any of the
candidate substances, and the individual values don't vary much from
low to high latitudes, or between 2012 and 2015.

The various determinations of the pressure boundaries of the main tropospheric cloud
layer are very similar for both years, both latitudes, and both particle-size solutions,
extending from a base near 2.5 bars to a top near 1.1 bar.  The optical depths
do differ substantially between large and small particle solutions because the
larger particles are more forward scattering and have a lower refractive index, both
differences reducing the back-scattering efficiency of the particles,
 requiring increased optical depth to make up for the losses.

\begin{figure*}[!htb]\centering
%\hspace{-0.15in}\includegraphics[width=3.2in]{derivatives_large-particle.eps}
\hspace{-0.15in}\includegraphics[width=3.2in]{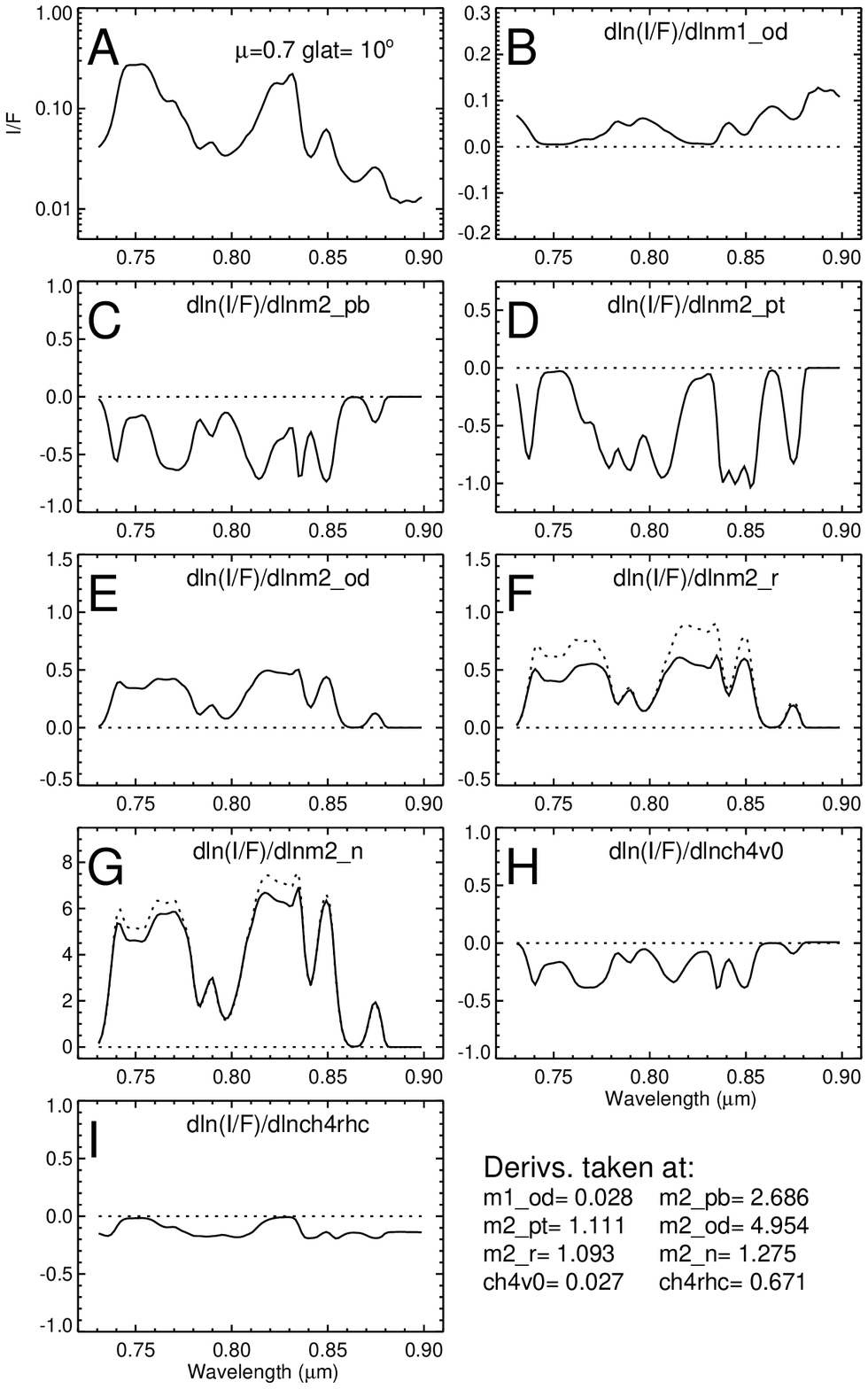}
%\hspace{-0.05in}\includegraphics[width=3.2in]{derivatives_small-particle.eps}
\hspace{-0.05in}\includegraphics[width=3.2in]{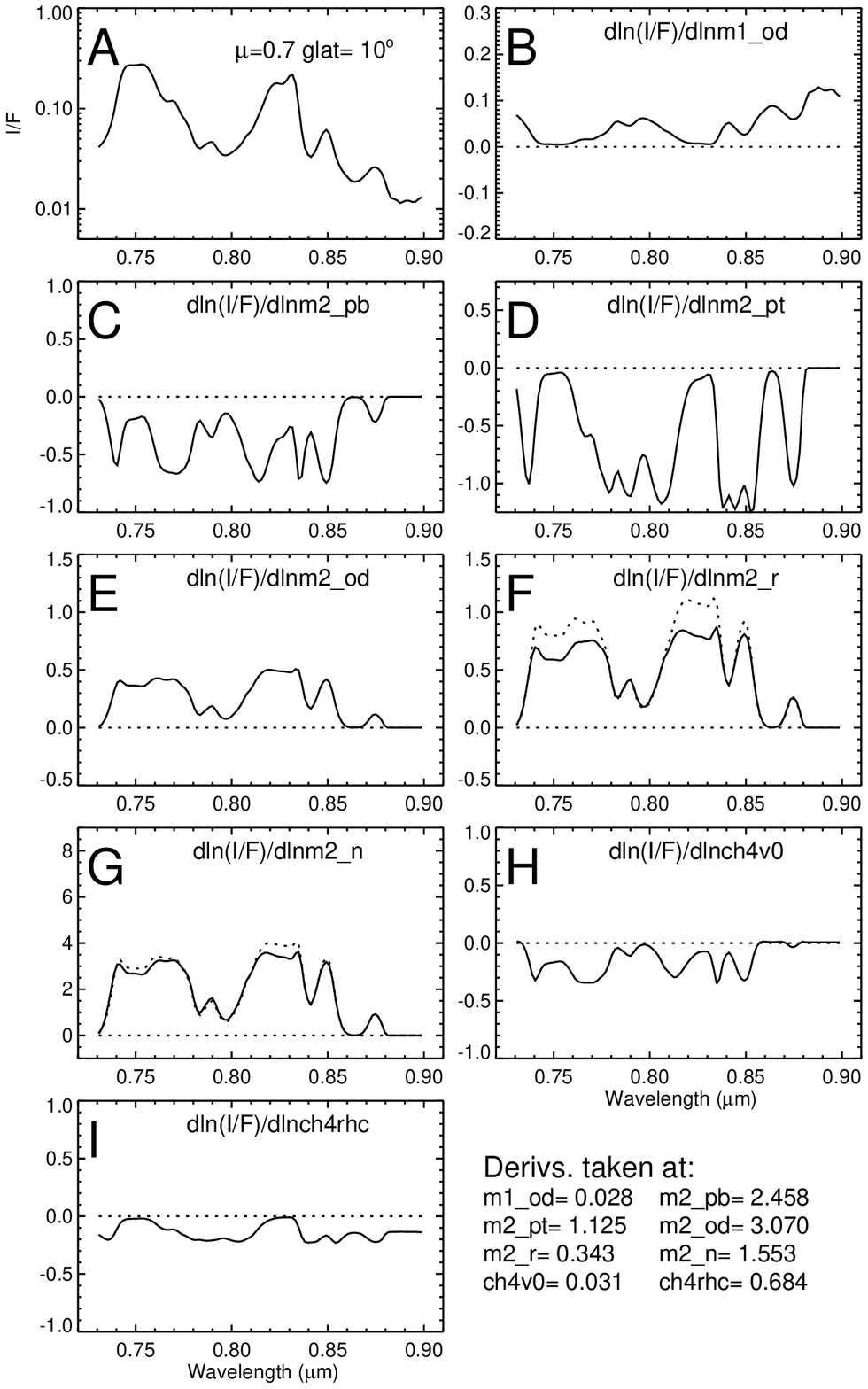}
%SOURCE: produced by running ir_ms_ura_lmgx_stisma_mpi.pro on puck with 'd' option
% then copying output files to triton:/home/home2/sro/uranus/paper_stis15 and
% running plot_deriv_spec17.pro in that directory.
%\vspace{-0.15in}
\caption{Derivative spectra for uniform mixing ratio models evaluated for
the large-particle solution (Left group) and small particle solution (Right group).
In each group we show I/F model spectrum (A) and derivatives of fractional changes in I/F
  with respect to fractional changes in parameters $m1\_od$ (B), $m2\_pb$
  (C), $m2\_pt$ (D), $m2\_od$ (E), $m2\_r$ (F), $m2\_nr$ (G), $ch4v0  \equiv \alpha_0$ (H) and
  $ch4rhc$ (I).  All the derivative panels
  are scaled the same, except for panel B, which has been expanded by
  a factor of 4, and panel G, which has been compressed by a factor of
  5 because of their unusually small and large effects, respectively, on
  the I/F spectrum.  In panels F and G, the dotted curve represents a
  version of the $m2\_od$ derivative spectrum scaled to match the
  lower features of the $m2\_r$ and $m2\_nr$ derivative spectra
  respectively, to illustrate their strong correlations but resolvable
  differences.
\label{Fig:miederiv}}
\end{figure*}

These results explain the brightening of the polar region at
pseudo-continuum wavelengths between 2012 and 2015. To understand how
influential these various parameters are on the observed spectrum, we
computed logarithmic spectral derivatives (Fig. \ref{Fig:miederiv}).
These have the useful property of showing the fractional changes in
the spectrum produced by fractional changes in the various parameters
used to model it.  For the small particle solution we see that between 2012
and 2015 m2\_od
increased by 32\% at 60\deg N, providing the main driver for the increase.
According to Fig.\ \ref{Fig:speccomp}, near 750 nm the I/F increased
by about 20\%, and according to the derivative spectra in
Fig.\ \ref{Fig:miederiv}, the optical depth increase would account
for about 11\%, while the increase in refractive index by just 2.8\%
would increase the I/F by an additional 10\%, accounting for the 20\%
total.  However, these derivatives were computed for a latitude of
10\degx N; somewhat different derivatives might be found at 60\degx N.
A similar increase of 33\% is seen in the optical depth derived for
the large particle solution, although in this case there is also an
increase in particle size by 22\%, which would also contribute
significantly.  Weighting these by respective factors of 0.42 and 0.45
(from Fig.\ \ref{Fig:miederiv}) we obtain from just these
parameters an I/F increase of about 24\%, which is again close to the
entire change observed.  The small changes in inferred methane mixing
ratios are increases of 6\% for the small particle solution and about
10\% for the large particle solution, which would yield I/F decreases
of 1.2\% and 1.5\% for the small and large particle solutions
respectively, both of which are well below uncertainties.  The fitting
errors at high latitudes, which are most evident in the 0.75-\mum
region are highlighted by blue dotted ovals in
Fig.\ \ref{Fig:specmie_ls}.

%\clearpage

\subsubsection{Fitting the 2-cloud non-spherical model assuming a uniform \chf distribution.}

We next consider fits in which the main tropospheric layer consists of a single
particle type characterized by the simplest possible Henyey-Greenstein function,
which is a one-term version of Eq.\ \ref{Eq:dhg} characterized by a single
asymmetry parameter. The vertical structure parameterization and stratospheric haze layer
parameterization are both unchanged from the spherical particle example used
in the previous section.  Because some wavelength dependence is required, we
introduce a wavelength dependent optical depth using a simple linear slope, which
is a parameter that is adjusted to optimize the fit.  Our model is given by \begin{eqnarray}
 \tau (\lambda) = \tau_o \times (1 + \mathrm{k_{OD}}\times (\lambda - \lambda_0))\label{Eq:wdep}
\end{eqnarray}
where $\lambda_0$ is taken to be 800 nm.  This also makes $\tau_o$ the
optical depth at 800 nm.  We could also have made the asymmetry
parameter wavelength dependent instead of, or in addition to, the optical
depth, but found excellent fits without adding any further complexity.
We will not be making any claims regarding the true source of
wavelength dependence in any case.  Our main objective is to find out
how this different kind of model affects the methane distribution, and
to determine the average asymmetry parameter of these particles.  We
will also try to infer a single-scattering albedo.

\begin{figure*}[!htb]\centering
\includegraphics[width=3.1in]{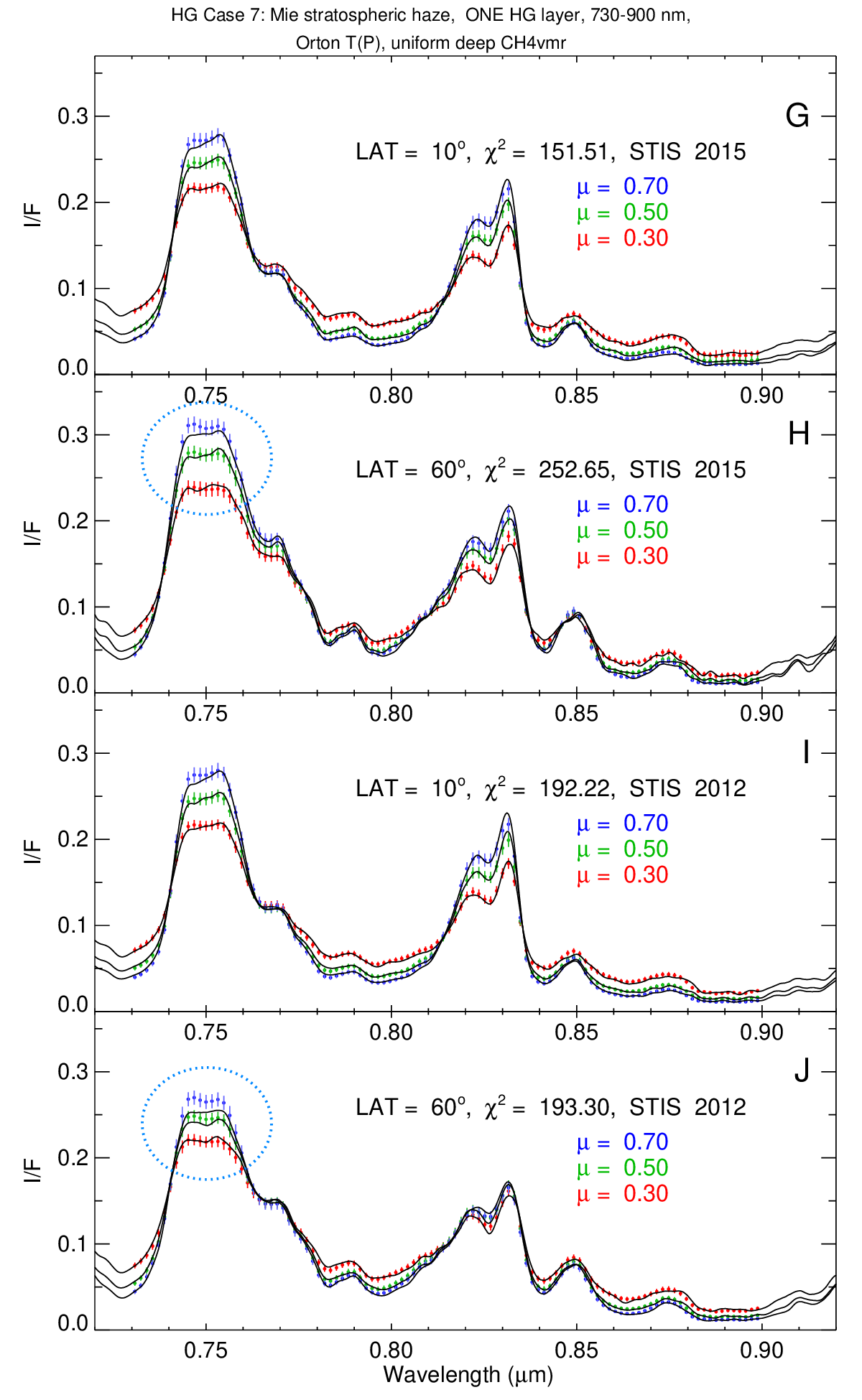}\hspace{-0.05in}
\includegraphics[width=3.1in]{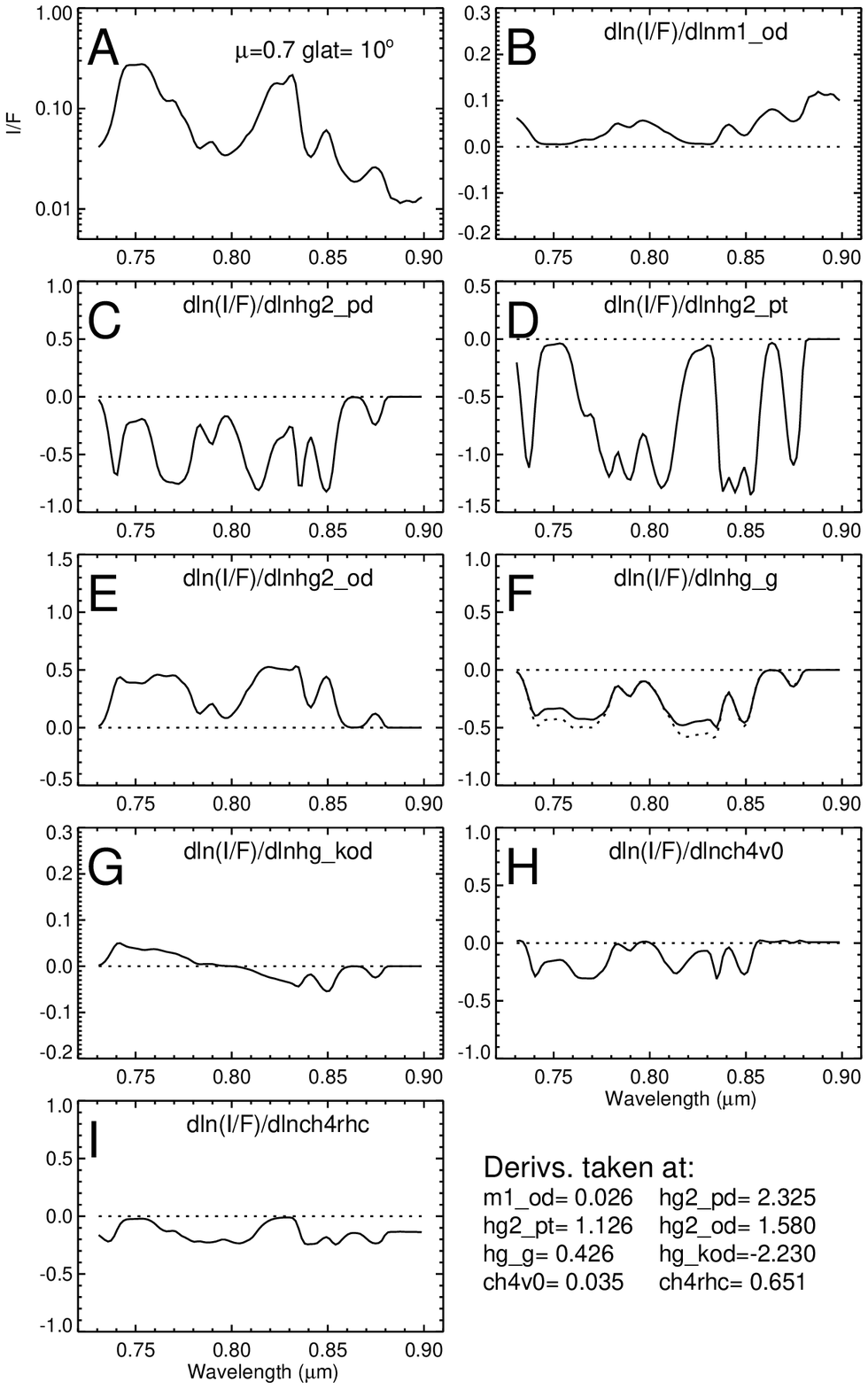}
\caption{Left: HG Model spectra compared to observations at 10\degx N
  and 60\degx N for 2012 (bottom pair) and 2015 observations (top
  pair), with models plotted as points with error bars.  Right:
  Derivative spectra showing the ratio of a fractional change in I/F
  to the fractional change in the parameter producing the
  change (here $ch4v0 \equiv \alpha_0$). The dotted curve in panel F of the derivative group is an inverted
plot of the curve in panel E, with minima scaled to match the solid curves. Note that
the maxima do not match, making them distinguishable.\label{Fig:hgspec}}
%SOURCE: Left is plotted with plot_fitctl2015_hg_6panel.pro on puck. Right is plotted with plot_deriv_spec17.pro on triton.
\end{figure*}

Best-fit parameter values and uncertainties for fits at 10\degx N and
60\degx N for 2012 and 2015 are presented in Table\ \ref{Tbl:hgpars}.
Best-fit model spectral are compared to observations in the left panel
of Fig.\ \ref{Fig:hgspec}, while fractional derivative spectra are
displayed in the right panel.  These fits are comparable in quality to
the spherical particle fits presented in the previous section, and
have the same problem fitting the high-latitude spectra, most notably
in the 750-nm region. This region senses more deeply than other parts
of this limited spectral range (see Fig.\ \ref{Fig:pendepth}), and
thus is most likely to be affected by vertical variations in the
methane mixing ratio.  According to the derivative spectra, an
increase in the methane mixing ratio with depth would reduce the I/F
in this region, which we would expect to produce a better fit,
 and we will later show that this does in fact improve
the spectral fit in this region.

The methane mixing ratio values for this model average somewhat higher than found for the model using
spherical particles, although all are within uncertainty limits for a given latitude, and all results
indicate an effective mixing ratio decrease by slightly more than a factor of three from 10\degx N to
60\degx N.

The best-fit asymmetry parameter for this model is generally near 0.4,
well below the commonly assumed value of 0.6 for near-IR analysis,
which is in part based on an analysis of limb-darkening measurements
by \cite{Sro2008grism}.  That analysis predates the significant
improvement in methane absorption coefficients seen in the last decade
\citep{Sro2012LBL} and may no longer be valid.  It seems unlikely that
this difference is merely a wavelength dependence. For the sizes
inferred for spherical particle solutions, the asymmetries either
decrease with wavelength (small particle solution), or remain
relatively flat (large particle solution).  While the asymmetry
parameter is highly negatively correlated with the optical depth
parameter, these two parameters do have sufficiently different ratios
between peaks and valleys to allow them to be independently
determined (shown in Fig.\ \ref{Fig:hgspec}F).  
The asymmetry was determined to within about 10\% and the
optical depth to within slightly better accuracy.  The optical depths
for this model appears to be considerably lower than for the spherical
particle models, which were at a shorter wavelength of 500 nm.  If we
convert those Mie scattering optical depths to a wavelength of 800 nm,
we find that the 0.3-\mum particle optical depth drops from 3.1 to 2.6
and the 1.54 \mum particle optical depth increases from 6.1 to 7.5.
Thus, the wavelength difference does not explain the low optical
depths of the non-spherical model.  It is more likely due to the
latter's more symmetric scattering.  The Mie particle models have
asymmetries of about 0.68 and 0.87 for the small and large particle
solutions respectively, both adjusted to a reference wavelength of 800
nm.  The particles of KT2009, which use a
DHG phase function, with their adopted values of $g_1$ = 0.7 and $g_2$=
-0.3, yield an asymmetry of 0.6, which is much closer to that of our
small particle solution for spherical particles, and much larger than
our HG particle solutions.  We tried to find an HG solution with larger
asymmetry by using a first guess with $g$ = 0.63, but the regression again
converged on $g$ = 0.43.  It is apparently the case that very different
scattering properties can lead to very nearly the same fit quality,
but very different optical depths and asymmetry parameters. At phase
angles near zero, there is a considerable ambiguity between more forward
scattering particles with larger optical depths and more backward scattering
particles with smaller optical depths.

%\hspace{-1in}
\begin{table*}[!hbt]\centering
\caption{Single tropospheric HG layer fits to 10\degx N and 60\degx N STIS spectra.\label{Tbl:hgpars}}
% SOURCE: This table is produced by running case 7 of plot_fitctl2015_hg_6panel.pro in ir_ms_ura on puck:
\small
\vspace{0.1in}
\setlength\tabcolsep{2pt}
\begin{tabular}{ c c c c c c c c c c c}
   Lat.    &      $m1\_od$     &           &      $hg2\_pt$ &    $hg2\_pb$ &         &     $hg2\_kod$      &      $\alpha_0$    &              &        & \\[-0.03in]
   (\degx) &  $\times$100 &  $hg2\_od$ &       (bar) &     (bar) &       $hg2\_g$ &   (/\mum)  &      (\%)     &       $ch4rhc$ & \chisq & YR\\[0.05in]
\hline
\\[-.15in]
%   PGLAT &    m1\_od &   hg2\_od &   hg2\_pt &   hg2\_pd &     hg\_g &   hg\_kod &    ch4v0 &   ch4rhc &    \chisq &  YEAR\\[0.05in]
10 &  2.6$\pm$0.5 &   1.58$\pm$0.13 &   1.13$\pm$0.03 &   2.33$\pm$0.15 &   0.43$\pm$0.04 &  -2.23$\pm$0.4 &   3.48$\pm$0.45 &   0.65$\pm$0.08 &   151.51 &  2015\\[0.05in]
60 &  0.0$\pm$0.0 &   0.97$\pm$0.05 &   1.01$\pm$0.02 &   2.53$\pm$0.12 &   0.26$\pm$0.02 &  -3.18$\pm$0.3 &   0.97$\pm$0.06 &   0.31$\pm$0.03 &   252.65 &  2015\\[0.05in]
%   PGLAT &    m1_od &   hg2_od &   hg2_pt &   hg2_pd &     hg_g &   hg_kod &    ch4v0 &   ch4rhc &    chisq &  YEAR\\[0.05in]
\hline
\\[-.15in]
10 &  2.7$\pm$0.5 &   1.57$\pm$0.14 &   1.07$\pm$0.03 &   2.47$\pm$0.17 &   0.42$\pm$0.04 &  -1.91$\pm$0.4 &   2.85$\pm$0.32 &   0.87$\pm$0.12 &   192.22 &  2012\\[0.05in]
60 &  2.0$\pm$1.2 &   0.71$\pm$0.06 &   1.01$\pm$0.02 &   2.04$\pm$0.08 &   0.39$\pm$0.04 &  -3.95$\pm$0.3 &   1.04$\pm$0.07 &   0.39$\pm$0.15 &   193.30 &  2012\\[0.05in]
\hline
\end{tabular}
\normalsize  % turned off for single-spaced text
\parbox{6.in}{\vspace{0.1in} NOTE: The optical depth is for a wavelength of 0.8 microns for $hg2\_od$, and for 0.5 \mum for the stratospheric
haze ($m1\_od$). These fits used 318 points of comparison and fit 8 parameters, for a nominal value of NF=310, for which the normalized \chisq /NF
ranged from 0.48 to 0.802.}
\end{table*}

The best-fit optical depth slope parameter $hg2\_kod$ is negative, as
generally expected, and is around -2/\mum at 10\degx N but -3.2
/\mum to -4/\mum at 60\degx N.  A spherical particle of radius 0.3
\mum and real index 1.4 would have a slope of about -2.4/\mumx.  For
spherical particles, some of the wavelength dependence in scattering
is provided by wavelength dependence in the phase function.  This
suggests a possible decrease in particle size at high latitudes.

There is better agreement between the HG solution and the small particle Mie solutions
regarding other parameters, including pressure boundaries, methane mixing
ratios and above cloud humidities.  Thus, the preponderance of evidence suggests that
the cloud particles can be roughly approximated by the small particle Mie solutions,
which is the solution type we will use to investigate the latitude dependent
characteristics in more detail.

%\clearpage

\subsubsection{Latitude-dependent fits}

%% A summary of the various fits as a function of latitude that are to be
%% discussed in the following sections is provided in Table\ \ref{Tbl:fitslist}.

%% \begin{table}[!hb]\centering
%% \caption{Summary of latitude dependent fits and links to results.}
%% \vspace{0.15in}
%% \begin{tabular}{c c c c}
%% \hline\\[-.15in]
%% Aerosol Model & CH4 Vertical distrib. & Figure reference & Table reference \\
%% \hline\\[-.15in]
%% 3-Mie layers &  uniform &  Fig.\ \ref{Fig:3miecompact730-900} & none \\
%% Mie + 2 DHG layers & uniform &  Fig.\ \ref{Fig:dhgcompact} & none\\
%% 3-Mie layers &  descended depletion &  Fig.\ \ref{Fig:3miedescendfits} & none \\
%% 3-Mie layers &  step depletion &  Fig.\ \ref{Fig:3miestepch4fits} & none \\
%% Mie + 2 DHG layers & descended depletion &  Fig.\ \ref{Fig:dhgdescend} & none\\
%% \hline
%% \end{tabular}\label{Tbl:fitslist}
%% \end{table}

\begin{figure*}[!hbt]\centering
\includegraphics[width=3.1in]{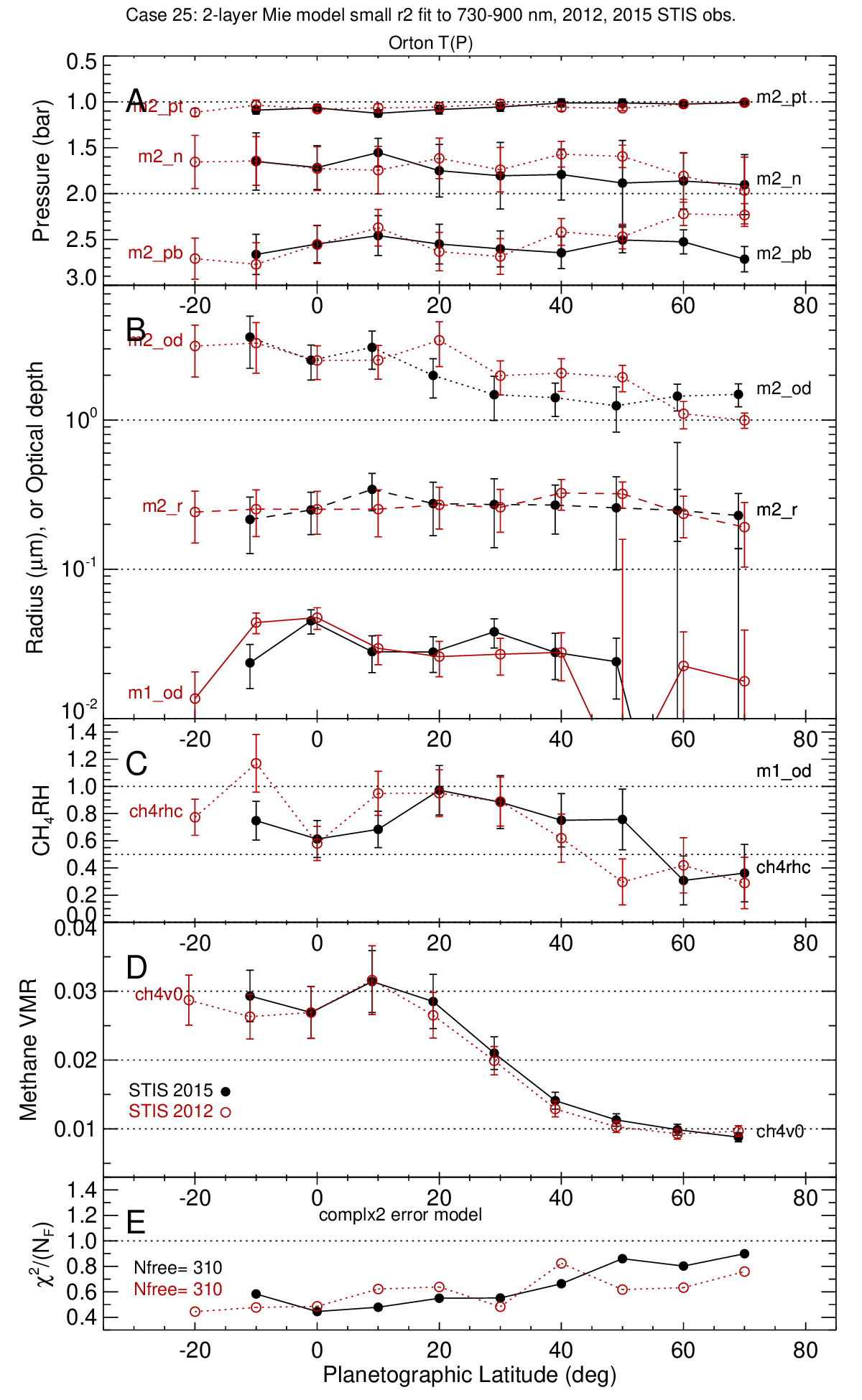}\hspace{-0.15in}
\includegraphics[width=3.1in]{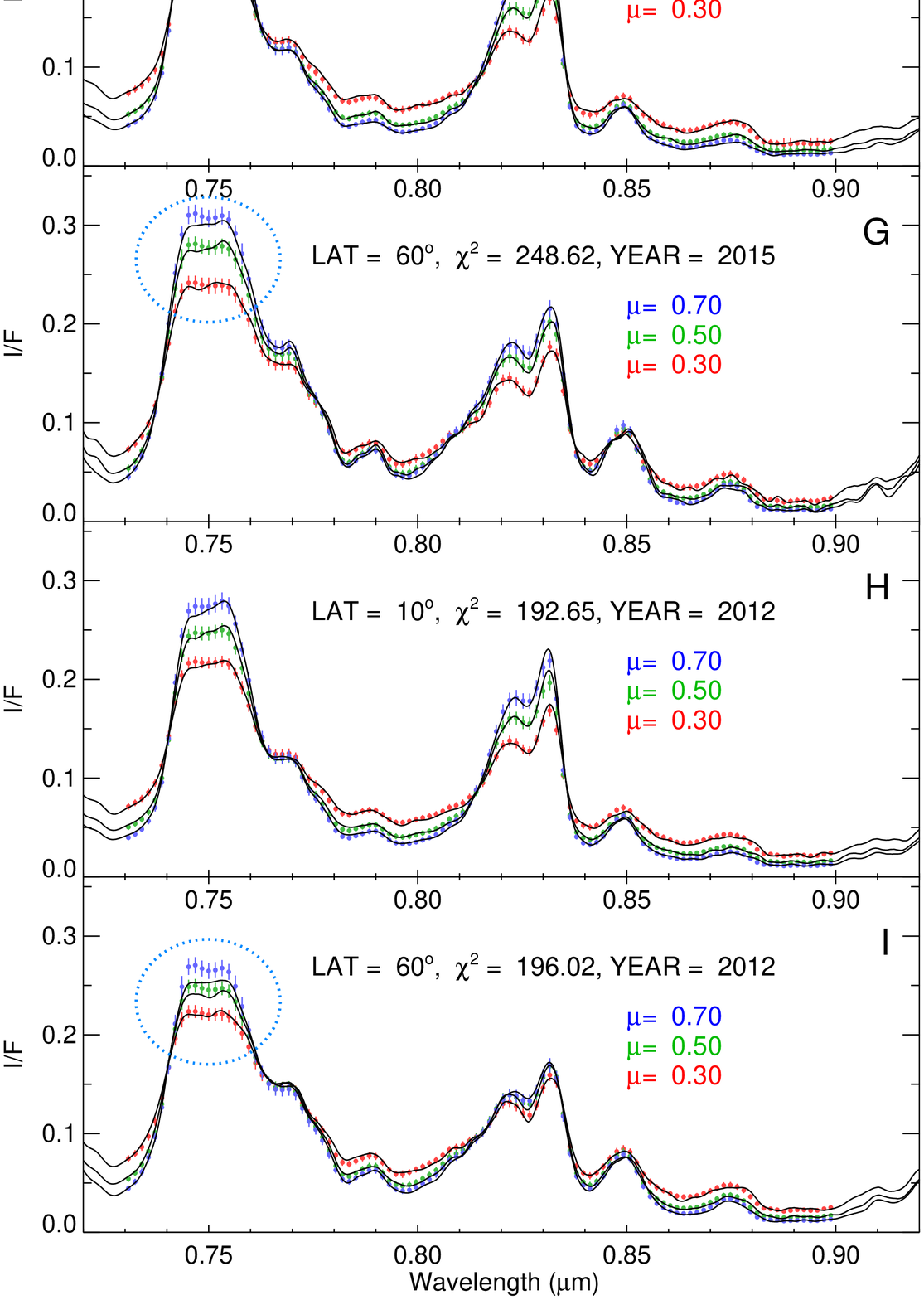}
%SOURCE: Figure produced by running case 25 plot_fitctl2015mie_5panel17.pro in ir_ms_ura on puck:
% See documentation on P 58-59 of Uranus Log L.
%\vspace{-0.2in}
\caption{Left: Single tropospheric Mie model fits as a function of latitude under the assumption that the methane VMR is
constant for pressures exceeding the condensation level. Parameter values are also
given in Table\ \ref{Tbl:mielat1}. Right: sample spectra, with blue dotted ovals identifying
regions of larger I/F errors. \label{Fig:mielat1}}
\end{figure*}

To illustrate the latitude dependence of the effective methane mixing ratio and the
inferred aerosol distribution we selected the simple 2-layer model using a compact
stratospheric haze and an extended diffuse layer of spherical tropospheric particles,
characterized by Mie scattering parameters of radius and refractive index.  
We also chose the small-radius solution set because
of their high quality fits and relative consistency between 2012 and 2015, as well as
their better agreement with HG fits as noted in the previous section.  Other
models show similar characteristics, except that they contain more variation between
years, as can be surmised from the table of fit parameters from fits at 10\degx N and 60\degx N,
shown in Table\ \ref{Tbl:mie_ls} for large Mie particle fits and in Table\ \ref{Tbl:hgpars} for HG model
fits.
We assumed a  methane profile
that has a vertically uniform fitted deep mixing ratio, a fitted relative humidity
immediately above the condensation level, a minimum relative humidity
of 30\%, with linear interpolation filling in values between the condensation level
and the tropopause.  Above the tropopause we assumed a mixing ratio equal to the
tropopause value.   From fitting spectra every 10\deg of latitude
for both 2012 and 2015 observations we obtained the best-fit
parameters and their formal uncertainties given in Table\ \ref{Tbl:mielat1}.  The
parameters are also plotted in  Fig.\ \ref{Fig:mielat1},
where panels A-E display the fit parameter values and their estimate errors, and
panels F-I display samples of model and observed spectra for 10\deg N and 60\deg N
for 2015 (F and G) and 2012 (H and I).    

\hspace{-1in}\begin{table*}\centering[!t]
\caption{Single tropospheric Mie layer fits to the 730-900 nm spectra
  as a function of latitude assuming vertically uniform \chf below the
  condensation level.\label{Tbl:mielat1}}
% SOURCE: This table is produced by running case 25 plot_fitctl2015mie_5panel17.pro in ir_ms_ura on puck:
% See documentation on P 58-59 of Uranus Log L.
\small
\vspace{0.1in}
\setlength\tabcolsep{2pt}
\begin{tabular}{ c c c c c c c c c c c}
   Lat.    &  $m1\_od$        &           &      $m2\_pt$ &    $m2\_pb$ &        $m2\_r$ &          &      $\alpha_0$    &              &        & \\[-0.03in]
   (\degx) &  $\times$100 &  $m2\_od$ &       (bar) &     (bar) &       (\mum) &   $m2\_nr$  &      (\%)     &       $ch4rhc$ & \chisq & YR\\[0.05in]
\hline
\\[-.15in]
%   PGLAT &    m1_od &    m2_od &    m2_pt &    m2_pb &     m2_r &     m2_n &    ch4v0 &   ch4rhc &    chisq &  YEAR\\[0.05in]
-10 &  2.4$\pm$0.8 &   3.60$\pm$1.37 &   1.09$\pm$0.04 &   2.66$\pm$0.22 &   0.22$\pm$0.09 &   1.65$\pm$0.31 &   2.93$\pm$0.37 &   0.75$\pm$0.14 &   180.61 &  2015\\[0.05in]
  0 &  4.5$\pm$0.8 &   2.52$\pm$0.67 &   1.07$\pm$0.04 &   2.55$\pm$0.20 &   0.25$\pm$0.08 &   1.72$\pm$0.24 &   2.69$\pm$0.38 &   0.61$\pm$0.14 &   137.93 &  2015\\[0.05in]
 10 &  2.8$\pm$0.8 &   3.07$\pm$0.88 &   1.13$\pm$0.04 &   2.46$\pm$0.22 &   0.34$\pm$0.10 &   1.55$\pm$0.16 &   3.14$\pm$0.45 &   0.68$\pm$0.13 &   148.39 &  2015\\[0.05in]
 20 &  2.8$\pm$0.7 &   1.99$\pm$0.58 &   1.08$\pm$0.05 &   2.55$\pm$0.22 &   0.28$\pm$0.11 &   1.75$\pm$0.29 &   2.85$\pm$0.39 &   0.97$\pm$0.18 &   170.32 &  2015\\[0.05in]
 30 &  3.8$\pm$0.8 &   1.48$\pm$0.49 &   1.06$\pm$0.05 &   2.60$\pm$0.20 &   0.27$\pm$0.13 &   1.81$\pm$0.36 &   2.10$\pm$0.24 &   0.88$\pm$0.19 &   170.86 &  2015\\[0.05in]
 40 &  2.8$\pm$0.9 &   1.41$\pm$0.36 &   1.01$\pm$0.04 &   2.65$\pm$0.17 &   0.27$\pm$0.10 &   1.79$\pm$0.28 &   1.41$\pm$0.12 &   0.75$\pm$0.20 &   205.54 &  2015\\[0.05in]
 50 &  2.4$\pm$1.1 &   1.25$\pm$0.42 &   1.01$\pm$0.04 &   2.51$\pm$0.14 &   0.26$\pm$0.16 &   1.88$\pm$0.46 &   1.13$\pm$0.09 &   0.76$\pm$0.22 &   266.49 &  2015\\[0.05in]
 60 &  0.1$\pm$70.7 &   1.45$\pm$0.29 &   1.02$\pm$0.02 &   2.53$\pm$0.13 &   0.25$\pm$0.09 &   1.86$\pm$0.30 &   0.99$\pm$0.08 &   0.31$\pm$0.18 &   248.62 &  2015\\[0.05in]
 70 &  0.4$\pm$13.4 &   1.49$\pm$0.26 &   1.01$\pm$0.03 &   2.71$\pm$0.14 &   0.23$\pm$0.09 &   1.90$\pm$0.33 &   0.88$\pm$0.07 &   0.36$\pm$0.21 &   278.56 &  2015\\[0.05in]
%   PGLAT &    m1_od &    m2_od &    m2_pt &    m2_pb &     m2_r &     m2_n &    ch4v0 &   ch4rhc &    chisq &  YEAR\\[0.05in]
\hline
\\[-.15in]
-20 &  1.4$\pm$0.7 &   3.14$\pm$1.19 &   1.11$\pm$0.04 &   2.71$\pm$0.22 &   0.24$\pm$0.09 &   1.66$\pm$0.29 &   2.87$\pm$0.36 &   0.77$\pm$0.13 &   137.66 &  2012\\[0.05in]
-10 &  4.4$\pm$0.7 &   3.28$\pm$1.22 &   1.04$\pm$0.05 &   2.77$\pm$0.23 &   0.25$\pm$0.09 &   1.64$\pm$0.27 &   2.63$\pm$0.32 &   1.17$\pm$0.21 &   147.88 &  2012\\[0.05in]
  0 &  4.7$\pm$0.8 &   2.51$\pm$0.64 &   1.08$\pm$0.04 &   2.56$\pm$0.21 &   0.25$\pm$0.08 &   1.73$\pm$0.24 &   2.69$\pm$0.37 &   0.58$\pm$0.12 &   150.60 &  2012\\[0.05in]
 10 &  3.0$\pm$0.7 &   2.52$\pm$0.64 &   1.07$\pm$0.04 &   2.37$\pm$0.20 &   0.25$\pm$0.09 &   1.74$\pm$0.26 &   3.16$\pm$0.50 &   0.95$\pm$0.16 &   192.65 &  2012\\[0.05in]
 20 &  2.6$\pm$0.7 &   3.42$\pm$1.14 &   1.06$\pm$0.05 &   2.63$\pm$0.21 &   0.27$\pm$0.08 &   1.62$\pm$0.22 &   2.65$\pm$0.33 &   0.95$\pm$0.17 &   197.68 &  2012\\[0.05in]
 30 &  2.7$\pm$0.7 &   1.98$\pm$0.51 &   1.02$\pm$0.05 &   2.69$\pm$0.19 &   0.26$\pm$0.08 &   1.74$\pm$0.24 &   1.99$\pm$0.21 &   0.89$\pm$0.18 &   149.64 &  2012\\[0.05in]
 40 &  2.8$\pm$1.0 &   2.07$\pm$0.51 &   1.06$\pm$0.04 &   2.42$\pm$0.15 &   0.32$\pm$0.08 &   1.57$\pm$0.14 &   1.29$\pm$0.12 &   0.62$\pm$0.18 &   255.27 &  2012\\[0.05in]
 50 &  0.4$\pm$1.55 &   1.94$\pm$0.39 &   1.07$\pm$0.03 &   2.47$\pm$0.13 &   0.32$\pm$0.06 &   1.59$\pm$0.12 &   1.03$\pm$0.08 &   0.30$\pm$0.17 &   191.32 &  2012\\[0.05in]
 60 &  2.2$\pm$1.6 &   1.10$\pm$0.23 &   1.02$\pm$0.04 &   2.22$\pm$0.13 &   0.24$\pm$0.07 &   1.81$\pm$0.25 &   0.93$\pm$0.08 &   0.42$\pm$0.20 &   196.02 &  2012\\[0.05in]
 70 &  1.8$\pm$2.1 &   1.00$\pm$0.12 &   1.01$\pm$0.03 &   2.23$\pm$0.12 &   0.19$\pm$0.09 &   1.97$\pm$0.36 &   0.97$\pm$0.08 &   0.29$\pm$0.19 &   235.18 &  2012\\[0.05in]
\hline
\end{tabular}
\normalsize
\parbox{6.in}{\vspace{0.1in} NOTE: The optical depths are for a wavelength of 0.5 \mumx. 
These fits used 318 points of comparison and fit 8 parameters, for a nominal value of NF=310, 
for which the normalized \chisq /NF
ranged from 0.44 to 0.90.}
\end{table*}

Most of the model parameters are found to have only weak variations
with latitude.  The top pressure of the sole tropospheric cloud
layer is surprisingly invariant from low to high latitudes as well
as from 2012 to 2013, even though there are substantial variations
in optical depth between years as well as with latitude.  This
boundary pressure is also very well constrained by the observations.
The bottom pressure of this cloud is more variable, but its variation
is not much more than its uncertainty which is much larger than that
of the cloud top pressure.  The larger uncertainty is consistent with
the derivative spectra given in Fig.\ \ref{Fig:miederiv}, which
shows that, compared to the top pressure, the bottom pressure has a smaller fractional effect
on the I/F spectrum for a given fractional change in pressure.

The most prominent latitudinally varying parameter is
the effective deep methane mixing ratio, which attains a low-latitude maximum of
about 3.15\%, dropping to about 2\% by 30\deg N, reaching a high-latitude
value of about 1\% at between 50\degx N and 60\degx N.  Close behind, is the
variation in methane humidity above the condensation level, which was
found to be 60-100\% at low latitudes, declining to about 30-40\%
for regions poleward of 50\deg N. This decline towards the north pole
is also seen in other model types as well.

 There is also close agreement, for this model, 
between between 2012 and 2015 results for both the extremes in the
 methane mixing ratio
and in its latitudinal variation.  The slight dip at the equator is 
also present in results for both years, as is the peak at 10\degx N.  The agreement of
 the 2012 and 2015 methane profiles (on both the deep mixing ratio and
the above cloud humidity) is close enough that we must look elsewhere
to explain the brightening of the polar region between 2012 and 2015.
The most likely aerosol change responsible for the polar brightening is
the increase in the bottom cloud layer optical depth ($m2\_od$) by about
60\% at latitudes north of 50\degx, a factor already discussed in
 Section \ref{Sec:2mieu}.  However, because multiple aerosol parameters
differ between 2012 and 2015, it is useful to show that the combined
effect of layer m2 parameter changes does indeed result in the increased
scattering that produced the observed brightness increase. This was
done by starting with  the model spectrum for 2012 and computed a new model spectrum in
which {\it only the layer-m2 parameters} were changed to match those of
2015, leaving other parameters unchanged.  We also computed the
spectrum change when only the optical depth of the m2 layer was
changed to the 2015 value.  We did this at latitudes of 50\degx N,
60\degx N, and 70\degx N.  The results are summarized in the following
figures.  The left-hand figure provides a sample spectral view at
60\deg N.  It shows the measured spectral difference between 2012 and
2015 as a shaded curve, with shading range indicating
uncertainties.  Also shown are the difference in model fits (+), the
difference due only to layer m2 differences ($\times$), and the
difference due only to the change in optical depth (o).
The right hand plot displays the latitude dependence for two
pseudo-continuum wavelengths.  Again are shown the measured differences
(shaded curves), the model difference ($\times$), and the brightness
change due only to layer m2 (+).  This figure shows that {\it layer m2 is clearly
responsible for the vast majority of the brightness increase between 2012 and 2015}, but
changes in the m2 layer optical depth are only responsible for about half of
the total scattering increases of that layer (as in the left hand
plot), except at 50\degx N, where even though the optical depth decreased, the
layer still brightened because of changes in particle size and
refractive index).

\begin{figure*}\centering
\includegraphics[width=3.1in]{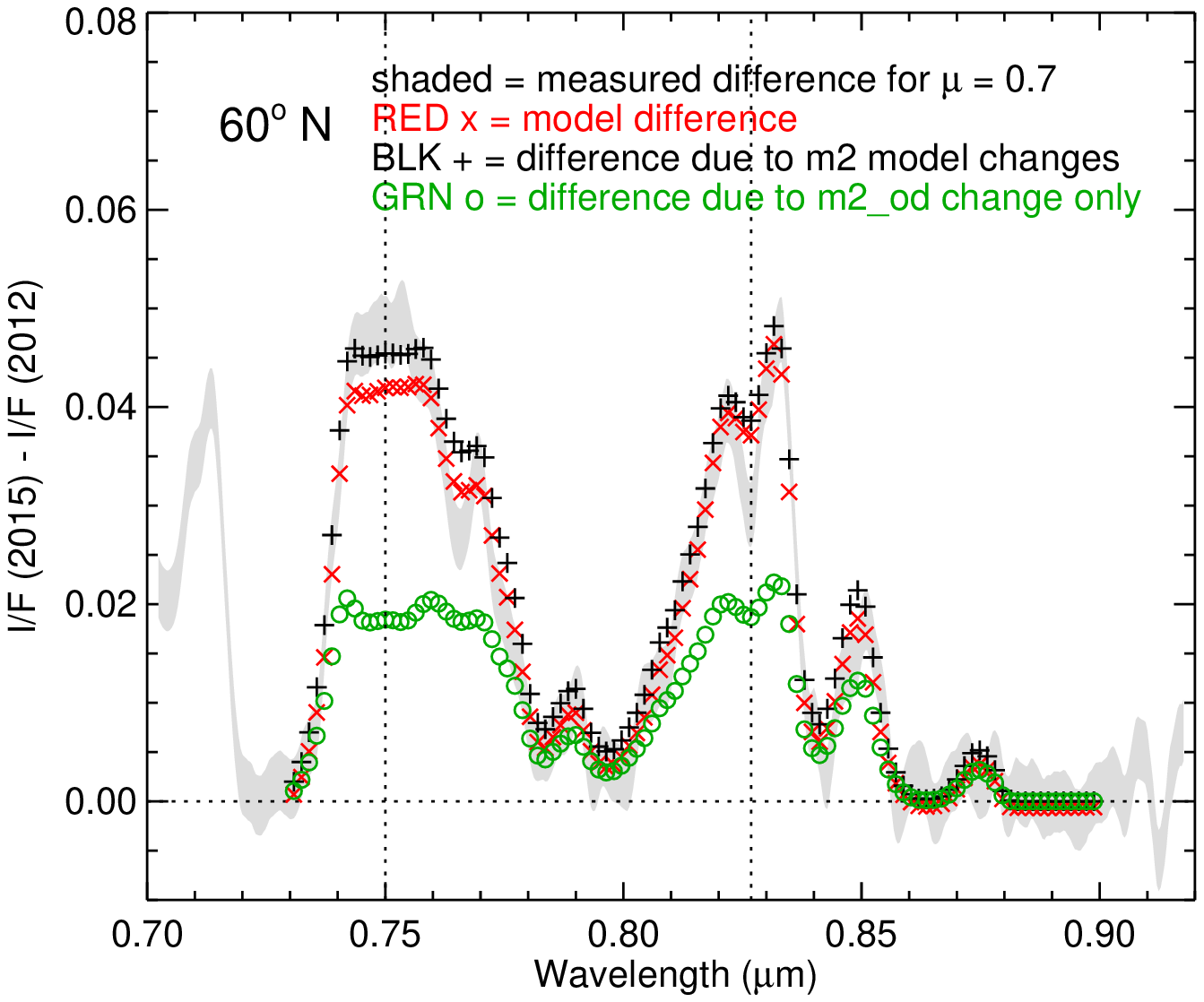}
\includegraphics[width=3.1in]{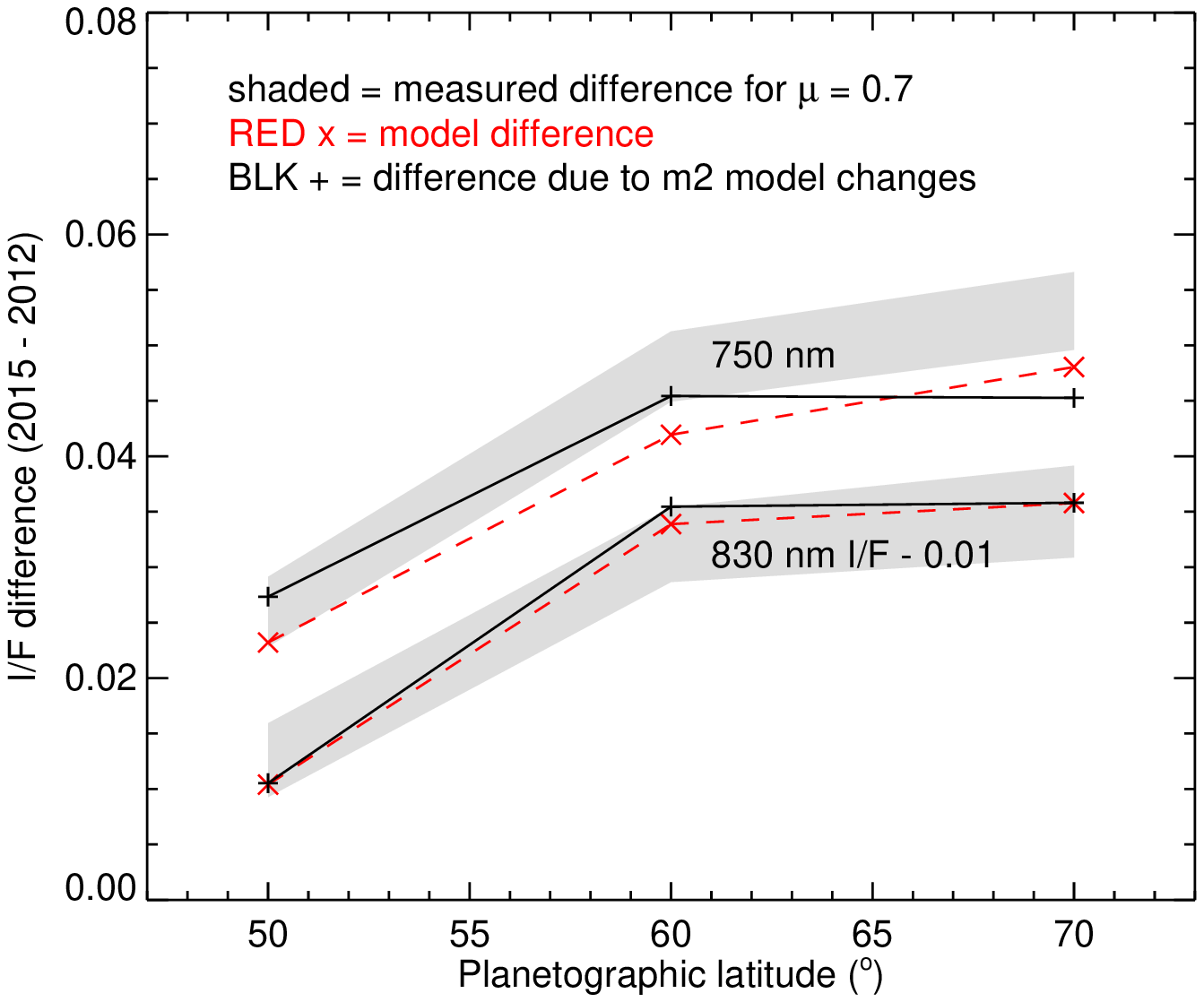}
\caption{Left: spectral difference at 60\degx N between 2012 and 2015 observations at a zenith angle cosine of 0.7 (shaded curve)
compared to all model differences ($\times$), to those contributed only by layer m2 (+), and
to those due only to the m2 optical depth change (o). Right: latitudinal variation of
observed temporal differences at 750 nm (upper shaded curve) and 830 nm (lower shaded curve offset
by 0.01), compared to total model differences ($\times$) and differences due to all changes
in layer m2 only (+).  This shows that increased scattering by layer m2 is primarily
responsible for the observed brightening of the polar region between 2012 and 2015.}\label{Fig:timediff}
\end{figure*}

At low latitudes, the fit quality for both years is better than
expected from our uncertainty estimates, but fit quality decreases significantly
at high northern latitudes, especially for the 2015 fits, which have
increased aerosol scattering. The high latitude fitting problem is
most obvious just short of 750 nm, as shown in panels G and I of
Fig.\ \ref{Fig:mielat1}, where the model values exceed the
measured values (note the encircled regions).  This problem can be
greatly reduced by using an altered vertical profile of methane, as
discussed in Section 9.3.

\subsubsection{Summary of uniform methane results}

\begin{figure*}[!hbt]\centering
\includegraphics[width=5.2in]{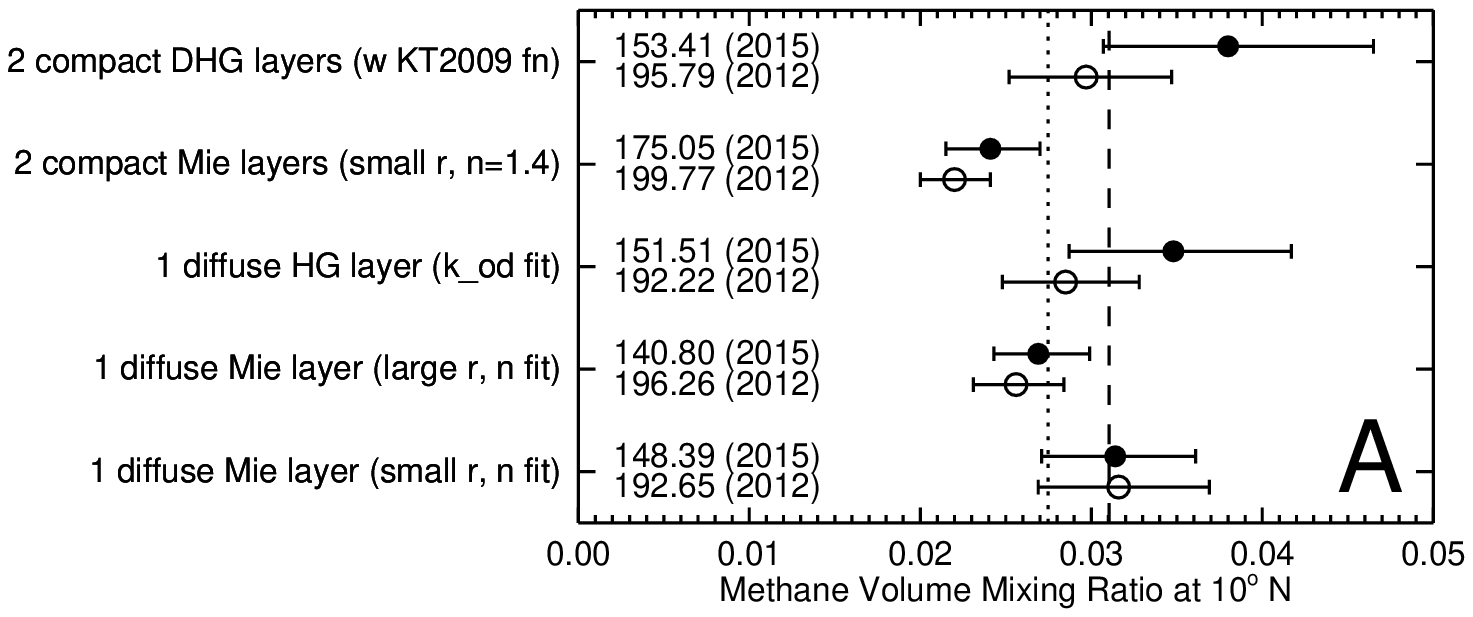}
\includegraphics[width=5.2in]{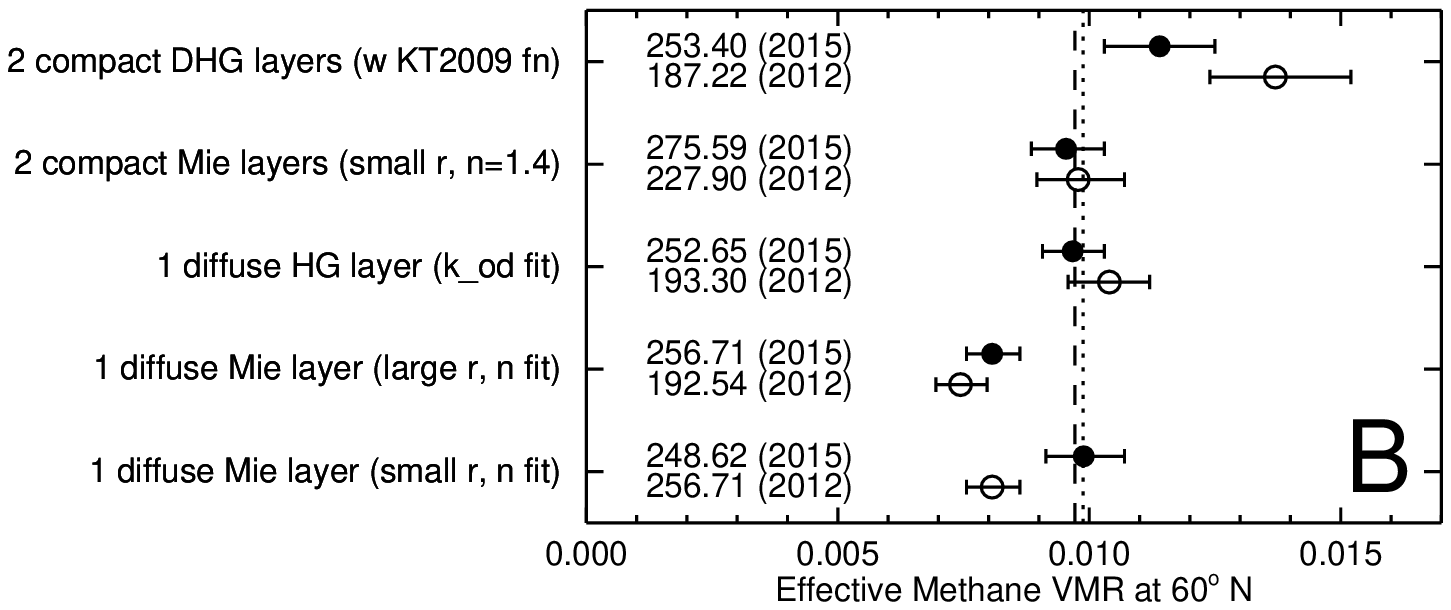}
%SOURCE Created on puck with plot_model_comparisons.pro using casenum=3 (top)
% and casenum=4 (bottom).  See Page 57, Uranus Log L
\caption{Effective deep methane VMR for different aerosol model parameterizations
at 10\deg N (A) and 60\deg N (B).  Vertical lines show unweighted
mean values for 2015 (dashed) and 2012 (dotted).}
\label{Fig:uniformch4}
\end{figure*}

Both spherical particle and HG models for the upper tropospheric layer lead to
declining effective methane volume mixing ratios with latitude by similar
factors, but are in some disagreement with respect to magnitudes,
as shown in greater detail in Fig.\ \ref{Fig:uniformch4}.  
The more detailed latitudinal fit results in Fig.\ \ref{Fig:mielat1} for the
small-particle solution, show that the effective methane mixing ratio peaks
near 10\deg N in both years, has a local minimum at the equator
and declines with latitude by more than a factor of two
by 50-60\deg N.  For each year, the two aerosol models lead
to similar shapes, and in the 50-70\deg range the two models
agree that there is a crossover in which the 2015 vmr declines
from 50\deg to 70\degx, while the 2012 vmr rises slightly over
the same interval.  

%Compute averages for 20\deg S to 20\deg N
%and averages for 50\deg N to 70\deg N and compare (perhaps tabulate).

The fitted values of the methane relative humidity just above the condensation level,
shown in  Fig.\ \ref{Fig:uniformch4}B, have considerable uncertainty.  
But both results indicate a peak near 20\deg N, a
clear local minimum near the equator, and a strong decline towards the
north pole. This is suggestive of rising motions near 20\deg and descending
motions near the equator and poles, with the latter being more significant.

\subsubsection{The effects of particle absorption on derived methane amounts}

The modeling results presented so far are for conservative particles
($\varpi$ = 1.0).  Particles that absorb some fraction of the incident
light will act to darken the atmosphere and reduce the amount of
methane needed to fit the spectrum.  This is true even if the
particles are not distributed vertically in the same fashion as
methane, and even though they lack the band structure of methane. The
aerosol optical depths and derived pressure locations of the layers
are also altered. To investigate the magnitude of these effects we did
fits of the 2 Mie layer model to the 2015 STIS observations, under the
assumption of vertically uniform methane, but with the imaginary index
of the tropospheric layer increased from zero to 0.0049, which, for a
0.3-\mum radius particle with a real refractive index of 1.8
corresponds to a decrease in single scattering albedo at 0.8 \mum from
$\varpi$ = 1.0 to $\varpi \approx$ 0.979.  This amount of absorption
in the 730 to 900 nm part of the spectrum, makes it possible to fit
the entire spectrum (down to 540 nm) if the particles are assumed to
be conservative at the shorter wavelengths (see Sec. \ref{Sec:wdep}
for more information).  Table \ref{Tbl:ssa} shows that adding this
amount of absorption changes the layer-2 top pressures by just
fractions of a percent, while increasing the bottom pressures by
18-22\%.  The optical depth of the layer changes in less consistent
directions. If the particle's refractive index and size did not change
much, then an increase in optical depth would be required to make up
for the lower single-scattering albedo produced by
absorption. However, the optical depth is seen to decrease at 10\degx
N, where $m2\_nr$ has increased by almost 10\% (and $m2\_nr$ - 1 by 28\%),
increasing the scattering efficiency substantially. For 60\degx N, the
changes in $m2\_r$, $m2\_nr$, and $m2\_od$ are all substantially smaller.
Most importantly, the effective deep methane mixing ratio is decreased
by 3\% at 10\degx N and 7\% at 60\degx N, which suggests that a fair
approximation of the latitudinal profile for absorbing cloud particles
can be obtained by scaling the profile we derived from conservative
scattering.  Whether the cloud particles are actually absorbing in the
730-900 nm region remains to be determined.

 For the large-particle solution, we made a similar comparison, but
just for 10\deg N and for 2012.  In this case the increase
of imaginary index needed to adjust the wavelength dependence (as
described above for the small-particle solution) is only from 0 to
6.2$\times 10^{-4}$, which decreases the single-scattering albedo for
a 1.535 \mum particle with real index 1.225 
to  $\varpi =$ 0.990 at 0.8 \mumx.  Although this seems like
a small change, it produces a 50\% increase in optical depth,
a 56\% increase in the cloud bottom pressure, and a 10.6\% decrease
in the best-fit methane mixing ratio, as shown in Table\ \ref{Tbl:ssalp}. 
For non-spherical particles in which wavelength-dependent optical depths
or wavelength dependent phase functions might be used to adjust
the wavelength dependent I/F spectrum, there may be no need for absorbing
particles, in which case the somewhat higher methane mixing ratios
may apply.

\begin{table*}\centering
\caption{Changes in small-particle best-fit parameter values derived from the STIS
  2015 observations, as a result of adding absorption to aerosol layer
  2 by increasing $m2\_ni$ from 0.0 to 0.005.}\label{Tbl:ssa}
%SOURCE: Table created on puck using plot_model_comparisons.pro with casenum=6
% See page 67, Uranus Log L
\vspace{0.15in}
\begin{tabular}{|c | c c c | c c c |}
\hline
      &\multicolumn{3}{| c|}{10\degx N Latitude} &\multicolumn{3}{| c|}{60\degx N Latitude} \\
Parameter  &  Value    &   Value    &           &  Value     & Value         &   \\
Name       & $m2\_ni$ = 0 & $m2\_ni$ = 0.005 & Difference &  $m2\_ni$ = 0 & $m2\_ni$ = 0.005 & Difference \\
\hline
%            m1\_od&  0.028&  0.029&  3.84\%&  0.000&  0.004&752.27\%\\
            $m2\_od$&  3.084&  2.864& -7.15\%&  1.445&  1.482&  2.54\%\\
      $m2\_pt$ (bar)&  1.126&  1.127&  0.14\%&  1.023&  1.032&  0.87\%\\
      $m2\_pb$ (bar)&  2.450&  2.993& 22.12\%&  2.519&  2.968& 17.79\%\\
     $m2\_r$ (\mumx)&  0.342&  0.307&-10.16\%&  0.248&  0.256&  2.89\%\\
     $m2\_nr$ &  1.554&  1.706&  9.77\%&  1.862&  1.900&  2.05\%\\
 $\alpha_0$ $\times$100&  3.160&  3.060& -3.16\%&  0.989&  0.916& -7.38\%\\
            $ch4rhc$&  0.687&  0.701&  2.04\%&  0.318&  0.355& 11.64\%\\
            \chisq& 148.03& 148.85&  0.55\%& 248.29& 246.28& -0.81\%\\
\hline
\end{tabular}
\end{table*}

\begin{table}\centering
\caption{Changes in large-particle best-fit parameter values derived from the STIS
  2015 observations, as a result of adding absorption to aerosol layer
  2 by increasing $m2\_ni$ from zero to 6.2$\times 10^{-4}$.}\label{Tbl:ssalp}
%SOURCE Created on puck with plot_model_comparisons.pro using casenum=7
% See page 100, Uranus Log L for documentation
\vspace{0.15in}
\begin{tabular}{|c | c c c |}
\hline
      &\multicolumn{3}{| c|}{10\degx N Latitude}  \\
Parameter  &  Value    &   Value    &          \\
Name       & $m2\_ni$ = 0 & $m2\_ni$ = 6.2$\times 10^{-4}$ & Difference \\
\hline
%            m1\_od&  0.028&  0.028&  1.36\% \\
            $m2\_od$&  6.088&  9.124& 49.87\% \\
      $m2\_pt$ (bar)&  1.094&  1.112&  1.64\% \\
      $m2\_pb$ (bar)&  2.675&  4.183& 56.39\% \\
     $m2\_r$ (\mumx)&  1.535&  1.597&  4.01\% \\
     $m2\_nr$ &  1.225&  1.243&  1.42\% \\
 $\alpha_0$ $\times$100&  2.560&  2.290&-10.55\% \\
            $ch4rhc$&  0.879&  0.877& -0.23\% \\
            \chisq& 196.26& 193.13& -1.59\% \\
\hline
\end{tabular}
\end{table}

%\clearpage

\subsection{Fitting latitude-dependent vertically non-uniform methane depletion models}

\subsubsection{Alternative models of vertically varying methane}

The fits discussed in previous sections have assumed that the methane
profile is vertically uniform from the bottom of our model atmosphere
all the way up to the methane condensation level.  We have already
noted problems with those fits in the 750 nm region of the spectrum,
which suggest that the methane mixing ratio likely increases with
depth at high latitudes. There are also independent physical arguments
suggesting the same characteristic.  \cite{Sro2011occult} pointed out that
extending the very low high latitude mixing ratios to great depths
would result in horizontal density gradients over great depths.  As a
consequence of geostrophic and hydrostatic balance, these gradients
would lead to vertical wind shears \citep{Sun1991}.  This would result
in an enormous wind difference with latitude at the visible cloud
level, which would be inconsistent with the observed winds of Uranus.
Thus, we would expect that the polar depletion would be a relatively
shallow effect, as we have inferred from our previous work
\citep{Kark2009IcarusSTIS,Sro2011occult,Sro2014stis}. As indicated by
KT2009, the 2002 spectral observations did not require that methane
depletions extend to great depths, and \cite{Sro2011occult} showed
that shallow depletions were preferred by the 2002 spectra.  This was
further supported by \cite{DeKleer2015}, who used our descended profile
parameterization, fixed the shape parameter at $vx=2$, and constrained
the depth parameter vs latitude using H band spectra.  They found
a clear latitude trend, with a low-latitude value of 1.7$\pm$0.2 bars,
increasing to 3.2$\pm$1 bars in the 40--50\degx N band, and
as deep as 26$^{+11}_{-18}$ bars in the 60--70\degx N band, although at that
extreme value the depth parameter is constrained more by the shape of the
profile at much lower pressures than by any direct sensing of sunlight
reflected from the 26-bar level.  

From the previous discussion, we
expect a reasonable physical model has some pressure value $P_d$
for which the methane mixing ratio is independent of latitude for $P >
P_d$, but allows a decline in mixing ratio with latitude for $P < P_d$.
We assume that the highest mixing ratio we observe at low latitudes
(which turns out to be at 10\degx N) is representative of the deep
mixing ratio and assume all of the variation with latitude is a
depletion relative to that level.  Here we describe the results of
fitting two alternative vertically varying depletion models: the
descended profile model described in Fig.\ \ref{Fig:descend}A and
Eq.\ \ref{Eq:deplete}, and the step function depletion described in
Fig.\ \ref{Fig:descend}B.  Both options result in improved fit quality
at high latitudes, with depletions confined to the upper troposphere.

\begin{figure*}[!htb]\centering
\includegraphics[width=5.25in]{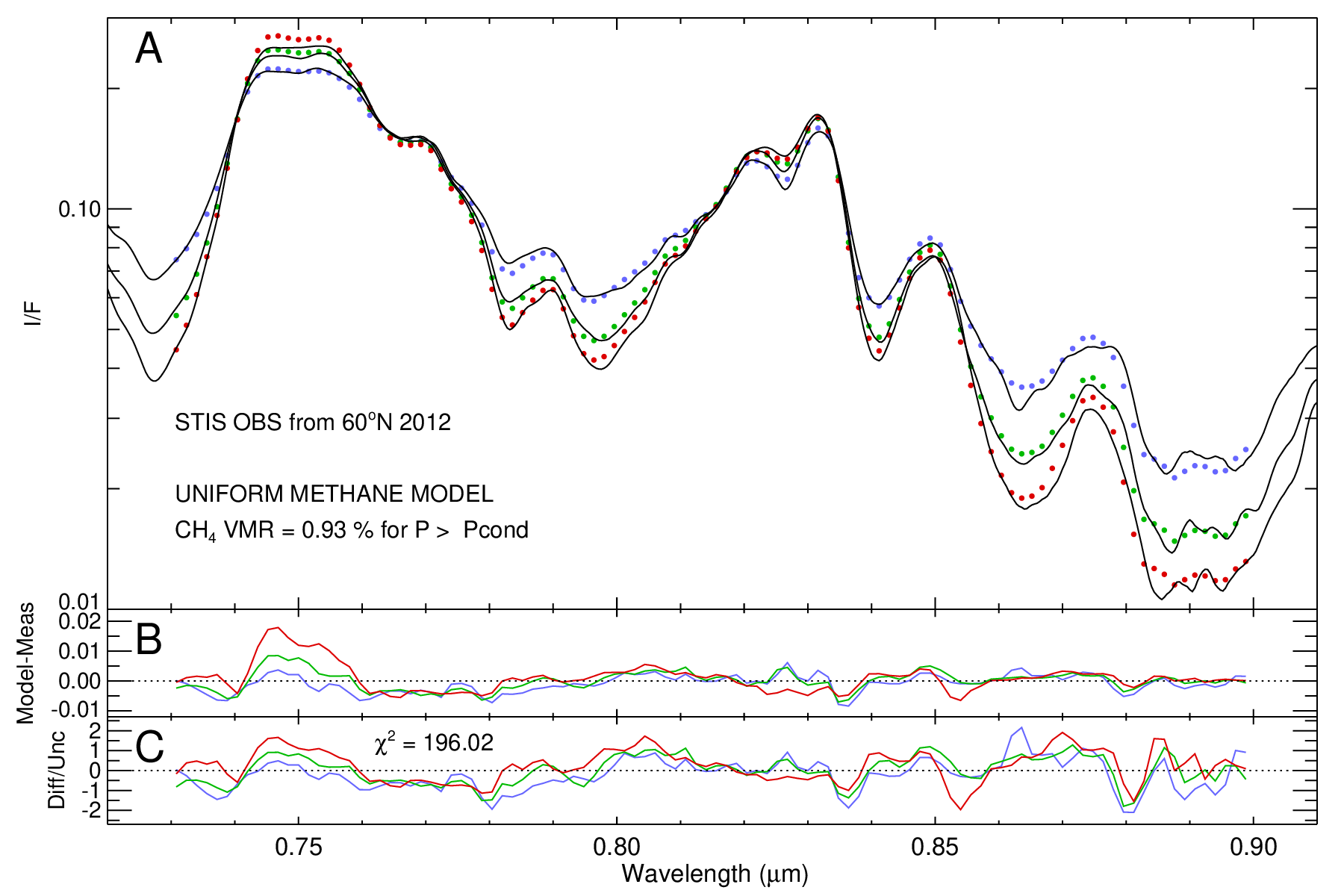}
\includegraphics[width=5.25in]{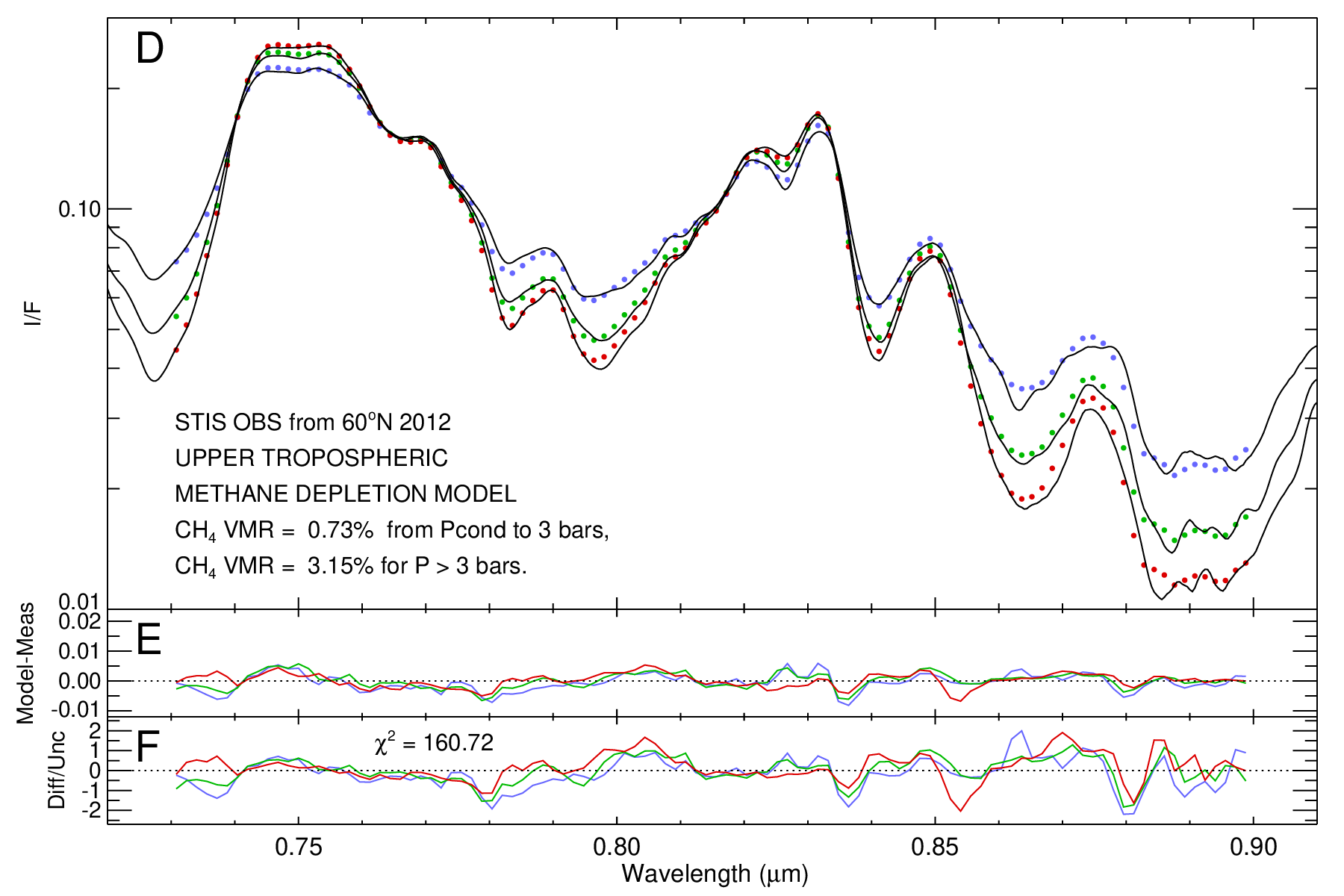}
%SOURCE: Plotted on puck with plot_spec_comparisons.pro using casenum = 2
% casenum 1 documented on page 53 Log L, casenum 2 on page 56
%\vspace{-0.15in}
\caption{Detailed comparison of 60\degx N 2012 STIS observations with  best-fit model spectra
assuming vertically uniform methane VMR (A-C) and with model calculations assuming
a step-function change in methane VMR (D-F), where observations are plotted as continuous
curves and models as colored points, using red, green, and blue for $\mu$ values of
0.3, 0.5, and 0.7 respectively. Below each spectral plot are plots of model minus
observation (B and E) and the same difference divided the expected uncertainty (C and F). Methane
profiles are described in the legends.
\label{Fig:step60}}
\end{figure*}

We first consider the stepped depletion model shown in
Fig.\ \ref{Fig:descend}B because it is easier to constrain
its bottom boundary at all latitudes.  A more detailed look at the 60\degx N
spectrum from 2012 in comparison with a model fit using a vertically
uniform methane mixing ratio is shown in Fig.\ \ref{Fig:step60}A-C,
while our best fit model for the stepped methane profile is displayed
in Fig.\ \ref{Fig:step60}D-F.  Here we assume that the deep mixing
ratio is equal to the 10\degx N best fit uniform VMR value of 3.14\%,
and optimize the depleted mixing ratio $\alpha_1$ and the depth of the
depletion $P_d$ to minimize \chisqx.  The result is seen to be a
substantial improvement of the fit in the 750-nm region, with minor
improvements in other areas, with an overall significant reduction in
\chisq for the entire fit from 196.02 to 160.72.  The fact that the
difference plots show strong features in the vicinity of large slopes
in the spectrum, particularly at 0.88 \mumx, suggests that there may
be a slight error in the STIS wavelength scale.  If we move the
observed spectrum just 0.24 nm towards shorter wavelengths, these
\chisq values can be reduced to 170.02 and 137.93 respectively. (Although
the STIS wavelengths are very accurate up to 653 nm because of the availability of
 numerous Fraunhofer calibration lines, longer wavelengths require
extrapolation that allows errors of this size.) 
 The best fit values for the methane profile parameters are $ch4vx$ =
0.73$\pm$0.08\% and $P_d$ = 3.0$^{+3.5}_{-1.5}$ bars.  The methane
value is a little below the 0.93$\pm$0.08\% for the uniform mixing
ratio model, as expected.  The methane relative humidity above the
condensation level was found to be 95$\pm$16\% for the uniform case
and 67$\pm$32\% for the upper tropospheric depletion case. The
uncertainty in the depth of depletion ($P_d$) is much larger on the
high side because the sensitivity to that parameter decreases with
depth.

We also tried fits with the descended depletion function described in
Fig.\ \ref{Fig:descend}A and Eq.\ \ref{Eq:deplete}, which is defined
by a shape parameter $vx$ and a depth parameter $P_d$.  We found the
depth parameter difficult to constrain because the rate of change
of mixing ratio with depth can be come quite small for large depths
due to the shape of the function.  However, fixing $P_d$ at 5 bars, and using just the
shape parameter to control the depletion, we
obtained a \chisq value of 167.34 and a shape parameter of $vx$ = 1.22$\pm$0.54,
fitting the same 2012 60\degx N observation as in Fig.\ \ref{Fig:step60}.
Thus, a descended depletion fit is also viable, and probably a more realistic
vertical variation than the step function.  The advantage of the step function
is that both parameters can be fit without too much difficulty.

%\clearpage

\subsubsection{Latitude dependent fits with a stepped depletion of methane}

\begin{figure*}\centering
\hspace{-0.15in}
\includegraphics[width=3.21in]{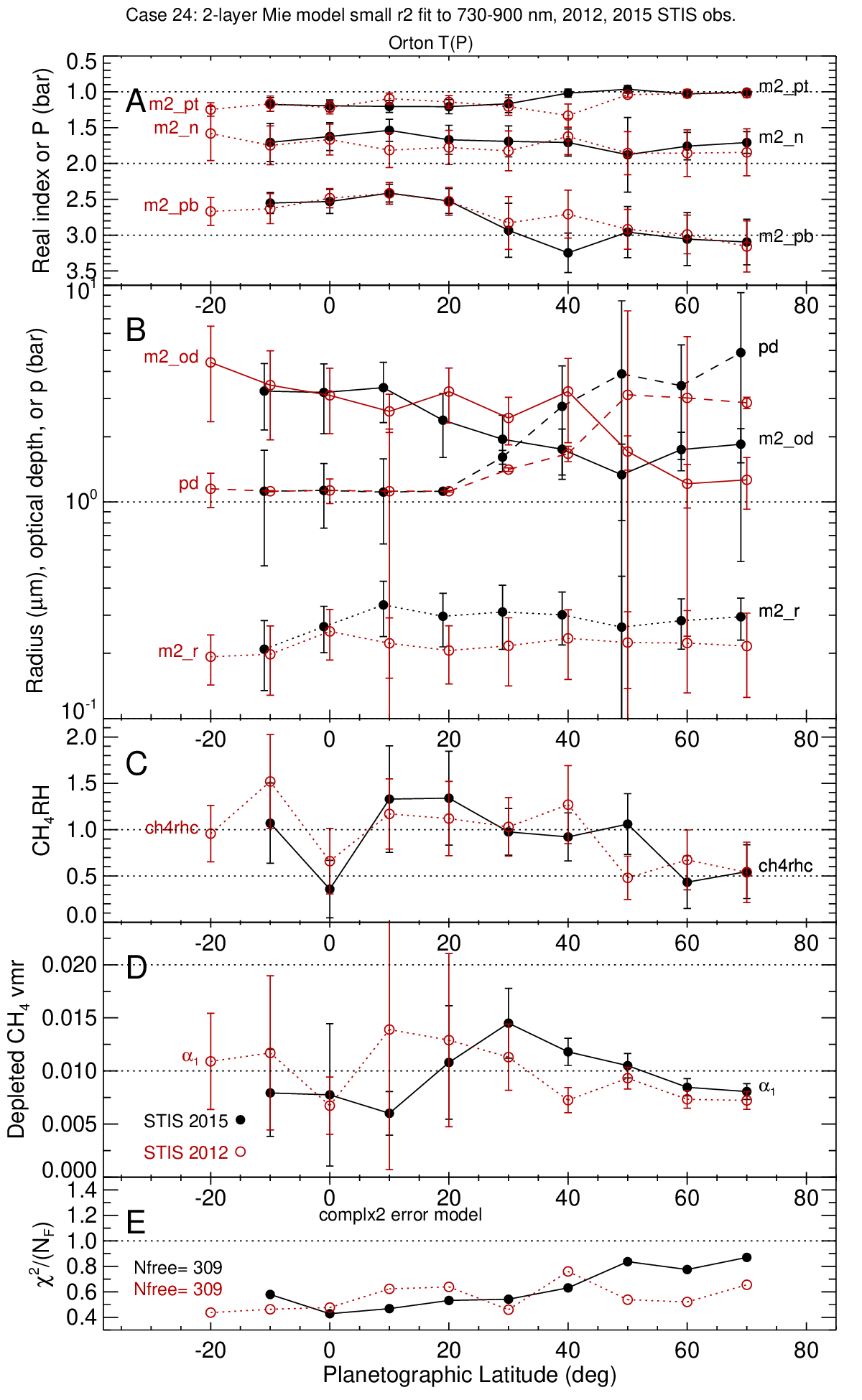}
\hspace{-0.17in}
\includegraphics[width=3.21in]{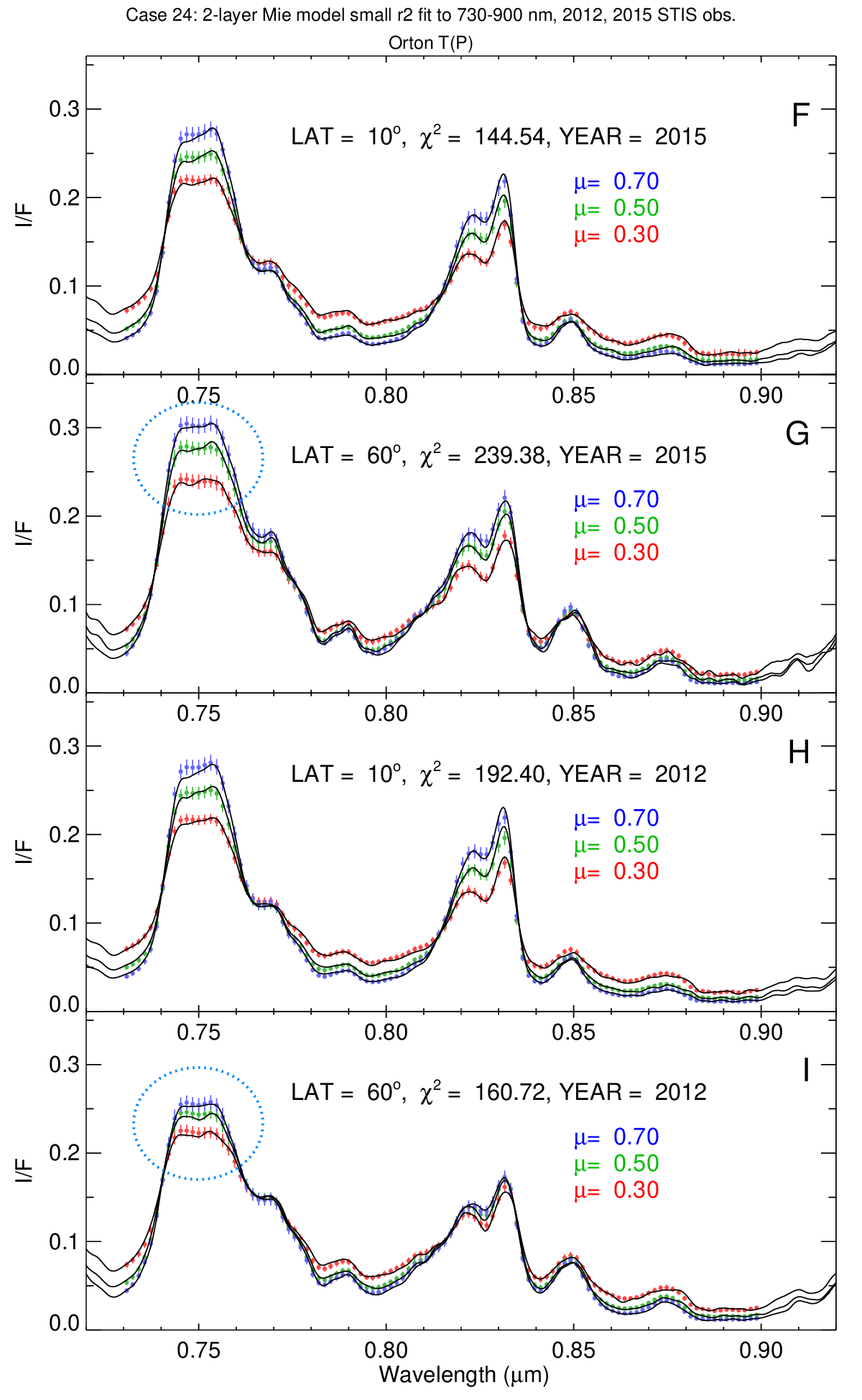}
%SOURCE Page 55 Log L, plotted with plot_fitctl2015mie.pro on puck using case=17 in ir_ms_ura
\caption{Stepped depletion model of vertical methane distribution
fit to STIS spectra from 2012 and 2015.  
  Conservative cloud model and gas profile parameters for a
  Mie-scattering haze above a single diffuse Mie-scattering tropospheric
  layer, assuming a deep mixing ratio of $\alpha_0$ = 0.0315, and a
  methane profile characterized by a pressure depth parameter $P_d$ and
  a depleted mixing ratio  $\alpha_1$ (defined in Fig.\ \ref{Fig:descend}) and
  constrained by spectral observations from 730 nm to 900 nm.  The
  parameter values are in panels A-E, with red (open circle) points displaying
  results of fitting 2012 STIS observations and black (filled circle) points
displaying the results of
  fitting 2015 observations.  Sample comparisons between measured and
  large-particle model spectra are in panels F-I. Note the great
  improvement in the high latitude fits near 745 nm, compared to
  results given in Fig.\ \ref{Fig:mielat1}. }
\label{Fig:steplat}
\end{figure*}

Here we describe the results of assuming a stepped depletion of
methane, parameterized by one fixed parameter ($\alpha_0$, the deep
methane VMR, which is set to 0.0315) and two adjustable parameters
($P_d$, the pressure at which the step occurs and $\alpha_1$, the
decreased mixing ratio between that level and the condensation level
(which is a function of the decreased methane VMR).  In addition to
fitting these two parameters, we fit the usual aerosol parameters and
the methane relative humidity above the condensation level, resulting
in a net increase of one adjustable parameter, for a new total of
nine.  The best-fit parameter values and their uncertainties are given
at 10\deg latitude intervals for both 2012 and 2015 in
Table\ \ref{Tbl:steplat}.  These are plotted versus latitude in the
left column of Fig.\ \ref{Fig:steplat} and comparisons of model and
observed spectra are displayed in the right column.

The best-fit methane depletion depth parameter values are shown by
dashed lines in panel B of Fig.\ \ref{Fig:steplat} for $P_d$ and by
dotted lines in panel D for $\alpha_1$.  At high latitudes the latter
is near 0.8\%, and increases somewhat at low latitudes, but becomes
very uncertain at low latitudes, which is a result of having less and
less influence on the spectrum as the depth of the depletion decreases
towards the condensation level.  As shown in panel D, the depletion
depth is in the 3-5 bar range from 70\degx N down to about 50\degx N,
and then declines to nearly the condensation level by 20\degx N, and
at low latitudes there is almost no depletion.  The improvement in fit
quality is significant at high latitudes.

\begin{table*}[!t]\centering
\caption{Single tropospheric Mie layer fits to the 730-900 nm spectra as a function of latitude assuming stepped depletion of \chf below the condensation level.\label{Tbl:steplat}}
% SOURCE: This table is produced by running case 25 plot_fitctl2015mie_5panel17.pro in ir_ms_ura on puck:
% See documentation on P 58-59 of Uranus Log L.
\vspace{0.1in}
\small
\setlength\tabcolsep{2pt}
\begin{tabular}{ c c c c c c c c c c c c}
   Lat.    &  $m1\_od$        &           &    $m2\_pt$ &    $m2\_pb$ &      $m2\_r$ &          &           &  $\alpha_1$   & $P_d$     &        & \\[-0.03in]
   (\degx) &  $\times$100 &    $m2\_od$ &       (bar) &     (bar) &     (\mum) &   $m2\_nr$  &    $ch4rhc$ & (\%) & (bar) & \chisq & YR\\[0.05in]
\hline
\\[-.15in]
%   PGLAT &   100 x m1\_od &    m2\_od &    m2\_pt &    m2\_pb &     m2\_r &     m2\_n &   ch4rhc &       vx &       pd &    chisq &  YEAR\\[0.05in]
-10 & 3.5$\pm$1.3 &   3.25$\pm$1.10 &   1.18$\pm$0.10 &   2.55$\pm$0.15 &   0.21$\pm$0.07 &   1.71$\pm$0.27 &   1.07$\pm$0.43 &   0.79$\pm$0.41 &   1.12$\pm$0.61 &   178.88 &  2015\\[0.05in]
  0 & 3.1$\pm$2.0 &   3.20$\pm$1.14 &   1.19$\pm$0.08 &   2.53$\pm$0.17 &   0.27$\pm$0.06 &   1.62$\pm$0.19 &   0.36$\pm$0.31 &   0.77$\pm$0.67 &   1.13$\pm$0.37 &   132.16 &  2015\\[0.05in]
 10 & 4.7$\pm$1.3 &   3.37$\pm$1.04 &   1.20$\pm$0.08 &   2.41$\pm$0.12 &   0.33$\pm$0.10 &   1.54$\pm$0.15 &   1.33$\pm$0.57 &   0.60$\pm$0.20 &   1.11$\pm$0.47 &   144.54 &  2015\\[0.05in]
 20 & 3.8$\pm$1.3 &   2.38$\pm$0.78 &   1.21$\pm$0.10 &   2.52$\pm$0.17 &   0.30$\pm$0.08 &   1.67$\pm$0.20 &   1.34$\pm$0.51 &   1.08$\pm$0.53 &   1.12$\pm$0.01 &   164.33 &  2015\\[0.05in]
 30 & 4.2$\pm$1.0 &   1.95$\pm$0.57 &   1.17$\pm$0.12 &   2.93$\pm$0.38 &   0.31$\pm$0.10 &   1.69$\pm$0.22 &   0.98$\pm$0.25 &   1.45$\pm$0.33 &   1.61$\pm$0.12 &   167.31 &  2015\\[0.05in]
 40 & 3.3$\pm$1.0 &   1.75$\pm$0.42 &   1.02$\pm$0.06 &   3.25$\pm$0.28 &   0.30$\pm$0.08 &   1.71$\pm$0.19 &   0.92$\pm$0.26 &   1.18$\pm$0.13 &   2.76$\pm$1.49 &   194.88 &  2015\\[0.05in]
 50 & 3.1$\pm$1.0 &   1.33$\pm$0.51 &   0.96$\pm$0.05 &   2.96$\pm$0.36 &   0.26$\pm$0.19 &   1.88$\pm$0.52 &   1.06$\pm$0.33 &   1.05$\pm$0.12 &   3.90$\pm$4.58 &   258.44 &  2015\\[0.05in]
 60 & 0.8$\pm$3.8 &   1.74$\pm$0.35 &   1.03$\pm$0.03 &   3.06$\pm$0.37 &   0.28$\pm$0.07 &   1.76$\pm$0.19 &   0.43$\pm$0.28 &   0.85$\pm$0.08 &   3.44$\pm$1.87 &   239.38 &  2015\\[0.05in]
 70 & 1.4$\pm$1.9 &   1.85$\pm$0.33 &   1.00$\pm$0.04 &   3.10$\pm$0.32 &   0.29$\pm$0.06 &   1.71$\pm$0.15 &   0.55$\pm$0.29 &   0.81$\pm$0.08 &   4.89$\pm$4.36 &   268.77 &  2015\\[0.05in]
\hline
-20 & 2.0$\pm$1.1 &   4.41$\pm$2.06 &   1.25$\pm$0.09 &   2.67$\pm$0.19 &   0.19$\pm$0.05 &   1.58$\pm$0.38 &   0.96$\pm$0.30 &   1.09$\pm$0.45 &   1.15$\pm$0.21 &   134.90 &  2012\\[0.05in]
-10 & 5.2$\pm$1.1 &   3.46$\pm$1.53 &   1.17$\pm$0.11 &   2.63$\pm$0.21 &   0.20$\pm$0.07 &   1.75$\pm$0.27 &   1.52$\pm$0.51 &   1.17$\pm$0.73 &   1.12$\pm$0.01 &   142.97 &  2012\\[0.05in]
  0 & 5.1$\pm$1.5 &   3.10$\pm$1.04 &   1.22$\pm$0.09 &   2.48$\pm$0.13 &   0.25$\pm$0.07 &   1.67$\pm$0.22 &   0.66$\pm$0.35 &   0.67$\pm$0.27 &   1.13$\pm$0.15 &   147.11 &  2012\\[0.05in]
 10 & 3.7$\pm$1.1 &   2.62$\pm$0.53 &   1.09$\pm$0.08 &   2.42$\pm$0.15 &   0.22$\pm$0.07 &   1.81$\pm$0.24 &   1.17$\pm$0.38 &   1.39$\pm$1.32 &   1.12$\pm$1.05 &   192.40 &  2012\\[0.05in]
 20 & 3.1$\pm$1.1 &   3.23$\pm$0.92 &   1.14$\pm$0.09 &   2.53$\pm$0.20 &   0.21$\pm$0.06 &   1.78$\pm$0.24 &   1.12$\pm$0.40 &   1.29$\pm$0.82 &   1.12$\pm$0.01 &   197.25 &  2012\\[0.05in]
 30 & 3.2$\pm$1.0 &   2.44$\pm$0.60 &   1.21$\pm$0.12 &   2.83$\pm$0.37 &   0.22$\pm$0.08 &   1.82$\pm$0.28 &   1.03$\pm$0.32 &   1.13$\pm$0.31 &   1.41$\pm$0.03 &   141.96 &  2012\\[0.05in]
 40 & 4.8$\pm$1.1 &   3.24$\pm$1.36 &   1.33$\pm$0.16 &   2.71$\pm$0.33 &   0.23$\pm$0.08 &   1.62$\pm$0.25 &   1.27$\pm$0.42 &   0.72$\pm$0.12 &   1.67$\pm$0.14 &   234.70 &  2012\\[0.05in]
 50 & 0.9$\pm$2.1 &   1.71$\pm$0.31 &   1.04$\pm$0.03 &   2.92$\pm$0.28 &   0.22$\pm$0.09 &   1.86$\pm$0.30 &   0.48$\pm$0.23 &   0.93$\pm$0.10 &   3.12$\pm$4.50 &   166.19 &  2012\\[0.05in]
 60 & 3.4$\pm$1.4 &   1.21$\pm$0.28 &   1.03$\pm$0.06 &   2.99$\pm$0.27 &   0.22$\pm$0.09 &   1.86$\pm$0.33 &   0.67$\pm$0.32 &   0.73$\pm$0.08 &   3.02$\pm$2.78 &   160.72 &  2012\\[0.05in]
 70 & 3.4$\pm$1.7 &   1.26$\pm$0.34 &   1.02$\pm$0.06 &   3.16$\pm$0.36 &   0.22$\pm$0.09 &   1.84$\pm$0.33 &   0.54$\pm$0.33 &   0.72$\pm$0.08 &   2.87$\pm$0.17 &   202.39 &  2012\\[0.05in]
\hline
\end{tabular}
\normalsize  % turned off for single-spaced text
\parbox{6.in}{\vspace{0.1in} NOTE: The optical depth is for a
  wavelength of 0.5 \mumx. These fits used 318 points of comparison
  and fit 8 parameters, for a nominal value of NF=310, for which the
  normalized \chisq /NF ranged from 0.426 to 0.87.}
\end{table*}

In comparison with the uniform methane fit results, we see only minor changes in most
of the other parameters.  The top pressure
of the tropospheric cloud layer is nearly the same for both models, although the stepped
depletion model results show a little more variability.  The retrieved bottom pressure
shows more significant changes.  The new results show much closer agreement between years,
but more change with respect to latitude, increasing from about 2.5 bars at low latitude
to 3 bars at high latitude.  The prior results showed no consistent trend with latitude, averaging
about 2.7 bars.  The optical depth for that layer shows about the same trend with latitude
and the same increase at high latitudes between 2012 and 2015.  The particle size generally
remains between 0.2 and 0.4 \mum for both models, but the descended model fits indicate
that particles in the northern hemisphere are about 40\% larger in 2015 than in 2012,
while the uniform model showed much less difference between years. All these particle
size differences are within uncertainties, however.  The relative humidity results
for methane are roughly similar for the two model types, with higher, near saturation levels
at low latitudes and a factor of two decline in the polar region.  Both find the methane
humidity depressed at the equator, with a slightly sharper decline seen in the descended
profile results.

The refractive index results differ a little.  For the descended depletion model
fits for 2012 and 2013 are in somewhat better agreement than for the uniform model,
and do not show as much trending towards slightly higher values at high latitudes.

\subsubsection{Latitude dependent fits with descended depletion of methane}

\begin{figure*}\centering
\hspace{-0.15in}
\includegraphics[width=3.2in]{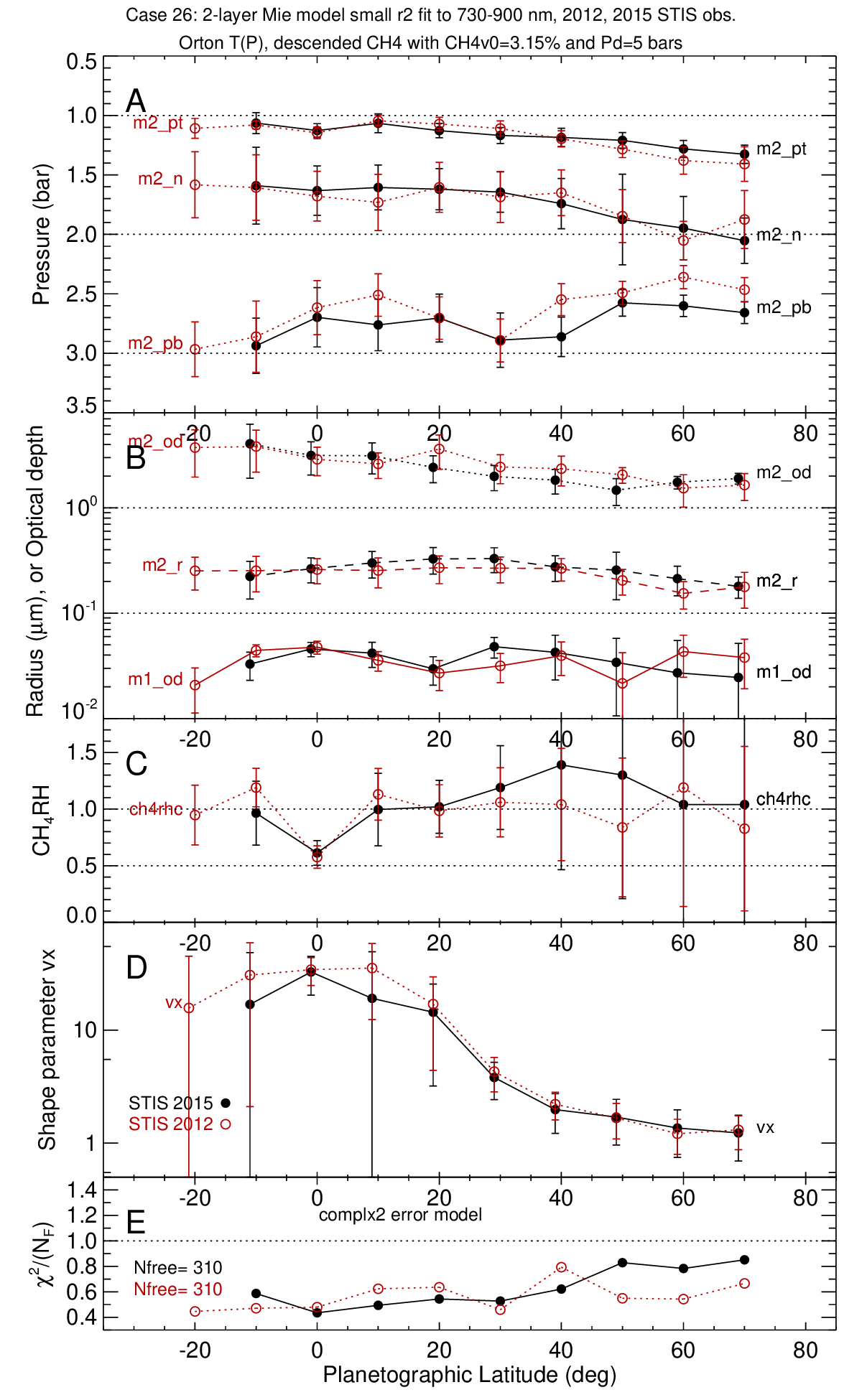}
\hspace{-0.2in}
\includegraphics[width=3.2in]{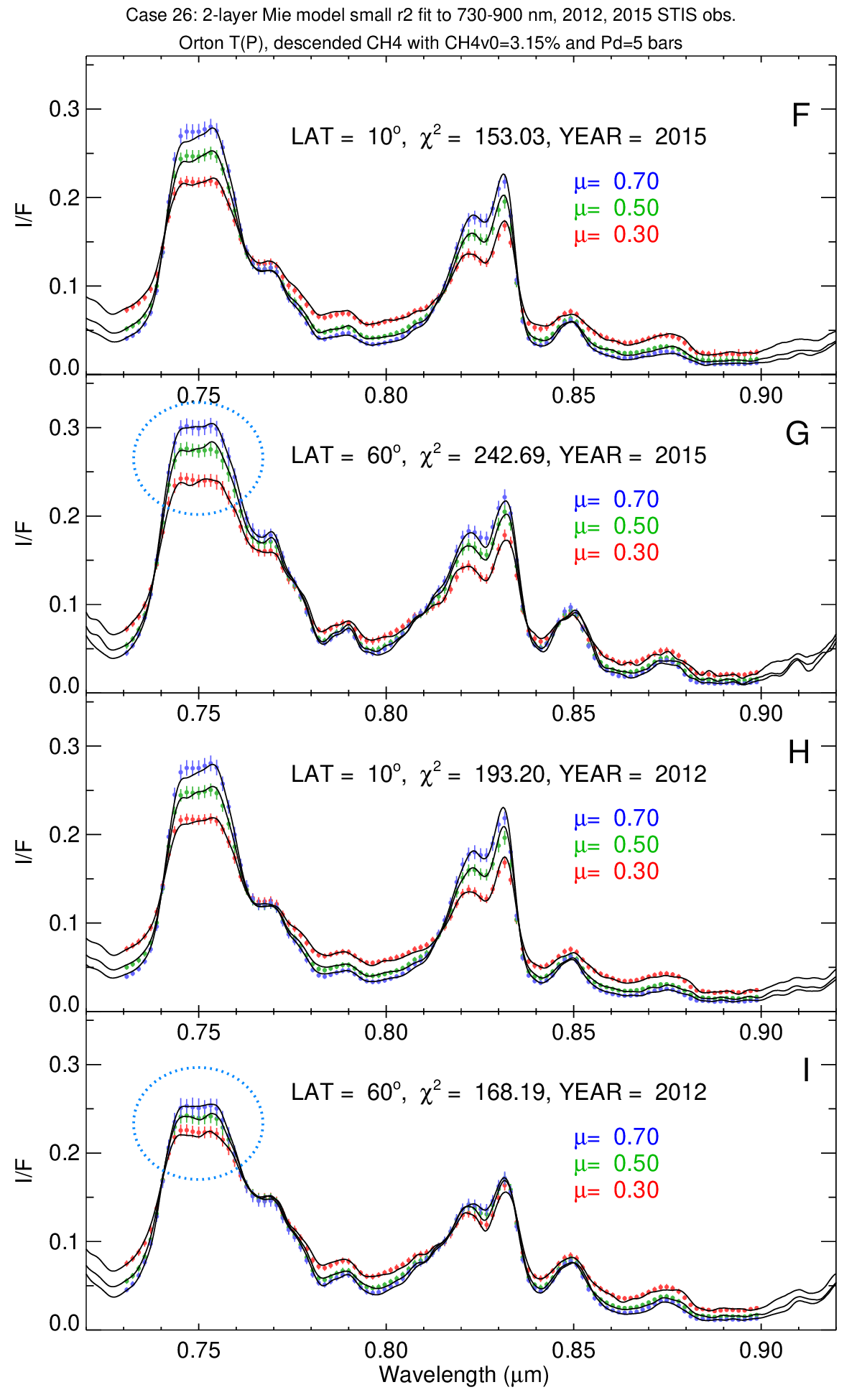}
%SOURCE Page 61 Log L, plotted with plot_fitctl2015mie_5panel17.pro on puck using case=26 in ir_ms_ura
\caption{Descended depletion model of vertical methane distribution
  fit to STIS spectra from 2012 and 2015.  Conservative cloud model
  and gas profile parameters for a Mie-scattering haze above a single
  diffuse Mie-scattering tropospheric layer, assuming a deep mixing
  ratio of 0.0315, and a methane profile characterized by a pressure
  depth parameter $P_d$ and a shape parameter $vx$ (defined in
  Eq.\ \ref{Eq:deplete} and illustrated in Fig.\ \ref{Fig:descend})
  and constrained by spectral observations from 730 nm to 900 nm.  The
  parameter values are in panels A-E, with red (open circle) points
  displaying results of fitting 2012 STIS observations and black
  (filled circle) points displaying the results of fitting 2015
  observations.  Sample comparisons between measured and
  large-particle model spectra are in panels F-I. Note the great
  improvement in the high latitude fits near 745 nm, compared to
  results given in Fig.\ \ref{Fig:mielat1}. }
\label{Fig:descendlat}
\end{figure*}

Because the descended depletion function approaches the deep mixing ratio
on a tangent, it is hard to constrain the depth parameter for this model at most latitudes.
Thus, from preliminary fits we found a $P_d$ value that worked well at high latitudes ($P_d$ = 5 bars) and
kept that constant, while using just the shape parameter ($vx$) as the additional
adjustable parameter in maximizing fit quality as a function of latitude.  The results
for best fit parameter values and uncertainties are given in Table\ \ref{Tbl:descendlat}
and plotted in Fig.\ \ref{Fig:descendlat}.

\begin{table*}[!t]\centering
\caption{Two-cloud spherical particle fits as a function of latitude
  assuming descended depletion of \chf as a function of
  latitude.\label{Tbl:descendlat}}
% SOURCE: This table is produced by running case 25 plot_fitctl2015mie_5panel17.pro in ir_ms_ura on puck:
% See documentation on P 58-59 of Uranus Log L.
\vspace{0.1in}
\small
\setlength\tabcolsep{2pt}
\begin{tabular}{ c c c c c c c c c c c c}
   Lat.    &  $m1\_od$        &           &    $m2\_pt$ &    $m2\_pb$ &      $m2\_r$ &          &           &          &        & \\[-0.03in]
   (\degx) &  $\times$100 &    $m2\_od$ &       (bar) &     (bar) &     (\mum) &   $m2\_nr$  &    $ch4rhc$ & $vx$  & \chisq & YR\\[0.05in]
\hline
\\[-.15in]
%   PGLAT &  100Xm1_od &    m2_od &    m2_pt &    m2_pb &     m2_r &     m2_n &   ch4rhc &       vx &    chisq &  YEAR\\[0.05in]
-10 &  3.3$\pm$1.0 &   4.06$\pm$2.15 &   1.06$\pm$0.09 &   2.94$\pm$0.23 &   0.22$\pm$0.09 &   1.59$\pm$0.32 &   0.96$\pm$0.28 &  16.90$\pm$31.60 &   181.84 &  2015\\[0.05in]
  0 &  4.6$\pm$0.7 &   3.14$\pm$1.09 &   1.13$\pm$0.06 &   2.70$\pm$0.25 &   0.26$\pm$0.07 &   1.63$\pm$0.21 &   0.61$\pm$0.11 &  32.70$\pm$12.30 &   134.79 &  2015\\[0.05in]
 10 &  4.2$\pm$1.1 &   3.12$\pm$1.02 &   1.07$\pm$0.08 &   2.76$\pm$0.22 &   0.30$\pm$0.08 &   1.61$\pm$0.19 &   1.00$\pm$0.32 &  19.10$\pm$30.30 &   153.03 &  2015\\[0.05in]
 20 &  3.0$\pm$0.9 &   2.42$\pm$0.69 &   1.13$\pm$0.06 &   2.70$\pm$0.20 &   0.33$\pm$0.09 &   1.62$\pm$0.17 &   1.02$\pm$0.23 &  14.40$\pm$11.20 &   168.49 &  2015\\[0.05in]
 30 &  4.8$\pm$1.1 &   1.99$\pm$0.54 &   1.17$\pm$0.07 &   2.89$\pm$0.23 &   0.33$\pm$0.09 &   1.64$\pm$0.17 &   1.19$\pm$0.37 &   3.81$\pm$ 1.38 &   163.38 &  2015\\[0.05in]
 40 &  4.2$\pm$1.9 &   1.83$\pm$0.48 &   1.18$\pm$0.08 &   2.86$\pm$0.17 &   0.28$\pm$0.08 &   1.74$\pm$0.21 &   1.39$\pm$0.93 &   1.98$\pm$ 0.76 &   192.65 &  2015\\[0.05in]
 50 &  3.4$\pm$2.4 &   1.47$\pm$0.42 &   1.21$\pm$0.07 &   2.58$\pm$0.11 &   0.26$\pm$0.12 &   1.87$\pm$0.38 &   1.30$\pm$1.09 &   1.70$\pm$ 0.74 &   256.78 &  2015\\[0.05in]
 60 &  2.7$\pm$2.8 &   1.75$\pm$0.24 &   1.28$\pm$0.07 &   2.60$\pm$0.09 &   0.21$\pm$0.07 &   1.95$\pm$0.27 &   1.04$\pm$1.14 &   1.36$\pm$ 0.61 &   242.69 &  2015\\[0.05in]
 70 &  2.5$\pm$2.7 &   1.90$\pm$0.22 &   1.33$\pm$0.08 &   2.66$\pm$0.09 &   0.18$\pm$0.04 &   2.05$\pm$0.19 &   1.04$\pm$1.18 &   1.23$\pm$ 0.54 &   263.79 &  2015\\[0.05in]
\hline
%   PGLAT &  100Xm1_od &    m2_od &    m2_pt &    m2_pb &     m2_r &     m2_n &   ch4rhc &       vx &    chisq &  YEAR\\[0.05in]
-20 &  2.1$\pm$0.9 &   3.73$\pm$1.77 &   1.11$\pm$0.08 &   2.97$\pm$0.23 &   0.25$\pm$0.09 &   1.58$\pm$0.28 &   0.95$\pm$0.26 &  15.70$\pm$29.40 &   138.18 &  2012\\[0.05in]
-10 &  4.4$\pm$0.6 &   3.82$\pm$1.64 &   1.08$\pm$0.03 &   2.86$\pm$0.30 &   0.25$\pm$0.09 &   1.61$\pm$0.28 &   1.19$\pm$0.17 &  30.70$\pm$28.60 &   145.93 &  2012\\[0.05in]
  0 &  4.8$\pm$0.7 &   2.89$\pm$0.87 &   1.15$\pm$0.05 &   2.62$\pm$0.23 &   0.26$\pm$0.07 &   1.68$\pm$0.21 &   0.58$\pm$0.10 &  34.30$\pm$ 9.49 &   148.60 &  2012\\[0.05in]
 10 &  3.6$\pm$0.7 &   2.61$\pm$0.71 &   1.04$\pm$0.05 &   2.51$\pm$0.18 &   0.25$\pm$0.08 &   1.73$\pm$0.24 &   1.13$\pm$0.23 &  35.40$\pm$23.00 &   193.20 &  2012\\[0.05in]
 20 &  2.7$\pm$0.9 &   3.62$\pm$1.30 &   1.07$\pm$0.06 &   2.70$\pm$0.18 &   0.27$\pm$0.08 &   1.60$\pm$0.21 &   0.98$\pm$0.23 &  17.00$\pm$12.60 &   197.04 &  2012\\[0.05in]
 30 &  3.2$\pm$1.0 &   2.44$\pm$0.75 &   1.11$\pm$0.06 &   2.89$\pm$0.18 &   0.27$\pm$0.07 &   1.69$\pm$0.21 &   1.06$\pm$0.31 &   4.29$\pm$ 1.45 &   142.42 &  2012\\[0.05in]
 40 &  3.9$\pm$1.4 &   2.35$\pm$0.74 &   1.20$\pm$0.07 &   2.55$\pm$0.13 &   0.27$\pm$0.06 &   1.65$\pm$0.19 &   1.04$\pm$0.50 &   2.21$\pm$ 0.61 &   245.82 &  2012\\[0.05in]
 50 &  2.2$\pm$2.1 &   2.06$\pm$0.35 &   1.28$\pm$0.07 &   2.49$\pm$0.10 &   0.20$\pm$0.06 &   1.85$\pm$0.22 &   0.84$\pm$0.61 &   1.67$\pm$ 0.58 &   170.15 &  2012\\[0.05in]
 60 &  4.3$\pm$1.9 &   1.54$\pm$0.52 &   1.38$\pm$0.11 &   2.36$\pm$0.10 &   0.15$\pm$0.04 &   2.05$\pm$0.16 &   1.19$\pm$1.05 &   1.21$\pm$ 0.41 &   168.19 &  2012\\[0.05in]
 70 &  3.8$\pm$1.9 &   1.65$\pm$0.47 &   1.41$\pm$0.15 &   2.47$\pm$0.10 &   0.18$\pm$0.07 &   1.88$\pm$0.24 &   0.83$\pm$0.73 &   1.31$\pm$ 0.44 &   206.35 &  2012\\[0.05in]
\hline
\end{tabular}
\normalsize  % turned off for single-spaced text
\parbox{6.in}{\vspace{0.1in} NOTE: The optical depth is for a
  wavelength of 0.5 \mumx. These fits used a fixed value of $Pd =$
  5 bars, $\alpha_0=$ 3.15\%. There were 318 points of comparison and
  8 fitted parameters, for a nominal value of NF=310, for which the
  normalized \chisq /NF ranged from 0.42 to 0.83.}
\end{table*}

\begin{figure*}\centering
\hspace{-0.15in}
\includegraphics[width=3.2in]{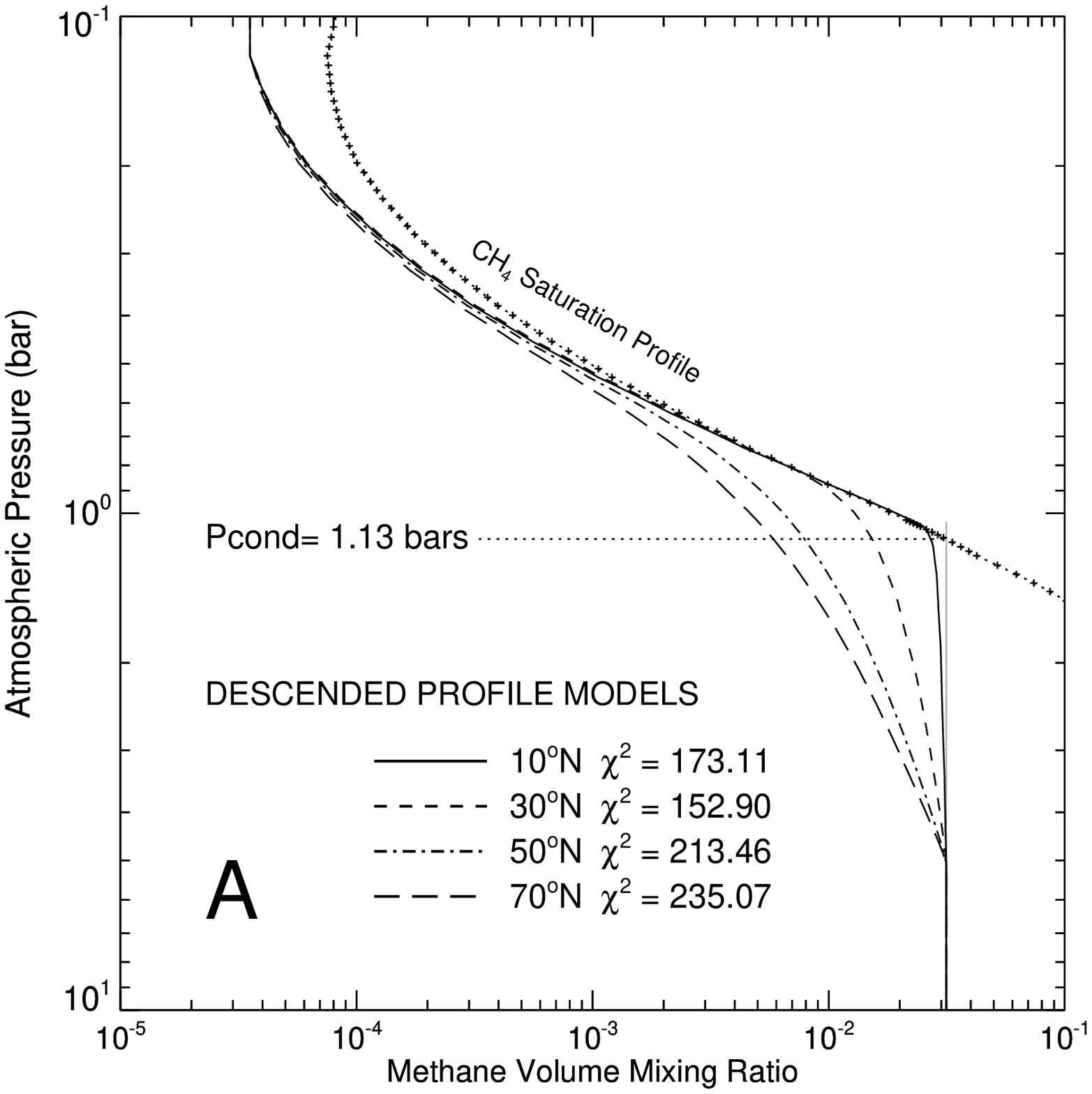}
\hspace{-0.15in}
\includegraphics[width=3.2in]{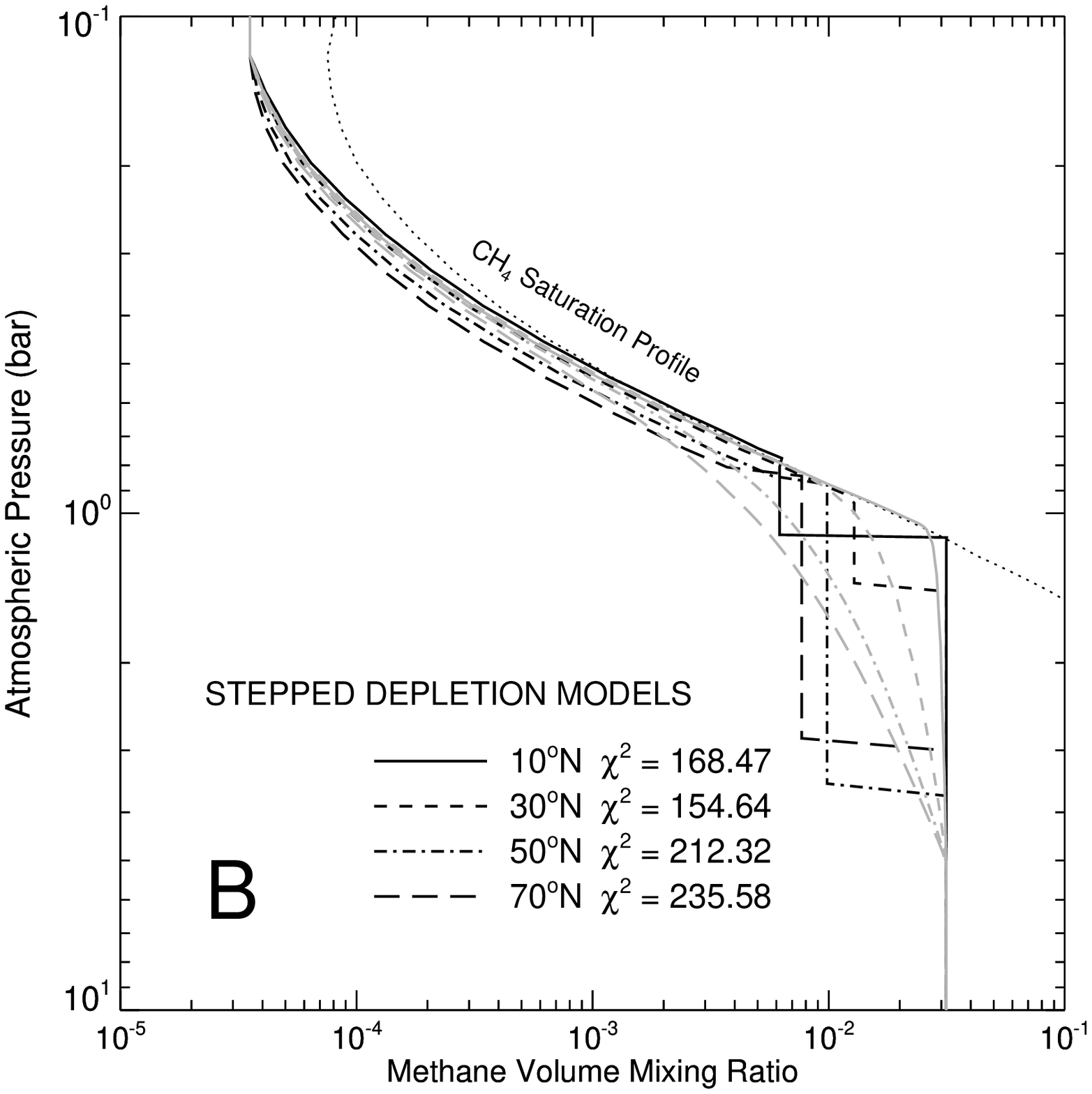}
%SOURCE Made with plotcldgasura_nonsub17.pro, see p 62-64, Uranus Log L.
%SOURCE Previous work on p 94-101, Uranus Log K.
\caption{A: Best-fit descended profiles at 4 latitudes, using average parameter fits for
2012 and 2015. B: Best fit stepped depletion profiles, using average parameter fits for
2012 and 2015. The profiles in A are overlain in light gray in B for reference. Both sets of 
fits show decreasing methane humidity with latitude above the 1 bar level, and both
indicate that the depletion is of limited depth ($\sim$5 bars or less). In both cases average
$\chi^2$ values for 2012 and 2015 are given in the legend. }
\label{Fig:vertch4mods}
\end{figure*}

\begin{figure*}\centering
%\hspace{-0.15in}\includegraphics[width=3.2in]{chisq_vs_lat_comp2012.eps}\hspace{-0.2in}
\hspace{-0.15in}\includegraphics[width=3.2in]{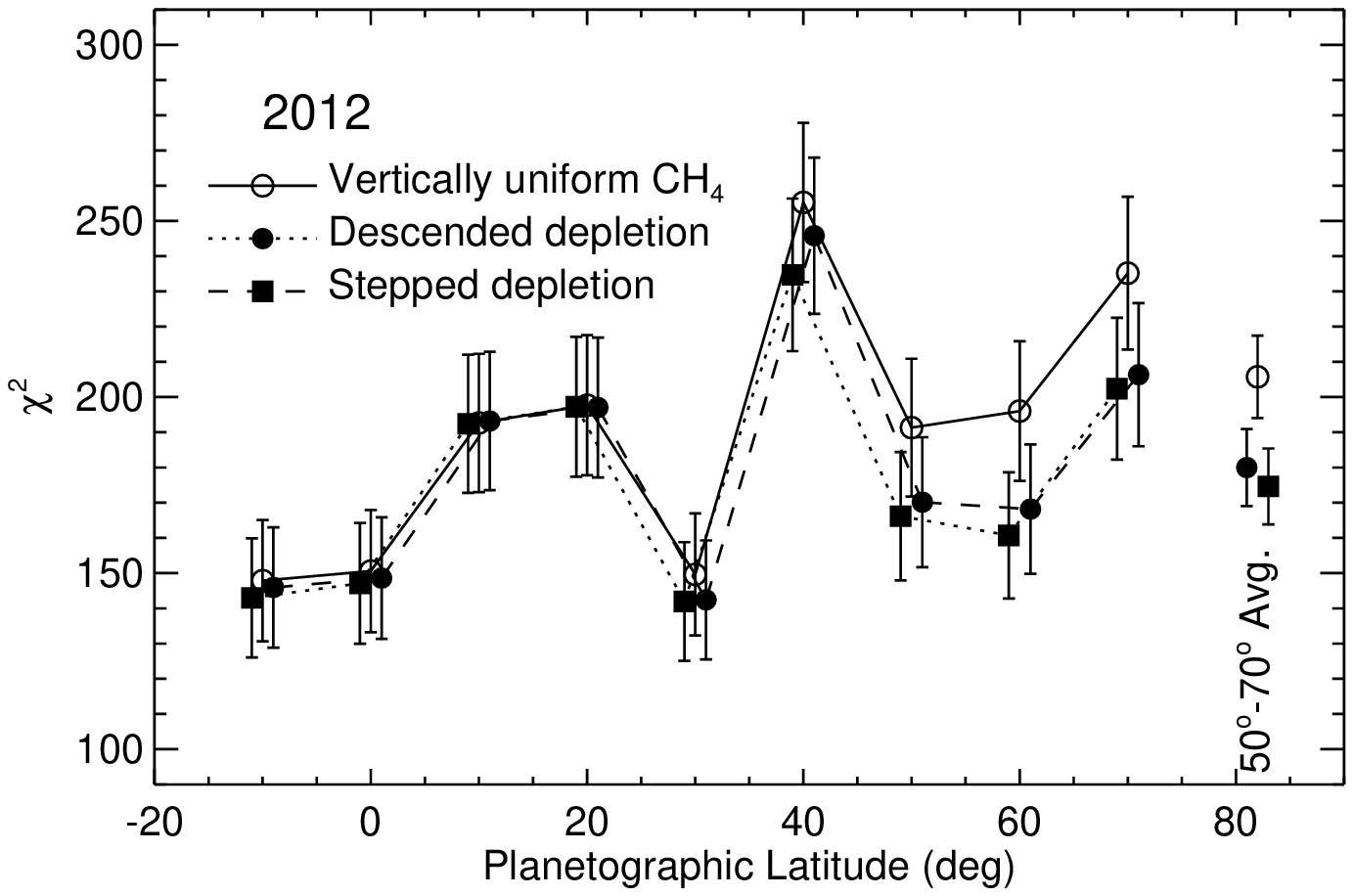}\hspace{-0.2in}
\includegraphics[width=3.2in]{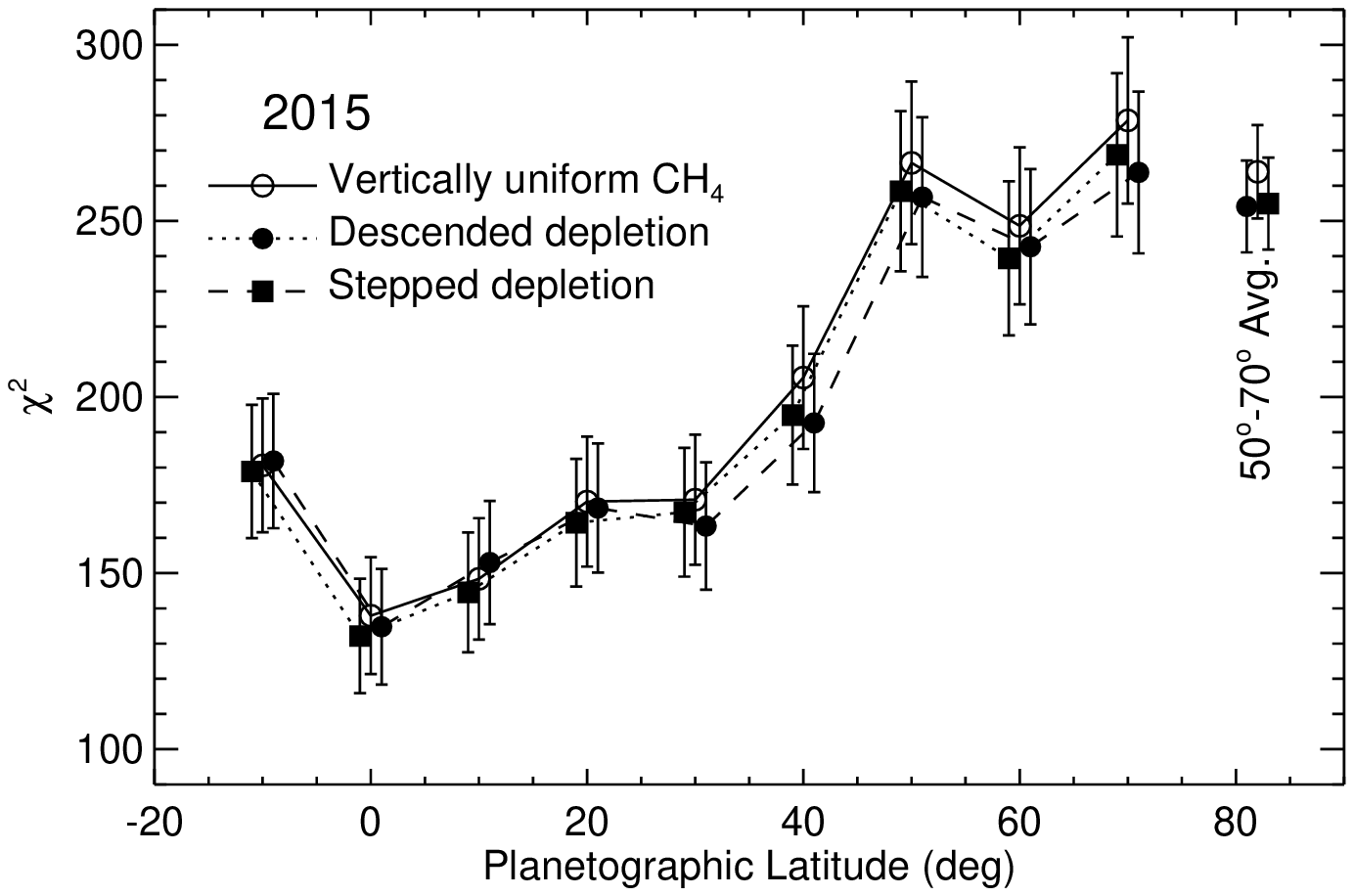}
%SOURCE Made with plot_chisq.pro, see p 65, Uranus Log L, or for earlier versions p 98, Uranus Log K.
\vspace{-0.15in}
\caption{Comparison of 2012 (left) and 2015 (right) \chisq versus
  latitude values for for three different methane vertical
  distribution models: uniform (solid line), descended depletion
  (dotted line), and stepped depletion (dashed line). Corresponding
  averages over the 50\deg -- 70\deg latitude range are also shown
  near 80\degx N in each panel.  The depleted profile values are
  slightly shifted in latitude to avoid error bar overlaps. This shows
  that the overall fit quality is improved by use of the descended
  profile, in addition to the more obvious improvement near 750 nm.
  The small overall improvement seen in fits to the 2015 observations
  is likely due to the increased noise level at high latitudes and low
  signal levels for this data set.}
\label{Fig:chi2comp}
\end{figure*}

These two depletion model fits are compared in
Fig.\ \ref{Fig:vertch4mods}, with descended model fits in panel A and
the stepped depletion model fits in panel B. The descended model fits
yield slightly lower $\chi^2$ values, especially at 70\deg N, although
even there the difference is smaller than the expected uncertainty of
$\sqrt{2\chi^2}$, which is 22 in this case. Both models imply that
the high latitude depletion is of limited depth, and both imply that
the methane humidity above the 1 bar level is near saturation at low
latitudes and decreases poleward.  Not only can we obtain good fits
with a shallow depletion of methane, they are preferred on the basis
of fit quality.  Not only does the high latitude fit near 745 nm
improve significantly when the vertically varying depletion models are
used, but the overall $\chi^2$ at high latitudes is also significantly
improved, as illustrated in Fig.\ \ref{Fig:chi2comp}.  This is especially
apparent at the higher latitudes and in comparing averages over the
50\deg -- 70\deg latitude range.  Although the stepped depletion
model is seen to yield slightly better \chisq values than the
 descended depletion models, the difference is less than the
expected uncertainty.  The virtue of the stepped depletion model
is that it can be well constrained at all latitudes, while the
virtue of the descended depletion model is that it makes more
sense physically.  We were able to extend the latitude range of
the descended model fits by fixing the value of the depth parameter $P_d$
to 5.0 bars.  We then found that both depletion models do not quite
yield zero depletion at low latitudes, which one might interpret
to mean that we should have chosen a slightly lower deep methane VMR value.
However, the \chisq values for the vertically uniform values are just as
good or slightly better than the depleted models at low latitudes.

\subsection{Wavelength dependence issues}\label{Sec:wdep}

\begin{figure*}\centering
\includegraphics[width=5in]{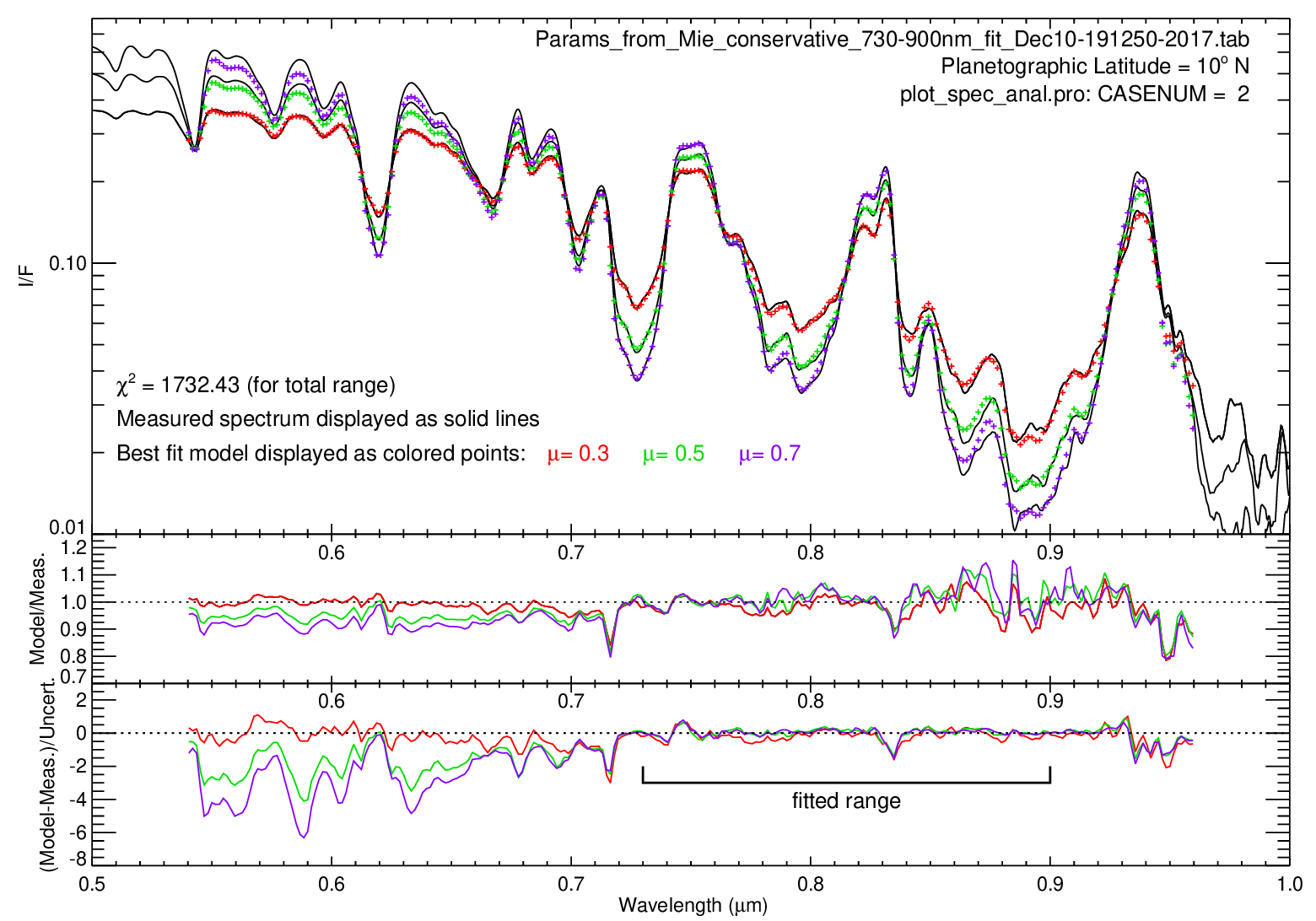}
%SOURCE produced by plot_spec_anal.pro on puck, see Page 19, Uranus Log J
\caption{As in Fig.\ \ref{Fig:initfit}, except 10\deg N spectral
  comparisons are shown for a conservative cloud model that provide
  the best match to the 10\deg N 2015 STIS spectra from 730 nm to 850
  nm, using the small particle solution. Note the significant model
  falloff at shorter wavelengths. See text for implications.}
\label{Fig:flawed}
\end{figure*}

Although the best-fit parameters given in Table\ \ref{Tbl:mie_ls}
provide great spectral matches over the fitted range (730--900 nm),
they do not provide good matches over the entire range. As expected,
and as illustrated in Fig.\ \ref{Fig:flawed}, the corresponding model
spectra fit even worse over the rest of the wavelength range than the
initial fit shown in Fig.\ \ref{Fig:initfit}.  The problem with
both the small-particle and large-particle models is that they do not
produce a large enough I/F at the short wavelength side of the
spectrum (from 0.54 \mum to 0.68 \mumx) for the two largest zenith
angle cosines, and produce too high an I/F in the deeply penetrating
region near 0.94 \mum for all three zenith angles.  The problem is
less extreme for the small-particle solution because it produces a
larger increase in I/F at shorter wavelengths.

One way to solve the short wavelength deficit problem is to abandon
spherical particles and use a wavelength-dependent phase function that
provides increased backscatter at short wavelengths, which is the
approach followed by KT2009, and one which we will return to in a
later section.  An alternative approach considered here is to use a
wavelength-dependent imaginary index that is small at short
wavelengths and larger at long wavelengths, an approach used by
\cite{Irwin2015reanalysis} to solve a similar problem in fitting
near-IR spectra. The utility of this approach is that the increased
optical depth required to compensate for the small absorption at long
wavelengths leads to a needed increase in the I/F at short wavelengths
where the absorption is absent.  To follow up on this approach we
added an adjustable imaginary index to cloud particles in the m2 
Mie layer, and then optimized model parameters to fit both the
730--900 nm region and the 540-580 nm region simultaneously, as
described in the following section.

%We also fit both the 2015 and 2012 spectra to look for temporal
%and latitude dependent changes in the fitted parameters.

\subsubsection{Controlling wavelength dependence with particulate absorption}

\begin{figure*}[!htb]\centering
\includegraphics[width=5.5in]{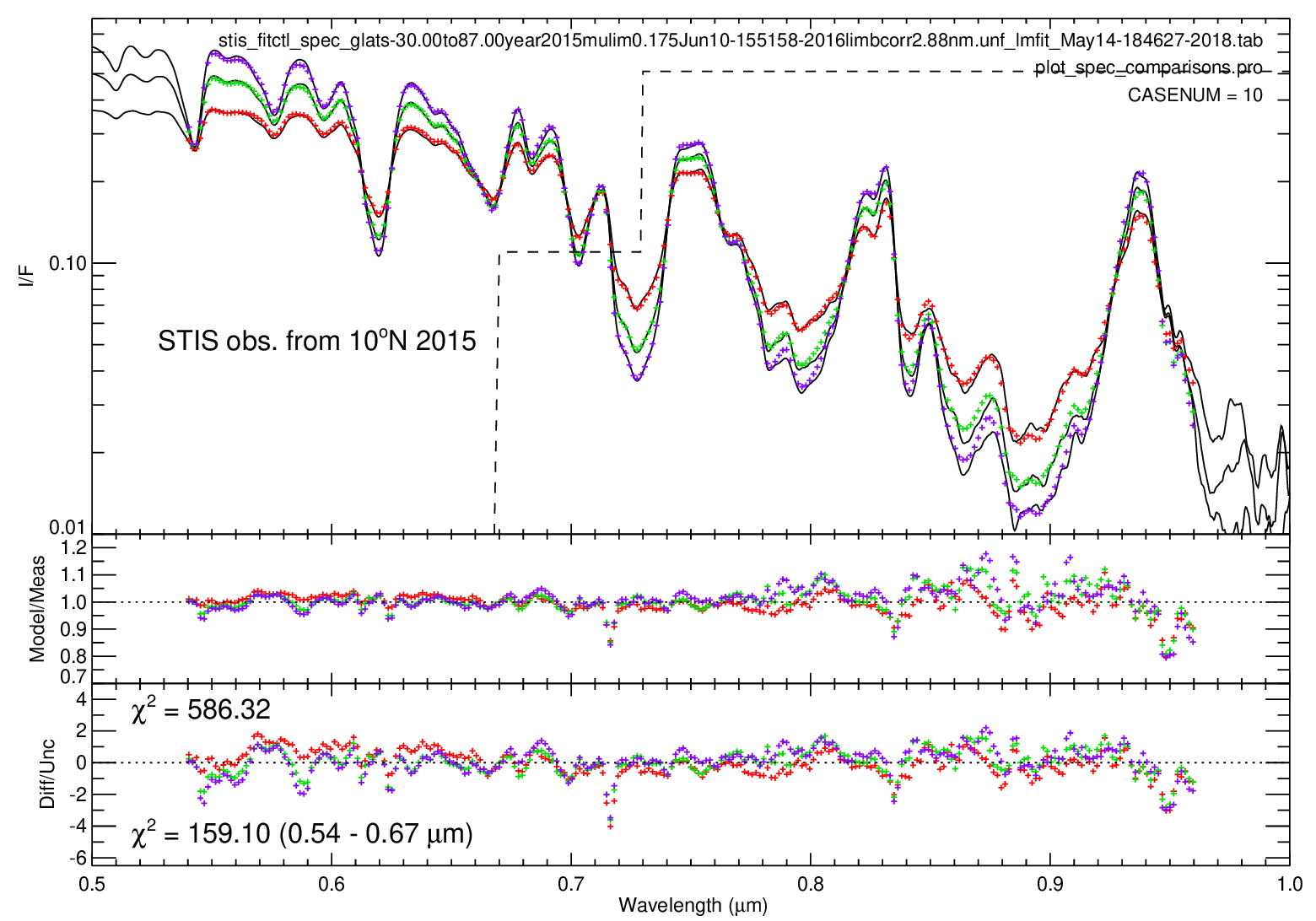}\par
%\vspace{0.05in}
\caption{Extended range wavelength dependent model, using imaginary
index variations to adjust the wavelength dependence.  The imaginary
index $m2\_ni$, multiplied by a factor of 100, is shown by the dashed curve.}\label{Fig:fullmie1}
%SOURCE: plotted by plot_spec_comparisons.pro on puck using casenum = 10
% See Uranus Log L, page 133, and Page 128 for documentation of the fit
% prior version plotted by plot_spec_comparisons.pro on puck using case 7
% See page 71, Uranus Log L
\end{figure*}

The first example of controlling wavelength dependence over a larger
spectral range is based on adjustment of particulate absorption.  For
this example, we assume two Mie scattering clouds, with the top layer
(m1) located at an arbitrary pressure of 50 mbar and containing
conservative particles with an assumed effective radius of 0.06 \mumx,
and an adjustable optical depth.  The top layer has a very small
optical depth and its particle size is not very well constrained by
our observations.  We chose a somewhat arbitrarily value based on
preliminary fitted values.  The \cite{Rages1991} haze model estimates
a particle size closer to 0.1\ \mum at 50 mbar.  The other Mie layer
is assumed to be composed of a non-conservative material,
characterized by a refractive index of $m2\_nr$ + 0 $\times$ $i$ for
$\lambda < $700 nm and n = $m2\_nr$ + $m2\_ni\times i$ for $\lambda >
$ 710 nm. The tropospheric Mie layers (m2) is characterized by three
additional fitted parameters: pressure, particle size, and optical
depth.  We then simultaneously fit just two sub regions of the
spectrum: the 540-580 nm region, where we assume the particles are
conservative, and the 730--900 nm region, where we assume a locally
wavelength-independent imaginary index that is adjusted to minimize
\chisqx.  We also allowed $m2\_nr$ to be adjustable.  This process
produced a best-fit value of (4.9$\pm$1.3)$\times 10^{-3}$ for the
imaginary index and 2.7$\pm$0.3\% for the deep methane mixing
ratio. However, this process slightly degraded the fit in the 730-900
nm region.  To better constrain the methane mixing ratio for the case
with absorbing aerosols we adopted the imaginary index obtained from
the dual fit, then refit the remaining parameters using the 730-900 nm
region for our spectral constraints, yielding the results given in
Table\ \ref{Tbl:ssa}.  Applying these parameters over the entire
spectral range from 540 nm to 960 nm, we then obtained a much improved
match to the observations, with a \chisq of 724.50. This was further
improved to 586.32 by optimizing values of $m2\_pt$ (1.09$\pm$0.01 bar), $m2\_pb$
(3.35$\pm$0.13 bar), $m1\_od$ (0.030$\pm$0.002), $m2\_od$
(3.91$\pm$0.34), $m2\_r$ (0.30$\pm$0.02), $m2\_nr$ (1.69$\pm$0.04), and
$m2\_nilw$ (0.0051$\pm$0.0003), after adding an intermediate
imaginary index of 0.0011 to the tropospheric aerosol particles in the
spectral interval from 670 nm to 730 nm, yielding the fit displayed in
Fig.\ \ref{Fig:fullmie1}.  Although the fit is good, it is not known
whether any plausible cloud material has this absorption
characteristic.  Complex hydrocarbons, such as tholins
\citep{Khare1993}, absorb more at shorter wavelength and have
declining absorption over the range where our example model shows
increased absorption.  Judging from frost reflection spectra obtained
by \cite{Lebofsky1976}, H$_2$S does not appear to exhibit such a trend
either.  Thus we have some motivation to consider other ways to
generate wavelength dependence.

\subsubsection{Controlling $\lambda$ dependence with optical depth variations}

Although Mie scattering calculations for spherical particles produce wavelength
dependent optical depth and scattering phase functions, if these do not yield
needed dependencies, and if particle size is constrained, and wavelength-dependent
absorption is not acceptable, then non-spherical particles need to be considered.

\begin{figure*}[!htb]\centering
\includegraphics[width=5.5in]{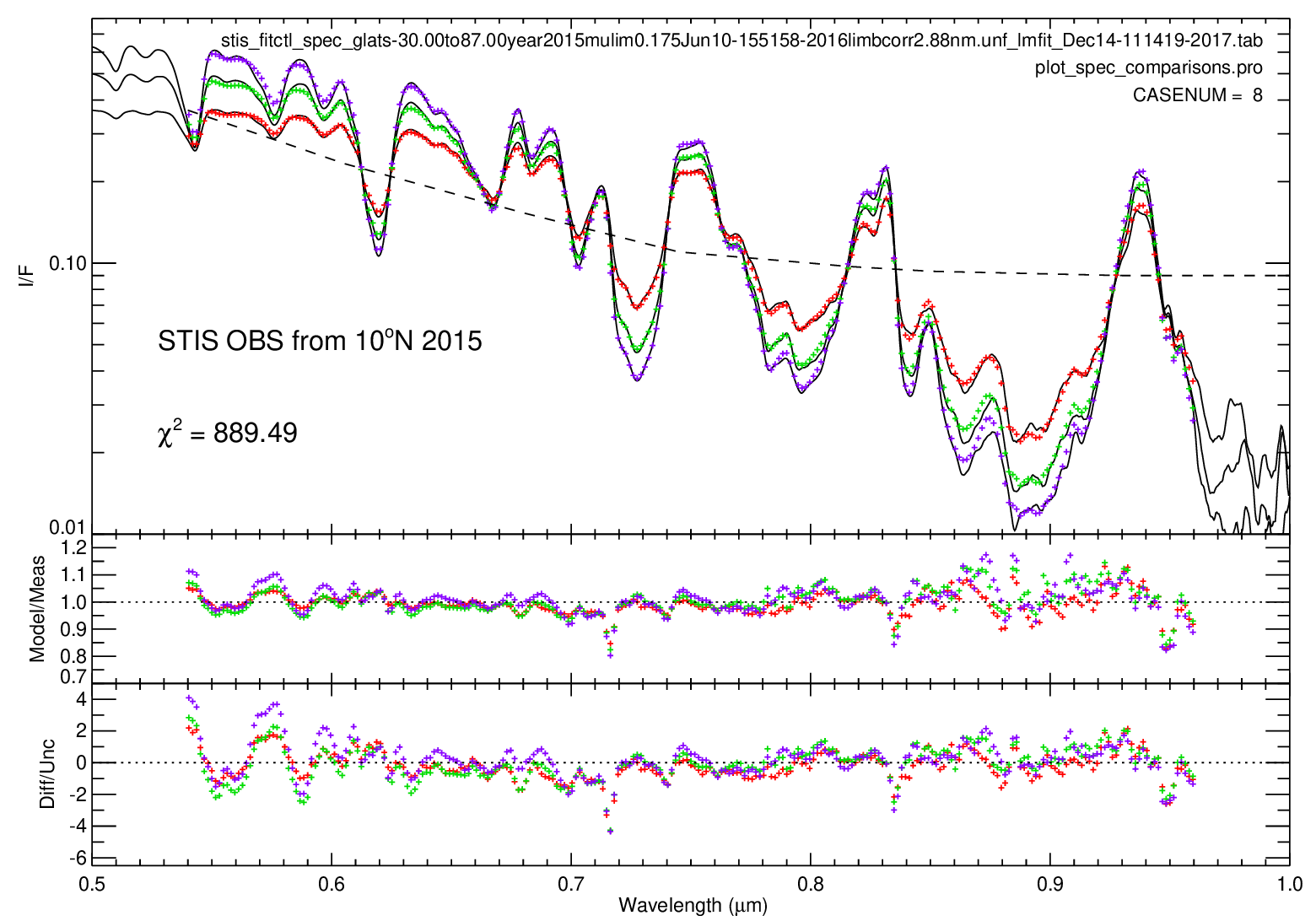}
%\vspace{-0.15in}
\caption{Extended range wavelength dependent model for HG particles, using
an optical depth variation with wavelength adjusted  to match the 2015 STIS observations
at 10\degx N.  The dashed curve displays the wavelength dependent optical depth normalized
by its value at 800 nm, then scaled downward by a factor of 10.}\label{Fig:fullhg}
%SOURCE: plotted by plot_spec_comparisons.pro on puck using case 8
% See page 73, Uranus Log L
\end{figure*}

The simplest option is to use a single HG scattering phase function
and simply adjust the wavelength dependence of the optical depth to
match the observed spectral variation.  An increase in optical depth
with size parameter (2$\pi r/\lambda$) is certainly a characteristic
shared by most particles and by aggregates in our trial calculations. 
 It also is plausible that a non-spherical
particle might exhibit a greater $\lambda$ dependence in optical depth
than a spherical particle for the case in which both particles satisfy
the other constraints in the 730-900 nm region.  To define the needed
$\tau (\lambda)$ function we began by taking our best fit vertical
structure and asymmetry fit for the 730-900 nm region, then computed a
series of model spectra with optical depths increasing until we could
find an optical depth at any wavelength that would match the observed
I/F at that wavelength.  But we found a problem with this approach.
At short wavelengths, the optical depth needed to match two successive
continuum regions (e.g. at 560 nm and 585 nm) was about 4-5 times the
value at 800 nm that was derived from fits to the 730-900 nm region.
But to match the intervening absorption feature at 576 nm would
require about half of that optical depth.   Thus a smoothly varying optical
 depth function could not
be created in this fashion, and a function that included wiggles at
all the methane features was completely implausible.  The fix to this
problem was to distribute the cloud particles over a greater
atmospheric depth.  This would not change the continuum I/F values
very much, but in the weakly absorbing regions, there would be
more absorption.  At longer wavelengths this required an increase
in the cloud's optical depth, which in turn required readjustment
of the optical depth ratio between 800 nm and 540 nm.  The
result of this process applied to our model of the 2015 STIS spectrum
at 10\degx N is shown in Fig.\ \ref{Fig:fullhg}.

\subsubsection{Controlling $\lambda$ dependence with  phase function variations}

KT2009 assumed that the main tropospheric cloud layer had a wavelength-independent
optical depth, which is a plausible assumption for large particles,
 and used a wavelength dependent phase function to match the observed
spectral variation.  A general form of their function can be written
as 
 \begin{eqnarray} f_1(\lambda)=a-b\times
  \sin^\alpha[\frac{\pi}{2}(\lambda_o-\lambda)/(\lambda_0 - \lambda_1)],\\ \nonumber
 \lambda_1 \le \lambda \le
  \lambda_o \label{Eq:f1wdep}
\end{eqnarray}
in which KT2009 assumed $\alpha = 4$, $a=0.94$, $b=0.427$, $\lambda_0
= 1$ \mumx, and $\lambda_1 = 0.3$ \mumx, which makes $f_1$ reach a
maximum of 0.94 at a wavelength of 1 \mum and a minimum of 0.513 at
0.3 \mumx.  They applied this to a double HG function with adopted
values of $g_1$ = 0.7 and $g_2$ = -0.3.  (Note that there is no basis
for applying this function to wavelengths greater than 1 \mum or less
than 0.3 \mumx.)  We found that this function was able to fit low
latitude spectra over the 730 nm to 900 nm range quite well, but that
the $a$ and $b$ constants needed to vary with latitude and that we
needed to increase $g_1$, leading us to adopt a new value of 0.8.
When applied to the extended spectral range, we needed to increase $\lambda_1$
to about 0.45 and $\alpha$ to 5.  The resulting spectral match was
intermediate between those shown in Figs. \ref{Fig:fullmie1} and \ref{Fig:fullhg}.
A problem with this formulation is that extending the idea to longer wavelengths
would require the particles to become more and more forward scattering at
longer wavelengths (in order to produce the same effect that absorbing
Mie particles produce, as discussed later).  This is not a plausible
trend.  For large enough wavelengths the particles must become less
forward scattering.

We also considered whether the HG model could use a wavelength-dependent
asymmetry parameter instead of a wavelength dependent optical depth to
match the observed spectrum over a wider spectral range.  
However, matching the shorter wavelengths
required a negative asymmetry parameter, which is an implausible condition,
and thus not an acceptable solution.

Thus over the longer spectral ranges it is most likely that optical
depth variation and possible particulate absorption will be needed
to model reflected spectra. Phase function variations will also be
present, but cannot be the sole way to produce the needed wavelength dependence in
scattering properties.

\subsection{Two-layer Mie model applied to Near-IR spectra}

To test whether our 2-layer Mie models would be capable of fitting
near-IR spectra, we extended model calculations to 1.6 \mum and
compared them to a central meridian SpeX spectrum covering the
0.8--1.65 \mum range.  [We obtained this spectrum from the Infrared
  Telescope Facility on 18 August 2013, using the cross-dispersed mode
  of the SpeX spectrometer. The spectrum was spatially averaged over
  the central 0.4 arcseconds of the central meridian covered by the
  0.15-arcsecond slit, corresponding to an average latitude of 24\degx
  N.  It was spectrally smoothed to the same spectral resolution as
  the smoothed STIS spectrum (a FWHM of 2.88 nm).  The spectrum was
  scaled to match the 1.09$\times 10^{-2}$ I/F center-of-disk H-band
  I/F from \cite{Sro2007struc}.] The initial small-particle model
parameters we used were from the 20\degx N spectrum and used an
imaginary index of 0.0046 at all wavelengths longer than 730 nm.  For
the initial large-particle model, we used a fit to the 10\degx N
spectrum and used an imaginary index of 6.2$\times 10^{-4}$ for
$\lambda >$ 730 nm.  The smaller index for the large particle solution
is a result of the lower real refractive index for the best-fit larger
particles.

\begin{figure*}\centering
\includegraphics[width=6in]{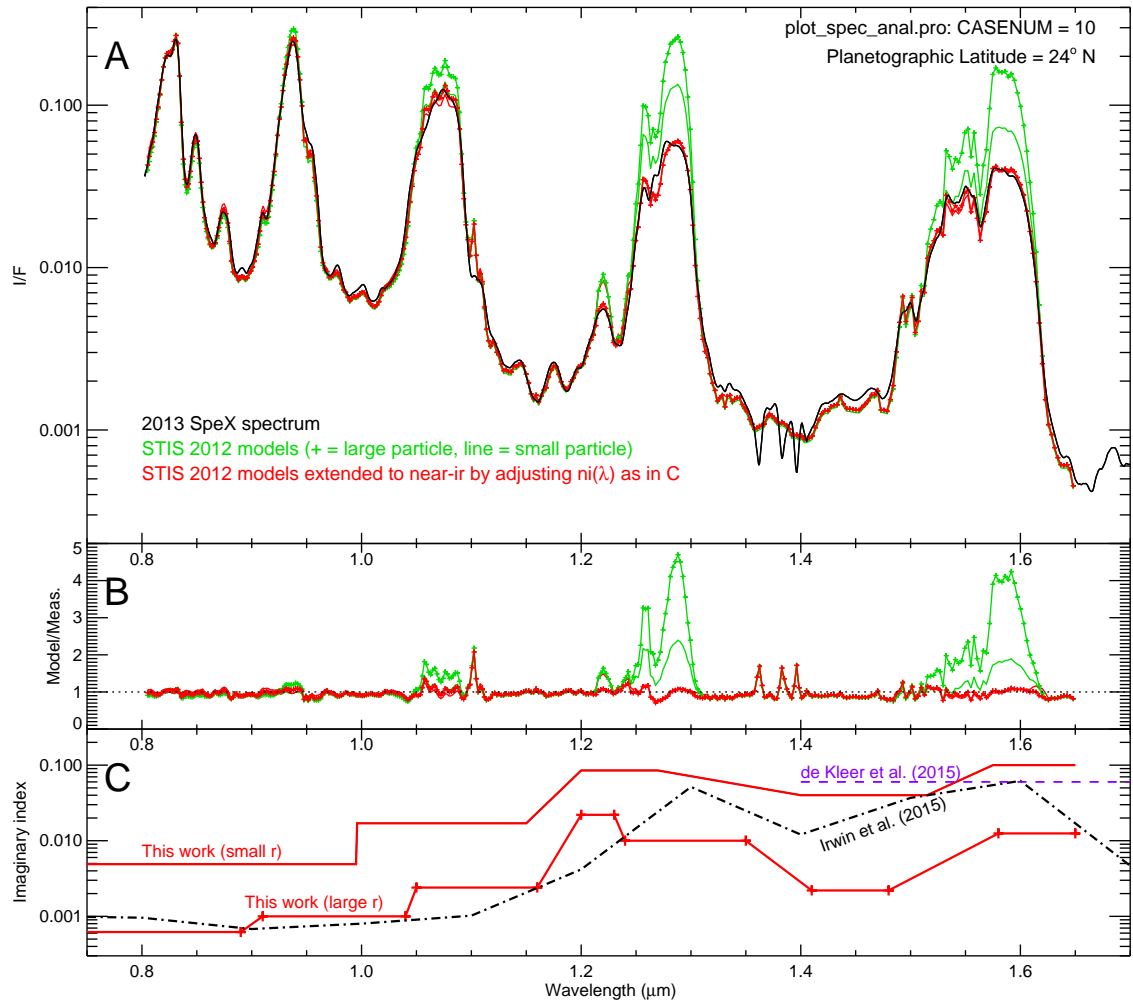}
%SOURCE Plotted by plot_spec_anal.pro, documentation in Log L, P88, 95-99
\caption{A: Our 2013 SpeX near-IR spectrum of Uranus from latitude
  24\deg N (black) compared to model spectra for the same observing
  geometry but using the gas and aerosol parameters from the 20\deg N
  two-layer Mie scattering model for two particle size solutions:
  small-particle (line only) and large-particle (lines with points).
  The same models, extended to the near-IR by adjusting the imaginary
  index of the cloud particles, are shown in red. In the first model
  set of models, a vertically uniform methane mixing ratio of 2.65\%
  was assumed up to the methane condensation level. In the second set
  (red curves) a deep mixing ratio of 3.15\% was assumed and a
  descended depletion profile shape was used, with vx = 7.34 an Pd = 5
  bars. B: ratio of model spectra to the SpeX observed spectrum. C:
  Imaginary index spectra assumed in the second set of models (red),
  compared to imaginary index values derived by
  \cite{Irwin2015reanalysis} and a value inferred from a single
  scattering albedo used by \cite{DeKleer2015}.}
\label{Fig:spex}
\end{figure*}

Fig.\ \ref{Fig:spex} shows that the extended models agree well in the
dark regions of the spectrum, indicating that little change in
stratospheric haze properties is needed, but is far too bright in the
longer wavelength continuum regions, indicating that the real cloud
particles have, at longer wavelengths, a lower optical depth or
greater absorption than the model particles.  The large particle
solution is the worst offender because its scattering efficiency is a
relatively weak function of wavelength, while the scattering
efficiency of the smaller particles declines substantially, though not
enough to match the falloff in pseudo continuum I/F values with
wavelength.

The excess model I/F at these wavelengths can be reduced by increasing
the imaginary index as indicated in the bottom panel of
Fig.\ \ref{Fig:spex}.  Our procedure for developing these solutions
was to start with a conservative solution constrained by the 730-900
nm spectrum.  We then used that as an initial guess for a split fit of
the 540-580 nm plus 730-900 nm region, assuming that the imaginary
index was zero for $\lambda \le$ 580 nm and had an adjustable value of
$m2\_nilw$ for for $\lambda \ge$ 730 nm. From that we obtained an
estimate for the imaginary index in the 730-900 nm region.  We
then fixed that imaginary index and did a new fit within the 730-900
nm region to get a revised estimate of the methane profile.  We then
fixed the methane profile and used a second split fit to improve the optical
depth and vertical aerosol distributions, as well as particle size and
real index.  We then adjusted the imaginary index in the 670-730 nm
range to optimize the fit to that part of the spectrum.  That provided
the parameters used for the initial near-IR calculations.  To match
the near-IR spectrum we did a suite of forward calculations with
different constant imaginary index values to find in each wavelength
region the imaginary index that provided the best model match to the
observations.  This was not done at a fine wavelength resolution in an
attempt to match every detail because the solid materials making up
the cloud particles would not likely have such fine-scale absorption
features.

This figure shows that the STIS-based model
with two layers of small spherical particles can match the observed infrared
spectrum out to 1.65 \mum by increasing the imaginary index with wavelength
as shown in Fig.\ \ref{Fig:spex}, reaching a maximum of 0.1 for the H-band region.
Our index is generally larger than the imaginary index estimated by
\cite{Irwin2015reanalysis} although of roughly similar shape. Our mean
value in the H band is similar to the adopted value of
\cite{DeKleer2015}.  Our fitted real index of 1.72$\pm$0.2 for the extended-wavelength
 small-particle model is
significantly larger than the value of 1.4 assumed by
\cite{Irwin2015reanalysis}. Since our particles are thus inherently
brighter, it is not surprising that we might need more absorption than Irwin et al. to
match the observations.  Our large-particle model, with a lower real index,
has an imaginary index profile of similar shape but lower amplitude. The imaginary index value for
\cite{DeKleer2015} we derived from their assumed single-scattering
albedo of 0.75, which corresponds to an imaginary index of 0.06 for
1-\mum particles. Irwin et al. suggested that the refractive index
spectrum would allow us to determine the composition of the cloud
particles. However, the most likely cloud material (H$_2$S) does not
have well characterized (quantitative) absorption properties, and
frost reflection spectra between 1.2 and 1.6 \mum \citep{Fink1982}
provide little qualitative evidence for significant absorption of the
type we seem to need to match the observed spectrum.

We could also have modeled the drop in I/F at longer wavelengths using
a HG particle scattering model, either by varying the single-scattering
albedo with wavelength, or by varying the optical depth as a function
of wavelength.  It is left for future work to evaluate which sort of
variation provides the best overall compatibility with the observations.

Although our modified imaginary index allows our two cloud model to
closely reproduce the observed spectrum in most regions, there are some
problems that need further work to address.  First, note that at the
1.08-\mum continuum peak, the model contains modulations that are not
observed in the measured spectrum.  This is also the case for model
calculations shown by \cite{Tice2013}, and is an indication of a possible flaw in
our commonly used absorption coefficients in this region.

There is also a relatively sharp feature at 1.1 \mum that is much
larger in the model than in the observations.  Further, the detailed shape of the
pseudo continuum peak near 1.27 \mum is not fit very well.

\section{Discussion}

\subsection{Why occultation constrained fits produced larger methane VMR values}

\begin{figure}[!htb]\centering
\includegraphics[width=3.2in]{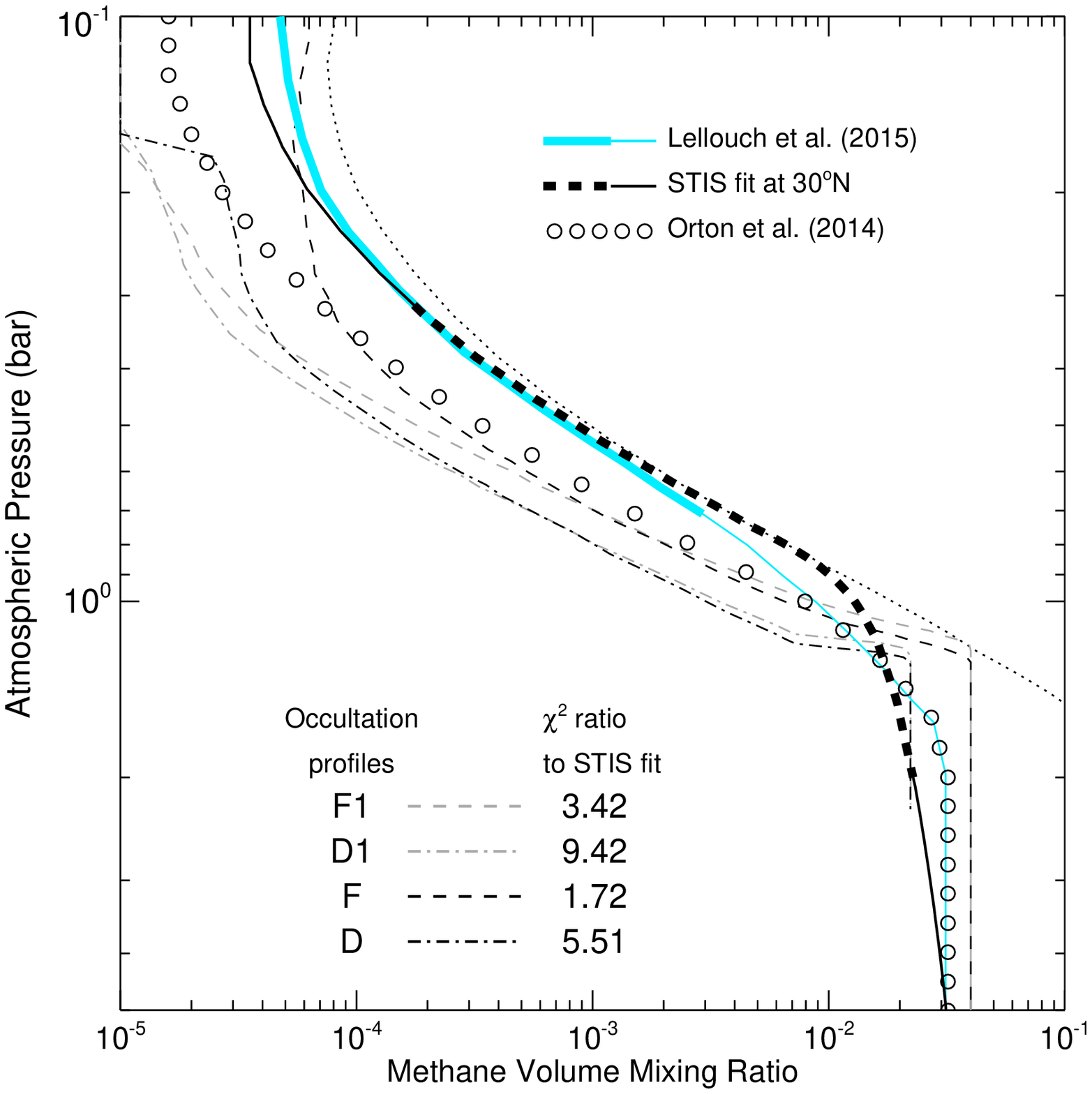}
%SOURCE  For latest documentation see P62-64, Uranus Log L.  
% Made with pltcldgas_ura_nonsub17.pro on puck. For prior plot see p 99,101, Uranus Log K.
\caption{Comparison of methane profiles derived from STIS-constrained and other
spectral observations by \cite{Lellouch2015} and \cite{Orton2014uracomp} compared
to those derived from occultation observations by \cite{Lindal1987} and \cite{Sro2011occult}.
The 4\% deep methane VMR occultation profiles provide better agreement with STIS-constrained
results in the upper troposphere.  But without occultation constraints, the preferred
deep mixing ratio is closer to 3\% for most aerosol models fit to the
730-900 nm spectrum.}
\label{Fig:ch4oldnew}
\end{figure}

Given the previous discussion of methane depletion profiles, this
might be a good point at which to compare the methane profiles in
Fig.\ \ref{Fig:vertch4mods} with those obtained from the occultation
analysis of \cite{Lindal1987} or \cite{Sro2014stis}.  This is provided
in Fig.\ \ref{Fig:ch4oldnew}, where we also show the results of
\cite{Orton2014uracomp} and \cite{Lellouch2015}.  The main regions of
sensitivity to the methane VMR values are indicated by thicker lines
for our current STIS results and those of \cite{Lellouch2015}.  Note
that our current STIS results at 30\deg N are in very good agreement
with the Lellouch et al. results where they have overlapping
sensitivity (roughly the 200--700 mbar range).  Both have relatively
high methane relative humidities compared to the saturation
vapor pressure profile computed for the \cite{Orton2014uratemp}
thermal profile.  The occultation results for methane are at much
lower levels at pressures less than the putative methane condensation
pressure (about 1.2 bars).  In the occultation analysis, temperature
and methane profiles are linked.  Both temperature and composition
affect density, which in turn affect refractivity versus altitude,
which is the main result produced from the radio measurements.  The
refractivity profile can be matched by a family of thermal and
corresponding methane profiles.  A hotter atmosphere is less dense,
and thus allows more methane to produce the same refractivity.
Because the occultation profiles have such low relative methane
humidities above the cloud level compared to what the STIS
spectra require to obtain the best fits, the hottest occultation
profile is favored.  If the only allowed adjustment of methane is
selection of the optimum occultation profile, as was the case for our
previous analyses \citep{Sro2011occult,Sro2014stis}, then we obtain a
deep mixing ratio that is relatively high (4\%) so that the methane
mixing ratio near and above the cloud level can approach closer to the
level needed to provide the best spectral match.  As an example of
this behavior, we carried out fits of STIS spectra at 10\deg N, using
STIS spectral fit quality as the only constraint, and compared that to
the best fits obtained for profiles with fixed occultation consistent
methane vertical profiles. The results, tabulated in the legend of
Fig.\ \ref{Fig:ch4oldnew}, show that all the occultation fits are much
worse than the STIS-only constrained fits, and that the best of the
occultation constrained fits (for the F profile) is for the hottest
profile, which provides the most upper tropospheric methane, even
though that has a deep methane VMR that is much higher than is needed
if one does not force the methane to fit an occultation profile.  Just
below the cloud level, the methane VMR at low latitudes is closer to
2\% at least in the region above the lower tropospheric clouds (near
2.5 bars) and perhaps deeper, although the STIS spectra are not
sensitive to values deeper than that.

%% \subsection{Latitude dependence for 2-layer diffuse conservative Mie-scattering models}

%% Given the arguments in favor of a photochemically generated haze as the main constituent
%% of the background aerosols on Uranus, we also investigated fits in which the cloud layers
%% were vertically extended.  Such models were used successfully by \cite{Kark2009IcarusSTIS},
%% and were also part of our original analysis of the 2012 STIS observations \citep{Sro2014stis}.
%% The questions we wish to address here are whether such models can provide better
%% fits to the observed spectral variations and whether such distributed models are consistent
%% with the idea that all the background aerosols in the atmosphere of Uranus are photochemically
%% generated.   According to photochemical models of \cite{Rages1991}, 

\subsection{Evidence for a deep cloud layer}

In our previous paper dealing with earlier STIS observations \citep{Sro2014stis}, we found that the
fit quality at short wavelengths was improved by adding a deep cloud layer, which
we fixed at the 5 bar level and assumed had the same tropospheric scattering parameters
as KT2009.  The only adjustable parameter for that layer was its wavelength-independent optical depth, which
we found to vary from about 4 at low latitudes to about half that at high latitudes.
It is possible in our current modeling that the more extended vertical extent of our upper
tropospheric cloud layer serves to reduce the need for the contribution of a 
deeper cloud.  The main function of the deeper cloud is to improve the fit in the 540 to 600 nm
range where matching weak methane band depth is easier if some of the aerosol
scattering is moved to higher pressures.  

%% To provide a better test of the existence
%% of a deeper cloud, we looked at spectra with more vertical view angles.  Choosing a
%% spectrum for 
%% $\mu$=0.9 at 10\deg N,  we did find that this spectrum with greater
%% penetration depth revealed greater spectral errors in matching the methane
%% band depths than is apparent from spectra at our chosen cosines of 0.3, 0.5, and 0.7 (these
%% were chosen to give more extended latitude coverage).  We found that adding an optically
%% thick cloud layer at 13.4$\pm$0.8 bars improved the fit quality even for our original
%% set of view angles. Adding that layer to the model plotted in Fig.\ \ref{Fig:fullmie1},
%% decreased \chisq from 625.15 to  591.03, when we also increased
%% slightly the imaginary index in the 730 nm - 960 nm range, from 4.9$\times 10^{-3}$ 
%% to 5.7$\times 10^{-3}$.  Although this decrease in \chisq is comparable to its
%% uncertainty, the change is concentrated in a small wavelength range that makes
%% it an obvious improvement.  Further investigation of the nature of this deep
%% cloud layer, including its latitude dependence, is left for future work.  A plausible
%% composition for such a cloud is \nhfshx.

\begin{figure}\centering
\includegraphics[width=3.2in]{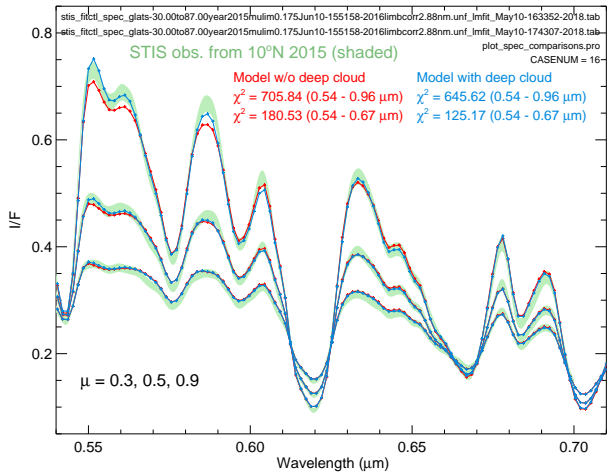}
\caption{STIS 2015 observations at 10\degx N (green shading indicating uncertainties),
compared to fitted model results without a deep cloud layer (red) and
with a deep cloud layer (blue). These are for non-standard, more deeply
penetrating, zenith angle cosines of 0.3, 0.5, and 0.7, with largest
cosines corresponding to largest I/F values at continuum wavelengths. 
The legend gives \chisq values for the entire spectral range that
was fitted (0.54 - 0.96 \mumx) and for the region most influenced
by the deep cloud (0.54 - 0.70 \mumx).}\label{Fig:deepcld}
\end{figure}

 To provide a better test of the existence of a deeper cloud, we
 looked at spectra with deeper penetration. Choosing a spectrum with a
 nearly vertical view ($\mu$ = 0.9 at 10\degx N), we computed
 simultaneous model spectra for view angle cosines of $\mu$ = 0.3,
 0.5, and 0.9, based on the fit we obtained using the standard set of
 view angle cosines ($\mu$ = 0.3, 0.5, and 0.7).  That model did not
 fit the weak methane bands very well even with the standard view
 angles and was even worse for this more deeply penetrating set.  The
 \chisq values rose from 586.32 to 714.59, with an expected \chisq
 uncertainty of 35-40.  This \chisq increase by 128.3 is about three
 times its uncertainty.  However, by refitting the same model (still
 without a deep cloud) to the new set of view angles, we reduced the \chisq value for this new set of
 angles to 705.84, and thus reducing the difference to 119.5, which is
 still about three times the expected uncertainty in \chisqx.  After
 inserting an optically thick deep cloud with an adjustable pressure,
 a new fit further reduced \chisq from 705.84 to 645.62, a decrease of
 59.52, which is about 1.6 times its uncertainty. A comparison of the
 latter and initial fits to the measured spectra is displayed in
 Fig.\ \ref{Fig:deepcld}.  The \chisq improvement is even more dramatic when
 computed just for the region from 540 nm through 670 nm.  In that
 case the \chisq change is from 180.53 to 125.17, a decrease of 75.46,
 which is over three times the expected uncertainty of about 22 for
 this more limited range that has 243 comparison points.  The model
 with a deep cloud also improved fits at the original set of view (and
 zenith) angles.  Adding that layer to the model plotted in Fig.\ 29
 and refitting, decreased \chisq from 586.32 to 529.95, a decrease of
 56.37, with most of this change taking place in the 540-670 nm region
 where \chisq dropped from 159.10 to 102.47, a decrease by 56.63,
 which is about 2.6 times the expected uncertainty.  Thus both sets of
 view angles lead to significant local fit improvements, with derived
 effective pressures of 10.6$\pm$0.4 bars for the more deeply
 penetrating view angles and 9.5$\pm$0.5 bars for our standard set.  A
 better estimate for the effective pressure of an optically thick deep
 cloud is probably 10$\pm$0.5 bars.  A lower pressure is likely if the
 cloud is not optically thick.  When we fixed the deep cloud pressure
 at 5 bars the best-fit optical depth of the cloud was 4.2$\pm$0.7
 (using the more deeply penetrating spectral constraints). This is
 quite consistent with the optical depth of the 5-bar deep cloud fits
 of \cite{Sro2014stis}.  Further investigation of the nature of this
 deep cloud layer, including its latitude dependence, is left for
 future work.  A plausible composition for such a cloud is \nhfshx.

\subsection{Comparison with other models of gas and aerosol structure on Uranus.}

\begin{figure*}[!htb]\centering
\includegraphics[width=5.75in]{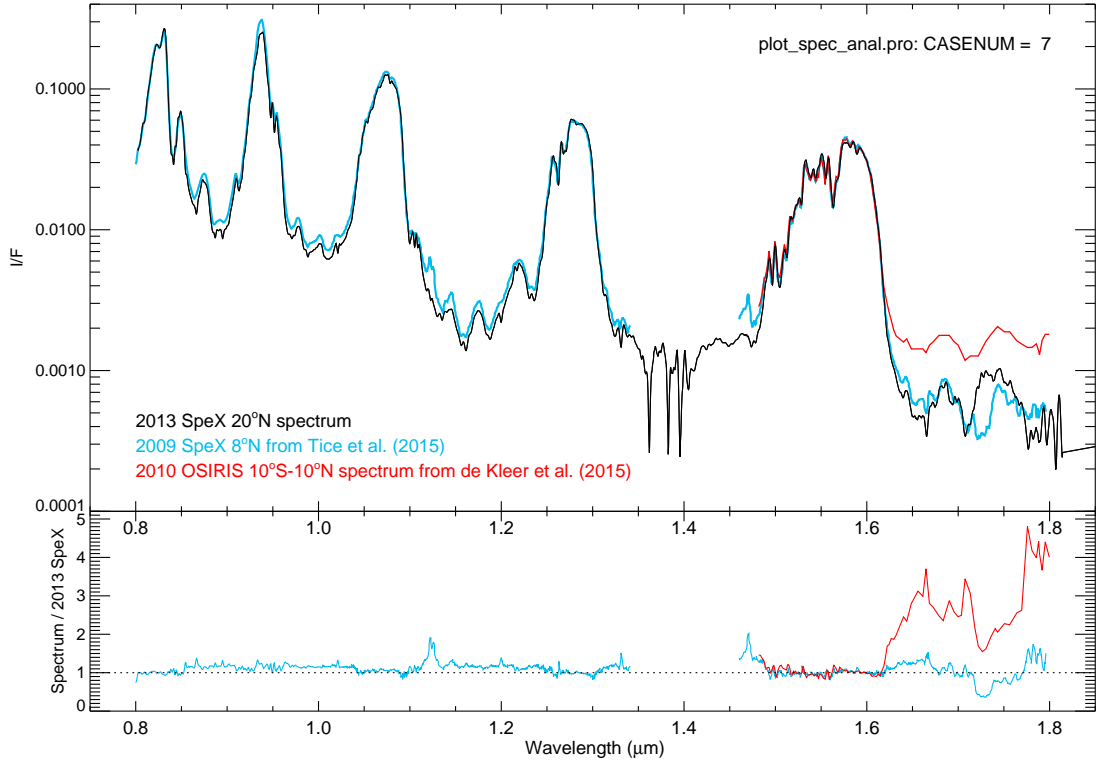}
%SOURCE: Case 7 plot by plot_spec_anal.pro on puck
% documentation on page XX Log L
\caption{Comparison of near-IR spectra of Uranus. Our 2013 central-disk spectrum is shown
in black. The 2009 SpeX central-disk spectrum of \cite{Tice2013} is shown in blue, and
the 2010 OSIRIS 10\degx S - 10\deg N spectrum of \cite{DeKleer2015} in red.  The bottom panel
plots the ratio of each spectrum to our 2013 SpeX spectrum.}
\label{Fig:nirtriple}
\end{figure*}

Models of 0.8-1.8 \mum SpeX spectra of Uranus by \cite{Tice2013} and
more recently by \cite{Irwin2015reanalysis} and recent models of
H-band (1.47-1.8 \mumx) spectra by \cite{DeKleer2015} present what
appear to be different views of the cloud structure from that derived
from our STIS observations.  Some fraction of the differences are due
 to different constraining
assumptions.  The other authors typically constrain the upper cloud
boundary pressure and fit the scale height ratio, while we have here
mainly assumed a unit scale height ratio (particles uniformly mixed with
gas) and treated the upper boundary pressure as adjustable.
 The differences are probably not due to very different conditions
on Uranus, as the spectral observations are generally very similar, as
illustrated in Fig.\ \ref{Fig:nirtriple}.  These spectra are all obtained near
the center of the disk, and all near latitude 20\deg N.  In most of
the spectral range they are all within 10\% of each other.  The main
exception is the \cite{DeKleer2015} spectrum, which is much brighter than the
other two spectra in the 1.63-1.8 \mum region.  This would presumably
lead to a model with much greater stratospheric haze contributions
than would be needed to match the other spectra.  To better
characterize these differences and better understand their origin, we
attempted to reproduce results from these near-IR analyses.

The first attempt was to match the \cite{Irwin2015reanalysis} retrieval
of a two-cloud structure from the \cite{Tice2013} 2009 SpeX central meridian data (0.8-1.8
\mum range). They used 1.6\% deep \chf with 30\% RH above condensation
level and the \cite{Lindal1987} Model D T(P) profile. They retrieved self-consistent
refractive indexes for their Tropospheric Cloud (TC) and Tropospheric
Haze (TH), similar to Tice et al. 2-cloud model. They retrieved
particle sizes, but used "combined H-G" phase function fits in the
forward modeling, rather than Mie calculations. An additional
complication was that their plots of optical depth vs pressure were
incorrect in the paper (estimated to be about an order
of magnitude too large, P.G.J. Irwin private communication). That
and the uncertain way double HG phase functions were obtained from
the Mie phase functions, led us to not attempt detailed quantitative
comparisons.

We decided to make our quantitative comparisons with 2-cloud results
of \cite{Tice2013}. This was more tractable, as the phase functions
for both TC (tropospheric cloud) and UH (Upper Haze, called TH in
Irwin et al.)  were simply H-G phase functions with an assumed
asymmetry parameter g = 0.7.  They also utilized wavelength-dependent
optical depths based on Mie calculations of extinction efficiency, but
they did not use the wavelength-dependent phase functions or
wavelength-dependent asymmetry parameters for either particle mode.
Although, for the larger particles in the TC, the asymmetry parameter
obtained from Mie calculations is close to their chosen value, the
0.1-\mum particle model has a very small asymmetry, which leads to a
backscatter phase function value about ten times that for an HG
function with g = 0.7.  Since their UH (or TH) particles have such
small optical depths, their contribution can be well approximated by
single-scattering, in which case the observed I/F contribution is
given by \begin{eqnarray} I/F = \frac{1}{4}\varpi P(\theta)
  \tau/\mu \label{Eq:singlescatt}
\end{eqnarray}
where $\theta$ is the scattering angle (about 180\deg in this case),
 $\tau$ is the vertical optical depth, and $\mu$ is the cosine
of the observer zenith angle.  This makes the modeled I/F strongly
dependent on the assumed phase function, specifically its backscatter
amplitude.  While there is substantial variation
in scattering efficiency with wavelength for a 0.1-\mum particle,
such a particle would not have such a strongly forward peaked
phase function, and would probably require roughly a factor of ten
lower optical depth than \cite{Tice2013} found for their UH layer.
However, using their peculiar scattering characterization for
this layer, and using their more plausible characterization for
the lower layer, and their chosen single-scattering albedos,
we were able to roughly match our own 2013 SpeX center-of-disk
spectra (which are quite similar to the spectra shown in the Tice
paper). Thus we have two different vertical structures that can
match the spectra.  Ours has a single tropospheric layer uniformly mixed
between 1.06 and 3.3 bars (small particle solution), while theirs
has a very strongly varying optical depth per bar between
their assumed cloud top of 1 bar and their fitted bottom at 2.3 bars.
We did not attempt to reproduce the more complex structures
based on \cite{Sro2011occult} three- and four-cloud models.

We also tried to reproduce \cite{DeKleer2015} results for a 2-cloud
model. Their retrievals were for a more limited H-band wavelength
range (H-band spectra). Their spectra were also similar to our 2013
SpeX results, except their dark regions were as much as 3-4 times brighter
(see Fig.\ \ref{Fig:nirtriple}). They used a two-stream radiative
transfer model, with wavelength dependent H-G parameters based on Mie
calculations. Using their retrieved optical depths, we roughly matched
their window I/F. However,  we used correlated-k
coefficients for Hartmann type line-shape wings, while de Kleer et
al. used the hybrid wing shape from \cite{Sro2012LBL} that produces more absorption in the H-band
window. Using these c-k coefficients, our I/F values in the methane
window were lower than those of de Kleer et al. by a factor of 2 or so.
The origin of these differences remain to be determined.  It is
likely that it is not entirely a result of very different
numbers of streams, as de Kleer et al. did trial calculations
showing that their approximation was good to within $\sim$10\%.

\begin{figure*}[!htb]\centering
\includegraphics[width=5.5in]{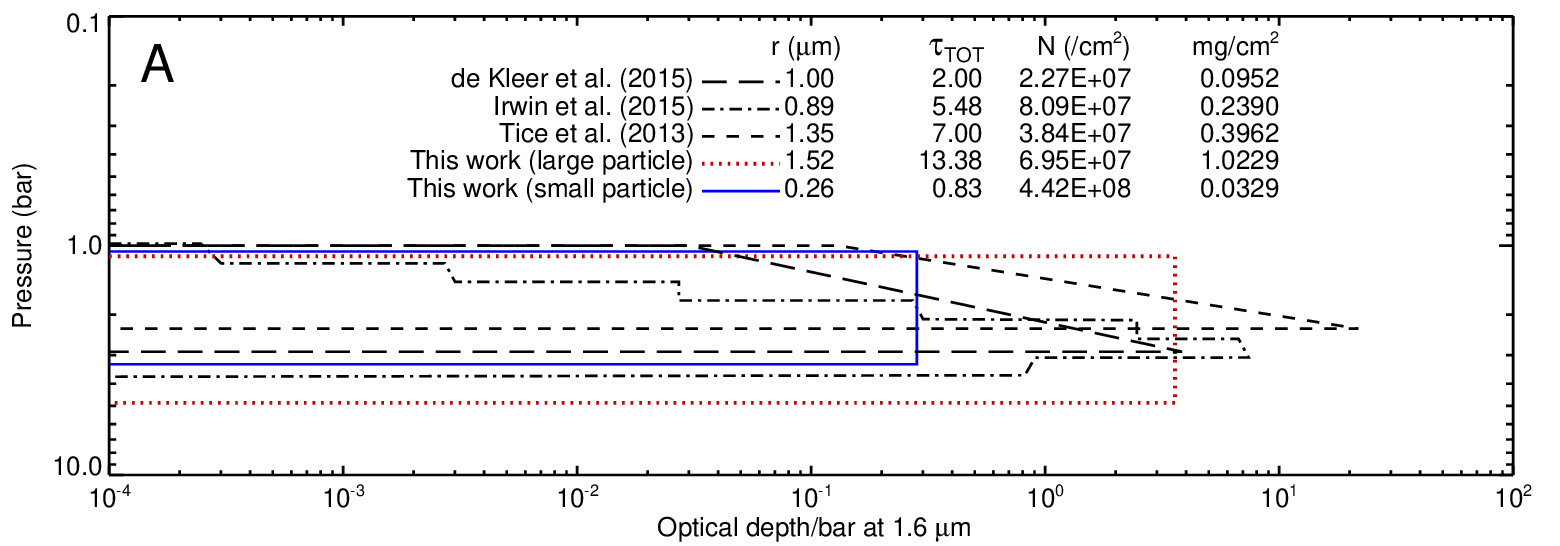}
\includegraphics[width=5.5in]{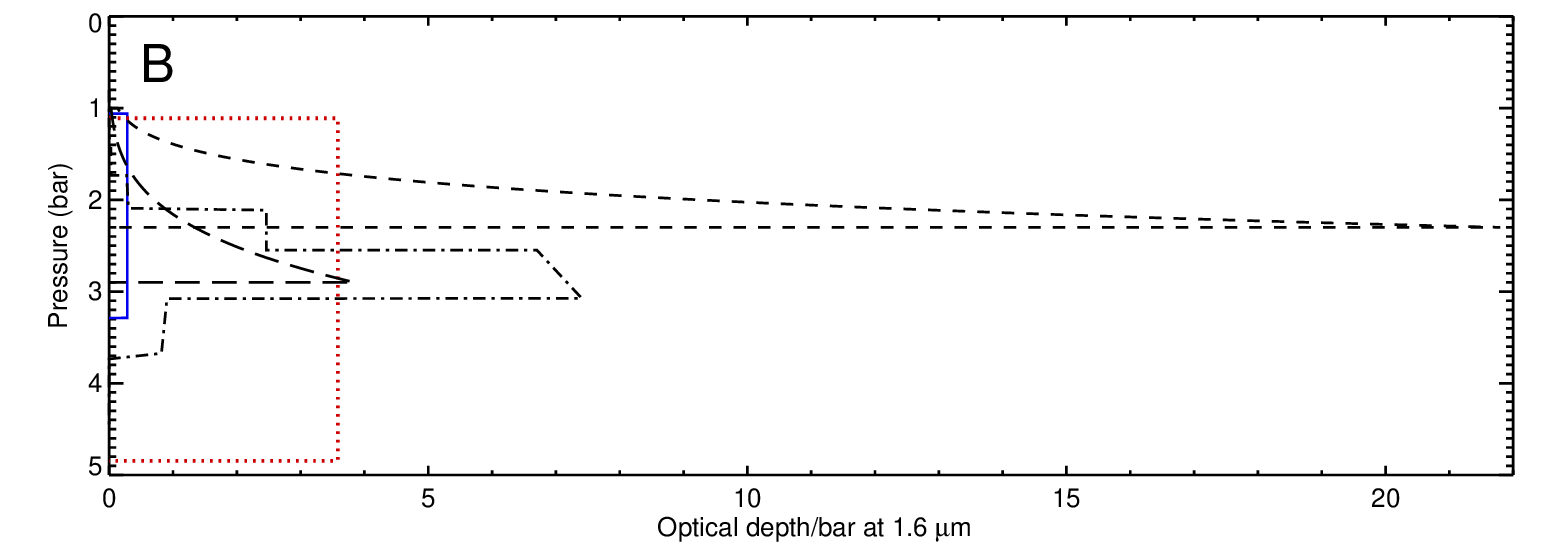}
%SOURCE Plotted by plot_cloud_profiles.pro on puck in ir_ms_ura
% See page 90-91, Uranus Log L for documentation
\caption{Comparison of tropospheric cloud density vertical profiles on log scales
(A) and linear scales (B).  Our small-particle fit is shown with solid 
lines in both panels, while results from other investigators are
shown using lines defined in the legend.  The \cite{Irwin2015reanalysis} result
has been scaled downward by a factor of 10, which is a rough correction
from what is shown in the left panel their Fig. 2, suggested by P.G.J. Irwin (personal
communication). Their profile was derived for a deep methane mixing ratio of 1.6\%
and would move upward by several hundred mbar for double that mixing ratio.
The \cite{Tice2013} profile was derived using a deep methane mixing ratio of 2.2\%,
which is also the case for the \cite{DeKleer2015} profile.
As noted in the legend, our small-particle model has much less optical
depth at 1.6 \mum and much less total mass than the other results shown. The estimated
total column cloud mass per unit area assumes a density of 1 g/cm$^3$.}
\label{Fig:cloudcomp}
\end{figure*}

A comparison of the characteristics of the tropospheric cloud models
from aforementioned references is displayed in Fig.\ \ref{Fig:cloudcomp}.
Although all of these models provide good fits to the spectra (ignoring
the fact that we could not reproduce all these results), they have
very different vertical structures and total optical depths and column masses.
In fact the widest variation in total cloud mass is between our own small-particle
and large-particle solutions.  In the log-log plot in Fig.\ \ref{Fig:cloudcomp}A,
the various cloud structures seem more similar than in the linear plot in panel B,
where the huge differences in optical depths and mass loading are more accurately
conveyed.  The column number density in particles per unit area is computed as $n = \tau/(\pi r^2 Q_{ext}$),
where $r$ is the particle radius and $Q_{ext}$ is the extinction efficiency (extinction
cross section divided by geometric cross section).  From $n$ the mass
loading (mass per unit area) is computed as $m$ = $n \rho \pi r^3$, assuming
that the particle density $\rho$ is 1 g/cm$^3$.  Our small-particle
tropospheric cloud is one of very low maintenance.  It needs very little
material to form, the particles fall slowly because they are small, and thus
probably a low level of mixing is needed to sustain it.  It also has the virtue
of having a refractive index similar to that of its potential main component, \htsx.
The large particle cloud is thirty times more massive, with larger particles
that fall much more quickly, needing much more vertical transport to be sustained.

\begin{figure}[!htb]\centering
\includegraphics[width=3.2in]{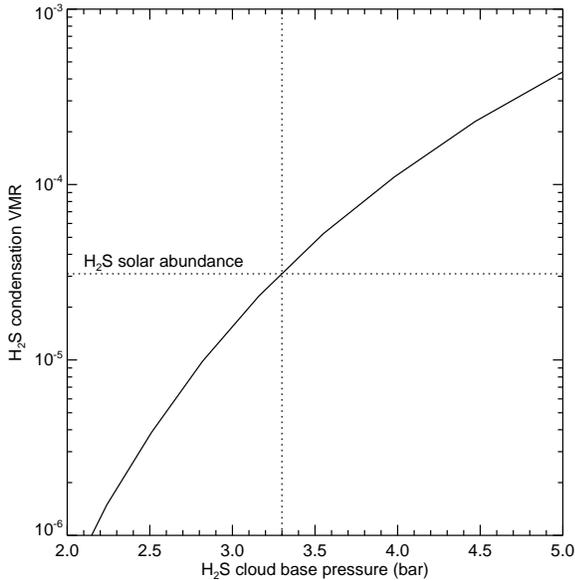}
%SOURCE Plotted plot_orton2014h2s.pro on triton in ../paper_stis15
% See page 105, Uranus Log L for additional documentation
\caption{Minimum H$_2$S VMR required to condense at the cloud base
versus cloud base pressure.}
\label{Fig:condensep}
\end{figure}

If these clouds are to be made of H$_2$S, it is worth considering whether 
there is enough H$_2$S available to make them.   For a  mixing ratio $\alpha_{H_2S}$, 
the mass per unit area of \hts between two pressures
separated by $\Delta$P would be (M$_{H_2S}$/M)$\alpha_{H_2S} \Delta P/g$, where
$g$ is gravity (9.748 m/s$^2$), and the ratio of molecular weights
of \hts to the total is given by 34/2.3 = 14.78.  For \hts to condense
at the tropospheric (layer-2) cloud base its mixing ratio must have a minimum
value that depends on base pressure as shown in Fig.\ \ref{Fig:condensep}.
To condense at the 3.3 bar level would require the \hts VMR to be equal
to its the solar mixing ratio of 3.1$\times 10^{-5}$ \citep{Lodders2003}.  About
10 times that VMR would be needed to condense as deep as the 5 bar level and about
ten times less would lead to condensation no deeper than the 2.4 bars.  Microwave
observations by \cite{DePater1991Icar} suggest \hts is at least a factor of ten above
solar.
Even for just a 10 ppm mixing ratio, this yields
an \hts mass loading of 169 mg/cm$^2$ per bar of pressure difference.  Thus,
condensing all the \hts in just a 1-bar interval would make 170 times the cloud 
mass that is inferred for the large-particle
solution and more than 5000 times the mass needed for the small-particle
cloud.  Thus, none of these clouds is immediately ruled out by lack of condensable
supply.  A more sophisticated microphysical analysis would be needed to evaluate
them, accounting for eddy mixing, coagulation, sedimentation, and other effects.
Another test would be to compare model spectra for these various distributions
with STIS spectra at CCD wavelengths.
We have verified that our STIS-based models can fit near-IR spectra, but
the reverse has not yet been demonstrated for near-IR based models.

\section{Summary and Conclusions}

We observed Uranus with the HST/STIS instrument in 2015, following the
same approach as in 2012 and 2002. We aligned the instrument's slit
parallel to the spin axis of Uranus and stepped the slit across the
face of Uranus from the limb to the center of the planet, building up
an image of half the disk with each of 1800 wavelengths from 300.4 to
1020 nm. The main purpose was to constrain the distribution of methane
in the atmosphere of Uranus, taking advantage of the wavelength region
near 825 nm where hydrogen absorption competes with methane absorption
and displays a clear spectral signature. Our revised analysis approach
used a considerably simplified cloud structure, relaxed the
restriction that methane and thermal profiles should be consistent
with radio occultation results, considered the new Uranus global mean
profile of \cite{Orton2014uratemp} that was inconsistent with radio
occultation results, and included parameters defining the methane
profile as part of the adjusted parameter sets in fitting observed
spectra.  This revised analysis applied to STIS observations of Uranus
from 2015 and comparisons with similar 2002 and 2012 observations, as
well as analysis of HST and Keck/NIRC2 imaging observations from 2007
and 2015, and IRTF SpeX spectra from 2013, have led us to the
following conclusions.

\setlength{\leftmargini}{6mm}
\setlength{\leftmarginii}{2mm}

\begin{enumerate}%[topsep=0pt,itemsep=20pt,parsep=0pt,partopsep=0pt]
\item TEMPORAL CHANGES
\newcounter{con}
\begin{list}{1.\arabic{con}}
{\usecounter{con}
\setcounter{con}{0}
}
\item A direct comparison of 2012 STIS spectra with 2015 STIS spectra
reveals no statistically significant difference at low latitudes.
At 10\degx N and a zenith cosine of 0.7, the spectra from the two years are within the noise
level of the measurements.

\item A different result is obtained by comparing 2012 and 2015 STIS spectra at high latitudes.
There we find significant differences at pseudo-continuum wavelengths beyond 500 nm, where
weaker methane bands are present, and where the 2015 I/F exceeds 2012 I/F values
by up to 0.04 I/F units (about 15-20\%). However, no difference is seen in the strong methane
bands that would be sensitive to changes in stratospheric aerosols.  

\item The brightening of high latitudes at pseudo continuum wavelengths between 2012 and 2015 
 is a result of increased scattering by tropospheric aerosols, and not due to a change in the
effective methane mixing ratio.  This is
shown by radiation transfer modeling as well as by direct comparisons
of imaging at wavelengths with different fractions of hydrogen and methane
absorption.

\item The polar brightening from 2012 to 2015 that we found in comparisons
of STIS spectra is part of a long-term trend evident from comparisons
of H-band images from the 2007 equinox and onward, including recent
images obtained in 2017 \citep{Fry2017DPS}.

%% \item The brightness of high latitudes relative to low latitudes is partly due
%% to the latitudinal decline in upper tropospheric methane mixing ratios, but
%% more recently a significant additional polar brightening has
%% been produced by the increase in aerosol scattering over time.
\end{list}

\item METHANE DISTRIBUTION:
%\begin{enumerate}

\begin{list}{2.\arabic{con}}
{\usecounter{con}
\setcounter{con}{0}
}

\item While the increased brightness of the polar region between 2012 and 2015
is due to increased aerosol scattering, the fact that the polar region is
much brighter than low latitudes in 2015 is due to the lower mixing ratio of
upper tropospheric methane at high latitudes.

\item We found that the STIS spectra from 2015 and 2012 can be well
  fit by relatively simple aerosol structures. We used a two-layer
  cloud structure with an optically thin stratospheric haze, and one
  tropospheric cloud, the latter extending from near 1 bar to several
  bars. This is similar to the 2-cloud model of \cite{Tice2013} except
  that we fit the upper boundary instead of fixing the upper boundary
  and fitting the scale height ratio. The particles in the tropospheric cloud were
  modeled either as spherical particles uniformly mixed with the gas
  and with a fitted real index, or as non-spherical particles using an
  HG phase function with a fitted asymmetry parameter. 

\item Our initial fits to the 2015 STIS spectra over the entire range
from 540 nm to 980 nm using either a two-cloud or three-cloud
model using spherical particles of real refractive index of n = 1.4
produced good overall fits that were especially bad near 830 nm, just where
the spectrum is especially sensitive to the methane to hydrogen ratio.
Much better fits were obtained by allowing the refractive index
of the tropospheric aerosols to be adjusted, which yielded
two solutions, one a large-particle low-index solution and second
small-particle high-index solution, the latter providing the better
fit and somewhat closer match to the refractive index of \htsx.

%% \item We decided to first focus more narrowly on fitting the 730 nm to
%%   900 nm region to better separate the determination of the methane
%%   distribution from the problem of constraining the wavelength
%%   dependent properties of the aerosol particles.

\item Our preliminary 2-cloud models using spherical particles
found little variation as a result of using different temperature profiles,
as long as we did not force the deep methane mixing ratio or the methane
humidity above the condensation level to be constrained either by occultation
results or by a prohibition against supersaturation.  We chose to use
the \cite{Orton2014uratemp} profile, even though it is inconsistent with
occultation results, because its higher upper tropospheric
temperatures allowed more methane without supersaturation.

\item For subsequent models containing a stratospheric haze and
 just a single tropospheric conservative Mie-scattering layer mixed
uniformly with the gas, we did preliminary fits to spectra at 10\deg N and
60\deg N over the 730 nm to 900 nm range, and for both 2012 and 2015, assuming that
methane was uniformly mixed below the condensation level. We found two classes of solutions, 
one with large particles  of 1.1-1.75 \mum in
radius and a real index of 1.22$\pm$0.05 to 1.28$\pm$0.07,  
and a second solution set with small particles about 0.24$\pm$0.07 \mum to 0.34$\pm$0.1 \mum in radius
with much larger real index values from 1.55$\pm$0.16 to 1.86$\pm$0.30. The small
particle index values are much closer to that of H$_2$S, a prime candidate
for the cloud's main constituent. 
 
\item The above preliminary fits with uniform methane mixing ratios found those
ratios ranged from 2.56\%$\pm$0.26\% to 3.16\%$\pm$0.5\% at 10\degx N, and
 from 0.74\%$\pm$0.05\% to 0.99\%$\pm$0.08\% at 60\degx N, with
lower values in both cases obtained from the large particle solutions, but
good agreement between 2012 and 2015 in both cases. 

\item Preliminary fits using non-spherical HG particles for the single
  tropospheric cloud layer produced similar results, with a methane
  mixing ratio from 2.85\%$\pm$0.3\% to 3.48\%$\pm$0.5\% at 10\degx
  N and from 0.97\%$\pm$0.06\% to 1.04\%$\pm$0.07\% at 60\degx N, and in this case
  differences between 2012 and 2015 are within estimated
  uncertainties.

\item All the above preliminary fits found methane humidities in the 68\% to 95\% at 10\degx N,
and 30\% to 56\% at 60\degx N, generally with uncertainties of 12-16\% and 18-26\% respectively.

\item  STIS results in the upper troposphere are in good
agreement with the \cite{Lellouch2015} results based on Herschel observations.
For 2015, the relative methane humidity above the nominal condensation level, which is roughly
at the 1-bar level, for the Orton et al. thermal profile is roughly 50\% north of 30\deg N
 but near saturation from 20\deg N and southward, 
but becomes supersaturated for the F1 and F0 profiles.  

%% \item 
%% \item There is no resolvable change in methane distribution between 2012 and 2015 in the north
%% polar region, even though the polar region is notably brighter at wavelengths of intermediate
%% methane absorption.

\item Latitude dependent fits assuming a uniform methane mixing ratio below the condensation level
show that a local maximum value of about 3\% is attained near 10\degx N latitude.  From that point the
effective mixing ratio smoothly declines by a factor of 2 by 45\degx N, and by
a factor of three by 60\degx N, attaining a value of about 1\% from 60\deg to 70\degx N.  However, if
particle absorption is present, the derived mixing ratios are lowered by
up to 10\% of their values, or possibly more, depending on models.  Thus, it is not possible to give
a firm value of the mixing ratio without a deeper understanding to the aerosols within the atmosphere.

\item For a vertically uniform methane mixing ratio, the high-latitude model fits failed
to accurately follow the observed spectra in the 750 nm region, suggesting that the
upper tropospheric methane mixing ratio increased with depth.  This was especially obvious
for the 2012 observations, probably because of reduced aerosol scattering in 2012. A model
profile containing a vertical gradient above the 5-bar level, using either what
\cite{Sro2011occult} called a descended depletion profile or a step decrease
at the 3 bar level made a substantial improvement in the fit quality.

\item When the methane depletion with latitude is modeled as a stepped
depletion, we find that the step change occurs at pressures between 3 and 5 bars, although
the uncertainty is typically 2 bars.  This level applies between about 50\deg and 70\degx N,
but moves to lower pressures between 50\deg N and 20\deg N, and remains near the condensation
level from that point to 20\degx S.  The mixing ratio above the break point pressure is
near 0.75\% in the 60\degx N to 70\degx N range, increasing to about 1.2\% at low latitudes,
although by that point the depleted layer is so thin that it is hard to distinguish
from the uniformly mixed case with a single mixing ratio up to the condensation level.

\item Because the shape of the descended profile makes the depth parameter of that profile
difficult to constrain with the spectral observations, we were guided by the stepped
depletion results to choose a fixed depth parameter of 5 bars, and fit just the
shape parameter $vx$ as a function of latitude.  The results show a relatively smooth
variation from slightly greater than 1 at high latitudes, increasing to about 4 by 30\degx N,
then rising to very high values at low latitudes, which yields a nearly vertical
profile that produces negligible depletion.  

%% \item The variation of the methane vertical distribution with
%%   latitude, when modeled as a ``descended'' version of the vertically
%%   uniform low latitude profile, displays an increasing depth of
%%   depletion with latitude, reaching 8$\pm$4 bars at 70\deg N in
%%   2015. The best-fit shape parameter ($vx$) is relatively large,
%%   ranging from $\ge$ 9 at low latitudes to 2.4$\pm$0.7 at 70\deg N,
%%   meaning that the pressure domain of the descended dry gas does not
%%   extend much above the 1 bar level.  The low pressure at which
%%   depletion begins decreases with latitude, starting near 1 bar and
%%   decreasing to nearly 800 mbar at 70\deg N.

\end{list}

\item AEROSOL PROPERTIES:

\begin{list}{3.\arabic{con}}
{\usecounter{con}
\setcounter{con}{0}
}

\item Preliminary fits with non-spherical particles with a simple HG
  phase function yielded asymmetry parameters that ranged from
  0.43$\pm$0.04 at 10\degx N to 0.26-0.39 at 60\degx N. These are
  smaller values than the commonly used value of $g$ = 0.7, e.g. by
  \cite{Tice2013}.  It is also smaller than the asymmetry parameters
  of even the small-particle solutions for the tropospheric aerosols, which ranged from
  about 0.4 at 1.6 \mum to 0.6 at 0.8 \mumx. The large-particle
  asymmetry values were near 0.86 at 0.8 \mum and 0.9 at 1.6 \mumx.

\item The cloud pressure boundaries varied with model structure.  When
  a vertically uniform methane profile is assumed, the top boundary of
  the cloud is precisely constrained and nearly invariant with
  latitude, moving from slightly greater than 1 bar at low latitudes
  to almost exactly 1 bar at high latitudes. The lower boundary is
  more uncertain varying about a mean near 2.6 bars.  The optical
  depth of the tropospheric cloud declines by roughly a factor of two
  from low to high latitudes, when 2012 and 2015 results are averaged.
  For the stepped depletion models, the top boundary behavior is
similar to that of the uniform model, but the bottom boundary moves
from 2.5 bars at low latitude to 3 bars at high latitude. 

\item A very different characteristic is seen for the descended
  methane fits as a function of latitude.  In this case the upper
  boundary of the tropospheric cloud moves significantly downward with latitude instead of
  slightly upward, with the pressure increasing from about 1.1 bar to
  1.3 bar. We also found that the refractive index increased with
  latitude, from 1.6 to about 2.0, perhaps a result of a low-index
  coating evaporating from a high index core as the cloud
  descends to warmer temperatures. The particle radius also decreases
  somewhat with latitude, which would be consistent with that
  speculation.  The tropospheric cloud optical depth is also seen to
  decline somewhat at high latitudes, as seen for other models.

\item The real refractive index of the main cloud has a relatively
flat latitude dependence for the stepped depletion model, but significant
increases with latitude are seen for uniform and descended depletion models.
Better agreement is obtained at low latitudes, where weighted averages
over 2012 and 2015 from 20\degx S to 20\degx N are 1.65$\pm$0.08, 1.66$\pm$0.07, and 1.63$\pm$0.07
for uniform, stepped depletion, and descended depletion models respectively,
which are all above the expected value of 1.55 for \hts by amounts that
are not much greater than combined uncertainties.

%% \item The temporal increase in brightness of the north polar region seen in
%%   H-band groundbased images, as well as in HST/STIS and HST/WFC3
%%   images sampling intermediate methane bands is due to increased
%%   aerosol scattering in 2015, and has a character similar to what was
%%   observed in NICMOS F165M images obtained in 1997.

\item The way aerosol contributions produce the increased polar
  brightness between 2012 and 2015 is simplest to understand within the context of the
  models assuming vertically uniform methane.  In these cases an
  increased amount of scattering in the main cloud layer produces
  the brightness increase. And at 60\deg N and 70\deg N this is due
  to a combination of increased optical depth and increased particle size.
  Similar effects are seen in the stepped depletion model
  model (small particle solution). In the descended depletion model it
  appears that an increase in the cloud top pressure over time may be
  a significant factor.  For the simple non-spherical HG particle
  cloud we found a 37\% increase in optical depth coupled with a 33\%
  decrease in the asymmetry parameter from 0.39 to 0.26.  These
  effects would produce a combined rise in pseudo continuum I/F of
  about 32\% (= 0.45$\times$(0.37+0.33)), which is comparable to the
  observed change.

\item The association of high-latitude methane depletions with
  descending motions of an equator-to-pole deep Hadley cell does not
  seem to be consistent with the behavior of the detected aerosol
  layers, at least if one ignores other cloud generation mechanisms
  such as sparse local convection.  Both on Uranus and Neptune \citep{DePater2014nep},
  aerosol layers seem to form in what are thought to be downwelling
  regions on the basis of the effective methane mixing ratio
  determinations.  

\item Models using conservative spherical particles in the tropospheric
cloud layer have significant flaws when fit to the wider
spectral range from 540 nm to 980 nm and assuming a real index of refraction of
$n$ = 1.4.  Much smaller flaws are seen with small
particles with a larger refractive index, but more accurate fits require additional
wavelength dependent scattering characteristics. 
This can be done by adding absorption in the longer wavelength regions, which
allows increasing optical depths enough to brighten the shorter wavelength regions.
For small particles with high real index values we needed to increase
the imaginary index from zero at short wavelengths to 1.09$\times 10^{-3}$
between 670 to 730 nm and to  4.9$\times 10^{-2}$ from 730 nm to 1 \mumx.
For non-spherical HG particles, we were able to match the same 
spectral region by creating an appropriate variation in optical depth
with wavelength.  It is also possible to produce a similar fit for
DHG particles by appropriate wavelength dependence in the phase function,
following an approach used by KT2009.

\item We were able to extend Mie model fits to the near-IR spectral range by further
adjustments of the imaginary index with wavelength.  For small particles the
imaginary index had to be elevated to 0.1 at 1.6 \mumx, where its single-scattering
albedo descends to 0.64.  For large particles, the needed imaginary index  
increase was to a level eight times less than for small particles, and the
single-scattering albedo was decreased to a more modest value of 0.90.

\item Our two solutions for cloud structures that can match spectra from
visible to near-IR wavelengths to at least 1.6 \mumx, require vast
differences in the total optical depth and cloud mass.  These solutions bound solutions
from other investigators, which have different vertical structures that
in most cases match spectra from 0.8 to 1.6 \mumx.  The column masses of
particles in these clouds range from 500 to 17 times smaller than the total
mass of \hts in a 1-bar pressure interval, and thus, even the most
massive of these clouds cannot be ruled out on the basis of insufficient
parent condensate.

\item We found evidence for a deep cloud layer in
  the 9 bar to 11 bar range if optically thick and possibly composed of \nhfshx.  Including
  this layer in our models has the main effect of improving
  our fits to the weak methane band structure at wavelengths from 540
  nm to 600 nm.  Placing the deep cloud at 5 bars yields an optical depth
near 4 but a worse fit to the spectra.  Further work is needed to better
constrain the properties of this cloud.

\end{list}

%\newpage
\item CONSTRAINTS ON H$_2$S:

\begin{list}{4.\arabic{con}}
{\usecounter{con}
\setcounter{con}{0}}
\setlength\leftmargin{0in}

\item If the tropospheric cloud is a condensation cloud, \hts is the
  likely main component.  This conclusion is based on the fact that
  virtually all of the model cloud mass is below the level at which
  methane can condense, but likely within the pressure range at which
  \hts can condense.  It is also the case that our preferred
  small-particle solutions are in rough agreement with the refractive
  index of \hts at low latitudes, although that agreement worsens at
  high latitudes and thus does not provide compelling support.  More
  compelling support for \hts as the main constituent of this cloud is
  the recent detection of \hts vapor at saturation levels above this
  cloud \citep{Irwin2018h2s}. What remains unclear is whether the 
spectrally varying imaginary index that seems to be required for this
cloud is compatible with the absorbing properties of condensed \htsx.

% The
%  small-particle cloud also has small particles with small
%  sedimentation speeds and minimal requirements for maintenance by
%  eddy mixing. 

\item Based on the estimated bottom boundary of the tropospheric
aerosol layer, if the small particle solution is to be consistent
with a composition of H$_2$S,  the mixing ratio of H$_2$S 
at the 3.3-bar level and immediately below must be at least $\sim$30 ppm.
To be consistent with the large particle solution would require
around an order of magnitude higher VMR near the 5-bar level.

\end{list}
\end{enumerate}

\noindent
Advancing our understanding of the distribution and composition of
Uranus' aerosols would be helped by good measurements of the optical
properties of \htsx, the most likely primary constituent of the most
visible tropospheric cloud layer.  Another helpful undertaking would
be microphysical modeling of photochemical haze formation and seasonal
evolution as well as microphysical modeling of condensation
clouds. The variety of vertical aerosol structures and mass loadings
that can produce model spectra matching the observations is
surprisingly large and it seems likely that not all of these options
would satisfy microphysical constraints.  A better understanding of
the aerosols is also the key to better constraints on the distribution
of methane because different aerosol models yield different methane
mixing ratios, with deep VMR values mostly falling between 2\% and
4\%.  As Uranus seems to be continuing to change, especially the
continued brightening of the north polar region through at least 2017,
and many uncertainties remain, continued observations are also
warranted.

%% noindent In the future, better constraints on the vertical profile of
%% methane as a function of latitude could be addressed by additional
%% modeling work with the 2012 STIS spectra, trying different functional
%% forms for vertical depletion profiles.  More detailed analysis at more
%% latitudes using full radiation transfer modeling for both 2002 and
%% 2012 would be useful in clarifying whether low-latitude changes
%% betweeen 2002 and 2012 are real. Additional STIS observations in
%% future cycles are also needed to confirm the northern high-latitude
%% modulations we found in the apparent mixing ratio.  Additional
%% quantitative constraints might also be derived from analysis of the
%% vertical wind shears that are implied by the horizontal density
%% gradients associated with latitudinal compositional gradients.
%% Additional work with numerical circulation modeling might also be
%% productive in understanding how the methane mixing ratio affects and
%% is affected by atmospheric circulation patterns.

\section*{Acknowledgments}\addcontentsline{toc}{section}{Acknowledgments}

This research was supported primarily by grants from the Space
Telescope Science Institute, managed by AURA. GO-14113.001-A supported
LAS and PMF. Partial support was provided by NASA Solar System
Observations Grant NNXA16AH99G (LAS and PMF). EK also acknowledges
support by an STScI grant under GO-14113. I.dP was supported by NASA
grant NNX16AK14G.  We thank staff at the W. M. Keck Observatory, which
is made possible by the generous financial support of the W. M. Keck
Foundation.  We thank those of Hawaiian ancestry on whose sacred
mountain we are privileged to be guests. Without their generous
hospitality none of our groundbased observations would have been
possible.

%\newpage
\addcontentsline{toc}{section}{References}

%\bibliographystyle{/home/home2/sro/uranus/paper3/elsart-harv-nomonth}
%%\bibliographystyle{newicarus}
%%\bibliographystyle{chicago}
%\bibliography{/home/home2/sro/uranus/paper3/outerplanets}

\begin{thebibliography}{48}
\expandafter\ifx\csname natexlab\endcsname\relax\def\natexlab#1{#1}\fi
\expandafter\ifx\csname url\endcsname\relax
  \def\url#1{\texttt{#1}}\fi
\expandafter\ifx\csname urlprefix\endcsname\relax\def\urlprefix{URL }\fi

\bibitem[{{Acton}(1996)}]{Acton1996}
{Acton}, C.~H., 1996. {Ancillary data services of NASA's Navigation and
  Ancillary Information Facility}. Planet. and Space Sci. 44, 65--70.

\bibitem[{{Borysow} et~al.(2000){Borysow}, {Borysow}, and {Fu}}]{Borysow2000}
{Borysow}, A., {Borysow}, J., {Fu}, Y., 2000. {Semi-empirical model of
  collision-induced absorption spectra of H$_2$-H$_2$ complexes in the second
  overtone band of hydrogen at temperatures from 50 to 500 K}. Icarus 145,
  601--608.

\bibitem[{{Colina} et~al.(1996){Colina}, {Bohlin}, and {Castelli}}]{Colina1996}
{Colina}, L., {Bohlin}, R.~C., {Castelli}, F., 1996. {The 0.12-2.5 micron
  Absolute Flux Distribution of the Sun for Comparison With Solar Analog
  Stars}. \aj 112, 307--315.

\bibitem[{{Conrath} et~al.(1987){Conrath}, {Hanel}, {Gautier}, {Marten}, and
  {Lindal}}]{Conrath1987JGR}
{Conrath}, B., {Hanel}, R., {Gautier}, D., {Marten}, A., {Lindal}, G., 1987.
  {The helium abundance of Uranus from Voyager measurements}. \jgr 92~(11),
  15003--15010.

\bibitem[{{Conrath} et~al.(1990){Conrath}, {Gierasch}, and
  {Leroy}}]{Conrath1990}
{Conrath}, B.~J., {Gierasch}, P.~J., {Leroy}, S.~S., 1990. {Temperature and
  circulation in the stratosphere of the outer planets}. Icarus 83, 255--281.

\bibitem[{{Conrath} et~al.(1991){Conrath}, {Pearl}, {Appleby}, {Lindal},
  {Orton}, and {B\'ezard}}]{Conrath1991urabook}
{Conrath}, B.~J., {Pearl}, J.~C., {Appleby}, J.~F., {Lindal}, G.~F., {Orton},
  G.~S., {B\'ezard}, B., 1991. {Thermal structure and energy balance of
  Uranus}. Uranus, pp. 204--252.

\bibitem[{{de Kleer} et~al.(2015){de Kleer}, {Luszcz-Cook}, {de Pater},
  {{\'A}d{\'a}mkovics}, and {Hammel}}]{DeKleer2015}
{de Kleer}, K., {Luszcz-Cook}, S., {de Pater}, I., {{\'A}d{\'a}mkovics}, M.,
  {Hammel}, H.~B., 2015. {Clouds and aerosols on Uranus: Radiative transfer
  modeling of spatially-resolved near-infrared Keck spectra}. Icarus 256,
  120--137.

\bibitem[{{de Pater} et~al.(2014){de Pater}, {Fletcher}, {Luszcz-Cook},
  {DeBoer}, {Butler}, {Hammel}, {Sitko}, {Orton}, and
  {Marcus}}]{DePater2014nep}
{de Pater}, I., {Fletcher}, L.~N., {Luszcz-Cook}, S., {DeBoer}, D., {Butler},
  B., {Hammel}, H.~B., {Sitko}, M.~L., {Orton}, G., {Marcus}, P.~S., 2014.
  {Neptune's global circulation deduced from multi-wavelength observations}.
  Icarus 237, 211--238.

\bibitem[{{de Pater} et~al.(1991){de Pater}, {Romani}, and
  {Atreya}}]{DePater1991Icar}
{de Pater}, I., {Romani}, P.~N., {Atreya}, S.~K., 1991. {Possible microwave
  absorption by H$_2$S gas in Uranus' and Neptune's atmospheres}. Icarus 91,
  220--233.

\bibitem[{{Fink} and {Sill}(1982)}]{Fink1982}
{Fink}, U., {Sill}, G.~T., 1982. {The infrared spectral properties of frozen
  volatiles}. In: {Wilkening}, L.~L. (Ed.), IAU Colloq. 61: Comet Discoveries,
  Statistics, and Observational Selection. pp. 164--202.

\bibitem[{{Friedson} and {Ingersoll}(1987)}]{Friedson1987}
{Friedson}, J., {Ingersoll}, A.~P., 1987. {Seasonal meridional energy balance
  and thermal structure of the atmosphere of Uranus - A
  radiative-convective-dynamical model}. Icarus 69, 135--156.

\bibitem[{{Fry} and {Sromovsky}(2017)}]{Fry2017DPS}
{Fry}, P.~M., {Sromovsky}, L.~A., 2017. {Uranus' post-equinox north polar
  brightening characterized with 2013 and 2016 IRTF SpeX observation}. In:
  AAS/Division for Planetary Sciences Meeting Abstracts. Vol.~49 of
  AAS/Division for Planetary Sciences Meeting Abstracts. p. 115.17.

\bibitem[{{Fry} et~al.(2012){Fry}, {Sromovsky}, {de Pater}, {Hammel}, and
  {Rages}}]{Fry2012}
{Fry}, P.~M., {Sromovsky}, L.~A., {de Pater}, I., {Hammel}, H.~B., {Rages},
  K.~A., 2012. {Detection and Tracking of Subtle Cloud Features on Uranus}.
  Astron. J. 143, 150--161.

\bibitem[{{Hanel} et~al.(1986){Hanel}, {Conrath}, {Flasar}, {Kunde}, {Maquire},
  {Pearl}, {Pirraglia}, {Samuelson}, {Horn}, and {Schulte}}]{Hanel1986Sci}
{Hanel}, R., {Conrath}, B., {Flasar}, F.~M., {Kunde}, V., {Maquire}, W.,
  {Pearl}, J., {Pirraglia}, J., {Samuelson}, R., {Horn}, L., {Schulte}, P.,
  1986. {Infrared observations of the Uranian system}. Science 233, 70--74.

\bibitem[{{Hansen}(1971)}]{Hansen1971JAScircpol}
{Hansen}, J.~E., 1971. {Circular polarization of sunlight reflected by clouds.}
  Journal of Atmospheric Sciences 28, 1515--1516.

\bibitem[{{Havriliak} et~al.(1955){Havriliak}, {Swenson}, and
  {Cole}}]{Havriliak1954}
{Havriliak}, S., {Swenson}, R.~W., {Cole}, R.~H., 1955. {Dielectric Constants
  of Liquid and Solid Hydrogen Sulfide}. J. Chem. Phys. 23, 134--135.

\bibitem[{{Hernandez} et~al.(2012){Hernandez}, {Aloisi}, {Bohlin}, {Bostroem},
  {Diaz}, {Dixon}, {Ely}, {Goudfrooij}, {Hodge}, {Lennon}, {Long}, {Niemi},
  {Osten}, {Proffitt}, {Walborn}, {Wheeler}, {York}, and
  {Zheng}}]{Hernandez2012}
{Hernandez}, S., {Aloisi}, A., {Bohlin}, R., {Bostroem}, A., {Diaz}, R.,
  {Dixon}, V., {Ely}, J., {Goudfrooij}, P., {Hodge}, P., {Lennon}, D., {Long},
  C., {Niemi}, S., {Osten}, R., {Proffitt}, C., {Walborn}, N., {Wheeler}, T.,
  {York}, B., {Zheng}, W., 2012. STIS Instrument Handbook, Version 12.0,
  (Baltimore: STScI). Space Telescope Science Institute, Baltimore, Maryland.

\bibitem[{{Irwin} et~al.(2010){Irwin}, {Teanby}, and {Davis}}]{Irwin2010Icar}
{Irwin}, P.~G.~J., {Teanby}, N.~A., {Davis}, G.~R., 2010. {Revised vertical
  cloud structure of Uranus from UKIRT/UIST observations and changes seen
  during Uranus' Northern Spring Equinox from 2006 to 2008: Application of new
  methane absorption data and comparison with Neptune}. Icarus 208, 913--926.

\bibitem[{{Irwin} et~al.(2015){Irwin}, {Tice}, {Fletcher}, {Barstow}, {Teanby},
  {Orton}, and {Davis}}]{Irwin2015reanalysis}
{Irwin}, P.~G.~J., {Tice}, D.~S., {Fletcher}, L.~N., {Barstow}, J.~K.,
  {Teanby}, N.~A., {Orton}, G.~S., {Davis}, G.~R., 2015. {Reanalysis of Uranus'
  cloud scattering properties from IRTF/SpeX observations using a
  self-consistent scattering cloud retrieval scheme}. Icarus 250, 462--476.

\bibitem[{{Irwin} et~al.(2018){Irwin}, {Toledo}, {Garland}, {Teanby},
  {Fletcher}, {Orton}, and {B{\'e}zard}}]{Irwin2018h2s}
{Irwin}, P.~G.~J., {Toledo}, D., {Garland}, R., {Teanby}, N.~A., {Fletcher},
  L.~N., {Orton}, G.~A., {B{\'e}zard}, B., 2018. {Detection of hydrogen sulfide
  above the clouds in Uranus's atmosphere}. Nature Astronomy 2, 420--427.

\bibitem[{{Karkoschka} and {Tomasko}(2009)}]{Kark2009IcarusSTIS}
{Karkoschka}, E., {Tomasko}, M., 2009. {The haze and methane distributions on
  Uranus from HST-STIS spectroscopy}. Icarus 202, 287--309.

\bibitem[{{Karkoschka} and {Tomasko}(2010)}]{Kark2010ch4}
{Karkoschka}, E., {Tomasko}, M.~G., 2010. {Methane absorption coefficients for
  the jovian planets from laboratory, Huygens, and HST data}. Icarus 205,
  674--694.

\bibitem[{{Karkoschka} and {Tomasko}(2011)}]{Kark2011nep}
{Karkoschka}, E., {Tomasko}, M.~G., 2011. {The haze and methane distributions
  on Neptune from HST-STIS spectroscopy}. Icarus 211, 780--797.

\bibitem[{{Khare} et~al.(1993){Khare}, {Thompson}, {Cheng}, {Chyba}, {Sagan},
  {Arakawa}, {Meisse}, and {Tuminello}}]{Khare1993}
{Khare}, B.~N., {Thompson}, W.~R., {Cheng}, L., {Chyba}, C., {Sagan}, C.,
  {Arakawa}, E.~T., {Meisse}, C., {Tuminello}, P.~S., 1993. {Production and
  optical constraints of ice tholin from charged particle irradiation of (1:6)
  C2H6/H2O at 77 K}. Icarus 103, 290--300.

\bibitem[{{Krist}(1995)}]{Krist1995}
{Krist}, J., 1995. {Simulation of HST PSFs using Tiny Tim}. In: {Shaw}, R.~A.,
  {Payne}, H.~E., {Hayes}, J.~J.~E. (Eds.), Astronomical Data Analysis Software
  and Systems IV. Vol.~77 of Astronomical Society of the Pacific Conference
  Series. pp. 349--352.

\bibitem[{{Lebofsky} and {Fegley}(1976)}]{Lebofsky1976}
{Lebofsky}, L.~A., {Fegley}, Jr., M.~B., 1976. {Laboratory reflection spectra
  for the determination of chemical composition of ice bodies}. Icarus 28,
  379--387.

\bibitem[{{Lellouch} et~al.(2015){Lellouch}, {Moreno}, {Orton}, {Feuchtgruber},
  {Cavali{\'e}}, {Moses}, {Hartogh}, {Jarchow}, and {Sagawa}}]{Lellouch2015}
{Lellouch}, E., {Moreno}, R., {Orton}, G.~S., {Feuchtgruber}, H.,
  {Cavali{\'e}}, T., {Moses}, J.~I., {Hartogh}, P., {Jarchow}, C., {Sagawa},
  H., 2015. {New constraints on the CH$_{4}$ vertical profile in Uranus and
  Neptune from Herschel observations}. \aap 579, A121.

\bibitem[{{Lindal} et~al.(1987){Lindal}, {Lyons}, {Sweetnam}, {Eshleman}, and
  {Hinson}}]{Lindal1987}
{Lindal}, G.~F., {Lyons}, J.~R., {Sweetnam}, D.~N., {Eshleman}, V.~R.,
  {Hinson}, D.~P., 1987. {The atmosphere of Uranus - Results of radio
  occultation measurements with Voyager 2}. J. Geophys. Res. 92, 14987--15001.

\bibitem[{{Lodders}(2003)}]{Lodders2003}
{Lodders}, K., 2003. {Solar System Abundances and Condensation Temperatures of
  the Elements}. \apj 591, 1220--1247.

\bibitem[{{Orton} et~al.(1987){Orton}, {Aitken}, {Smith}, {Roche}, {Caldwell},
  and {Snyder}}]{Orton1987spectra}
{Orton}, G.~S., {Aitken}, D.~K., {Smith}, C., {Roche}, P.~F., {Caldwell}, J.,
  {Snyder}, R., 1987. {The spectra of Uranus and Neptune at 8-14 and 17-23
  microns}. Icarus 70, 1--12.

\bibitem[{{Orton} et~al.(2014{\natexlab{a}}){Orton}, {Fletcher}, {Moses},
  {Mainzer}, {Hines}, {Hammel}, {Martin-Torres}, {Burgdorf}, {Merlet}, and
  {Line}}]{Orton2014uratemp}
{Orton}, G.~S., {Fletcher}, L.~N., {Moses}, J.~I., {Mainzer}, A.~K., {Hines},
  D., {Hammel}, H.~B., {Martin-Torres}, F.~J., {Burgdorf}, M., {Merlet}, C.,
  {Line}, M.~R., 2014{\natexlab{a}}. {Mid-infrared spectroscopy of Uranus from
  the Spitzer Infrared Spectrometer: 1. Determination of the mean temperature
  structure of the upper troposphere and stratosphere}. Icarus 243, 494--513.

\bibitem[{{Orton} et~al.(2014{\natexlab{b}}){Orton}, {Moses}, {Fletcher},
  {Mainzer}, {Hines}, {Hammel}, {Martin-Torres}, {Burgdorf}, {Merlet}, and
  {Line}}]{Orton2014uracomp}
{Orton}, G.~S., {Moses}, J.~I., {Fletcher}, L.~N., {Mainzer}, A.~K., {Hines},
  D., {Hammel}, H.~B., {Martin-Torres}, J., {Burgdorf}, M., {Merlet}, C.,
  {Line}, M.~R., 2014{\natexlab{b}}. {Mid-infrared spectroscopy of Uranus from
  the Spitzer infrared spectrometer: 2. Determination of the mean composition
  of the upper troposphere and stratosphere}. Icarus 243, 471--493.

\bibitem[{{Rages} et~al.(1991){Rages}, {Pollack}, {Tomasko}, and
  {Doose}}]{Rages1991}
{Rages}, K., {Pollack}, J.~B., {Tomasko}, M.~G., {Doose}, L.~R., 1991.
  {Properties of scatterers in the troposphere and lower stratosphere of Uranus
  based on Voyager imaging data}. Icarus 89, 359--376.

\bibitem[{{Rannou} et~al.(1999){Rannou}, {McKay}, {Botet}, and
  {Cabane}}]{Rannou1999}
{Rannou}, P., {McKay}, C.~P., {Botet}, R., {Cabane}, M., 1999. {Semi-empirical
  model of absorption and scattering by isotropic fractal aggregates of
  spheres}. \planss 47, 385--396.

\bibitem[{{Sromovsky}(2005{\natexlab{a}})}]{Sro2005raman}
{Sromovsky}, L.~A., 2005{\natexlab{a}}. {Accurate and approximate calculations
  of Raman scattering in the atmosphere of Neptune}. Icarus 173, 254--283.

\bibitem[{{Sromovsky}(2005{\natexlab{b}})}]{Sro2005pol}
{Sromovsky}, L.~A., 2005{\natexlab{b}}. {Effects of Rayleigh-scattering
  polarization on reflected intensity: a fast and accurate approximation method
  for atmospheres with aerosols}. Icarus 173, 284--294.

\bibitem[{{Sromovsky} and {Fry}(2007)}]{Sro2007struc}
{Sromovsky}, L.~A., {Fry}, P.~M., 2007. {Spatially resolved cloud structure on
  Uranus: Implications of near-IR adaptive optics imaging}. Icarus 192,
  527--557.

\bibitem[{{Sromovsky} and {Fry}(2008)}]{Sro2008grism}
{Sromovsky}, L.~A., {Fry}, P.~M., 2008. {The methane abundance and structure of
  Uranus' cloud bands inferred from spatially resolved 2006 Keck grism
  spectra}. Icarus 193, 252--266.

\bibitem[{{Sromovsky} and {Fry}(2010)}]{Sro2010iso}
{Sromovsky}, L.~A., {Fry}, P.~M., 2010. {The source of 3-{$\mu$}m absorption in
  Jupiter's clouds: Reanalysis of ISO observations using new NH$_{3}$
  absorption models}. Icarus 210, 211--229.

\bibitem[{{Sromovsky} et~al.(2012{\natexlab{a}}){Sromovsky}, {Fry}, {Boudon},
  {Campargue}, and {Nikitin}}]{Sro2012LBL}
{Sromovsky}, L.~A., {Fry}, P.~M., {Boudon}, V., {Campargue}, A., {Nikitin}, A.,
  2012{\natexlab{a}}. {Comparison of line-by-line and band models of near-IR
  methane absorption applied to outer planet atmospheres}. Icarus 218, 1--23.

\bibitem[{{Sromovsky} et~al.(2009){Sromovsky}, {Fry}, {Hammel}, {Ahue}, {de
  Pater}, {Rages}, {Showalter}, and {van Dam}}]{Sro2009eqdyn}
{Sromovsky}, L.~A., {Fry}, P.~M., {Hammel}, H.~B., {Ahue}, W.~M., {de Pater},
  I., {Rages}, K.~A., {Showalter}, M.~R., {van Dam}, M.~A., 2009. {Uranus at
  equinox: Cloud morphology and dynamics}. Icarus 203, 265--286.

\bibitem[{{Sromovsky} et~al.(2012{\natexlab{b}}){Sromovsky}, {Fry}, {Hammel},
  {de Pater}, and {Rages}}]{Sro2012polar}
{Sromovsky}, L.~A., {Fry}, P.~M., {Hammel}, H.~B., {de Pater}, I., {Rages},
  K.~A., 2012{\natexlab{b}}. {Post-equinox dynamics and polar cloud structure
  on Uranus}. Icarus 220, 694--712.

\bibitem[{{Sromovsky} et~al.(2011){Sromovsky}, {Fry}, and
  {Kim}}]{Sro2011occult}
{Sromovsky}, L.~A., {Fry}, P.~M., {Kim}, J.~H., 2011. {Methane on Uranus: The
  case for a compact CH$_4$ cloud layer at low latitudes and a severe CH$_4$
  depletion at high latitudes based on re-analysis of Voyager occultation
  measurements and STIS spectroscopy.} Icarus 215, 292--312.

\bibitem[{{Sromovsky} et~al.(2014){Sromovsky}, {Karkoschka}, {Fry}, {Hammel},
  {de Pater}, and {Rages}}]{Sro2014stis}
{Sromovsky}, L.~A., {Karkoschka}, E., {Fry}, P.~M., {Hammel}, H.~B., {de
  Pater}, I., {Rages}, K.~A., 2014. {Methane depletions in both polar regions
  of Uranus inferred from HST/STIS and Keck/NIRC2}. Icarus 238, 137--155.

\bibitem[{{Sun} et~al.(1991){Sun}, {Schubert}, and {Stoker}}]{Sun1991}
{Sun}, Z., {Schubert}, G., {Stoker}, C.~R., 1991. {Thermal and humidity winds
  in outer planet atmospheres}. Icarus 91, 154--160.

\bibitem[{{Tice} et~al.(2013){Tice}, {Irwin}, {Fletcher}, {Teanby}, {Hurley},
  {Orton}, and {Davis}}]{Tice2013}
{Tice}, D.~S., {Irwin}, P.~G.~J., {Fletcher}, L.~N., {Teanby}, N.~A., {Hurley},
  J., {Orton}, G.~S., {Davis}, G.~R., 2013. {Uranus' cloud particle properties
  and latitudinal methane variation from IRTF SpeX observations}. Icarus 223,
  684--698.

\bibitem[{{Tomasko} et~al.(2005){Tomasko}, {Archinal}, {Becker}, {B{\'e}zard},
  {Bushroe}, {Combes}, {Cook}, {Coustenis}, {de Bergh}, {Dafoe}, {Doose},
  {Dout{\'e}}, {Eibl}, {Engel}, {Gliem}, {Grieger}, {Holso}, {Howington-Kraus},
  {Karkoschka}, {Keller}, {Kirk}, {Kramm}, {K{\"u}ppers}, {Lanagan},
  {Lellouch}, {Lemmon}, {Lunine}, {McFarlane}, {Moores}, {Prout}, {Rizk},
  {Rosiek}, {Rueffer}, {Schr{\"o}der}, {Schmitt}, {See}, {Smith}, {Soderblom},
  {Thomas}, and {West}}]{Tomasko2005}
{Tomasko}, M.~G., {Archinal}, B., {Becker}, T., {B{\'e}zard}, B., {Bushroe},
  M., {Combes}, M., {Cook}, D., {Coustenis}, A., {de Bergh}, C., {Dafoe},
  L.~E., {Doose}, L., {Dout{\'e}}, S., {Eibl}, A., {Engel}, S., {Gliem}, F.,
  {Grieger}, B., {Holso}, K., {Howington-Kraus}, E., {Karkoschka}, E.,
  {Keller}, H.~U., {Kirk}, R., {Kramm}, R., {K{\"u}ppers}, M., {Lanagan}, P.,
  {Lellouch}, E., {Lemmon}, M., {Lunine}, J., {McFarlane}, E., {Moores}, J.,
  {Prout}, G.~M., {Rizk}, B., {Rosiek}, M., {Rueffer}, P., {Schr{\"o}der},
  S.~E., {Schmitt}, B., {See}, C., {Smith}, P., {Soderblom}, L., {Thomas}, N.,
  {West}, R., 2005. {Rain, winds and haze during the Huygens probe's descent to
  Titan's surface}. Nature 438, 765--778.

\bibitem[{{Tomasko} et~al.(2008){Tomasko}, {Doose}, {Engel}, {Dafoe}, {West},
  {Lemmon}, {Karkoschka}, and {See}}]{Tomasko2008}
{Tomasko}, M.~G., {Doose}, L., {Engel}, S., {Dafoe}, L.~E., {West}, R.,
  {Lemmon}, M., {Karkoschka}, E., {See}, C., 2008. {A model of Titan's aerosols
  based on measurements made inside the atmosphere}. Plan. \& Space Sci. 56,
  669--707.

\end{thebibliography}

%\vspace{1in}

\section*{Supplemental material.}
The calibrated hyperspectral STIS cubes are archived at the Mikulski
Archive for Space Telescopes (MAST) as High Level Science Products
(HLSPs). They can be found at\newline
https://archive.stsci.edu/prepds/uranus-stis/;
https://dx.doi.org/10.17909/T9KQ4N.
 The 2015 cube is named
hlsp\_uranus-stis\_hst\_stis\_uranus-2015\_g430l-g750l\_v1\_cube.fits;
2002 and 2012 cubes are named analagously.  
The
hyperspectral cubes contain calibrated I/F values as a function of
wavelength and location, with navigation backplanes that provide
viewing geometry and latitude-longitude coordinates for each pixel.  A
detailed explanation of the file contents is provided in the
file README\_SUPPLEMENTAL.TXT.  A sample IDL program that reads a cube
file, plots a monochromatic image, extracts data from a particular
location on the disc, and plots a spectrum, is provided in the file
stis\_cube\_example.pro.  The IDL astronomy library will be needed to
run the sample program.

\vspace{0.15in}
\noindent
NOTE: Until the MAST archive submission is finalized, which is
underway at this writing and expected to be complete by the end of June
2018, the materials will be available for review at
http://www.ssec.wisc.edu/planetary/
uranus/ \newline onlinedata/ura2015stis/.

\newpage

%\tableofcontents

\end{document}